\documentclass[12pt,a4paper]{article}
\usepackage{amsmath,amssymb}
\usepackage[french]{babel}
\usepackage[utf8]{inputenc}
\usepackage{color}
\usepackage{graphicx}        
\usepackage{xcolor}

\begin{document}

\author{Daniel Bennequin \thanks{Institut de Math\'ematiques de Jussieu,
Universit\'e Denis Diderot Paris VII, France, bennequin@math.univ-paris-diderot.fr}}

\title{The unfolded universe of elementary particles. A geometric explanation of the
standard model structure.}
\date{}
\maketitle \tableofcontents

\section{Introduction}

Our universe seems to be made of assemblies of interacting elementary particles; the best known description of
small assemblies is Quantum Field Theory, relying on quantum corrections to a set of classical evolution equations.
A very particular set of elementary particles were experimentally discovered in the past: a photon, three massive bosons $W_+$, $W_-$, $Z_0$, eight gluons,
an electron, a muon, a tau, their three respective neutrinos, and six families of three quarks, plus, recently,
a Higgs boson. These particles belong to the Standard Model of Glashow, Salam and Weinberg (see \cite{weinberg1996quantum},\cite{itzykson1980quantum},
\cite{peskin1995introduction},\cite{t1994under}). The predictions of
this model are very well verified by experiments (cf. \cite{schorner2015large}, \cite{gardi2015lhc}, \cite{plehn2015lectures}).
But why the table has this form? Why three times the same structure of fermions?
Why masses are as they are? much larger for bosons than for fermions, and having a sort of geometric growth for the leptons.
The present text pretends to give an answer based on Differential Geometry to these four questions, and to suggest new physics after the standard model.\\
\indent The proposed answer is a generic quantum field theory, with local covariance. It is the theory in four dimension generated by the
Einstein and Dirac Lagrangian in a universe of twelve Lorentzian dimensions, where our four dimensional space-time is embedded
as a Lorentz sub-manifold, and supposed flat in comparison to the transversal curvature. This four dimensional theory is obtained
by developing along our world the transversal Taylor series of the gravitation field and the Dirac fermion field in dimension twelve. In this way, we
get a Yang-Mills field for the group $Spin_8$, a large collection of Higgs-Hitchin vector fields and scalar fields associated
to transversal curvature, plus several families of Dirac spinors in dimension four.\\
\indent The main mechanism which gives birth to
the standard model is a certain gauge fixing (a choice of coordinates respecting general covariance, as done by Brout, Englert and Higgs)
which permits to identify the multiplicity of fermions, as seen from the four dimensional world, with the eight unseen dimensions
of the generating universe. We prove that this gauge fixing kills half of the fermionic degrees of freedom and makes apparent
the standard structure of the surviving one. In fact, half of the quarks appearing in this process behave like anti-particles,
thus we need to replace them by particles, ensuring the experimentally observed multiplicity $M_8$ of fermions. \\
When looking at the bosons, we see that this replacement forbids the action on fermions for all the bosonic fields,
with the exception of the generators of an $su_3$
connection (gluons), a real three dimensional space of abelian connections (photon, baryon number and lepton number) and the generators $W_+, W_-,Z$ of a $su_2$ gauge,
in the broken form predicted by Weinberg.\\
The fields in four dimensions correspond to the coefficients of series in a virtual displacement $v$. Then, the gauge fixing, completed by
a choice of diffeomorphism, selects a privileged vector $v$; this explains why several
generations of fields appear, and in particular for the fermions, why the structure at each order is the same.\\
\indent The first somehow fantastic thing is the interpretation of $v$ as the Higgs boson (divided by a constant in $GeV$). It stays at the right place for giving the expected
masses to the other particles; in particular it gives no mass to the gluons and the photon. We propose that its dynamical degrees of freedom come by bosonization (in the sense of A.Okninski, 2014, \cite{okninski2014neutrino})
of a fixed direction of spinor, that is
the odd (Right) chirality of the neutrino, which consequently looses its freedom, as it is asked in the standard model.\\
This Higgs field $H=M_0v$ gives their masses to all particles. The decoupled bosons could have masses in this way, possibly constituting the dark matter.\\
A particular way fermions could get masses with the help of $H$ is by using the transversal Riemann sectional curvatures of the twelve dimensional metric
of the universe, in
plane sections containing the direction $v$. Continuing in this way, we explain why the first neutrino has theoretically a mass zero.
Quantum corrections and higher order terms in our Taylor series can give it a small mass, as observed (cf. \cite{suekane2015neutrino}).
In addition, the manner fermions appear in the development
implies that eigenvectors of masses and eigenvectors of flavors are not the
same, as soon as the generating universe is generic. This is coded in $PMNS$ and $CKM$ matrices for neutrinos and quarks respectively.
Cf. \cite{weinberg1996quantum}, \cite{itzykson1980quantum},
\cite{peskin1995introduction}, \cite{suekane2015neutrino}, \cite{kuhr2013flavor}.\\
From here, using the known masses of $W,Z, H$ and the ratio of masses of leptons, we deduce the scales of the other masses, and we find a good
agreement with the observed size orders of masses.\\
Our theory does'nt pretend to give any new information on mass gap and confinement for
gluons and quarks, which are certainly due to non perturbative effects of chromodynamics in strong coupling, cf. \cite{jaffe2006quantum}.\\
Note that,
\emph{a priori} the exact values of the masses of new particles at each order in $v$, are unpredictable. However some hope is permitted
to get more precise expectations, using the
constraint equations between transverse curvature and matter fields, which are not taken in account in the four dimensional classical Lagrangian.\\
\indent The electromagnetic field appears in this approach as a subtle choice of a certain circle $U_1$ in a quotient of the subgroup $U_6\times U_2$
in $U_8$ acting naturally to the multiplicity $M_8$. It is worth mentioning that exactly the same group supports the masses distribution of fermions.\\
This is a timid sign of supersymmetry in our effective theory, after spontaneous symmetry breaking. A second sign is the computation
of bosonic and fermionic degrees of freedom, which gives a perfect equilibrium at $128$.
However, the spectrum of unitary representations of the
Poincaré group, is not made by super-multiplet. Anyway, recent results on $N=8$ super-gravity (coming
from dimension eleven super-gravity)
have many flavor in common with ours, cf. Meissner and Nicolai \cite{meissner2015standard}, and
Kleinschmidt and Nicolai \cite{kleinschmidt2015}. However, at the moment this super-gravity does'nt
offer a natural $SU_2$ and allows other broken sectors than $SU_3\times U_1$; to the contrary, our approach gives a symmetry $SU_2$,
and for it $SU_3\times U_1$ is the only possibility.\\
Then we note that introduction of a cosmological constant in the initial Lagrangian could explain the paradox of dark energy,
cf. \cite{henneaux1989cosmological}, \cite{bousso2008cosmological}, \cite{dupays2013can}.\\

In the first part of the methods section we describe the four dimensional field theory obtained by developing
the Einstein-Dirac energy-moment tensor along $W^{4}$ in $X^{12}$; we do not pretend here to a great originality, but we do not have find
this development in the literature. In fact, membrane theory (cf. \cite{maartens2015standard}) assumes $W$ more curved than the transverse
directions, and Kaluza-Klein theories (cf. \cite{coquereaux1983geometry}, \cite{bailin1987kaluza}) rely on the action of a group on $X$, and describe
a quotient, not a submanifold.\\
The second part is the heart of the model: fixing a \emph{triality} gives birth to leptons and quarks. Triality is a particular beauty of dimension eight,
linking the two sorts of spinors and the vectors together.\\
The third part deduces the $W,Z$, the Higgs boson and the photon.\\
The fourth part discusses fermion masses and their electric charges.\\
\indent In the results section, we compare qualitatively predicted masses to their experimental values, and
discuss the $PMNS$ and $CKM$  matrices, which describe respectively neutrino and quark flavors and masses departures.
We count bosons and fermions degrees of freedom, we describe the hidden bosons and we explain the origin of
the photon in more details. We also suggest testable prediction beyond the Standard Model.\\
\indent The discussion is short and mostly  dedicated to the quantum corrections and effective renormalization, plus a comparison
with some others theories. We hope to be able to present soon a more precise discussion of the perturbative quantum theory.\\
\indent More details and proofs are given in a paper written in French and diffused at the same time, entitled "L'univers déplié des particules élémentaires", $UDPE$.
Presently, the interested reader can  go to this
French text to get more information.
The author apologizes for the strange English he is using in the present text, he hopes that this will be corrected in a near
future, and that he should offer soon a translation of the more extended paper.\\

\section{Methods}

\subsection{Fields and classical Lagrangian}

We start with one Dirac spinor field in twelve dimensions coupled to Einstein gravitation. The universe is a smooth
manifold $X$ of dimension $12$, equipped with a Lorentzian vector bundle $E_V$ and a spin bundle $E_S$. The
variable quasi-classical fields are a vielbein $\epsilon:T(X)\rightarrow E_V$ and a section $\psi$ of
$E_S$; the metric $g_X$ is the pull-back of the Lorentz metric $\eta_V$ on $E_V$; the gravitational field
$\nabla_X$ is the Levi-Civita connection of $g_X$, i.e. the unique linear connection on the tangent bundle $T(X)$
which is compatible with
the metric and has zero torsion. The action $S_c$ for the quasi-classical dynamic is the sum of the Dirac action
and the Einstein-Hilbert action, with minimal coupling.
We do not consider a Lorentz connection on $E_V$ as an independent field, only the pull-back of the Levi-Civita
plays a role. The group of local covariance is the cartesian product of the group of diffeomorphisms of $X$
with the gauge group of spin automorphisms of $E_S$.\\
\indent However, the theory we aim to consider is not quantum gauge theory coupled to quantum gravity in twelve dimensions,
it is the quantum field theory in four dimension, which is given by developing the infinite jets of fields along a four dimensional
smooth Lorentz manifold $W$ in $X$, and considering local covariance for a sub-group of jets of diffeomorphisms
along $W$. More precisely, given $W$, we can choose a tubular neighborhood of $W$ parameterized by a vector
bundle $E_F$ of rank eight, and we can assume that along $W$, the metric makes the fibers of $E_F$ orthogonal to $W$
with an Euclidian metric along the fibers fixed at the first order. Those assumptions are allowed by applications
of diffeomorphisms of $X$, and reduce the covariance without loosing general covariance, as was explained
by Einstein \cite{einstein1952principle}, when discussing the coordinates where the discriminant of the metric is equal to minus one.\\
\indent We choose to write the vielbein and the spinors in a moving frame $(e_F,e_W)$ that
contains eight vectors tangent to the fiber directions and four vectors perpendicular to these directions. The distribution of four planes which are orthogonal to the
fibers is denoted by $M_W$. Greek indices are attributed to vectors tangent to $M_F$ and latin indices from the middle of the
alphabet are attributed
to vectors tangent to the fibers of $E_F$. If $x^{\mu};\mu=0,...,3$
are local coordinates on $W$, the fibration $E_F$ can be used to extend them uniquely in four independent functions,
also denote by $x^{\mu}$; and local coordinates $v^{i};i=1,...,8$ along the fibers of $E_F$ permit to define
a complete basis of tangent vectors to $X$, $\partial_\mu;\mu=0,...,3$, $\partial_i;i=1,...,8$. Lifting the $\partial_\mu$
to $M_W$ gives four vectors $e_\mu;\mu=0,...,3$, which can be written
\begin{equation}
e_\mu=\partial_\mu+\sum_i a_{\mu}^{i}(x,v)\partial_i.
\end{equation}
Then the coordinate functions representing the potential metric form three groups
denoted by $g_{\mu\nu};\mu,\nu=0,...,3$, $g_{ij};i,j=1,...,8$ and $a_\mu^{i};\mu=0,...,3;i=1,...,8$.
The natural $g_{\mu i}$ are zero in the chosen frame.\\
We will
note $D_\mu$ the Lie derivatives along the vectors $e_\mu$, which in general differ from the vectors $\partial_\mu$.
We decide to also note $e_i$ the vectors $\partial_i$ for $i=1,...,8$ and sometimes $D_i$ for the asociated derivations,
to unify latin and greek notations. For the Lie brackets, we get
\begin{align}
[e_i,e_\mu]&=-[e_\mu,e_i]=\sum_k\partial_ia_\mu^{k}e_k\\
[e_\mu,e_\nu]&=-[e_\nu,e_\mu]=\sum_j(\partial_\mu a_\nu^{j}-\partial_\nu a_\mu^{j}+\sum_k(a_\mu^{k}\partial_ka_\nu^{j}
-a_\nu^{k}\partial_ka_\mu^{j}))e_j\\
[e_i,e_j]&=[e_j,e_i]=0
\end{align}
Latin indices from the beginning of the alphabet are used to unify $\mu$ and $i$; practically $a=0,...,3$ correspond
to $\mu$ and $a=4,...,11$ to $i+4$. (Majuscule letters will be reserved to spinor indices.) Then the identities
\begin{equation}
[e_a,e_b]=\sum_cM_{ab}^{c}e_c,
\end{equation}
imply
\begin{align}
M_{\mu i}^{j}=-M_{i \mu}^{j}&=-\partial_i a_{\mu}^{j},\\
M_{\mu \nu}^{j}=-M_{\nu \mu}^{j}&=\partial_\mu a_\nu^{j}-\partial_\nu a_\mu^{j}+\sum_k (a_\mu^{k}\partial_i a_\nu^{j}-a_\nu^{k}\partial_i a_\mu^{j});
\end{align}
all the other coefficients $M_{ab}^{c}$ being zero.\\
The Christoffel symbols $\Gamma_{ab}^{c}$ denote the coordinates of the Levi-Civita connection in the basis $e_a;a=0,...,11$. We also
introduce
\begin{equation}
\Gamma_{abc}=\sum_d \Gamma_{ab}^{d}g_{dc},\quad M_{abc}=\sum_d M_{ab}^{d}g_{dc}.
\end{equation}
It is easy to prove that
\begin{equation}
\Gamma_{abc}=\frac{1}{2}(D_ag_{bc}-M_{bca}+D_bg_{ac}-M_{acb}-D_cg_{ab}+M_{abc}).
\end{equation}
As explained by Einstein \cite{einstein1952principle}, these functions constitute the coordinates of the gravitational field.
See also \cite{hawking1973Largescale}, \cite{cartan1946leccons}. They can be
distributed in six groups.  The first one, $\Gamma_F$ corresponds to indices $i,j,k$ along fibers and is named
\emph{vertical gravitation field}, and the second one, $\Gamma_H$ corresponds to indices $\mu,\nu,\kappa$ along $M_W$
and is named \emph{horizontal gravitation field}. The four others are true tensors fields: the \emph{vertical deformation
field} $D_F$, defined by $D_i g_{\mu\nu}$ is a section of $E_F^{*}\otimes S^{2}M_W^{*}$; the \emph{horizontal deformation
field} $D_H$, made by the $D_\mu g_{ij}$ is a section of $M_W^{*}\otimes S^{2}E_F^{*}$; the \emph{connection field} $C_F$,
of components $M_{i\mu}^{j}=D_ia_\mu^{j}$ is a section of $M_W^{*}\otimes End(E_F)$; the \emph{monodromy field} $M_H$, whose
components are
\begin{align*}
M_{\mu\nu}^{j}&=D_\mu a_\nu^{j}-D_\nu a_\mu^{j}\\
&=\partial_\mu a_\nu^{j}-\partial_\nu a_\mu^{j}+\sum_ka_\mu^{k}\partial_ka_\nu^{j}-\sum_ka_\nu^{k}\partial_ka_\mu^{j},
\end{align*}
is a section of $\Lambda^{2}M_W^{*}\otimes E_F$.\\
To signal the field $C_F=C$, we will also write $C_{\mu i}^{j}=M_{i\mu}^{j}=-M_{\mu i}^{j}$, i.e. $C_{\mu i}^{j}=\partial_{i}a_{\mu}^{j}$.\\
The above division is made to understand fields from the point of view of the four dimensions of $W$, however it is a bit
arbitrary; for instance inside $\Gamma_H$ we have $D_\lambda g_{\mu\nu}$, where $D_F$, $C_F$ and $M_H$ are present through the
expressions $a_\lambda^{i}D_ig_{\mu\nu}$, and inside $D_H$ we have $a_\mu^{k}\partial_kg_{ij}$, where
$C_F$, $M_H$ and $\Gamma_F$ are hidden. The only indepenent degrees of freedom are represented by $g_{\mu\nu}$, $a_\mu^{i}$
and $g_{ij}$, that we will denote respectively $G_H$, $M_W$ and $G_F$.\\
The pair $(D_F,M_H)$ corresponds to the sets of Chistoffel symbols with two times more greek indices than latin indices,
that is
$(\Gamma_{\mu\nu}^{i}, \Gamma_{\mu i}^{\nu},\Gamma_{i \mu}^{\nu})$, and the pair $(C_F,D_H)$ corresponds to the sets of Christoffel
symbols with two times more latin indices than greek indices, that is $(\Gamma_{i j}^{\mu}, \Gamma_{i \mu}^{j},\Gamma_{\mu i}^{j})$.\\
More precisely, we have
\begin{equation}
M_{\mu\nu}^{i}=\Gamma_{\mu\nu}^{i}-\Gamma_{\nu\mu}^{i},
\end{equation}
\begin{equation}
D_i g_{\mu\nu}=\Gamma_{\mu\nu i}+\Gamma_{\nu\mu i}=\sum_jg_{ij}(\Gamma_{\mu\nu}^{j}+\Gamma_{\nu\mu}^{j}),
\end{equation}
in such a way that $M_H$ and $D_F$ describe respectively the anti-symmetric and symmetric parts of
$\Gamma_{\mu\nu}^{i}$. And we have
\begin{equation}
D_\mu g_{ij}=\Gamma_{\mu i j}+\Gamma_{\mu j i},
\end{equation}
\begin{equation}
C_{\mu i}^{j}=\Gamma_{i\mu}^{j}-\Gamma_{\mu i}^{j},
\end{equation}
in such a way that $D_H$ corresponds to the symmetric part of $\Gamma_{\mu i j}$, and
$C_F$ corresponds to the anti-symmetric part of $\Gamma_{\mu i j}$.\\
\indent The field $C_F$ of components $C_{\mu i}^{j}$ determines the distribution $M_W$ of components
$a_\mu^{j}$ by integration, thus it determines the derivations $D_\mu$, and it determines
the field $M_H$ through the following formula
\begin{multline}\label{courburedecf}
M_{\mu\nu k}^{i}(x,v)=D_\mu C_{\nu k}^{i}(x,v)-D_\nu C_{\mu k}^{i}(x,v)\\
+\sum_j(C^{j}_{\mu k}(x,v)C_{\nu j}^{i}(x,v)-C^{j}_{\nu k}(x,v)C_{\mu j}^{i}(x,v)).
\end{multline}
In intrinsic manner $C_F$ is an $1$-form on $X$ with values in $End(E_F)$.\\

The curvature tensor $R_{abcd}$ can be expressed according to this division, then we obtain
formulas for the Ricci tensor $R_{a d}=\sum_b R_{ a b d}^{b}$ and the scalar curvature
$R=\sum_a R_a^{a}$ (cf. \cite{hawking1973Largescale}, \cite{cartan1946leccons}). The Einstein tensor
\begin{equation}
S_{ab}=R_{ab}-\frac{1}{2}Rg_{ab}
\end{equation}
is divergence
free, and the Einstein-Dirac equation in $X$ equals this tensors to the energy-moment tensor of the
spinor field
\begin{equation}\label{energiemomentfermion}
T^{(s)}=(Id_{T^{*}(X)}\otimes ^{t}\gamma)(\psi^{*}\otimes\nabla \psi
+(\nabla \psi)^{*}\otimes \psi).
\end{equation}
multiplied by a coupling constant $c_0$.\\
Here $\gamma$ designates the Dirac morphism from $T(X)$ to $End(E_S)$, and its transpose is applied
to the spinor part of the $1$-form $\psi^{*}\otimes\nabla \psi
+(\nabla \psi)^{*}\otimes \psi$.\\
The Lagrangian is
\begin{equation}\label{lagrangienhilbertdirac}
\mathcal{L}=R + c_0\widetilde{\psi}D_S\psi,
\end{equation}
where the tilde denotes the conjugation
of a spinor, seen as a linear form on spinors (because $\psi\in S^{\pm}$ implies $D_S\psi\in S^{\mp}$, and
$S^{\mp}=\overline{S^{\pm}}$).\\
\indent As $g_{\mu i}=0$ there are two separate parts in the Lagrangian, one involving $R_{\mu\nu}$ and
one involving $R_{ij}$, giving two sets of semi-classical equations, which develop in two sets
of Dyson-Schwinger equations for quantum dynamics; only the first set can be considered as a field theory in
four dimensions, the other one being considered as constraint equations.\\
\indent A computation with Christoffel symbols gives that $R_{\mu\nu}$ is the sum of two expressions
$R^{(4)}_{\mu\nu}$ and $T^{(4)}_{\mu\nu}$ that we interpret as \emph{inertia} and \emph{energy-moment} of bosonic
matter respectively; they are given by:
\begin{multline}
R^{(4)}_{\mu\nu}=\frac{1}{4}(D_\mu D_\nu+D_\nu D_\mu)\ln det G_H+\frac{1}{2}\sum_{\lambda\kappa jk}D_k g_{\mu\nu}g^{j k}D_j \ln det G_H\\
+\frac{1}{2}\sum_{ij}(D_i(g^{ij}D_j g_{\mu\nu})+\sum_{k}\Gamma_{i k}^{k}g^{ij}D_jg_{\mu\nu})-\frac{1}{2}\sum_{\kappa\lambda i j}g^{\lambda\kappa}g^{ij}D_ig_{\nu\lambda}D_jg_{\mu\kappa}\\
+\frac{1}{4}\sum_{\lambda\rho\kappa\omega}(D_\lambda g_{\mu\rho}-D_\rho g_{\mu\lambda})g^{\rho\kappa}g^{\lambda\omega}(D_\kappa g_{\nu\omega}-D_\omega g_{\nu\kappa})\\
+\frac{1}{4}\sum_{\lambda\rho\kappa\omega}D_\mu g_{\lambda\rho}g^{\rho\kappa}D_\nu g_{\kappa\omega}g^{\omega\lambda}-\frac{1}{4}\sum_{\lambda\rho\kappa\omega}g^{\rho\kappa}g^{\lambda\omega}D_\kappa g_{\lambda\omega}(D_\mu g_{\rho\nu}+D_\nu g_{\rho\mu});
\end{multline}
and
\begin{multline}
T^{(4)}_{\mu\nu}=\frac{1}{2}\sum_{\kappa \lambda i}g^{\kappa\lambda}M_{\mu\kappa}^{i}M_{\nu\lambda i}
+\frac{1}{4}\sum_{i j k l}(D_\mu g_{lk})g^{il}(D_\nu g_{ij})g^{jk}\\
+\frac{1}{4}(D_\mu D_\nu+D_\nu D_\mu+2\sum_\kappa\Gamma_{\mu\nu}^{\kappa}D_\kappa)(\ln det G_F)\\
+\frac{1}{2}\sum_{i j k}(C_{\mu i}^{k}(D_\nu g_{kj})g^{ij}
+C_{\nu i}^{k}(D_\mu g_{jk})g^{ij})
+\frac{3}{4}\sum_{i j k l}C_{\mu j}^{l}g_{lk}C_{\nu i}^{k}g^{ji}\\+\frac{1}{4}\sum_{i k}C_{\mu i}^{k}C_{\nu k}^{i}
+\frac{1}{2}\sum_i(D_\mu C_{\nu i}^{i}+D_\nu C_{\mu i}^{i}+\sum_{\lambda}\Gamma_{\mu\nu}^{\lambda}C_{\lambda i}^{i}),
\end{multline}
Remind that $D_\mu$ involves $a_\mu^{i}$, then inertia contains other fields than $\Gamma_H$, and the decision to
put $D_F$ amost entirely on the inertial side is not very well founded, except perhaps in the approximation which neglects
$D_F$ and consider $G_H$ as almost flat. This shows how justified was Einstein when he questioned the geometrical meaning
of $T_{\mu\nu}$.\\

\indent To write the Dirac Lagrangian (\cite{dirac1967principles}), we need the vielbein, $\epsilon_\alpha^{a}$, where the greek letters of the beginning
of the alphabet concern the local trivialisation of the Lorentz auxiliary bundle $E_V$; the potential metric tensor is deduced
from the auxiliary Lorentz metric through the equation
\begin{equation}
g^{ab}=\epsilon^{a}_\alpha g^{\alpha\beta}\epsilon_\beta^{b}.
\end{equation}
If large letters $A,B,...$ refer to a trivialisation of the spinor bundle $E_S$, we can write $\Gamma_{aA}^{B}$ the
Christoffel symbols of the unique spinor connection $\omega_V=\epsilon^{-1}(\omega)\epsilon+\epsilon^{-1}d\epsilon$
which lifts the pull-back by $\epsilon$ of the Levi-Civita connection $\omega$ on $T(X)$;
then the Dirac operator is
\begin{equation}
D_S(\psi)^{C}=\sum_{\alpha,a,b,B}\gamma_{\alpha,B}^{C}
(\epsilon^{-1})^{\alpha}_b\eta^{ab}(\partial_a\Psi^{B}+\Gamma_{a,A}^{B}\Psi^{A}).
\end{equation}
The decomposition of $T(X)$ along $W$ can be reflected in a decomposition of $E_V$ and $E_S$, i.e. we decompose
the fibers of $E_V$ in such a way that the vielbein is made by two blocs; this gives rise to
four sectors. If the large and fat letters $\mathbf{I},\mathbf{J},\mathbf{K},\mathbf{H},...$ and
$\mathbf{A},\mathbf{B},\mathbf{C},...$ concern respectively the part dedicated to the fibers of $E_F$ and the orthogonal
part, we get
\begin{multline}\label{diracdevlop}
(D_S(\psi))^{\mathbf{C},\mathbf{K}}=\sum_{\alpha,\mu,\nu,\mathbf{B}}\gamma_{\alpha,\mathbf{B}}^{\mathbf{C}}
(\epsilon^{-1})^{\alpha}_\nu \eta^{\mu\nu}(\partial_\mu\Psi^{\mathbf{B},\mathbf{K}}+
\Gamma_{\mu,\mathbf{A}}^{\mathbf{B}}\Psi^{\mathbf{A},\mathbf{K}})\\
+\sum_{\alpha,\mu,\nu,\mathbf{J}}\gamma_{\alpha,\mathbf{J}}^{\mathbf{K}}
(\epsilon^{-1})^{\alpha}_\nu \eta^{\mu\nu}(\partial_\mu\Psi^{\mathbf{C},\mathbf{J}}+
\Gamma_{\mu,\mathbf{I}}^{\mathbf{J}}\Psi^{\mathbf{C},\mathbf{I}})\\
+\sum_{\iota,i,j,\mathbf{B}}\gamma_{\iota,\mathbf{B}}^{\mathbf{C}}
(\epsilon^{-1})^{\iota}_jg^{ij}(\partial_i\Psi^{\mathbf{B},\mathbf{K}}+
\Gamma_{i,\mathbf{A}}^{\mathbf{B}}\Psi^{\mathbf{A},\mathbf{K}})\\
+\sum_{\iota,i,j,\mathbf{J}}\gamma_{\iota,\mathbf{J}}^{\mathbf{K}}
(\epsilon^{-1})^{\iota}_jg^{ij}(\partial_i\Psi^{\mathbf{C},\mathbf{J}}+
\Gamma_{i,\mathbf{I}}^{\mathbf{J}}\Psi^{\mathbf{C},\mathbf{I}})
\end{multline}
Remark that the connection which couples the metric and the spinor in \eqref{diracdevlop} is negligible
except for the sums number two and four.\\

\indent The constant $c_0$ in front of the Dirac action is the constant $\kappa$
of Einstein; in the usual units of particle Physics where $\hbar=c=1$, the unit of length
is in $GeV^{-1}$, the constant $\kappa$ is in $GeV^{-2}$. This accords with the fact that
curvature is in $GeV^{2}$ and spinors $\psi$ are in $GeV^{3/2}$: the Dirac energy-momentum
has dimension $L^{-3}L^{-1}$, which gives $L^{-2}$ when multiplied by $\kappa$, as for
the curvature. However, the tradition is to also put $\kappa$ in front of $T_{\mu\nu}$
for the bosonic part of matter. This will appear naturally in the following development.\\
In fact, $\kappa$ is very very small, about $10^{-38}$, corresponding to the Planck length
around $10^{-19} GeV^{-1}$. We interpret this fact by saying that the horizontal curvature, in the
direction of $M_W$, almost parallel to $W$, is very small compared to the curvature in the
directions of the fibers $F$, thus we will neglect the fields $\Gamma_W$ and $D_F$. This justifies
the application of particle notions, from unitary representations of the Poincaré group in four
space-time dimensions.\\

Now we develop the fields $\psi$, $D_H$, $C_F$, $M_H$ and $\Gamma_F$ transversally to $W$,
by using a virtual displacement, which is a section $v$ of $E_F$, and we get a Taylor development
of the Lagrangian accordingly.\\
The unique spinor section $\psi_X$ becomes a series of Dirac spinors in four dimension along
$W$ with values in the fiber bundle $S^{*}(E_F)\otimes S_8$, of polynomials on the fibers
with coefficients Euclidian spinors in dimension $8$, which have $32$ real dimensions:
\begin{equation}
\psi(x,v)=\psi^{(0)}(x)+\sum_iv^{i}\psi_i^{(1)}(x)+\sum_{ij}v^{i}v^{j}\psi_{ij}^{(2)}(x)+...
\end{equation}
Here, $x$ denotes the point in $W$, and $v$ our exploratory section leaving the four dimensional universe.\\
As we will see in next sections, the vector $v$ will be fixed, thus we will get a simpler sequence of generations
of spinors in $S_8$; moreover half of them will be killed, which will give generations of $8$ fermions in $S_8^{+}$.
Today experiment has shown three generations, this is a good reason to stop the series at the order two
in $v$; however in the near future, the order three and more will hopefully been observed.\\
For the metric $g_{\mu\nu}(x,v)$, our hypothesis is that it could be taken as a constant.\\
For the metric $g_{ij}(x,v)$, using diffeomorphisms, we choose the coordinates in such a manner
that $g_{ij}(x,0)$ is the standard Euclidian structure $\delta_{ij}$ and such that all the symbols $\Gamma_{ij}^{k}(x,0)$
are zero, which is equivalent to tell that all the derivatives $\partial_kg_{ij}(x,0)$ are zero. Then the Taylor development is
\begin{equation}\label{metriqueverticalecourbure}
g_{i j}(x,v)=g^{(0)}_{i j}(x)-\frac{1}{3}R_{ikjl}(x)v^{k}v^{l}-\frac{1}{6}(D_mR_{ikjl})(x)v^{m}v^{k}v^{l}+...;
\end{equation}
where $R_{ijkl}=\sum_nR_{ijk}^{n}g_{nl}$, and where $g^{(0)}_{i j}(x)=\delta_{ij}$ is the identity matrix for every $x$.
Cf. \cite{cartan1946leccons}, \cite{green1987superstring}.\\
For the equation of the four plane field $M_W$, we have
\begin{multline}\label{devtmw}
a_\mu^{i}(x,v)=\sum_ka_{\mu k}^{i}(x)v^{k}+\sum_{k,l}a_{\mu k l}^{i}(x)v^{k}v^{l}+...\\+\sum_{k_1,...,k_n}a_{\mu K}^{i}(x)v^{K}+o(\| v\|)^{n}.
\end{multline}

To obtain the series for the connection $C_F$, given by $C_{\mu k}^{i}=\partial_k a_\mu^{i}$, we just have to delete a component
$v_k$ from \eqref{devtmw}.\\
The series for $D_F$, given by the $D_\mu g_{ij}$, are easily deduced from \eqref{metriqueverticalecourbure} and \eqref{devtmw};
only at the order $3$ and more in $v$ we have to take care of \eqref{devtmw}.\\
The development of $\Gamma_F$ is obtained from the development \eqref{metriqueverticalecourbure} by writing the Levi-Civita formulae;
here the vertical derivatives $D_mR_{ikjl}$ intervene at the order two in $v$.\\
And the monodromy $M_H$ is given by
\begin{multline}\label{courburedecf}
M_{\mu\nu}^{i}(x,v)=\sum_kv^{k}(\partial_\mu a_{\nu k}^{i}-\partial_\nu a_{\mu k}^{i}+\sum_j(a^{j}_{\mu k}a_{\nu j}^{i}-a^{j}_{\nu k}a_{\mu j}^{i}))\\
+\sum_{k_1,k_2}v^{k_1}v^{k_2}(\partial_\mu a_{\nu k_1 k_2}^{i}-\partial_\nu a_{\mu k_1 k_2}^{i}
+\sum_j(a_{\mu k_1 k_2}^{j}a_{\nu j}^{i}-a_{\nu k_1 k_2}^{j}a_{\mu j}^{i})\\
+\sum_j(a^{j}_{\mu k_1}a_{\nu j k_2}^{i}-a^{j}_{\nu k_1}a_{\mu j k_2}^{i})
+\sum_j(a_{\mu k_2}^{j}a_{\nu j k_1}^{i}-a_{\nu k_2}^{j}a_{\mu j k_1}^{i})).
\end{multline}
At the order one, we get the formula of the curvature of a linear connection; at the second order we get
an exterior $2$-form with value in the bundle of morphisms from a vector bundle in another one, then it is
not precisely a curvature.\\

\indent For the energy-momentum tensor, again assuming that $\partial_ig_{\mu\nu}$ and $\partial_\lambda g_{\mu\nu}$ are negligible,
at the order two in $v$ we have
\begin{multline}\label{energymomentsecondorder}
T^{(4)}_{\mu\nu}\thicksim \frac{1}{2}\sum_i(D_\mu C_{\nu i}^{i}+D_\nu C_{\mu i}^{i})
+\frac{3}{4}\sum_{i j k l}C_{\mu j}^{l}g_{lk}C_{\nu i}^{k}g^{ji}+\frac{1}{4}\sum_{i k}C_{\mu i}^{k}C_{\nu k}^{i}\\
+\frac{1}{2}\sum_{i j k}(C_{\mu i}^{k}(D_\nu g_{kj})g^{ij}
+C_{\nu i}^{k}(D_\mu g_{jk})g^{ij})
+\frac{1}{2}\sum_{\kappa \lambda i}g^{\kappa\lambda}M_{\mu\kappa}^{i}M_{\nu\lambda i}.
\end{multline}
For the fully developed formula we refer to the companion paper $UDPE$.\\

We will show in the following sections that a gauge fixing by the spin transformations, joined to a rotation
in the fiber bundle $E_F$, justifies the introduction of a special section $v_0$, and the identification of this
section to a multiple $H/M_0$
of the Higgs field $H$. Up to now, the scalar fields come from $\Gamma_F$ at order one given
by the curvature of the transversal metric. The Higgs field will result from the bosonisation of a special
spinor component.\\
From the point of view of the four dimensional space-time $W$, at the order two, the other bosonic fields
appearing in the energy-moment formula are the $1$-form $C_{\mu i}^{j}(x)$, and its first and second vertical derivatives
$C_{\mu i k}^{j}(x)$, $C_{\mu i k l}^{j}(x)$, components of $C_F$, its curvature $M_{\mu\nu i}^{j}(x)$, the component
of $M_H$, the curvature $R_{ikjl}(x)$, component of $\Gamma_F$, and the horizontal derivative of the vertical metric through
$\partial_\mu R_{ikjl}(x)$, the component of $D_H$. Just next, at the order three, we must add more vertical derivatives,
as $D_m R_{ikjl}(x)$.\\
We have to decompose the principal field $C_{\mu i}^{j}(x)$, as its higher order jets, and the curvature, in three parts:
the antisymmetric one, giving $A^{(0)}$, $A^{(1)}$, $A^{(2)}$,...,
$M_A^{(0)}$, ..., the trace $B^{(0)}$, $B^{(1)}$, $B^{(2)}$,...,
$M_B^{(0)}$, ..., and the part which is symmetric without trace $S^{(0)}$, $S^{(1)}$, $S^{(2)}$,...,
$M_S^{(0)}$, .... Only the matrix $A$ of $1$-forms $A^{(0)}$ can be interpreted as a true connection form in a
gauge theory, all the others must be interpreted as Higgs-Hitchin fields (see \cite{atiyahhitchin1988geometrymonopoles}) i.e. sections of associated
bundles to the principal bundle in $spin_8$ associated to $E_V$ and $E_S$.\\

\noindent The group of local covariance for the resulting Field Theory in four dimension is the product of
the group of infinite jets of diffeomorphims preserving globally $W$ and the fibration $E_F$ which
preserve the metric of the fibers at order one, with the group of spin gauge transformation of the
bundles $E_S$ and $E_V$.\\

\noindent Another part of the Einstein-Dirac equation relates the Ricci curvature $R^{F}_{ij}$ along the fibers of $E_F$
to the $4D$ fields; at the order zero, the only terms that appear are a quadratic monomial in $C_F$
and the energy-moment of the first generation of fermions; the following terms are of order two in $v$,
they involve $C_F$, $D_H$, $M_H$ and $\Gamma_F$.\\

\noindent The link between $T_{\mu\nu}$ and the Lagrangian density $\mathcal{L}$ is generally given by
\begin{equation}
T_{\mu\nu}=\frac{1}{\sqrt{-g}}\frac{\delta(\mathcal{L}\sqrt{-g})}{\delta g^{\mu\nu}}.
\end{equation}
cf. \cite{landau1971theoriedeschamps}, chapter $11$. This equation simplifies in coordinates where the determinant of $G_W$ is constant; the square root
of the determinant of the metric disappears. As Einstein recommended,
we choose to work in such coordinates. Then, the quadratic term in $M_F$ in $T^{(4)}_{\mu\nu}$, expressed around
the mean value $H_0$ of $H$, gives the ordinary Yang-Mills Lagrangian density.\\
For the Higgs-Hitchin fields, from the same Yang-Mills term, it is known that we recover the  Hitchin  equation for
associated vector fields.\\
The Lorentz condition
\begin{equation}\label{lorentzjauge}
\sum_\mu \partial_\mu A^{\mu}=0.
\end{equation}
appears naturally; we introduce its square in the Lagrangian with a Lagrange multiplier.\\

With the exception of $A$ or $H$ the bosonic fields are collectively written $\phi$; then the bosonic Lagrangian given by
developing the curvature $R$ at the order two has the following shape:
\begin{equation}
\mathcal{L}_B(A,\phi)=YM(A)+\epsilon (div_W(A))^{2}+d^{A}\phi.d^{A}\phi+L(d^{A}\phi)+V(\phi);
\end{equation}
where $L$ is linear and where, at order two, $V(\phi)$ is polynomial of degree two in $\phi$.\\
At this order two the potential $V$ has the following content:
\begin{multline}
V(\phi)=HB+HBB+HHB+HHBB+HAA+HSS+HHAA\\+HHSS+HHARS+HHB\nabla R+HHA\nabla R+HHS\nabla R\\
+HHM_AM_A+HHM_SM_S+HHM_BM_B;
\end{multline}
where $M_A,M_S,M_B$ denote the monodromy fields associated to the connection fields $A,S,B$ respectively,
that are there curvatures.\\

We see on this expression that the quadratic terms in the fields involving $HH$ are coming from the
Taylor series, thus they give
masses to all these fields except to the components which annihilate the displacement. Those components preserve
$H$; this is  the crucial point of the mechanism of Brout, Englert and Higgs, cf \cite{weinberg1996quantum},\cite{itzykson1980quantum},
\cite{peskin1995introduction}. Which gives a first justification
of our identification of displacement direction and Higgs field direction.\\

At higher orders the classical fields polynomials are of order less than four in $\phi$, but there is no limit for
the degree in $H$, thus they do not satisfy the renormalization criteria of Dyson, but, as we will see
in the discussion, we are confident they can be renormalized in the effective sense of Wilson and Weinberg,
cf. \cite{anselmi1994removal},\cite{anselmi2015renormalization},
\cite{gomis1996nonrenormalizable}.\\

However there is no need to enumerate the precise terms in $\mathcal{L}$, because all the terms involving the
fields and compatible with the local covariance will appear for quantum corrections and renormalization.\\

\subsection{Choice of triality and fermion structure}

Now the idea is to use the spin gauge transformations to kill half of the spinors and make
apparent the decomposition of the resting half in four sub-spaces which can be identified
with the neutrino direction, the electron direction, the three dimensional space of quarks up and
the three dimensional space of quarks down.\\

The particularity of dimension $8$ is that the even and odd spin representations
$\sigma^{+}$  in $S_8^{+}$ and $\sigma^{-}$ in $S_8^{-}$ are real and externally isomorphic to the standard vector representation
$\rho$ in $\mathbb{R}^{8}$, i.e. there exists an automorphism $j$ of the group $Spin_8$ or order $3$,
such that $\sigma^{+}$ is equivalent to $\rho\circ j$ and $\sigma^{-}$ to $\rho\circ j^{2}$.
The \emph{triality} associated to $j$ is the triple of isometries $\tau^{+}:\mathbb{R}^{8}\rightarrow S_8^{+}$,
$\tau^{-}:S_{8}^{+}\rightarrow S_8^{-}$, and $\tau_0:S_8^{-}\rightarrow \mathbb{R}^{8}$,
such that, for every element $g$ in the group $Spin_8$, every vector $v$ in $\mathbb{R}^{8}$, every
spinor $s_+$ in $S_{8}^{+}$ and every spinor $s_-$ in $S_{8}^{-}$,
we have
\begin{multline}
\sigma^{+}(g)(\tau^{+}(v))=\tau^{+}(\rho(g)(v)),\\
\quad \sigma^{-}(g)(\tau^{-}(s_+))=\tau^{-}(\rho(g)(s_+)),\\
\rho(g)(\tau_0(s_-))=\tau_0(\sigma^{-}(g)(s_-)).
\end{multline}
\indent In other terms, a \emph{triality} is an automorphism of the sum $\mathbb{R}^{8}\oplus S_8^{+}\oplus S_8^{-}$
that permutes the factors in cyclic manner and commutes with the action of $Spin_8$ after twisting by $j$.\\
\indent In some sense, the existence of triality goes back to Cayley, however the clear exposition is
due to Chevalley (cf. \cite{chevalley1997algebraic}). The book of Chevalley on spinors details the complex case, or more generally the case
where the quadratic form in dimension eight is hyperbolic, as in signature $(4,4)$ over real numbers. But
it is not difficult to adapt his study to the positive definite case, and to prove that, given any three
representations of the type $\rho$, $\sigma^{+}_{\mathbb{C}}$ and $\sigma^{-}_{\mathbb{C}}$, a triality is determined
by a pair of spinors of different chirality, a choice of spin invariant metric on $S_8\otimes \mathbb{C}$
and a choice of spin invariant real structure on $S_8\otimes \mathbb{C}$, the odd spinor being pure and
the even one being of norm one.\\

\noindent More precisely: let $F$ be a real vector space of even dimension $2m$, $\Phi$ a positive definite
quadratic form on $F$, and $S=S^{+}\oplus S^{-}$ a space of spinors for $\Phi$, of complex dimension $2^{m}$;
we write $\gamma$ the Dirac map from $F$ to $End(S)$, as well as its extension to $F\otimes \mathbb{C}$; we have
$\gamma(x)\gamma(x)=\Phi(x)Id_S$ for every $x$ in $F$.\\
A \emph{pure spinor} is an element $s$ of $S$ such that the vectors $v$ in $F\otimes \mathbb{C}$ satisfying
$\gamma(v)(s)=0$ form a maximal isotropic subspace $F'$ in $F\otimes \mathbb{C}$ (cf. \cite{cartan1966theory}, \cite{chevalley1997algebraic}).
Such a $s$ belongs necessarily
to $S^{+}$ or $S^{-}$. The space $F'$ does'nt intersect $F$ nor $iF$, then its conjugate $F"$ is another maximal
isotropic subspace of $F\otimes \mathbb{C}$; we have $F\otimes \mathbb{C}=F'\oplus F"$ and the pair $(F',F")$
is a \emph{complex polarization} of the quadratic space $(F, \Phi)$. We put on $F$ the orientation given
by the projection of $F'$ to $F$ parallel to $F"$ or to $iF$. Another manner to describe this situation is to
consider the complex structure $J$ on $F$ which has $F'$ (resp. $F"$) as eigenspace for $i$ (resp. $-i$).
Note that only the complex direction of $v$ is used to get the polarization or the complex structure.\\
The orbit of $s$ in $S^{+}$ (for instance) is isomorphic to the symmetric space $SO_{2m}/U_m$ of complex
structures on $\mathbb{R}^{2m}$, compatible with the orientation and the Euclidian metric.\\
The subgroup of $Spin(\Phi)$ that fixes a given pure spinor direction is a two-fold covering of the
unitary group of the vector space $F$ endowed with the complex structure $J$ and the hermitian product
deduced from $J$ and $\Phi$. Cf. A.Weil \cite{weil20varietes}.\\
When this choice is made, the two spinor representations $S^{+}$ and $S^{-}$ can be identified (this
fixes the phase of the pure spinor) with the even exterior sum $\Lambda^{ev}(F")$ and the odd exterior
sum  $\Lambda^{odd}(F")$ respectively; the map $\gamma_\mathbb{C}$ being given by the sum of creation and annihilation
operators: $\gamma(v'+v")(s)=\iota(v')s+v"\wedge s$. Thus we get a model of complex spinors based on the
space $F$. On $F"$, the natural hermitian structure is given by
\begin{equation}
\langle v"| w"\rangle=\frac{1}{2}(v|w)-\frac{i}{2}(Jv|w)
\end{equation}
where $v"=(v+iJv)/2,w"=(w+iJw)/2$. This extends to hermitian structures on $\Lambda^{ev}(F")$ and $\Lambda^{odd}(F")$.\\
When a unit of complex volume $\Omega$ is chosen with norm equals to one, an anti-$\mathbb{C}$-linear \emph{Hodge star} $*"$
is defined on $S$ by the formula
\begin{equation}
*"s\wedge u=\langle s| u\rangle \Omega.
\end{equation}

\indent Now suppose $m=4$, $F$ has real dimension $8$. The space $\Lambda^{ev}(F")$ splits in three summands,
the constants, the $6$ dimensions of bi-vectors and the rank one space of $4$-vectors. The choice of a non-zero $4$-vector
$\Omega$ of norm $1$ gives a complex quadratic form $B$ on $S^{+}$ and $S^{-}$ which is spin invariant (cf. \cite{chevalley1997algebraic}):
if $s$ belongs to $S$, $B(s)$ is the component over $\Omega$ of the Clifford product $\beta(s)s$, where $\beta$ denotes the
principal anti-involution of the Clifford algebra $C(\Phi)$, sending $\gamma(v_1)...\gamma(v_k)$ on $\gamma(v_k)...\gamma(v_1)$.\\
In dimension $4$ for $F"$, $*"\circ*"$ is $Id$ on $S^{+}$, and
$-Id$ on $S^{-}$. We define the anti-$\mathbb{C}$-linear map $I_0$, by
taking $*"$ on multi-vectors of degrees $1$ and $2$, and taking the opposite $-*"$ on multi-vectors
of degrees $0$, $3$ et $4$. In such a way that $I_0\circ I_0=Id$ defines a \emph{real structure} on $S$.\\
\indent Computations with creations and annihilations (cf. $UDPE$ for them) prove that
$I_0$ commutes with the spin representation of the Lie algebra $o(\Phi)$ of infinitesimal Euclidian rotations.
Consequently, the map $I_0$ defines the conjugation operator of a spin invariant real structure on $S$.
The self-dual multivectors $s+I_0s$ in $\Lambda^{odd}(F")$ (resp. $\Lambda^{ev}(F")$)
form a real subspace $S_8^{+}$ (resp. $S_8^{-}$) which is spin invariant. \\
All the other invariant real structures $I$ are obtained by multiplying $I_0$ by a complex number
of module one; this goes with changing the arbitrary phase of $\Omega$, the constraint being that
$\Omega$ must be definite positive on the real spinors spaces $S_8^{+}$ and $S_8^{-}$.\\
\indent If in addition, we choose a spinor $t$ in $S_{-}$ such that $B(t)=1$, we obtain a unique complex triality
$\tau$ sending $s$ to $-t$. In fact, $t=u"+t"$ where $u"$ is a vector and $t"$ a trivector; the vector $u_0=(u"+\overline{u"})/\|u"+\overline{u"}\|_F$
belongs to the real space $F$ and has Euclidian norm $1$; by applying $\gamma(u_0)$ to $t$ we get an element $s$ in $S^{+}$
such that $B(s)=1$. Then we follows the process described by Chevalley (\emph{up cit}): first define $\gamma^{+}:F\rightarrow S^{-}$
by $\gamma^{+}(v)=\gamma(v)s$; it is a bijection; then consider the involution
$\mu$ of $\Delta_\mathbb{C}=(F\otimes\mathbb{C})\oplus S^{+}\oplus S^{-}$ which coincides with $\gamma^{+}$ on $F\otimes\mathbb{C}$ and
with its inverse on $S^{-}$, and which is the reflection in the hyperplane orthogonal to $s$ in $S^{+}$. In the same manner, replacing the
spinor $s$ by the spinor we get an involution $\nu$ of $\Delta_\mathbb{C}$ exchanging $F$ and $S^{+}$ and preserving $S^{-}$.
Then the map $\tau=\mu\circ \nu$ is the announced triality.\\
If the vector $u_0$ is chosen, we decompose it according to $F'$ and $F"$ in $u_0=u'_0+u"_0$ and we decompose
the unit oriented $3$-vector $V_0$ orthogonal to $u_0$ in the same manner $V_0=V'_0+V"_0$, then
every possible choices of $t$ with $u"=u"_0$ are given by the following formula
\begin{equation}\label{spinort}
t=u"_0+e^{-i\beta}V"_0+u_0"\wedge a",
\end{equation}
where $\beta$ is real number modulo $2\pi$ and $a"$ a bi-vector in $S^{-}$. In this case the unit complex volume $\Omega$
satisfies $\Omega=2e^{-i\beta}u"_0V"_0$, and we have
\begin{equation}
s=\frac{1}{2}(\Omega+1+a").
\end{equation}
The necessary and sufficient condition such that $\tau$ preserves an invariant real structure and an invariant
positive definite norm on $\Delta_\mathbb{C}$ is $a"=0$. In all what follows we assume this condition.\\
If the unit volume defining $B$ and $I_0$ is $\Omega_0=2u"_0V"_0$, and if $t$ is defined by \eqref{spinort}, the
invariant conjugation for which the spinors $s$ and $t$ are real is $-e^{-i\beta}I_0$. Cf. $UDPE$.\\
\indent We choose a direct orthonormal basis $e_i;i=1,...,8$ of $F$ such that $e_8=u_0$ and $e_7=-Je_8$.\\
Then $\tau(e_8)=-s$,
\begin{equation}
\tau(e_7)=\frac{1}{2}(-i\Omega+i.1),
\end{equation}
and for $j=1,...,6$,
\begin{equation}
\tau(e_j)=e"_i\wedge u"_0+e^{-i\beta}\iota(e'_i)V"_0.
\end{equation}
Then we observe a natural decomposition of the complex subspace $Q=\Lambda^{2}(F")$ in $S^{+}$ in two subspaces
$Q'$ and $Q"$ of dimension three, respectively made of the bi-vectors of the form
\begin{equation}
q'(v')=2e^{-i\beta}\iota(v')V"_0,
\end{equation}
for every $v'$ in the $3$-plane $V'_0$ in $F'$ orthogonal to $u'_0$, and by the bi-vectors of the form
\begin{equation}
q"(v")=2\lambda v"\wedge u"_0
\end{equation}
for every $v"$ in the $3$-plane $V"_0$ in $F"$ orthogonal to $u"_0$.\\
We name the first ones magnetic quarks and the second ones electric quarks (in fact we will see they are
anti-quarks).\\

Thus the choice of the triple $J,\Omega, u_0$ (or equivalently $J,\Omega, t$) not only gives a triality but a decomposition
of $S^{+}$ in four summands $\Lambda^{0}F"=\mathbb{C}.1$, $\Lambda^{4}F"=\mathbb{C}.\Omega$, $Q'$ and $Q"$, where it is
possible (or evident) to recognize the \emph{neutrino}, the \emph{electron} and the two first flavors of \emph{quarks}.\\
If we forget the phase of $\Omega$, we get the same decomposition, only the identification of $Q'$ with $V'_0$ is lost.\\

We saw that the gauge fixing of the complex structure $J$ has a residual gauge group $\widetilde{U_4}$ which is a covering of the
unitary group $U_4$ in dimension four of $F$. The center of the corresponding Lie algebra has no effect on the quarks but it induces multiplication
by non-zero imaginary numbers on the leptons $1$ and $\Omega$, consequently the sub-algebra fixing $\Omega$ is $su_4$; the subgroup generated by
$su_4$ in $U_4$ being simply connected, the residual group in $Spin_8$ is $SU_4$. In this group the gauge fixing of $u_0$ or
$t$ has the same residual gauge group which is $SU_3$, corresponding to quantum chromodynamics. On the side of boson fields, the
components of the connection $A$ in the directions of the Lie algebra $su_3$ of this group are the \emph{gluons}.\\

\noindent Note that the representation of $Spin_8$ in $F$, $S^{+}$ and $S^{-}$ are not equivalent, they are only externally isomorphic,
thus it is non-trivial that $SU_3$ is the same for fixing $u_0$ and $t$, but it is true.\\
\indent In fact, every triality $\tau$ determines a unique automorphism $j$ of order $3$ of the group $Spin_8$, such that the sum $\sigma$
of the three representations $F$, $S_8^{+}$, $S_8^{-}$ in $\Delta$ verifies
\begin{equation}
\tau\circ \sigma(g)=\sigma(j(g))\circ \tau.
\end{equation}
The induced automorphism $j_*$ of the Lie algebra can be computed from its action on the commutative sub-algebra
$\mathfrak{t}_4=i\mathfrak{h}_4$ of $o_8$ which corresponds to diagonal matrices. The standard basis of $\mathfrak{t}_4$
is made by the nodal vectors of $SO_8$, that is $(K_j;j=1,3,5,7)$, where $K_j=2i\pi H_j$ and $H_j=\frac{2}{i} e_j\wedge e_{j+1}$.
In the infinitesimal spin representation $\sigma_*$, we have $2\sigma_*(e_j\wedge e_k=\gamma(e_k)\gamma(e_j)$.\\
Then, in the ordered basis $K_7$, $K_5$, $K_3$, $K_1$, the matrix of the restriction of $j_*$ to $\mathfrak{t}_4$ is
$\frac{1}{2}^{t}T_8$, where
\begin{equation}\label{matricejetoile}
^{t}T_8=
\begin{pmatrix}
-1&\quad 1&\quad 1&\quad 1\\
-1&\quad 1&-1&-1\\
-1&-1&\quad 1&-1\\
-1&-1&-1&\quad 1
\end{pmatrix}
\end{equation}\\
The action on the full Lie algebra $so_8$ follows by considering the associated Chevalley generators
(see , \cite{bourbaki1981groupes456}, \cite{bourbaki1982groupes9}). In particular the usual simple co-roots of $SU_3$, which are $H_1-H_3$ and
$H_3-H_5$ are invariant by $j_*$, thus $j$ preserves $SU_3$ and the three representations of $SU_3$
on $F$, $S_8^{+}$ and $S_8^{-}$ are equivalent.\\
However $j_*$ does'nt preserve $u_4$ or $u_3$. Also, the automorphism $j$ does'nt act on $SO_8$ or $SU_4$.
But it acts on the quotient $PSO_8$ of $SO_8$ by its center $\{\pm Id\}$ (cf. \cite{cartan1966theory}).\\
Fixing a chosen $t$ of spinorial norm $1$ in $S_8^{-}$ is always possible and defines the group $SU_3$ in
$SU_4$.\\

The group $SU_3$ fixes point by point the leptons and acts separately on the two sets of quarks $Q'$, $Q"$. As this
action is the complexification of an action on the real spinors, the two actions on $Q'$ are $Q"$
are not equivalent, the first one is standard on $\mathbb{C}^{3}$, noted traditionally $D(1,0)$ and the
second one is the conjugate on $\overline{\mathbb{C}}^{3}$, noted traditionally $D(0,1)$.
The two representations $D(1,0)$ and $D(0,1)$ are only
externally equivalent, through the conjugation automorphism of $SU_3$, which sends $g$ to $\overline{g}=^{t}g^{-1}$.\\
However, it is an experimental fact in Particle Physics that the creation operators on the two sorts of quarks are
equivalent, both being of type $D(1,0)$.\\
\indent The solution of this paradox is offered by anti-particles: every particle of spin $1/2$ in even
Lorentz dimension comes with a conjugate anti-particle, with reverse propagation (see \cite{dirac1967principles}, \cite{feynman1962quantumelectrodymaics},
\cite{weinberg1995quantum});
thus on the side of $S_8^{+}$ (resp. $S_8^{-}$) we have to consider the conjugate
spaces $\overline{S_8^{+}}$ (resp. $\overline{S_8^{-}}$). On the leptons this does'nt make great difference,
we keep them as they come for particles, but on the quarks we must take $\overline{Q"}$ for the particles
and $Q"$ for the anti-particles, keeping $Q'$ as it comes for the particles, the space $\overline{Q'}$ describing
its anti-particles.\\
Consequently we redefine the eight-dimensional spaces of multiplicities of Lorentz four dimensional Dirac spinors, by taking
\begin{multline}
L=\mathbb{C}1+\mathbb{C}\Omega;\quad \overline{L}=\overline{\mathbb{C}}\overline{1}+\overline{\mathbb{C}}\overline{\Omega};\\
Q=Q'\oplus \overline{Q"};\quad \overline{Q}=\overline{Q'}\oplus Q";\\
M_8^{+}=L\oplus Q;\quad \overline{M_8^{+}}=\overline{L}\oplus \overline{Q}.
\end{multline}
The spaces without overline give the directions of the particles, the ones with barre give the directions of the associated anti-particles.\\
There exists an analog redefinition for $S_8^{-}\oplus \overline{S_8^{-}}$ giving rise to $M_8^{-}$ and $\overline{M_8^{-}}$.
The letter $M$ is for multiplicity. We consider them as complex vector spaces; this makes no difference when tensoring
by the space of four dimensional Dirac spinors  which is a four dimensional complex space.\\

\subsection{W,Z, Higgs and dark matter}

In the pool of bosons, we not only have generators of $spin_8=o_8$, we also have the symmetric matrices
in $gl_8$. A natural way for allowing them to act on fermions, by Yukawa coupling, is to transform the
vector space $\mathfrak{p}_8$ in $i\mathfrak{p}_8$, and to consider $o_8\oplus i\mathfrak{p}_8$
as the Lie algebra $u_8$, which acts naturally on $M_8^{+}$
and $M_8^{-}$. This action has to be compatible with the action of $o_8$. The same for the
anti-particles in the conjugate multiplicities.\\
\indent The more evident operators correspond to the subalgebra $u_2$ orthogonal to the
$u_6$ of the quarks $Q$. \emph{A priori} it acts only on $L$ in the usual manner, however the reversing
of complex structure in $Q"$ makes possible (if not necessary) an action on $Q$ which commutes with the action of
the $u_3$ in $o_6$. In particular the off diagonal operators generate the isomorphisms between the two
representations on $Q'$ and $\overline{Q"}$ in $M_8^{+}$.\\
First we define the $\mathbb{C}$-linear operators $W_+$
and $W_-$ in $gl_2(\mathbb{C})=u_2\otimes \mathbb{C}$ by the following formulas, for every real vector $v$ in $F$,
decomposed as $v'+v"$ in $F'\oplus F"$,
\begin{gather}
W_+(\overline{v"}\wedge \overline{u"_0})=\exp(-i\beta)\iota(v')V"_0,\quad W_+(\iota(v')V"_0)=0\\
W_+(1)=-\exp(-i\beta)\Omega,\quad W_+(\Omega)=0,\\
W_-(\overline{v"}\wedge \overline{u"_0})=0,\quad W_-(\exp(-i\beta)\iota(v')V"_0)=-v"\wedge u"_0;\\
W_-(1)=0,\quad W_-(\exp(-i\beta)\Omega)=1.
\end{gather}
\indent The bracket $[W_+,W_-]=W_+W_--W_-W_+$ satisfies
\begin{multline}
[W_+,W_-](\overline{q"})=\overline{q"},\quad [W_+,W_-](q')=-q',\quad [W_+,W_-](\nu")=-\nu",\\ [W_+,W_-](\varepsilon")=\varepsilon".
\end{multline}
Thus, if we denote by $Z'_1$ the operator on $M_8^{+}$ which multiplies by $-i/2$ on $\Lambda^{0}F"$ and $Q'$
and  multiplies by $i/2$ on $\Lambda^{4}F"$ and $\overline{Q"}$, we have
\begin{equation}\label{crochetzcomplexe}
[W_+,W_-]=2iZ'_1.
\end{equation}
This operator $Z'_1$ comes from $u_4$ and generates the commutator of $u_3$ in $u_4$. Remark
its sign would have the same on $Q'$ and $Q"$, then ignores the reversing of $Q"$; is it a kind of
twist? Anyway we have no choice at this point.\\
\noindent The associated anti-hermitian operators are :
\begin{equation}
W_1=\frac{W_++W_-}{\sqrt{2}},\quad W_2=\frac{i(W_+-W_-)}{\sqrt{2}},
\end{equation}
in such a way that $\sqrt{2}W_+=W_1-iW_2$ and $\sqrt{2}W_-=W_1+iW_2$.\\
\noindent It follows
\begin{equation}\label{crochetzunitaire}
[W_1,W_2]=2Z'_1.
\end{equation}
Thus $W_1,W_2,Z'_1$ form a Lie alegbra $su_2$, and $W_+,W_-,H_7$ a Lie algebra $sl_2(\mathbb{R})$, in their
usual presentations.\\
On $M_8^{+}$ we also have the operator $Z_4=i(H_1+H_3+H_5+H_7)$, which generates the center of $u_4$;
it coincides with $4Z'_1$ on $L$ and gives zero on $Q$. To generate the center of $u_3$ we choose $Z_4-Z'_1$.\\
Coming from $\mathfrak{p}_8$, we have $B_0$ (baryon number) which is the identity of $M_8^{+}$ and $B_1$
(lepton number) which is identity on $L$ and zero on $Q$.\\
All these elements, identifiable in $gl_8$, have a natural action on $M_8^{-}$ as well: we copy the preceding
actions, replacing $1$ by $u"_0$, $\Omega$ by $V"_0$, $Q'$ by the hermitian orthogonal of $V"_0$ in $\Lambda^{3}(F")$,
and $\overline{Q"}$ by the conjugate space of the hermitian orthogonal of $u"_0$ in $F"$.\\

Let us denote by $\rho^{+}$ (resp. $\rho^{-}$) the representation of $\mathfrak{u}_3\times \mathfrak{u}_2\times \mathfrak{u}_1$
constructed as above on $M_8^{+}$ (resp. $M_8^{-}$), with the standard $s\mathfrak{u}_3$, the $Z_3$ in $\mathfrak{u}_3$, the $s\mathfrak{u}_2$ of $W_1,W_2,Z_1$,
$iB_1$ in $\mathfrak{u}_2$ and $iB_0$ generating $\mathfrak{u}_1$, and denote by $\mathfrak{u}_8$ the Lie algebra $spin_8\oplus i\mathfrak{p}_8$
that naturally contains this algebra $\mathfrak{u}_3\times \mathfrak{u}_2\times \mathfrak{u}_1$.\\
In $UDPE$ the following proposition is proved:\\
There exists no extension of $\rho^{+}$ (resp. $\rho^{-}$) in a representation of a sub-algebra of $\mathfrak{u}_8$ which
contains $\mathfrak{u}_3\times \mathfrak{u}_2\times \mathfrak{u}_1$, and which is stable by the adjoint action of
$s\mathfrak{u}_3$ in $\mathfrak{u}_8$.\\
This implies that in our theory, on the $64$ independent vector fields coming from the connection field $C_F$ at the order one,
$52$ of them cannot act on fermionic matter (the same will be true at any order, when specified by the Higgs field). Only
the eight gluons, the two bosons $W_+$, $W_-$ and the four abelian vectors $Z'_1$, $Z_4$, $iB_0$, $iB_1$ can do.\\
Within the four abelian fields, only two linear combination $Q_0$ and $Z_0$ are observable particles; we will discuss
why it should be so later.\\
\indent This proposition offers a natural explanation of \emph{dark matter}, because those $52$ fields certainly have non-zero masses,
comparable to the masses of $W_+$, $W_-$ and $Z_0$. Remark that in our theory, the dark matter will be abundant, because
each generation of bosons will reproduce this structure, with higher and higher masses, today inaccessible, at the
energy scale we are exploring. Remark that at first sight, this seems paradoxical that most of the masses is not accessible,
but it is a consequence of the dependency of quantum corrections on the energy scale; in the theory we suggest more and more
energies are needed to discover next generations, which is compatible with the fact that big masses correspond to rare events.  \\

Let us show now how the exact Weinberg mechanism of symmetry breaking appears in this framework:\\
\indent Once $\beta$ is chosen, fixing $t$ is equivalent to fix $u"_0$.
Consider the two dimensional complex vector subspace $S_4^{-}$ of $S_8^{-}$ (or $M_8^{-}$) which is
generated by the vector $u"_0$ in $\Lambda^{1}(F")=F"$ and by the tri-vector $V"_0$ in $\Lambda^{3}(F")$.\\
In this subspace, $W_{\pm}$ satisfy
\begin{align}\label{actionwimpaire}
W_+(\overline{u"_0})=\exp(-i\beta)V"_0,\quad &W_+(\exp(-i\beta)V"_0)=0,\\
W_-(u"_0)=0,\quad &W_-(\overline{\lambda}\exp(-i\beta)V"_0)=-\lambda u"_0.
\end{align}
The commutator $[W_+,W_-]$ acts as $Id$ on $u"_0$ and $-Id$ on
$V"_0$.\\
In the following we simply write $Z_1$ for $Z'_1$, cf. $UDPE$ for this point.\\
The operators $B_0$, $B_1$ act as $id$ on $S_4^{-}$, but $Z_1$ and $Z_4$ both induce
direct rotation on $u"_0$ and indirect one on $V"_0$. Consequently there exist linear
combinations of $Z_1$, $Z_4$, $iB_0$, $iB_1$ which give $0$ on $u"_0$ and other combinations which
maximize its change. In fact on $S_4^{-}$, the operator $Z'_1$ coincides with $Z_4$, and the
operator $iB_1$ coincides with $iB_0$, thus a pair should \emph{a priori} be sufficient, for instance $Z_1$ and
$iB_1$; but we prefer to conserve all the operators, for taking in account the actions on quarks.\\
It is easy to prove that the combinations
\begin{equation}
Q=i\beta_0B_0+i\beta_1B_1+2\alpha_1Z_1+\alpha_4\frac{Z_4}{2};
\end{equation}
which preserve $u"_0$ fixed are the ones for which $\beta_0+\beta_1=\alpha_1+\alpha_4$.\\
However, only one is correct if we want that $Q$ gives the known charges to the known quarks, i.e.
$-1/3$ in $\overline{Q"}$ ($d$, $s$ or $b$) and $+2/3$ in $Q'$ ($u$, $c$ or $t$); it is
\begin{equation}\label{photon}
Q_0=\frac{i}{6}(B_0-4B_1)-Z_1.
\end{equation}
Then we put
\begin{equation}\label{bosonneutre}
Z_0=\frac{i}{6}(B_0-4B_1)+Z_1.
\end{equation}
The operator $B'=(B_0-4B_1)/3$ is $-Id$ on the lepton space $L$ and $1/3.Id$ on the quarks
space $Q$. This combination $B'$ of $B_0$ and $B_1$ is the one that we must pair with
$Z_1$ to reproduce the Weinberg scenario in $S_4^{-}$.\\
In the results section, we will propose a geometric explanation of the formulas \eqref{photon}
and \eqref{bosonneutre} for the \emph{photon} and for the \emph{neutral boson} respectively, which suggests
an explanation of the known charges of the quarks.\\
\indent Let us signal the problem of the fields, which satisfy the
condition to be invariant, but are not elected as a photon, for instance $iB_0+Z_4/2$ and $iB_1+2Z_1$. They coincide on $L$ and on $Q$
the first is $iB_0$ and the second is $2Z_1$. They do not acquire masses from the Higgs mechanism. They could acquire masses from another
kind of gauge fixing; for instance the total trace $iB_0$ corresponds to the determinant of the transversal
metric, and $Z_4$ corresponds to the complex determinant of the transversal space. However it is more honest to say
that we do not know what these fields become. They are allowed to act on fermions. But they could simply decouple.\\

\indent The auxiliary gauge group $SU_2\times U_1$,
is made from the $SU_2$ generated by $W_{\pm}$ et de $Z_1$, and the $U_1$ generated by $iB'$ acting
on $L$, $Q$ and $S_4^{-}$, exactly as in the Standard Model of Glashow, Salam and Weinberg. Cf. \cite{weinberg1996quantum}.\\

\indent But, where is the Higgs field?\\
It cannot be a spinor, experiment is net on that. We propose it is a constant multiple of the section $u_0$ in $E_F$.\\
\indent Remind that the gauge group does'nt act on $E_F$. In fact until now, there exists no field here, all
the bosonic fields come from components of the metric. But the diffeomorphism group, even preserving $W$ and preserving
the metric on the fibers of $E_F$ along $W$ to the first order included, can transform any section of fixed norm
in any other one, thus we can use diffeomorphisms as gauge transformations to privileged $u_0$ given by our spinorial
gauge fixing. This does'nt solve the problem, that still $u_0$ is not a bosonic field. For that, another mechanism
is necessary, and we propose it is the \emph{bosonization of four degrees of freedom among the fermions}.\\
This bosonization has to offer the same structure we had in $S_4^{-}$, with two complex components $\varphi^{0}$ et $\varphi^{+}$.\\
The odd sector, $S_8^{-}$ or $M_8^{-}$ is now impossible to use, because it is frozen by the symmetry breaking.
However the orbit under $Spin_8$, or $\widetilde{U_4}$, or $SU_4$, of $t$ in $S_8^{-}$ is a sphere $S^{7}$, the same for
$u"_0$ in $F"$, and the same is true for $u_0$ in $E_F$ under the action of the constrained diffeomorphisms group.
Then we can use the even sector; this sector is active after symmetry breaking,
but after the choice of $J$, or $1$ in $S_8^{+}\otimes \mathbb{C}$, the complex direction of
the neutrino is fixed by the residual gauge group $SU_3\times U_1$. From the point of view of the four dimensional theory, over $W$, this corresponds to a complete
Dirac spinor of mass zero, four complex dimensions, which can be decomposed in two chiralities, $E^{+}$ and $E^{-}$, also named \emph{left}
and \emph{right}. And it is a characteristic of the standard model, issued from experimental observations, that only the
left component of the neutrino and the right component of the anti-neutrino have a physical presence. Thus we propose to transform the right component into the
Higgs boson. Andrzej Okninski recently remarked that the Dirac equation for a spinor $\nu$ of fixed direction gives
for its two coordinates $\varphi$ a diffrential system of the type of  Duffin-Kemmer-Petiau, $DKP$, of spin zero.
\cite{okninski2011supersymmetric}, \cite{okninski2014neutrino}.\\
We use a similar transformation, in three steps:\\
First a scalar field $\varphi=(\varphi',\varphi")$ is defined from the phase variations
of $\nu"$ and $\overline{\nu"}$:
\begin{equation}\label{bosonfermion}
\nu(x)=\varphi'(x)\nu"_0+\varphi"(x)\overline{\nu"}_0;
\end{equation}
Secondly, we identify the plane of $\overline{\nu"}_0$ and $\overline{\nu"}_0$ with the vectors $u"_0$ et $V"_0$
in the subspace $S_4^{-}$ of $M_8^{-}$. The first component
$\varphi'(x)$ gives the coordinate along $u"_0$, the second one
$\varphi"(x)$ gives the coordinate along $V"_0$, with perhaps a phase beta.\\
And we identify $\varphi'$ to $\varphi^{0}$ and $\varphi"$ to $\varphi^{+}$.\\
Finally we interpret $u"_0$ as $e"_8$ and $V"_0$ as $e'_8$ in the complexification of the real plane $(e_7,e_8)$,
which has a natural quaternionic structure due to $J$.\\
This gives the spaces for creation and annihilation of the Higgs field $H=M_0u$.\\
The modulus of $\varphi^{0}$ was considered by Weinberg as the only physical field in this sector, due to the
action of $SU_2\times U_1$, it corresponds to the radius of the sphere of diemnsion seven where
is varying $v$ and gives the scale  in $GeV$ for the theory.\\
Note that our three steps for the bosonisation follows the arrows of the triality, from even spinor to odd spinors to vectors.\\

\indent In $UDPE$ we compute the differential operator coming from the Dirac operator; there exists an orthonormal Lorentz frame where
\begin{equation}
(\varphi',\varphi")\mapsto
\begin{pmatrix}
-(\partial_1+i\partial_2)\varphi"\\
i(\partial_0+\partial_3)\varphi"\\
(\partial_1+i\partial_2)\varphi'\\
-i(\partial_0+\partial_3)\varphi'
\end{pmatrix}
\end{equation}
Up to the sign, it is the same operator $P(\partial)$ on $\varphi'$ and $\varphi"$.
A solution of $P(\partial)\varphi'=0$ is a function of $(x_0-x_3,x_1,x_2)$ which is
holomorphic in $x_1+ix_2$, which selects a chirality.\\
All solutions satisfy the ordinary wave equation, which becomes the Klein-Gordon
equation as soon as we take in account a mass for $H$.\\
\indent The mass of $H$ is supposed to come from a quartic term $C_0\overline{t} HH t$, in
the Taylor development of the Dirac Lagrangian,
where $t$ is the odd fixed spinor; the factor $C_0$ being necessary here to have a dimension
$L^{-4}$ in the Lagrangian. The dynamical term, coming from bosonization of $\nu"_R$, is
$M_1H\nabla H$, with a constant $M_1$ in $L^{-1}$; then the square of the mass of $H$ is expressed through the
mean value of $t_0$ as $M_H^{2}=C_0\overline{t_0}t_0$. The experimental value of $M_H$ is $125 GeV$.
Nothing a priori in our model can predict this value. However it is natural to expect from the triality
that $t_0$ correspond to the typical length $v_0$ by taking the exponent $-3/2$, and if we take $C_0$ of the
order of unity this gives the correct size of $M_H$. To the contrary, if we decide to adapt $t_0$ from
$H_0$ by taking its value to the exponent $3/2$, the known value of $M_H$ implies that $C_0$ is of the order $10^{-3}$.\\
\indent Terms of arbitrary degrees in $H$ appear in the same way, and give the Higgs potential; for instance
$\lambda \overline{t} H^{4} t$. It is preferable for the stability of the system, that, as supposed by
the standard model, the mean value $H_0$ is made for attaining a minimum of this potential; this
gives a probable value for $\lambda$.\\
\indent Observe that relaxing the gauge, $H$ departing from its mean value, the right component of the neutrino
could revive transiently, giving rise to a massive fermion, which is good with respect to neutrino oscillations,
cf. \cite{suekane2015neutrino}.\\

\noindent Except perhaps for $B_0$ and $Z_4$ all masses for bosons and fermions come from the Higgs field $H$.\\
\noindent In the case of bosons $\phi$, this
corresponds in the Lagrangian to quadratic terms in $H$ and $\phi$, like $HHSS$, $HHBB$ and $HHAA$ that are of
dimension $L^{-4}$. This is for the fields of the first order. For higher orders new fields appear and include the mean value
$v_0$ of $u=\|u\| u_0$, i.e. $v_0=H_0/M_0$, of dimension zero. Their masses correspond to terms $HH\phi^{(n)}\phi^{(n)}$.
In the boson case the Lagrangian term is proportional to the square of the mass. In particular the masses $M_W$ and $M_Z$
are of first order in $H_0$; they also depend on the coupling of dimension zero for $SU_2$ and the $U_1$ of $B=B_0-4B_1$; cf. \cite{weinberg1996quantum},
\cite{itzykson1980quantum}, \cite{peskin1995introduction}. $M_W=gH_0/2$, $M_Z=M_W/\cos \theta_W$. Today, the experimental values are
$M_W=80.4 GeV$,$M_Z=91.2 GeV$. The constant $H_0$ is experimentally given by the Fermi constant
$G_F=1/v^{2}\sqrt{2}\approx 1.166 10^{-5} GeV^{-2}$; this gives an approximate value
of $250 GeV$ for $H_0$; then $g$ et $g'$ are of the order of unity.\\
By construction, the gluons and the photon are of mass zero, in agreement with the identification of $H$ with a
multiple of $v$.\\
For $B_0$ and $Z_4$, there are two possibilities, either they have a very high mass, either they are decoupled or
even nonphysical. The reason to defend this last position is the fact they represent variations of volumes; for $B_0$
it is the determinant of $g_{ij}$, which can be considered as minus the determinant of $g_{ab}$ because $g_{\mu\nu}$
can be considered as a constant in this analysis; for $Z_4$ it is the complex determinant of $F$, identifiable with
the phase of $\Omega$.\\
For the $52$ dark bosons, there is no reason that their masses should be far from the masses of $W$ and $Z$, because the coupling
constants are also $g$ or $g'$ for them. They should interact normally with the Higgs field, but not with the fermions or the other
bosons, because of the previously mentioned exclusion.\\
If we omit the trace $B_0$, and neglect the horizontal derivatives of $R$, the bosonic Lagrangian at the order two in $v$ is given by
\begin{multline}\label{energielorentzcouplagedegredeux}
\mathcal{L}_2(C,\nabla C,\nabla\nabla C)
=\frac{3}{4}\sum_{i j k l}v^{k}v^{l}(\partial_k\partial_lC_{\mu i}^{j}C_{i}^{\mu j}+C_{\mu i}^{j}\partial_k\partial_lC_{i}^{\mu j})\\
+\frac{1}{4}\sum_{i j k l}v^{k}v^{l}(\partial_k\partial_lC_{\mu i}^{j}C_{j}^{\mu i}+C_{\mu i}^{j}\partial_k\partial_lC_{j}^{\mu i})\\
+\frac{3}{2}\sum_{i j k l}v^{k}v^{l}\partial_kC_{\mu i}^{j}\partial_lC_{i}^{\mu j}
+\frac{1}{2}\sum_{i j k l}v^{k}v^{l}\partial_kC_{\mu i}^{j}\partial_lC_{j}^{\mu i}\\
+\frac{1}{4}\sum_{i j k l m}v^{k}v^{l}R_{ i k j l}C_{\mu i}^{m}C_{j}^{\mu m}-\frac{1}{4}\sum_{i k l m n}v^{k}v^{l}R_{ m k n l}C_{\mu i}^{m}C_{i}^{\mu n}\\
+\frac{1}{2}\sum_{\kappa \lambda i k l}v^{k}v^{l}g^{\kappa\lambda}M_{\mu\kappa k}^{i}M_{\lambda l}^{\mu i}.
\end{multline}

\noindent We already wrote the fermionic Lagrangian, but we do not discussed Yukawa couplings, giving masses from $H$; we will do that in the next section.\\

\section{Results}

Our main result is a geometrical deduction of all the elements of the standard model of Weinberg and Salam for
elementary particles,
by developing in Taylor series the Einstein and Dirac equations in dimension twelve, along a flat Lorentz
four dimensional sub-manifold $W$, and using only general covariance for diffeomorphims and spin gauge transformations.
The diffeomorphisms are used to introduce a vector bundle $E_F$ of real rank eight, and adapted coordinates on it, permitting to
represent the metric by three tensors, giving six series of bosonic fields, and one series of fermionic fields; the gauge group
is used to identify the multiplicities of four dimensional spinors with vectors in $E_F$, by fixing a \emph{triality}.\\
At this stage, the color group $SU_3$ and the decomposition in neutrino, electron and two sets of quarks ar obtained,
and the odd half of the spinors is eliminated. \\
Then, to go further, two ingredients are injected: a conjugation of half of the quarks, selecting $W_+$, $W_-$ and $Z_1$,
and a bosonization of the right neutrino $\nu_R$, giving a new dynamic field, the Higgs boson $H$. Importantly, to reproduce the
properties of the masses, this field
is identified with a multiple $M_0v$ of the section of $E_F$ which is used in the development of fields along $W$.\\
Then the Yang-Mills theory for $SU_3\times SU_2\times U_1$ is recovered, with the precise mechanism of symmetry breaking of
Brout-Englert-Higgs, which was described by Weinberg. A unique photon $Q_0$ and a unique neutral current emerged, if we
admit the values of charges for the leptons and the quarks.\\
The existence of several families of leptons and quarks is natural in this theory, they simply correspond to the different
orders of the development in $v$. However, each generation will have higher and higher masses, thus, at energies before the Planck mass,
specially if we take in account the quantum non-perturbative augmentation of masses, the model does'nt work, due to black holes,
and another model is necessary.\\
As in most unified models, other bosons appear in this approach, but we have shown how unitarity forbids their interactions with fermions and
other bosons; this is a quite noticeable characteristic of our model, which suggests a natural interpretation of \emph{dark matter}.\\
Note that computing the number of bosons, $64$, and comparing to the number of fermions, $64$ real ones, we have a kind of
equilibrium, which was considered as a good reason to explain \emph{dark energy} by a non-zero cosmological constant. Cf. \cite{bousso2008cosmological},
\cite{henneaux1989cosmological}. However,
even a small departure from perfect super-symmetry is known to generate a false prediction by more that ten orders, thus the best
property of our theory in the direction of understanding dark energy is probably its origin, in pure gravitation coupled
with a Dirac spinor of mass zero.\\

\subsection{Fermion masses. CKM and PMNS matrices.}

We make now the further hypothesis that the Yukawa couplings of the fermions with $H$ are coming from the curvature tensor
in the directions of the fibers of $E_F$.\\
\noindent We decide to identify $M_8^{+}$ and $F\otimes \mathbb{C}$ in such a manner that $\nu"$ corresponds to
$e_8$, $\varepsilon"$ to $e_7$, the complex space generated by $e_1,e_3,e_5$ to
$Q'$, and the complex space generated by  $e_2,e_4,e_6$ to $\overline{Q"}$, in a compatible
manner with the action of $SU_3$. This correspondence is determined up to a phase on $L$ and a phase on $Q$.\\
\indent Then, for each generation $\psi$, we can \emph{a priori} introduce a real constant $C$,
and the Yukawa term:
\begin{equation}\label{lagrangienmassefermions}
\mathcal{L}_R(\psi)=\frac{C}{i}\widetilde{\psi}^{a}\sum_{k,l}v^{k}v^{l}R_{kalb} \psi^{b}.
\end{equation}
It is natural in our unified context to suppose that $C$ does'nt depend strongly on the generation.\\

\noindent In this scheme, what happens from one generation to the next one?\\
We saw the multiplicity $M_8^{+}$ has a natural norm. At the generation number $n$, the four dimensional spinor appears
with a factor $v_0^{n}$, then to find a spinor corresponding to a norm one element of $M_8^{+}$, we have do divide it by
$v_0^{n}$. The spinors have dimension $L^{-3/2}$, the division by $v_0^{n}$ can be interpreted as a multiplication of
lengths in space-time by $v_0^{2n/3}$, which implies a multiplication of the curvature by a factor $v_0^{-4n/3}$.\\
From one generation to the next, the masses are multiplied by $v_0^{-4/3}$.\\
This is a prediction experimentally well verified for the electron, muon and the tau. For the electron $m_e\approx 0.5 MeV$,
for the muon $m_\mu\approx 100 MeV$, and for the tau $m_\tau\approx 20 GeV$. Thus we can take $v_0^{-4/3}\approx 200$,
which implies a value $v_0\approx 1/53$.\\
\indent For $H=M_0v$, it is known that the mean value is $H_0\approx 250GeV$, thus the constant $M_0$ is around
$13250 GeV= 13.25 TeV$.\\
For the first generation of fermions $m\approx CRv_0^{2}\approx 4.10^{-4}CR$. Experimentally for the electron
the mass $m_e$ is near $0.5 MeV$, ou $5. 10^{-4}GeV$, and for the quarks $u,d$ we have few $MeV$. This accords
with values of the quantity $CR$ of the size of units.\\

The real matrix $8\times 8$ with indices $a,b$ given by the curvature is symmetric, then the imaginary factor $1/i$ is here to deduce
an anti-hermitian operator $T_R$ in $u_8$, acting on $M_8^{+}$, and respecting the chosen identification between $M_8^{+}$ and
$F$. This operator is diagonalisable; its eigenvalues (once multiplied by $i$)
give the fermions masses of this generation (the masses in the Dirac equation are naturally multiplied by $-i$ for respecting
the conventions on the energy).
However, nothing tells us that these eigenvalues are positive, they correspond to the sectional curvatures of the fibers $F_x$;
then we will interpret the modules as the masses.\\
\noindent In accord with our hypothesis
of zero curvature for $W$, we will suppose in what follows that $R_{kalb}$ varies very slowly in function of $x$, in
comparison with its transversal variations.
But it is useful to consider the development of $R_{kalb}$ in function of $v$:
\begin{equation}\label{devcourburesectionnelle}
R_{kalb}(v)=R_{kalb}+\sum_jR_{kalb;j}v^{j}+\sum_{j h}R_{kalb;j h}v^{j}v^{h}+...;
\end{equation}
which permits to get higher order corrections to masses.\\
At first order, the direction of $v$ is $e_8$. But, we can also introduce higher orders or the vector $v$:
\begin{equation}
v=v_0e_8+v_0^{2}\sum_j\alpha^{j}e_j+v_0^{3}\sum_,\beta^{j}e_j+...
\end{equation}
which implies a deformation of the mass operator:
\begin{multline}\label{massedeformation}
(T_R)_{ab}=\frac{C}{i}v_0^{2}(R_{8a8b}+v_0(\sum_k(R_{8akb}+R_{ka8b})\alpha^{k}+R_{8a8b;8})\\
+v_0^{2}(\sum_{j h}R_{8a8b;j h}\alpha^{j}\alpha^{h}+\sum_j(R_{8ajb}+R_{ja8b})\beta^{j}+R_{8a8b;88}+...)
\end{multline}
Note this choice of higher jets is compatible with general covariance, it makes appeal to diffeomorphisms
constrained by fixing the metric and $u_0$ at the first order, an even corresponding to the gauge
group $SU_3$ under our identification of $F\otimes \mathbb{C}$ and $M_8^{+}$. We enter in a kind of
non-linear gauge theory.\\

We have $R_{8a88}=0$ (resp. $R_{888b}=0$), then in first approximation the mass of the neutrino is zero, which
is nice, but not fully exact, because for $\nu_e$ it is near $2.5 eV=2.5\times 10^{-6} MeV$, for $\nu_\mu$
it is $170 keV= 0.17 MeV$ and for $\nu_\tau$ or the order of $18 MeV$. (The rule of
$200$ is not verified, but not too far.) The higher orders corrections have to be taken in account. This gives three
other eigen-vectors in the space $M_8$ for the smallest eigen-values, which do not necessarily coincide with
the theoretical direction $e_8$.\\
The smallest non-zero eigenvalue of $T(R)$ is orthogonal to $e_8$, it is always possible to take it
for $e_7$, which restricts the choice of the complex structure $J$ on $F$. As before, higher orders corrections
\emph{a priori} give deformed eigenvectors for this mass.\\
In the same manner, the other basis vectors can be chosen as orthogonal eigenvectors for larger eigenvalues, in the order
$e_6,e_4,e_2,e_5,e_3,e_1$. For a generic metric there is no reason they have multiplicities. However, apparently,
experiment gives two eigenvalues with multiplicity three, one for magnetic quarks, one for electric quarks. It could
be that experiment attains only a mean eigenvalue, but more probably, we can suppose that the constraint given by the
Einstein and Dirac equations on the transversal Ricci curvature $R_{ij}$ implies that the $SU_3$ symmetry of fermions,
that come from the gauge fixing of $t$ and $H$ to their mean values, influences the metric in the same way, converging
to masses of multiplicity three.\\

\indent \emph{A priori} the basis of masses eigenvectors could mix all kinds of fermions; however, the masses of quarks
being significatively larger than the masses of leptons, we can assume an absence of mixing between both kinds; moreover, the
direction of electron, muon an tau is experimentally the same for flavors and masses, cf. \cite{bernstein2013charged}.
Then from orthogonality we deduce the existence of two separated oscillations, one for neutrinos, the other one for
quarks (with $SU_3$ symetry). They are respectively described by the $3\times 3$ matrices $PMNS$ (Pontecorvo, Maki, Nakagawa and Sakata) and $CKM$
(Cabibbo, Kobayashi and Maskawa). To describe them in our context, we use the theory of operators perturbation (cf. \cite{kato2012perturbation}),
the small parameter being $v_0\approx 0.02$. The experimental results accord qualitatively with the order of the values we find
(see $UDPE$). In particular, the fact that $0$ has a degenerate multiplicity for neutrinos implies that the series
which  give eigenvectors and eigenvalues are in $v_0^{1/2}$, predicting stronger oscillations for them than for quarks.
Also the existence of phases, responsible of $CP$ breaking, can be explained by the ambiguity on phases in the correspondance
between $M_8$ and $F\otimes \mathbb{C}$.\\

\subsection{Bosons, structures and masses}

With the exception of the Higgs field, alls the bosonic fields come from the Levi-Civita connection of the
metric in dimension $12$. In the first generation the most important of them for the four dimensional dynamics are the
components of the connection $C_{\mu i}^{j}$ and the monodromy $M_{\mu\nu i}^{j}$, which is the curvature of $C$. They form a vector field
with value in the Lie algebra $gl_8$.\\
\indent The $spin_8$ gauge contains $28$ component vector fields, $20$ of them being
massive, $12$ of these $20$ constitute the orthogonal of $u_4$, and $7$ over $20$ constitute the orthogonal of $u_3$ in $u_4$,
the last component, generated by $Z_4$, being the center of $u_4$.\\
Under the action of $SU_3$ this gives the following decomposition in irreducible factors:
\begin{equation}
spin_8=su_3\oplus 2\mathbb{C}^{3}\oplus \overline{\mathbb{C}^{3}}\oplus 2\mathbb{R}.
\end{equation}
With dominant weights notations, this can be written
\begin{equation}
so_8=D(1,1)_{\mathbb{R}}\oplus 2 D(1,0)_{\mathbb{C}}\oplus  D(0,1)_{\mathbb{C}}\oplus  2D(0,0)_{\mathbb{R}}
\end{equation}
The $12$ form a $D(1,0)\oplus D(0,1)$, the $7$ a $D(0,0)_{\mathbb{R}}\oplus D(1,0)_{\mathbb{C}}$.\\
\indent The orthogonal of $spin_8$ is made by symmetric matrices, of $36$ dimensions, which are decomposed under $SU_3$ as follows
\begin{equation}
\mathfrak{p}_8(\mathbb{R})=su_3\oplus S^{2}\mathbb{C}^{3}\oplus 2\mathbb{C}^{3}\oplus \mathbb{C}\oplus 2\mathbb{R};
\end{equation}
which gives
\begin{equation}
\mathfrak{p}_8(\mathbb{R})=D(1,1)_{\mathbb{R}}\oplus D(2,0)_{\mathbb{C}}\oplus 2 D(1,0)_{\mathbb{C}}\oplus  D(0,0)_{\mathbb{C}}\oplus  2D(0,0)_{\mathbb{R}}
\end{equation}
\indent For comparison, each one of the representations $F$, $S_8^{+}$ and $S_8^{-}$ has the type $D(0,0)_{\mathbb{C}}\oplus D(1,0)_{\mathbb{C}}$.
For the retained fermionic multiplicity, this give the same type; as we have four complex dimensions for each Dirac field in Lorentz dimension $(1,3)$,
this gives $8D(0,0)_{\mathbb{C}}\oplus 8D(1,0)_{\mathbb{C}}$.\\
\indent The algebra $su_2\times u_1$ is made by the complex inert component of $\mathfrak{p}_8$, with real generators $X,Y$ giving
the complex fields $W_{\pm}$, by the real inert component of $\mathfrak{p}_8$ generated by $B_0=Id$, and by the real inert component of
$spin_8$ generated by $Z_1$, from the seven fields of second breaking.\\
\indent We have mentioned the problem with the fields $iB_0+Z_4/2$ and $iB_1+2Z_1$, that can \emph{a priori} act on fermions,
and do not get masses directly from $H$, but apparently are not observed. It is possible they get masses by another way, or decouple,
or appear as auxiliary fields; we do not know.\\

In our theory,  there exist at the beginning, for each generation, $64$ real directions of bosons, which gives
$128$ degrees of freedom when multiplied by two for non-massive particles, and also $64$ complex degrees of freedom for fermions,
because the $16$ dimensions of spinors have to be multiplied by the $4$ complex dimensions of Dirac spinors, two for
a particle and two for an antiparticle, which gives $128$ real degrees of freedom for fermions.\\

In addition to the vector fields there is a scalar four dimensional field $H=M_0v$, given by the displacement field $v$ around
the four dimensional universe, and acquiring its dynamical degrees of freedom through bosonization of
the right neutrino. Its mass comes from the gauge fixing of an odd spinor $t$ at its mean value $t_0$.\\
For $H$, $C_0\overline{t} HH t$, $M_H^{2}=C_0\overline{t_0}t_0$, if $t_0$ is deduced from $v_0$, considered as typical length, by taking exponent
$-3/2$, with  $C_0=1$, this gives $386 GeV$, the experiment gives $M_H=125 GeV$, thus $C_0\approx 1/3$.\\

All the couplings between bosons and $H$ are of the same type in our Lagrangian, then the simplest gess is that
all their masses at the firts generation are of the order of the masses of $W$ and $Z$, around $80 GeV$.\\

For the following generations, we propose a scaling argument, similar to the one made for the fermions: the field
of generation $n$ has norm multiplied by $v_0^{n}$, but bosons are of dimension $L^{-1}$, then the lengths are
multiplied by $v_0^{n}$ in the scaling, and the curvature is multiplied by $v_0^{-2n}$, consequently the part of
the square of mass which is explained by curvature is multiplied by $v_0^{-2n}$, as the other part does'nt change its order. Thus, from a generation to
the following one a factor $v_0^{-1}=53$ is applied to the mass. This would give $5 TeV$ for the next bosons generation, which
will be accessible fairly soon.\\

The known particles represent $20$ per cent of the mass of the observed universe. There are $55=64-9$
massive bosons. The
visible massive bosons plus the heavy fermions represent $9/55$ of the massive particles, that is not
far from $20$ per cent. Thus our suggestion for the \emph{dark matter} is not too absurd.\\

Another open question is the nature of \emph{dark energy}, which represents $70$ per cent of the energy of the accessible
universe, according recent estimations. The argument against the simplest explanation by a non-zero cosmological constant
is the theoretical computation of the energy generated by quantum fluctuation of the vacuum, that gives a gigantesque value,
mainly due to the excess of fermions with respect to bosons. It is known that a perfect super-symmetry, restoring the balance
between fermions and bosons, would treat the problem by compensating the  fluctuations. However the super-symmetry is
broken. In our theory, before symmetry breaking there is the same number of bosons and fermions, $64$ real vectors and
$64$ complex spinors, giving $128$ real dimensions on each side. This counting is modified by symmetry breaking,
distribution of masses and transfer of four degrees of freedom
from a fermion to the Higgs boson, all that attributing more dimensions to bosons than to fermions, however, at the level
of fundamental excitations the equilibrium is achieved. Then a positive cosmological constant could have an
influence on the universe's inflation.\\
Let us mention that models involving supplementary dimensions of the universe, as does our model, were proposed
for explaining the dark energy, cf. \cite{appelquist1983quantum}, \cite{dupays2013can}.\\

\subsection{Photons, electric charges values}

\indent In this section we prove that the values of electric charges have a natural geometric
explanation.\\
\indent Let us begin by giving a more geometric expression of the operator $Q_0$ representing the photon.\\
\indent For every integer $n$, we consider the canonical basis $H_k;k=1,...,n$ of the Euclidian space $\mathbb{R}^{n}$
and we denote by $x_k;k=1,...,n$ the dual basis; the canonical \emph{root system} of the Lie algebra $\mathfrak{su}_n$ is
the set of the $x_j-x_k;j\neq k$; its canonical simple roots are given by $x_{k+1}-x_k;k=1,...,n-1$.\\
The lattice $L$ generated by the $H_k$ in $\mathbb{R}^{n}$ is the fundamental group of a maximal torus of $U_n$ (also
named the nodal group of $U_n$, cf. \cite{bourbaki1981groupes456}, chapter $VI$); to get the sub-torus in $\mathfrak{su}_n$ we impose that the
sum of coordinates $x_k$ is zero. We note $y_n=x_1+x_2+...+x_n$ this sum and $Y_n=H_1+H_2+...+H_n$ the corresponding vector.\\
The dual lattice $P_n^{\vee}$ of the lattice $Q_n$ generated by the roots contains $L$, it is generated by the elements
$H'_k=H_k-Y_n/n$. The center of $U_n$ is the group of matrices $zId$ where $z$ is any complex number of modulus $1$,
the center $C_n$ of $SU_n$ is the cyclic subgroup of $n$-th roots of the unity; any element $\exp(2i\pi H'_k)$ is a generator of
$C_n$. This gives an isomorphism between $C_n$ and $P_n^{\vee}/L$.\\
\indent On the complex vector space $M_8^{+}$, the generator $H_7$ in $u_8$ acts by $Z'_1/i$, giving
$1/2$ on $\overline{Q"}$ and $\Omega$, $-1/2$ on $Q'$ and $1$; the operator $B_0$ is the identity and $B_1$
is the projector onto the leptons space $L_2$. \\
\indent In the group $U_8$ acting on $M_8^{+}$, we consider the two subgroups $U_6$ and $U_2$ of isometries of quarks and leptons
respectively; the corresponding elements $Y_n$ are $Y_6=H_1+H_2+...+H_6$ and $Y=H_7+H_8$ respectively.\\
It is natural to identify $Y_8$
with $B_0$, then $Y_6$ is $Id$ on $Q^{+}$ and is $0$ on $L$, while $Y$ is $0$ on $Q^{+}$ and is $Id$ on $L_2$.\\
We have $Q_0/i=(B_0-4B_1)/6-H_7$ and $Z_0/i=(B_0-4B_1)/6+H_7$; then replacing $B_0$ by $Y_8=Y_6+Y$ and $B_1$ by $Y$,
we obtain on $S_8^{+}$,
\begin{align}
\frac{Q_0}{i}&=\frac{Y_6}{6}-\frac{Y}{2}-H_7,\\
\frac{Z_0}{i}&=\frac{Y_6}{6}-\frac{Y}{2}+H_7.
\end{align}\\
A remarkable thing in these formulas is that the three present operators are the highest weights of the
fundamental representations of $U_6$, $U_2$ and $U_1$ respectively, the last $U_1$ acting by
$\sqrt{z}$ on $\Omega$ and $\overline{Q"}$, as $1/\sqrt{z}$ on $1$ and $Q'$.\\
\indent The circular group generated by $2\pi Q_0$ is not contained in $U_8$; only a $6$-fold covering
of it is contained in $U_8$. The natural group where $i(Y_6/6-Y/2)=iB'$ can be integrated is the product
$PU_6\times PU_2$ of $PU_6=U_6/C_6$ and $PU_2=U_2/C_2=SO_3\times U_1$, whose fundamental group is
$(\mathbb{Z}/12\mathbb{Z})\times \mathbb{Z}$.\\

\noindent Due to the ambiguity on the phases of the isomorphisms between $M_8$ and $F\otimes \mathbb{C}$, this quotient group
$PU_6\times PU_2$ is also the only possible one for containing a principal operator of masses.\\

\indent Now we establish that this formula is minimal for a gauge generator $Q$ which survives after the gauge fixing
giving $SU_3$ and a broken $SU_2\times U_1$, as soon as we assume that $(i)$ $W_+$ and $W_-$ have well defined charges, and $(ii)$
$Q$ generates a well defined subgroup in a quotient of $U_6\times U_2$. Two somewhat natural conditions for a quantum theory.\\
\indent Remember that the gauge fixing giving does'nt determine $Q_0$ without knowledge of the charges values for $Q_0$; it only
impose that $Q_0$ is a linear combination $Q=i\beta_0B_0+i\beta_1B_1+2\alpha_1Z_1+\frac{1}{2}\alpha_4 Z_4$ de $iB_0$, $iB_1$, $Z_1$ et $Z_4$,
verifying the zero mass condition $\beta_0+\beta_1=\alpha_1+\alpha_4$.\\
As we saw another natural basis of the space generated by $iB_0$, $iB_1$, $Z_1$, $Z_4$ is $Y=B_1$, $Y_6=B_0-B_1$, $Z_4/2$, $Z_6=2Z_1-Z_4/2$.
Let us write
\begin{equation}
Q=iz\frac{Y_6}{6}+tZ_6+iy\frac{Y}{2}+x\frac{Z_4}{2},
\end{equation}
the zero mass condition becomes
\begin{equation}
y=2x.
\end{equation}
This condition tells that the neutrino has charge zero. The coefficient $y$
is the charge of the electron. Note that we do not suppose it is equal to $-1$.\\
Admitting $(i)$, it follows that the difference between the common charge
of the $q'$ and the common charge of the $\overline{q"}$ must be $y$, then we get $t=-y/2$.\\
Only two arbitrary constants remind
\begin{equation}
Q=iz\frac{Y_6}{6}-y\frac{Z_6}{2}+iy\frac{Y}{2}+y\frac{Z_4}{4}.
\end{equation}
The hypothesis $(ii)$ is
equivalent to the requirement that $Q$ belongs to the weights lattice of this group, i.e. that $z$ and $y$ are rational integers.
The minimal solution is $z=\pm1$, $y=\mp1$.\\
Up to the sign this gives two possible solutions: $z=1$,$y=-1$ and $z=-1$, $y=-1$.\\
The second one corresponds to the exchange of $Q'$ and $\overline{Q"}$ and $W_{+}$ with $W_-$, which would give
in place of $Q_0$ the operator $Q=Q'_0$ such that
\begin{equation}
\frac{Q'_0}{i}=\frac{Y_6}{6}+\frac{Z_6}{2}-\frac{Y}{2}-\frac{Z_4}{4}=\frac{Y_6}{6}-\frac{Y}{2}-\frac{Z_4}{2}+Z_1.
\end{equation}
Then it is a matter of convention to chose between $Q_0$ and $Q'_0$, they are not truly different solutions.\\

Let us mention another interesting formula for $Q_0$, showing its relation to ambiguities of phases that are
taken in account by the electromagnetic gauge group.\\
\indent The phase parameters $\lambda$ and $\beta$ conditioned the choice of a triality. In particular $\beta$ determines
the real structure of $S_8^{+}$ and $S_8^{-}$, then the phase of $\Omega$; it corresponds to a generator $Z_\beta$ which acts by
$-1$ on $\Omega$ (or $\varepsilon"$), $-1$ on the quarks $q'$, and $0$ on $1$ (or $\nu"$) and
on the quarks $\overline{q"}$. There is in addition a graduation operator on $\Lambda^{*}(F\otimes \mathbb{C})$; its
generator $N'$ gives $0$ on $\nu"$, $-1$ on vectors $F"$, $+1$ on $F'$, etcetera. This gives
$-4$ for the electron, $-1$ for $\overline{q"}$ et $+1$ for the $q'$.\\
We have:
\begin{equation}\label{chargetiers}
Q_0=\frac{1}{3}N'-\frac{1}{3}Z_\beta
\end{equation}\\

\section{Discussion}

\subsection{Beyond the Standard Model}

The results we just exposed are mostly in agreement with the Standard Model of Glashow, Weinberg et Salam,
however they depart on some points, as the fermionic origin of symmetry breaking, the nature of the weak $SU_2$
and the complexity of the Higgs field, and they add possible explanations of the constituents of matter, the size orders of masses
and the flavors oscillations. Thus we can deduce from our approach some predictions to go beyond the Standard Model:\\
\noindent 1) We must expect new generations of fermions, with identical structure, two leptons and two quarks families.
The rule of geometric ratio $200$ predicts approximatively their masses: the next analog of the electron would weight $350 GeV$,
the next analog of the quark up $34 TeV$, of the quark down $8 TeV$. We suspect that this automatic prediction is not very original.\\
\noindent 2) At least a second generation of $W$ and $Z$ has to be awaited, with expected masses around $5$ or $6 TeV$.\\
\noindent 3) Many other massive bosonic particles do exist without interaction with the fermions, and constitute the
major part of dark matter; for the first generation they correspond to generators of $gl_8$, some antisymmetric, orthogonal to $u_3$
inside $so_8$, some symmetric orthogonal to $sl_2$ inside $\mathfrak{p}_8$. Their masses would be of order $100 GeV$.
A second generation will have masses of order $6 TeV$.\\
\noindent 4) Taking into account these bosons, or better, coming back to the pure gravitation coupled to one Dirac fermion
in dimension twelve, the dark energy could be explained by a positive cosmological constant (cf. \cite{bousso2008cosmological}, \cite{dupays2013can}).\\
\noindent 5) The suggested nature of the birth of the Higgs boson, i.e. gauge fixing of a fermion in the new invisible sector
given by odd eight dimensional spinors,
dynamical bosonization of the right chirality of the neutrino, and finally, identification of the resulting boson
to a displacement field in dimension twelve, should possess an identifiable experimental signature.\\
\noindent 6) In the same manner, the mechanism that is responsible of light propagation, i.e. the photon $Q_0$, appearing as a
combination of two traces in $gl_8$ and two roots of $u_4$ and $u_2$, should be experimentally identifiable,
perhaps under the form of two particles $iB_0$ and $Z_4$, to the side of the known $Z_0$. Those particles would be
neutral and massive, their masses coming from another origin than the Higgs boson. The charge of each fermion under $B_0$
should be $1$, and $Z_4$ should act as $Z_0$ on the leptons, but should be uncoupled with the quarks.\\

\subsection{Quantum Theory}

The context for interpreting the above structure is a four dimensional Quantum Field Theory. The Lagrangian of this theory is obtained by
developing the sum of Einstein and Dirac Lagrangian density in twelve dimensions, transversally to a Lorentz four dimensional
sub-manifold, then it contains \emph{a priori} an infinite number of fields, however after at most ten orders of generations,
the gravitational effects become so strong that $QFT$ could not be the right description. With respect to the four dimensional theory, the twelve
dimensional theory is a kind of generatrix function.\\
\indent In $QFT$, the states of independent free elementary particles are elements of Hilbert spaces equipped with irreducible representations
of the Poincaré group; these states are solutions of special partial differential equations on vector bundles over the Minkowski space of signature $(1,3)$,
like Dirac, Maxwell or Klein-Gordon equations. The states for several particles of the same type belong to the derived Fock spaces,
infinite symmetric for bosons and infinite antisymmetric for fermions. On these Fock spaces there exist natural operations
given by creations and annihilations operators. Cf. \cite{bogoliubov1960introduction}, \cite{itzykson1980quantum}, \cite{peskin1995introduction},
\cite{weinberg1995quantum}. The evolution of the fields in interactions is described by ordinary differential
equations in this infinite dimensional setting; the coefficients of the equations are polynomials of creations and
annihilations operators which correspond to the terms of the classical energy-impulsion tensor \cite{tomonaga1946relativistically},
\cite{koba1950note}. Practically,
these equations must be considered from a perturbative point of view, by computing series of integrals associated to Feyman graphs,
\cite{feynman1962quantumelectrodymaics}, \cite{thooftveltman1973}.
However these integrals are generically divergent and a process of renormalization is necessary to produce effective amplitudes
for the solutions, cf. \cite{t1994under}, \cite{polyakov1987gauge}, \cite{delamotte2012introduction}. Renormalization theory, which appeared in the forties, under the hands of Feynman, Tomonaga, Schwinger, Dyson,
evolved during the second half of the last century, in particular under the influence of Kadanoff and Wilson, connecting
to the study of phase transitions in Statistical Physics, until it was possible to compute radiative corrections and to prove the
renormalizability of the gauge theories that describe known particles in the Standard Model, in particular by 't Hooft, Veltman, Lee, Zinn Justin,
Becchi-Rouet-Stora-Tyutin, Kluberg-Stern and Zuber, Anselmi, cf. \cite{itzykson1980quantum}, \cite{peskin1995introduction}, \cite{t1994under},
\cite{weinberg1996quantum}, \cite{anselmi2014adler},\cite{anselmi2015adler}.
A concept or
effective renormalization, which permits to adapt the quantum corrections to the scales of energy, even in non-renormalizable theories
for the initial definition of Dyson, was elaborated; cf. Anselmi \cite{anselmi1994removal},\cite{anselmi2015renormalization},
Gomis and Weinberg \cite{gomis1996nonrenormalizable}. A crucial role in these results is played by
geometrical covariance, as it is apparent in the proof of annulation of anomalies, which not only save the symmetries at the quantum
level but also permit the compensations of infinite quantities.\\

From the mathematical point of view many progresses were made recently for a better understanding of the renormalization process,
in particular by
Connes and Kreimer \cite{conneskreimer1999}, cf. Connes and Marcolli, \cite{connes2008noncommutative}.\\
\indent To take in account the gauge invariance or general covariance, the best approach, from the point of views of concepts as from the point of
views of computations, is to consider that quantum states and amplitudes in such a theory are of co-homological nature;
the reason being that quantum states are invariants of constrained objects, that is a known source of co-homological constructions,
and that obstructions to form invariant quantities, like anomalies, also rely to homological algebra.
This idea was revealed by Becchi-Rouet-Stora-Tyutin in $1975$, cf. \cite{becchi1976renormalization}, and immediately followed by the first concrete calculations in this
context by Kluberg-Stern and Zuber and Zinn-Justin, cf. \cite{kluberg1975renormalization}, \cite{kluberg1975bisrenormalization},\cite{zinn1975renormalization}.
This theory is named today $BRST$ co-homology, and the generalization which
is more and more used is the Batalin-Vilkovyski co-homology, $BV$ theory
cf. \cite{weinberg1996quantum}, \cite{dewitt2003global}, \cite{cataneo1995}, \cite{roger2009gerstenhaber}.
Important reports and new works on $BRST$, $BV$ were made by Henneaux, 1985, 1990,
\cite{henneaux1985hamiltonian}, \cite{henneaux1990lectures}, Barnich, Brandt and Henneaux 1995 \cite{barnich1995local}, Henneaux
and Teitelboim \cite{henneauxteitelboim1992quantizationgauge}, the
scope from algebraic topology is well shown in
J.Stasheff, \cite{stasheff1997homological}, \cite{stasheff2005poisson}.\\
all that leads to a beautiful and rigourous theory; let us cite Kevin Costello \cite{costello2011renormalization},
Kasia Rejzner thesis \cite{rejzner2011batalin}, and  R.Brunetti, K. Rejzner and K.Fredenhagen, \cite{brunetti2013quantum}. For the full Standard Model, coupled to the Einstein Gravitation,
D. Anselmi recently succeeded to construct a finite effective theory at all orders in $\hbar$, for every energy scale  \cite{anselmi2014adler},\cite{anselmi2015adler}.\\

The companion text in French, referred as $UDPE$,  contains a summary of the renormalization method, following the $BV$ approach,
without any original contributions, the aim being only to indicate the method that can be suited for computing quantum corrections in our model.
The hope is that the achievements made for the Standard Model, for instance in the deep recent works of Anselmi, can be realized too
with the changes we have proposed, to go  beyond the Standard Model.\\

\subsection{Weak supersymmetry and gravitation}

The key of our approach is the definition of spinor's multiplicity by space geometry.
The generating theory for that is Einstein plus Dirac theory in twelve dimensions. The Lagrangian
of Dirac appears naturally in super-symmetry and super-gravity.
Scherk and Schwartz found that in string or super-string theory, Einstein equation is a first order condition
of renormalization cf. \cite{green1987superstring}, \cite{deligne1999quantum}. Could it be that the twelve
dimensional theory appears as a certain limit of a super-string theory?\\
\indent As we have shown, a kind of symmetry between bosons and fermions results in our model from a special choice of coordinates,
compatible with general covariance; we got the same number $128$ of degrees of freedom for bosons and
fermions before spontaneous symmetry breaking. Another property, which is in general a signature
of extended super-symmetry, is the close relation between electric charges and masses, here through
the group $PU_6\times PU_2$.\\
\indent Many aspects we met here are also present in eleven dimensional super-gravity: an effective local
$U_8$, an unbroken $SU_3\times U_1$,
as observed by B.DeWit, H.Nicolai and N.P.Warner in the eighties, \cite{nicolai19853}, \cite{de1986d}. Recent developments
ar even closer,
cf. \cite{meissner2015standard}, \cite{kleinschmidt2015}.
However, the symmetry $SU_2$ does'nt appear naturally in these developments, and $SU_3\times U_1$ is only one possible sector. \\
A common point of any super-symmetric theory and our theory is the fundamental role of the application $\Gamma$ from the tensor product of spinor spaces
$S\otimes S$ to the Lorentz space $V$, which is the origin of Haag-Lopuszanski-Sohnius discovery of super-symmetry, cf. \cite{wess1992supersymmetry},
\cite{weinberg2000quantum}, \cite{deligne1999quantum}. \\
However, ordinary super-symmetry uses the positive real version of $\Gamma$, corresponding to
Majorana spinors, and we use only Dirac spinors.\\
The main difference is not in a necessary use of anti-commuting variables in Susy, cf. a recent paper
with Michel Egeileh (\cite{egeileh2016}), but in the nature of super-multiplets, we have no peticle of spin $3/2$ and have an
excess of spin $1$ with respect to expected spin $0$.\\

The more general problem, to deduce Einstein theory or a modification of it from matter equations, retained the
attention of many researchers in the past (De Broglie, Heisenberg, Sakharov, ..., cf. \cite{wetterich2004gravity}). We can consider that the Twistor
Theory of R.Penrose is an attempt in this direction, \cite{penrose1984spinors}, \cite{penrose1988spinors}. Another approach was developed by Amati et Veneziano \cite{amati1981dynamical},\cite{amati1982unified},
who have shown that the universe's metric can be generated by qunatum fluctuations of fermions interacting with
an assembly of auxiliary bosons. More recently Wetterich \cite{wetterich2004gravity}, constructed a close theory,
where the metric is a composite object, and where Lorentz invariance is only global. In another circle of ideas,
the theory proposed by Chamseddine and Connes, \cite{chamseddine1997spectral}, \cite{connes2008noncommutative}, is based on
a principe of spectral action which concerns the Dirac operator of a Riemannian manifold, and deduces the Lagrangians of Einstein and Weyl,
plus a series of other terms derived from the curvature. This theory was adapted by the authors to contain the Standard Model;
a point in common with the theory we present is to consider a sequence of generations of fermions. However, for Chamseddine and Connes,
the fermionic spectrum, corresponding to a non-commutative part of space-time, is posed as an \emph{a priori} data; it is
not deduced from a more generic situation, as it is in our theory, by spontaneous symmetry breaking, which
identifies the multiplicities of the spinors with ordinary supplementary dimensions, and the masses of spinors with
ordinary Riemannian curvatures.\\
\indent That such a structure, two leptons, two families of three quarks, which is reproducing itself several times,
is generic, if we start with a Dirac field propagating in eight transversal dimensions, is due to the
miracle of triality. The non-abelian gauge theory in dimension $4$ comes as a direct consequence
of the Einstein equations in dimension $12$. This evokes the Kaluza-Klein hypothesis
(cf. \cite{coquereaux1983geometry}, \cite{bailin1987kaluza}); without contest the idea of extending the universe with supplementary dimensions comes from them,
and this idea gave wonderful theories as superstrings of  Neveu, Schwartz (see \cite{green1987superstring}, \cite{deligne1999quantum}) and supergravities of Cremmer, Julia, Scherk \cite{cremmer1978supergravity}, or McDuff,
or Horava, Witten and A.Sen, \cite{witten1995stringdimensions}, \cite{sen1998nonperturbativestring}. Moreover, most of those theories accord a major role to spinors. However an important difference with
our proposition is that we do not assume any \emph{a priori} symmetry hypothesis. We start and stay in  generic universe, in the sense
of René Thom, and this does'nt forbid the appearance of a rigid algebraic structure for fermions and of a precise balance between bosons and fermions.\\
\indent The reduction of fermions from $12$ to $4$ dimensions of the universe relies on the modulo $8$ periodicity, discovered by R. Bott for the stable homotopy of
orthogonal groups, in close relation with the periodicity of Clifford algebras and spinors (cf. \cite{atiyah1964clifford}).
This periodicity is an important face of the topological theory of differentiable manifolds of all dimensions, localized at the odd prime
numbers, as shown by D. Sullivan (\cite{sullivan2005geometric}), who suggested more recently that Quantum Fields and strings should give an access to
a direct definition of this type of localization of manifolds. It is tempting to go to explore beyond twelve, from eight to eight,
what could reveal a universe possessing an infinite number of dimensions. \\

\section{Aknowledgments}

The author want to thank Jean-Bernard Zuber for his interest, his questions and pertinent
and benevolent suggestions. Thanks equally to Bernard Julia for his very helpful remarks.
The support of Michèle Bompard-Porte, Alain Chenciner and
Sylvie Paycha was extremely precious all along this research, warm thanks to them.

The author is conscious of many flaws of the work he is presenting now. The main part was done exactly one year ago,
and the rest was almost completed since six months, thus he decided to publish it, hoping this work, as it is, could
inspire other researchers for a better understanding
of the domain of particles.

\bibliographystyle{abbrv}
\bibliography{biblio_unfolded}
\nocite{*}

\end{document}


\author{Daniel Bennequin \thanks{Institut de Math\'ematiques de Jussieu,
Universit\'e Denis Diderot Paris VII, France, bennequin@math.univ-paris-diderot.fr}}

\title{L'univers déplié des particules élémentaires}
\date{}
\maketitle \tableofcontents

\section{Résumé}

Ce travail présente une théorie géométrique de l'univers physique qui part de principes simples et prédit
la structure complète des particules élémentaires aujourd'hui connues, plus quelques
nouveautés qui pourraient être observés dans le futur. En ce sens la théorie présentée va au-delà du modèle standard,
mais en plus, certains aspects du modèle standard y sont modifiés. Le cadre adopté est celui de la théorie quantique
des champs, où une forme de renormamisation doit être appliquée à des interactions entre particules associées à des
représentations unitaires irréductibles du groupe de Poincaré. Mais l'accent, dans cette première partie du travail,
est mis sur la façon d'engendrer les particules à partir de principes géométriques. La dynamique de la théorie
engendrée sera une autre histoire. Le point de départ
pour cela est la covariance générale des équations d'Einstein couplées avec les équations de Dirac; ces équations
sont prises en dimension Lorentzienne douze, mais elles sont vues comme génératrices d'équations en dimension
Lorentzienne quatre, où les effets gravitationnels sont faibles aux échelles d'énergie considérées. L'idée
est d'appliquer un mécanisme de Higgs à une partie des spineurs, tout en récupérant
effectivement un champ de Higgs scalaire, capable de donner des masses aux bosons comme aux fermions.\\
\indent Partant de là, nous retrouvons en particulier que les seuls bosons sans masse sont ceux d'une théorie de jauge non-brisée de type
$SU_3\times U_1$, et que tous les fermions appartiennent à des familles comportant deuxs leptons (et leurs anti-particules) où $SU_3$ n'agit pas,
un lepton massif de charge électrique $\pm 1$, comme
l'électron et le positron,
et un lepton neutre possédant également éventuellement une masse, comme le neutrino et l'antineutrino, mais ne possédant qu'une
seule chiralité (la gauche), plus
des quarks massifs,
venant en deux familles de trois couleurs qui répondent à la même représentation standard de $SU_3$, la première famille ayant une charge électrique
$-1/3$ et la seconde
une charge électrique $2/3$. Il peut apparaître ainsi plus de six générations de fermions avant que les effets gravitionnels
forts ne les absorbent. La théorie prédit également l'existence de trois bosons
vecteurs $W_+,W_-, Z_0$ possédant avec le photon
la structure connue $SU_2\times U_1$ de la théorie standard de Weinberg et Salam. Eux aussi se présenteront en plusieurs générations.
Leur spécificité est que $SU_3$ les laisse invariants.\\
\emph{A priori}, à côté de ceux-ci, il devrait apparaître
plusieurs autres familles de bosons vecteurs massifs, mais nous démontrons que celles-ci n'interagissent pas avec
les fermions, contrairement aux générateurs de $SU_3\times SU_2\times U_1$. Ce qui expliquerait qu'ils ne soient pas
encore observés; ils sont des candidats pour la matière noire.\\
Sur ces $52$ bosons, deux sont des traces et sont neutres pour $SU_3$ comme pour $U_1$; les $50$ bosons restant appartiennent
à des représentations non-triviales de $SU_3$: on trouve cinq fois la représentation fondamentale
$D(0,1)=\mathbb{C}^{3}$, une fois la représentation symétrique $D(0,2)=S^{2}\mathbb{C}^{3}$ et une fois la représentation
adjointe $D(1,1)_{\mathbb{R}}=su_3$. Certaines particules associées auraient également des charges pour $U_1$; celles-ci pouvant
être égales, comme pour les fermions, à $0,\pm 1,\pm1/3, \pm 2/3$.\\
Toutes satisfont à des équations du type de Higgs vectoriel, dites Higgs-Hitchin, et se présentent en plusieurs générations.\\
\indent Une matrice PMNS (Pontecorvo-Maki-Nakagawa-Sakata) et une matrice CKM
(Cabbibo-Kobayashi-Maskawa)
apparaissent naturellement, qui expli-quent la différence entre les états propres des masses et les saveurs pures
pour les neutrinos et les quarks.\\
\indent La théorie inclut une explication d'une partie des effets du groupe de renormalisation, opérant sur les particules et leurs interactions,
en prenant en compte l'apport des générations nouvelles de fermions, toujours accompagnées par de nouveaux champs bosoniques, dont
la structure est précisément décrite à l'avance mais pas les valeurs numériques. Ce que nous présentons est donc une
suite de théories effectives au sens de Wilson et Weinberg; la  théorie à très haute énergie reposant sur
une forme de gravitation quantique, comme celle qui est décrite dans la première partie du texte.\\
\indent Du point de vue du groupe de renormalisation lui-même, traitant en particulier les corrections radiatives,
la chose la plus importante est l'existence d'un nouveau
groupe d'invariance, issu du modèle géométrique de départ: il s'agit du groupe des germes de difféomorphismes de
l'espace-temps de dimension douze le long de notre espace-temps de dimension quatre, étendu par le groupe des
automorphismes du fibré des spineurs; ce groupe contient les groupes
de jauge usuels mais il en offre une version non-linéaire étendue. En suivant l'approche
$BRST$ et $BV$ pour construire le Lagrangien des interactions et les états quantiques, guidée par ce groupe, nous pouvons
espérer définir à tous les ordres d'énergie, en-dessous de celle de Planck, des théories effectives
sans termes infinis. Cf. \cite{gomis1996nonrenormalizable}, \cite{anselmi1994removal}, \cite{anselmi1995more}, \cite{anselmi2015adler}.\\
\indent Chaque ordre correspond à un facteur d'environ $200$ pour les
masses des fermions. Nous prédisons qu'il existera également plusieurs générations de bosons, avec un facteur
de $50$ environ; ce qui donnerait des nouvaux $W,Z$ de l'ordre de $5 TeV$.\\
Un des intérêts principaux de notre approche est qu'elle rend compte naturellement de tous ces ordres de grandeur:
du fait que les fermions aient des masses beaucoup plus petites que celles des bosons, du fait qu'elles possèdent
une sorte de croissance géométrique, et du fait que les états propres des masses ne sont pas forcément égaux
aux saveurs, dans les proportions obtenues expérimentalement.\\

\indent La théorie que nous présentons est certainement non-renormalisable au sens de Dyson et probablement
renormalisable au sens effectif de Wilson et Weinberg (cf. dernière section), et nous comptons bien développer
le calcul perturbatif plus loin dans une suite à cette recherche. Cependant, pour rejoindre les propriétés des particules
existantes, il est certain que, comme pour le modèle standard lui-même, une approche non-perturbative serait nécessaire,
et extrêmement difficile d'un point de vue mathématique, avec les méthode connues aujourd'hui, cf. \cite{jaffe2006quantum}.
Par exemple, les ordres de grandeur des masses qui apparaissent dans ce travail, ne sont qu'une base de départ, elles
ne tiennent pas compte de la dynamique spontanée de la théorie. Il est bien établi aujourd'hui, par l'expérience
et par le calcul sur machine des théories de jauges sur réseau, que les masses des hadrons, ou les corrections des
masses des leptons et des quarks, dépendent des effets non-perturbatifs. Par exemple la plus grande partie de
la masse du proton et de celle du neutron n'est pas due aux masses des quarks, elle provient d'une configuration compliquée de gluons et de
quarks, or l'essentiel des masses autour de nous viennent de là. En particulier nous ne disons rien d'interessant sur
le mass gap ou le confinement des gluons et des quarks, qui relèvent de la dynamique non-perturnative.\\
Rien, dans notre approche, pour le moment, ne permet de dire que les trois facteurs du groupe de jauge ont les ordres de
grandeur de couplages que l'on connait: $1$ pour $su_3$, $1/137$ pour $U_1$ et $10^{-6}$ pour $SU_2$. En particulier
les propriétés originales du couplage fort, qui pourraient résulter de dualités de champs, échappent à la description
structurelle que nous avons développée.\\
\indent Pourtant, un des aspects du confinement semble apparaître dans la structure des charges et des masses des quarks.
En effet, le groupe $U_1$ qui explique l'électromagnétisme à notre échelle, doit agir sur les
phases des particules observables, pas seulement à traver son algèbre de Lie, mais en tant que groupe. Or notre approche
prédit (ou plutôt préfère) un unique groupe compatible avec l'électromagnétisme de Maxwell pour les fermions, et ce groupe
attribue (en accord avec l'expérience) des charges
fractionnaires aux quarks; par conséquent, il est nécessaire de passer au revêtement (unique) à trois feuillets de $U_1$
pour avoir un groupe de jauge des quarks, ce qui fait une obstruction d'ordre
trois à l'observabilité à notre échelle. Cette obstruction est levée pour les mélanges de quarks dans certaines particules
(produits tensoriels) relativement stables toutes de charges entières, comme le proton et le neutron. Notons que
la théorie $SU_3$ étant asymptotiquement libre, cela n'interdit pas de mettre en évidence des
quarks semi-classiques aux très hautes énergies, associés à l'algèbre de Lie plus qu'au groupe.
La relation entre ce fait électromagnétique et le confinement dynamique provenant du groupe de jauge $SU_3$ de
la chromodynamique n'est pas claire du tout.\\

\indent L'hypothèse de départ est que l'univers $X$, au moins en première approximation, possède douze dimensions,
quatre que l'on expérimente tous les jours,
avec une métrique Lorentzienne variable sur une variété $W$ et huit dimensions transverses, qu'on analyse ordre par ordre
en introduisant un fibré vectoriel $E_F$ de base $W$, orthogonal à $T(W)$ dans $T(X)$. Le choix du fibré est libre, la théorie
est co-variante de bout en bout. Dans ce texte nous nous contentons
d'étudier en détail les trois premiers ordres
transverses pour la métrique d'univers le long de $W$, ce qui impose une structure à tous les ordres suivant. De cette
manière, toutes les particules connues traduisent la géométrie des huit dimensions transverses à l'ordre deux, et leur dynamique
provient d'équations aux dérivées partielles en dimension quatre.\\
La métrique transverse le long de $W$ peut être fixée
en utilisant la liberté de reparamétrisation; elle donne une métrique définie positive de fibré vectoriel sur $E_F$.
Pour décrire l'ensemble du champ de gravitation en dimension douze, nous introduisons la distribution $M$ des $4$-plans
perpendiculaires aux fibres de $E_F$; ceci fait apparaître une division en six assemblées de champs, le champ de gravitation
horizontal $\Gamma_H$, suivant $M$, le champ de gravitation vertical $\Gamma_F$, suivant les fibres, leurs champs de
deformation respectifs $D_F$ et $D_H$ (verticale pour $\Gamma_H$ et horizontale pour $\Gamma_F$), la courbure $M_H$ de $M$
et son champ de déformation vertical $C$, qui apparaît au premier ordre comme une connexion linéaire.\\
Mais en plus l'espace-temps n'est pas supposé être vide, il y a un unique champ $\psi_X$ de spineurs de Dirac sans masse en dimension douze.
Pour ce champ, l'équation de Dirac en dimension douze est retenue. \\
\indent Nous faisons l'hypothèse que les équations couplées d'Einstein
et de Dirac en dimension douze fournissent une série
génératrice des équations des particules découvertes à partir de la dimensions quatre.
La Physique vue de notre point de vue à quatre dimensions résulte alors de l'exploration de champs effectifs, ordre par
ordre suivant les sections arbitraires $v$ du fibré $E_F$. \\
\indent Une génération de fermions est un ensemble de champs de
spineurs de Dirac en dimension quatre à valeurs dans les spineurs Euclidiens en dimension huit.
Un seul spineur en dimension douze
comporte $64$ degrés de libertés complexes. On sait que c'est le double de ce qui est autorisé par l'expérience, mais, comme on va le voir,
le point essentiel du modèle proposé est de faire disparaître $32$ des degrés de libertés fermioniques.\\
\indent A l'ordre zéro en $v$, il n'y a qu'une seule génération de fermions et on peut faire agir les difféomorphismes en
dimension douze pour qu'il n'apparaîsse que quatre familles de champs bosoniques:
1) le champ de gravitation $\Gamma_W$ à quatre dimensions; 2) la dérivée verticale $d_Fg$ de la métrique sur $M$, qui est le $D_F$ d'odre zéro;
3) une connexion linéaire $\nabla_F$ sur $E_F$; 4) une famille de
champs de Higgs vetoriels $\Phi_{F}$ associés à $E_F$.
Les équations d'Einstein en dimension douze induisent sur ces champs
les équations d'Einstein sur la métrique $\eta_W$ en dimension quatre, les équations de Yang-Mills avec
des courants associés à $d_Fg$ pour les connexions linéaires, et des équation de Higgs avec des potentiels
jusqu'au degré quatre pour $\Phi_{F}$.\\
Les ordres de grandeur respectifs mesurés des intéractions du type jauge et de la gravitation en dimension quatre font
que pendant au moins six générations, probablement plus, la courbure de notre espace-temps n'intervient pas. Par contre
la courbure transversale intervient dès l'ordre deux et va avoir un rôle important.\\
Remarque: comme nous avons surtout en vue les effets quantiques, aucune des équations n'est vraiment imposée aux champs.
Ces équations apparaissent comme dans les équations de Dyson-Schwinger; elles servent à décrire les degrés
de liberté quantique des champs, c'est-à-dire à quelles représentations unitaires
irréductibles d'énergie positive du groupe de Poincaré ils appartiennent lorqu'on réussit à les observer.\\

La structure algébrique qui semble expliquer les champs physiques en dimension $4$ est l'identification des spineurs et des vecteurs en dimension huit,
nommée \emph{trialité} et découvert par Cayley au XIXème siècle (cf. les livres de Elie Cartan et de Claude Chevalley, \cite{cartan1966theory} et \cite{chevalley1997algebraic}).\\
\indent L'espace des spineurs en dimension huit transverse se décompose en deux sous-espaces irréductibles
$S_8^{+}, S_8^{-}$ (pair et impair, ou gauche et droite).
Or dans cette dimension, les transformations de jauges, venant de l'ambiguité des solutions de l'équation
d'Einstein sous l'action des difféomorphismes, permettent d'identifier les espaces de spineurs $S_8^{+}$ et $S_8^{-}$ à
l'espace vectoriel de la fibre $F$ isomorphe aux vecteurs $\mathbb{R}^{8}$. Notons que cela n'arrive
qu'en dimension $8$, d'où la particularité de $12$ si on part du $4$ de tous les jours.\\
Le chemin qui mène à cette identification consiste à choisir d'abord une structure d'espace vectoriel complexe
sur $F$, interprétée comme un spineur pur imaginaire dans $S_8^{+}\otimes \mathbb{C}$ donnant un vecteur de vide,
puis à choisir un vecteur spécial $t$ de norme $1$ dans l'autre espace de
spineurs $S_8^{-}\otimes \mathbb{C}$; ceci suffit à déterminer une trialité permutant les trois espaces. Notons qu'\emph{a priori} il y a plusieurs
choix de trialités possibles, nous établissons qu'il est possible d'en fixer une en faisant
agir le groupe de jauge pour la théorie $Spin_8$, et nous décrivons l'ambiguité sur cette trialité. En particulier
il apparaît une ambiguité de phase. Très précisément, l'ambiguité de phase que nous rencontrons est celle qui décrit les choix
possibles d'une \emph{structure réelle} sur les spineurs. Nous constaterons qu'elle correspond, en un certain sesn, au groupe circulaire de
l'électromagnétisme. \\
Le mécanisme qui
fixe la direction d'un spineur pur dans le complexifié $S_8^{+}\otimes \mathbb{C}$ ramène le groupe structural compact $Spin_8$
à un sous-groupe
$\widetilde{U_4}$ de $Spin_8$, revêtement à deux feuillets du groupe unitaire $U_4$, et il apparaît aussi $12=28-16$ bosons
de jauge intermédiaires. Le spineur pur détermine une structure complexe $J_0$ sur $F$.
Puis, choisissant une métrique invariante sur $S_8$ on applique la transformation de jauge résiduelle $\widetilde{U_4}$
pour fixer une trialité.
Alors le groupe de jauge résiduel non-brisé est un revêtement à deux feuillets $\widetilde{U_3}$ du groupe
unitaire $U_3$, son algèbre de Lie est $su_3\times u_1$. Mais uniquement $SU_3$ est invariant par l'automorphisme de
$Spin_8$ qui correspond à la trialité.\\
\indent Une riche structure se révèle alors sur $S_8^{+}$, qui s'identifie naturellement à
l'espace $\Lambda^{ev}(F")$ des formes extérieures de degrés pairs sur le sous-espace complexe $F'$ de $F\otimes \mathbb{C}$
naturellement isomorphe à $F$ muni de $J_0$:
le vecteur $1$ s'identifie au vecteur fixé par le premier choix de jauge, ce sera la direction du neutrino, et
la droite des formes de degré $4$ sera la direction de l'électron, les formes de degré deux se divisent canoniquement
en deux sous-espaces invariants
pour $SU_3$, un espace $Q'$ où la représentation est isomoprphe à la fondamentale $Q'=D(1,0)$ et un espace $Q"$ où la représentation
est isomorphe à la conjugué $D(0,1)=\overline{D(1,0)}$, ces deux espaces vont correspondre aux deux types de quarks.\\
\indent Or la représentation expérimentalement bien établie de $SU_3$ sur les six quarks
est la somme de deux $D(1,0)$. La solution (unique ?) du paradoxe consiste à renverser la structure complexe du facteur $Q"$, en définissant $\overline{Q"}$
comme espace de multiplicité des particules quarks (on le verra, de charges $-1/3$) et en interprétant $Q"$
comme la multiplicité des anti-particules associées. Nous voyons ici une première manifestation de la dynamique
en dimension quatre, la nécessité des antiparticules.\\
\indent En plus des gluons de $SU_3$ et de la trace $Z_t$ du $u_3$ fixant $t$, les bosons au premier ordre de la théorie sont de deux
types, les opérateurs antisymétriques, appartenant à l'orthogonal de $u_3$ dans $spin_8$ et les opérateurs symétriques,
appartenant à l'orthogonal de $o_8$ dans $gl_8$.
\emph{A priori} les premiers agissent sur les fermions à travers la représentation spin, et pour les autres
une action possible est offerte par le "truc unitaire" de Weyl, en identifiant les matrices symétriques aux
matrices anti-hermitiennes imaginaires pures. Or ce schéma n'est pas facilement compatible avec le renversement de $Q"$ en $\overline{Q"}$;
les seuls opérateurs qui étendent l'action naturelle de $SU_3$ sont 1) deux opérateurs $W_+$ et $W_-$ qui échangent
neutrino et électron et qui échangent les quarks des deux types, 2) leur crochet $2iZ_1$, 3) l'identité $B_0$,
4) la projection $B_1$ sur l'espace $L$ des leptons et 5) la trace
$Z_4$ de l'algèbre $u_4$. Les opérateurs $W$, $B_0$ et $B_1$ viennent de la partie symétrique du commutant de $SU_3$
dans $gl_8$. Les opérateurs orthogonaux à tous ceux là ne peuvent pas agir de façon cohérente sur les fermions.
Le seul boson avec lequel \emph{a priori} ils interagiront sera le
boson de Higgs (qui ne va pas tarder à se montrer).\\

Parmi les combinaisons linéaires réelles $Q$ des quatre opérateurs diagonaux indépendants, $iB_0$, $iB_1$, $Z_1$ et $Z_4$ ($Z_t$ étant
égal à $Z_4+2Z_1$), un sous-espace de dimension trois satisfait à la condition du fixage de jauge, donc n'acquière pas de masse
par ce mécanisme. Autrement dit, tous les éléments de cet espace attribuent la charge zéro au neutrino.
Nous invoquons alors deux conditions supplémentaires: la première est que $W_+$ et $W_-$ possèdent une
charge (i.e. la différence entre la charge des $q'$ et celle des $\overline{q"}$ est égale à la charge de l'électron),
la seconde est que $Q$ engendre un cercle dans un sous-quotient de $U_8$, en l'occurence un quotient de $U_6\times U_2$
opérant naturellement séparément sur l'espace $L$ des leptons et sur l'espace $Q$ des quarks. Nous trouvons alors un
réseau épointé de rang deux. La nature a choisi le plus petit cercle possible, qui donne
\begin{equation}
\frac{Q_0}{i}=\frac{B_0-4B_1}{6}-Z_1.
\end{equation}
En fait, comme nous allons le voir, ce choix est aussi le seul
qui soit compatible avec le groupe contenant les opérateurs donnant les masses des fermions à l'ordre un. Il
apparaît donc une relation profonde entre charge électrique et masse des fermions (qui ne sera pas le seul écho de supersymétrie
comme nous allons le voir bientôt).\\

Les seuls degrés de liberté spinoriels qui survivent aux deux brisures de symétrie sont dans $S_8^{+}$,
puisque ceux de $S_8^{-}$ ont été fixés par la jauge, sauf peut-être la phase qui détermine la structure
réelle. \\
De façon remarquable vu le détour, la figure des
actions des champs $W$ et $Z$ dans le secteur $S_8^{-}$ reproduit exactement le mécanisme décrit par Weinberg, cf. \cite{weinberg1996quantum}. \\
Cependant, on ne peut pas prétendre que dans notre approche le groupe $SU_2\times U_1$ ait vécu comme un groupe
de jauge avant d'avoir été brisé.\\

\indent Le mécanisme utilisé est donc celui d'une
brisure spontanée de symétrie de jauge locale telle
qu'elle a été décrite par Brout, Englert et Higgs, et précisée par Weinberg dans ce cadre, mais appliquée à des spineurs
et non à des champs scalaires. Or l'expérience
nous apprend depuis 2012 que le champ de Higgs n'a pas de spin, cf. \cite{schorner2015large}.\\
La solution de ce paradoxe est un brin fantastique, il faut l'admettre, mais elle vient à nouveau de la trialité, et
semble surmonter élégament toutes les difficultés.\\
En effet, grâce à la trialité, il apparaît naturellement une section privilégiée $v_H$ de $E_F$, c'est celle qui correspond
au spineur fixé; elle peut être considérée comme un champ
scalaire auxiliaire $H=M_0v_H$, et interprétée comme un champ de Higgs ($M_0$ est une constante
dont la présence est nécessaire pour transfomer des nombres en $GeV$).
Afin de donner un terme cinétique au champ de Higgs, et avoir une authentique particule dans notre espace-temps,
nous réalisons le passage d'un spineur au vecteur par
bosonisation en suivant un modèle proche de celui qui a été décrit par A.Okninski (\cite{okninski2011supersymmetric}, \cite{okninski2014neutrino}),
en s'appuyant sur un spineur de direction constante.
Nous proposons que ce
spineur de direction constante, offrant ses quatre degrés de liberté réelle au boson de Higgs, soit la composante droite du neutrino;
celle-ci étant fixée par la première condition de jauge et liée à la composante $(1,0)$ de $t$ par trialité.
Nous obtenons ainsi un champ scalaire et faisons disparaître du même coup la moitié du neutrino. Autre caractéristique
du modèle standard, et autre intervention dans notre approche de la dynamique en dimension quatre.\\
\indent Le fait que $v$ corresponde à la direction fixée par le $SU_3$ fixant $t$ et par le $U_1$ de l'électromagnétisme
implique que les gluons et le photon n'acquièrent pas de masse suivant le Lagrangien developpé suivant $v$.
Par contre tous les autres bosons acquièrent ainsi une masse naturelle. Tous, en fait non, car les combinaisons de
$iB_0+Z_4/2$ et $iB_1+2Z_1$ sont épargnés;
pour elles un autre mécanisme pourrait apporter une masse. Par exemple, $iB_0$ est contraint par le choix d'Einstein, de choisir des
coordonnées où le déterminant de la métrique est égal à $-1$, et $Z_4$ est lié pareillement au déterminant complexe de la
structure unitaire sur les fibres de $E_F$ le long de $W$. Mais il faut avouer que ces champs là restent un peu mystérieux.\\

Remarquons que nous avons fondamentalement (avant brisure de symétrie, à l'ordre zéro)
$64$ directions réelles bosoniques non massives, les $9$ de $SU_3\times U_1$ plus les trois de $SU_2$ et les $52$
étranges, ce qui correspond évidemment de la dimension de $gl_8$, ce qui donne $128$ degrés de libertés réels bosoniques,
puisqu'un vecteur sans masse possède deux degrés de liberté, et que, par ailleurs, nous avons aussi $128$ directions réelles fermioniques, puisque
les quatre dimensions complexes de Dirac en dimension quatre (deux pour les particules, deux pour les antiparticules)
se multiplient par les $8$ dimensions réelles des secteurs pair et impair en dimension huit,
pour donner $64$ dimensions complexes.\\
Il y a donc la base d'une symétrie entre fermions, sans évidence de supersymétrie. En effet la structure algébrique évidente
sur les bosons est celle de $V^{*}\otimes gl_8$, ou de $\mathfrak{u}_8=\mathfrak{o}_8\oplus i\mathfrak{p}_8$, alors que celle des fermions
est $S_4^{*}\otimes S_8$.\\
Mais ceci décrit la situation avant brisure spontanée de symétrie; après brisure,
la moitié des degrés de liberté fermioniques qui a disparu devrait se retrouver du côté des bosons, cependant neuf d'entre eux
n'ont pas acquis de masse, les huit gluons et le photon; le photon étant un cas un peu à part, il faut au moins retrouver huit degrés
de liberté manquant. Et en effet, si on transfert quatre degrés de liberté du neutrino vers le champ de Higgs, le défaut de huit est supprimé.\\

Les masses des fermions sont également apportées par le champ $H$, ou de façon équivalente, par le déplacement $v$, mais en plus,
le mécanisme fait intervenir les courbures
sectionnelles $R_{a8b8}$ de la métrique dans les fibres de $E_F$ suivant les plans contenant la direction de $v$, prise comme
vcteur $e_8$. La matrice anti-hermitienne $iR_{a8b8}$ définit une transformation de l'espace complexe de dimension
huit comptant la multiplicité des fermions. Cette définition introduit deux ambiguités de phase, une sur $L$, une
sur $Q$; d'où le groupe projectif $PU_6\times PU_2$ que nous avons vu intervenir pour la charge $Q_0$.\\
De là nous tirons d'abord l'annulation au premier ordre de la masse du neutrino; par conséquent dans cette théorie, ce n'est pas par hasard
que le neutrino n'a ni charge ni masse, cela correspond à un choix de géométrie, plus pratique qu'un autre. Ensuite viennent les
résultats sur les ordres de grandeur des masses des fermions.\\
L'échelle est déterminée par la valeur moyenne $H_0$, qui est environ $250 GeV$, la valeur moyenne du déplacement $v_0$ est déduite à l'aide d'arguments
d'échelle des rapports approximatifs des masses de l'électron, du muon et du tau, elle est environ $1/53$. La
constante $M_0$ est donc de l'ordre du $TeV$. La masse $M_H$ du boson de Higgs, qui vaut $125 GeV$ en vrai,
est prédite égale à $115 GeV$ si on
attribue la valeur $1$ à la norme du spineur fixe $t$, mais on peut aussi voir seulement dans ce résultat une justification
de ce choix de $1$ pour $\| t\|$.\\
Nous montrons également pourquoi les masses des neutrinos suivants $\nu_\mu$ et $\nu_\tau$ ne sont pas forcément nulles,
et d'où vient la matrice $PMNS$. Du côté des quarks,  \emph{a priori} ils pourraient tous avoir des masses différentes, trois pour les $u$,
trois pour les $d$, cependant il n'est pas exclu que les contraintes dynamiques sur la courbure transversale de la métrique en
douze dimensions, fassent converger les masses des quarks vers des valeurs communes. Par contre les états propres de masse peuvent
différer des états propres de saveur, si bien que nous trouvons ainsi une matrice $CKM$ d'un ordre de grandeur raisonnable,
sans toutefois prédire ses coefficients exacts, puisque
le choix naturel dans ce cadre est plutôt celui d'un matrice générique.\\

Nous précisons ainsi par cette approche la nature d'une partie du flot de renormalisation suivant l'échelle d'énergie,
puisque celui, en plus des effets quantiques, doit réintégrer ordre par ordre les équations complètes en dimension douze.
On voit que de nouveaux paramètres indépendants surgissent à chaque itération
en $v$. La partie restante du
flot peut être attribuée aux fluctuations quantiques. Pour la décrire il faut revenir au schéma de renormalisation
et analyser les transferts entre échelles. Ce point de vue se rattache donc à celui de Wilson, faisant intervenir une variété croissante de Lagrangiens effectifs.
Un problème grave survient certainement au bout d'environ six à huit générations de
fermions, peut être même dix, car alors les trous noirs dans $W$ empêcheront de voir les nouveaux champs. Voir l'idée de 't Hooft
sur le role des trous noirs (\cite{t1994under}).\\
La possibilité de la renormalisation au sens de Wilson et Weinberg devrait être assurée par la covariance de la théorie proposée
vis à vis du groupe des difféomorphismes de $X$ préservant $W$. La ligne à suivre est celle qui est exposée par
J. Gomis et S. Weinberg  \cite{gomis1996nonrenormalizable} et D. Anselmi \cite{anselmi1994removal}, \cite{anselmi1995more}.
La théorie complète reposera donc sur une version de la co-homologie $BRST$ (Becchi-Rouet-Stora-Tyutin) ou son extension $BV$
(Batalin-Vilkovisky). Nous pourrons nous appuyer sur les remarquables travaux récents de Damiano Anselmi,
en particulier \cite{anselmi2014adler}, \cite{anselmi2015adler}.\\
\indent Remarquons que nous n'avons pas eu besoin de poser
\emph{a priori} un champ de Higgs parmi les variables de la théorie, puisque celui-ci apparaît comme un corollaire de la
possibilité (nécessité?) de fixer une trialité, afin de supprimer la grande ambiguité sur ce que sont les spineurs.
Après brisure de symétrie les polynomes à valeurs champs du développement sont restreints aux puissances de $H$; c'est ce qui fait que les générations
successives de fermions auront la même structure algébrique qu'à l'ordre zéro. Avec des masses plus élevées, déformées par les
champs de bosons plus compliqués à chaque ordre.\\
Par conséquent, si on admet notre schématisation,
il est possible de réconcilier deux raisons également satisfaisantes de l'apparition des
masses: les rayons critiques de sphères dans l'espace et les valeurs moyennes des champs de Higgs.\\

\indent Notre proposition théorique se rattache
à une démarche
ayant déjà fait ses preuves en Physique: que dire de la forme de ce qu'on ne peut pas savoir sans faire des choix arbitraires?
Ici la forme en question est l'ambiguité sur les trialités possibles, "cousant" les spineurs aux vecteurs. Le plus fondamental
du point physique (ou métaphysique) est que tous les choix soient équivalents, comme le choix d'une solution des équations
d'Einstein, à difféomorphismes près, ou le choix d'une jauge en théorie de Yang-Mills. L'instrument idéal pour mettre
en forme la théorie quantique est alors donné par une théorie de co-homologie. C'est la voie suivie par la théorie $BV$.\\

\indent L'idée que l'espace-temps de dimension quatre est comme une membrane dans un espace-temps
de plus grande dimension n'est pas originale, cf. par exemple Maartens, \cite{maartens2015standard}. Mais dans notre
approche, la
membrane n'est pas prise en compte comme un champ fondamental au départ, il s'agit seulement d'explorer l'univers à partir
d'une situation de co-dimension huit, un enlacement avec une petite sphère autour de dimension sept. Pourtant le champ
scalaire de Higgs prend la place du vecteur d'exploration de façon dynamique, après bosonisation. Le degré selon
ce vecteur représente la quantité d'énergie mise en jeu pour aller explorer l'univers autour, de la même façon
que le degré dans une série de Taylor permet de capturer les non-linéarités d'ordre de plus en plus élevées.\\
\indent Plusieurs théories très belles ont été proposées en augmentant la dimension de quatre à cinq,
dix, onze ou douze, en commençant par Kaluza-Klein, cf. \cite{bailin1987kaluza}, puis Neveu, Schwartz, cf. \cite{green1987superstring},
Cremmer, Julia, Scherk, cf. \cite{cremmer1978supergravity}, McDuff, Horava, Witten et A.Sen , \cite{witten1995stringdimensions},
\cite{sen1998nonperturbativestring}.
A partir de douze des structures non-Lorentziennes sont considérées, alors qu'ici c'est un douze de Lorentz que nous considérons.\\
La plupart de ces théories font déjà jouer un rôle clé au spineurs, par exemple à travers la supersymétrie, qui n'est
pas tout à fait étrangère à l'espace que nous avons présenté, puisque la trialité en est une des plus belles
expressions. Remarque: l'interdiction énoncée d'aller au-delà de la dimension $11$ si on tient compte des spins en dimension
quatre est levée par le fait que la brisure de symétrie, contrairement au scénario habituel fait disparaître
la moitié des fermions  (la composante impaire en dimension $8$). Mais il semble que l'idée d'exploiter l'ambiguité
fondamentale sur les spineurs soit nouvelle.\\
Notre approche est différente de celle de Kaluza-Klein et de celle des membranes, car nous ne
passons pas au quotient par des symétries, nous ne supposons aucune symétrie particulière, au contraire plus
les conditions sont génériques mieux c'est, et nous ne considérons pas un plongement particulier, tous
les champs de l'espace ambient entrent en jeu. L'équation des champs considérée
est celle d'Einstein et Dirac en dimension $12$, et nous ne faisons que choisir des coordonnées sur
ces deux sortes de champs en profitant de l'action des difféomorphismes. C'est donc uniquement ce choix d'un changement de variables et de
coordonnées qui fait émerger le modèle standard (un peu augmenté, pour aller au-delà). Il reste beaucoup à faire pour
comprendre la Physique des générations suivantes, et encore plus à faire pour intégrer au fur et à mesure les effets non-perturbatifs,
jusqu'au moment critique où le choix éléctromagnétique entrera peut-être en
conflit avec la théorie génératrice à douze dimension. Une éventualité plus optimiste est que les progrès en
Théorie quantique des champs non-perturbative incluant la Gravitation quantique, en particulier les propriétés
quantiques des trous noirs, feront apparaître une théorie nouvelle plus satisfaisante à cette échelle de Planck.\\

\section{La gravité génératrice}

\subsection{Les très grands champs de l'univers déplié}

Un voisinage d'une sous-variété Lorentzienne $W$ de dimension $4$ dans une variété Lorentzienne $X$ de dimension $12$
est difféomorphe à un voisinage tubulaire de la section nulle dans un fibré vectoriel $E_F$ de rang réel $8$ de base $W$.
En faisant agir les difféomorphismes on peut supposer que pour toute métrique de $X$ voisine d'une
métrique donnée, les fibres sont perpendiculaires à $W$ le long de $W$, et que la métrique sur les
fibres le long de la section nulle est une métrique définie positive fixe $g_F$.\\
Les directions des $4$-plans perpendiculaires
aux fibres de $E_F$ en chaque point $x$ au-dessus de $w\in W$ ne sont pas déterminées \emph{a priori}, elles
définissent une \emph{connexion de Ehresmann} $M$, capable de relever les chemins tracés sur $W$.\\
Pour chaque choix de cette connexion, la métrique sur $X$ est entièrement déterminée par un champ de métriques sur
la distribution $M$, qui implique en particulier le long de la section nulle, une métrique sur $W$,
et par une section au-dessus de $W$ du fibré des métriques Riemaniennes sur les fibres de $E_F$.\\
Nous ne nous intéressons dans cette note qu'à la série de Taylor de ces objets variables le long de $W$; si on choisit
une trivialisation locale de $E_F$ sur un ouvert $U$ de $W$, la série complète correspond à une section
au-dessus de $U$ du fibré vectoriel
\begin{equation}
N_F=S^{*}(F^{*})\otimes (S_{+}^{2}(F^{*})\oplus (T^{*}(W)\otimes F)\oplus S^{2}(T^{*}(W)).
\end{equation}
produit tensoriel de l'algèbre symétrique sur $F$ avec l'espace des formes bilinéaires symétriques
définies postives sur $F$,
l'espace des applications linéaires de $T(W)$ dans $F$ et l'espace des métriques du type de Lorentz
sur $T(W)$, qui sont relevées à $X$ en métriques sur $M$.\\
Pour les paramétriser et calculer les symboles de Christoffel de la métrique sur $X$ auprès
de $W$ on choisit un repère mobile $(e_W,e_F)$ fait par des champs de vecteurs $\{e_\mu;\mu=0,1,2,3\}$ dans
$M$ qui relèvent des dérivations suivant des coordonnées locales de $W$ et une trivialisation $\{e_i;i=1,...,8\}$
du fibré $E_F$. On peut toujours choisir un tel repère de sorte qu'en un point donné il soit orthonormé.
Avec un tel repère les composantes de la métrique en dimension $12$ sont des $g_{\mu\nu}$ et des $g_{ij}$
car les $g_{\mu i}$ sont tous nuls.\\

Dans le repère choisi de $T(X)$ la matrice de la forme de connexion de Levi-Civita s'écrit
\begin{equation}
\nabla e=\sum_{\mu=0}^{3}\nabla_\mu e+\sum_{k=1}^{8}\nabla_k e
\end{equation}
avec les deux groupes de matrices scalaires
\begin{equation}
\nabla_\mu e=
\begin{pmatrix}
_\lambda(\nabla_\mu e)_\nu & _\lambda(\nabla_\mu e)_j\\
_i(\nabla_\mu e)_\nu & _i(\nabla_\mu e)_j
\end{pmatrix}
\quad
\nabla_i e=
\begin{pmatrix}
_\lambda(\nabla_k e)_\nu & _\lambda(\nabla_k e)_j\\
_i(\nabla_k e)_\nu & _i(\nabla_k e)_j
\end{pmatrix}
\end{equation}
où les indices $\lambda,\nu,i,j$ varient où ils doivent.\\
Les symboles de Christoffel sont donnés par huit types, qui représentent les différents champs bosoniques en présence
\begin{gather}
\Gamma_{\mu\nu}^{\lambda}=_\lambda(\nabla_\mu e)_\nu,\quad \Gamma_{\mu\nu}^{i}=_i(\nabla_\mu e)_\nu,\\
\Gamma_{\mu j}^{\lambda}=_\lambda(\nabla_\mu e)_j,\quad \Gamma_{\mu j}^{i}=_i(\nabla_\mu e)_j\\
\Gamma_{k\nu}^{\lambda}=_\lambda(\nabla_k e)_\nu,\quad \Gamma_{k\nu}^{i}=_i(\nabla_k e)_\nu,\\
\Gamma_{k j}^{\lambda}=_\lambda(\nabla_k e)_j,\quad \Gamma_{k j}^{i}=_i(\nabla_k e)_j
\end{gather}
La notation suivante est plus traditionnelle en Géométrie.\\
\begin{equation}
\omega_\mu=\nabla_\mu e.dx^{\mu},\quad \omega_i=\nabla_i e.dx^{i}.
\end{equation}

\noindent La représentation de la distribution $M$ en coordonnées locales $\{x^{\mu},x^{i}\}$, au point
$(x,v)$, $x\in U$, $v\in F$, en identifiant l'espace tangent à $\mathbb{R}^{4}\times \mathbb{R}^{8}$,
est donnée par l'équation :
\begin{equation}
x^{i}=\sum_\mu a_{\mu}^{i}(x,v) x^{\mu}.
\end{equation}
Si bien que les vecteurs générateurs $e_\mu$ de $M$ au-dessus des générateurs $\partial_\mu$ de $T_x(W)$
sont donnés par
\begin{equation}
e_\mu=\partial_\mu+\sum_i a_{\mu}^{i}(x,v)e_i.
\end{equation}
Pour les indices $i=1,...,8$ on pose $e_i=\partial_i$. Les formules des crochets de Lie seront utiles:
\begin{align}
[e_i,e_\mu]&=-[e_\mu,e_i]=\sum_k\partial_ia_\mu^{k}e_k\\
[e_\mu,e_\nu]&=-[e_\nu,e_\mu]=\sum_j(\partial_\mu a_\nu^{j}-\partial_\nu a_\mu^{j}+\sum_k(a_\mu^{k}\partial_ka_\nu^{j}
-a_\nu^{k}\partial_ka_\mu^{j}))e_j\\
[e_i,e_j]&=[e_j,e_i]=0
\end{align}
En écrivant que $g_{i\mu}=g_{\mu i}=0$, on trouve
\begin{equation}
\widetilde{g}_{j \mu}=-\sum_ka_\mu^{k}g_{kj}
\end{equation}
ou inversement
\begin{equation}
a_\mu^{i}=-\sum_k\widetilde{g}_{k \mu}g^{ki}
\end{equation}
Par ailleurs, en tout point de $X$ au voisinage de $W$, on a
\begin{align}
g_{\mu\nu}&=(e_\mu|e_\nu)=\tilde{g}_{\mu\nu}+\sum_ia_\mu^{i}\tilde{g}_{\nu i}+\sum_ja_\nu^{j}\tilde{g}_{\nu j}
+2\sum_i\sum_j a_\mu^{i}a_\nu^{j}g_{ij}\\
&=\tilde{g}_{\mu\nu}+2\sum_i\sum_j a_\mu^{i}a_\nu^{j}g_{ij}-\sum_i\sum_ja_\mu^{i}a_\nu^{j}g_{ji}
-\sum_j\sum_ia_\nu^{j}a_\mu^{i}g_{ij}\\
&=\tilde{g}_{\mu\nu}
\end{align}

\noindent Remarque: les symboles de Christoffel $\Gamma_{ab}^{c}$ attachés au repère mobile des $e_i$ et des $e_\mu$ sont reliés à ceux qui viennent des coordonnées $x^{i}$ et
$x^{\mu}$, dont la formule est bien connue (\cite{cartan1946leccons}, \cite{hawking1973Largescale}):
\begin{equation}
\widetilde{\Gamma}_{ab}^{c}=\frac{1}{2}\sum_d (\partial_a \tilde{g}_{bd}+\partial_b \tilde{g}_{ad}-\partial_d \tilde{g}_{ab})\tilde{g}^{dc}
\end{equation}
\noindent En développant $\nabla_\mu (e_\nu)$ on trouve que
\begin{equation}
\Gamma_{\mu \nu}^{\lambda}=\widetilde{\Gamma}_{\mu \nu}^{\lambda}+\sum_j (a_\mu^{j}\widetilde{\Gamma}_{j \nu}^{\lambda}
+a_\nu^{j}\widetilde{\Gamma}_{j \mu}^{\lambda})+\sum_i\sum_ja_\mu^{i}a_\mu^{j}\widetilde{\Gamma}_{i j}^{\lambda},
\end{equation}
\begin{multline}
\Gamma_{\mu \nu}^{k}=D_{e_\mu}a_\nu^{k}+\sum_j (a_\mu^{j}\widetilde{\Gamma}_{j \nu}^{k}
+a_\nu^{j}\widetilde{\Gamma}_{j \mu}^{k})\\+\sum_i\sum_ja_\mu^{i}a_\mu^{j}\widetilde{\Gamma}_{i j}^{k}
-\sum_\lambda a_\lambda^{k}\sum_i(a_\mu^{i}\widetilde{\Gamma}_{i \nu}^{\lambda}+a_\mu^{i}\widetilde{\Gamma}_{i \nu}^{\lambda})\\
-\sum_i\sum_j\sum_ka_\mu^{i}a_\nu^{j}a_\lambda^{k}\widetilde{\Gamma}_{i j}^{\lambda}.
\end{multline}
De même en développant $\nabla_\mu (e_i)$ on trouve que
\begin{equation}
\Gamma_{\mu i}^{\lambda}=\widetilde{\Gamma}_{\mu i}^{\lambda}+\sum_ja_\mu^{j}\widetilde{\Gamma}_{j i}^{\lambda},
\end{equation}
\begin{equation}
\Gamma_{\mu i}^{k}=\widetilde{\Gamma}_{\mu i}^{k}+\sum_ja_\mu^{j}\widetilde{\Gamma}_{j i}^{k}-\sum_\nu a_\nu^{k}\widetilde{\Gamma}_{\mu i}^{\nu}.
\end{equation}
Pour calculer les symboles où $\mu$ et $i$ sont permutés, on peut utiliser l'absence de torsion de la connexion
qui dit que
\begin{equation}
\nabla_i e_\mu=\nabla_\mu e_i+[e_i,e_\mu];
\end{equation}
d'où
\begin{equation}
\Gamma_{i \mu}^{\lambda}=\Gamma_{\mu i}^{\lambda},
\end{equation}
\begin{equation}
\Gamma_{i \mu}^{k}=\Gamma_{\mu i}^{k}+\partial_ia_\mu^{k}.
\end{equation}
Enfin, le calcul de $\nabla_i e_j$ donne
\begin{equation}
\Gamma_{i j}^{\lambda}=\widetilde{\Gamma}_{i j}^{\lambda},
\end{equation}
\begin{equation}
\Gamma_{i j}^{k}=\widetilde{\Gamma}_{i j}^{k}-\sum_\nu a_\nu^{k}\widetilde{\Gamma}_{i j}^{\lambda}.
\end{equation}\\

\noindent Cependant il est plus convenable de calculer directement les symboles $\Gamma_{ab}^{c}$ à partir des trois
champs de potentiels géométriques, $a_\mu^{i}$, $g_{\mu \nu}$ et $g_{ij}$, en exprimant l'unitarité et l'absence de torsion.\\
Nous notons $D_a$ l'opérateur de dérivation ordinaire dans la direction de $e_a$, où $a$ peut être un $i$ allant de $1$ à $8$ ou
un $\mu$ allant de $0$ à $3$.
Et nous introduisons les symboles des coordonnées des crochets définis par:
\begin{equation}
[e_a,e_b]=\sum_cM_{ab}^{c}e_c.
\end{equation}
Si bien que
\begin{align}
M_{\mu i}^{j}=-M_{i \mu}^{j}&=-\partial_i a_{\mu}^{j},\\
M_{\mu \nu}^{j}=-M_{\nu \mu}^{j}&=\partial_\mu a_\nu^{j}-\partial_\nu a_\mu^{j}+\sum_k (a_\mu^{k}\partial_i a_\nu^{j}-a_\nu^{k}\partial_i a_\mu^{j});
\end{align}
les autres coefficients $M_{ab}^{c}$ étant nuls.\\
\noindent Nous poserons aussi
\begin{equation}
\Gamma_{abc}=\sum_d \Gamma_{ab}^{d}g_{dc},\quad M_{abc}=\sum_d M_{ab}^{d}g_{dc}.
\end{equation}
Il est facile de vérifier que les seuls symboles $\Gamma_{ab}^{c}$ qui correspondent à une connexion
affine de torsion nulle respectant la métrique sont donnés par
\begin{equation}
\Gamma_{abc}=\frac{1}{2}(D_ag_{bc}-M_{bca}+D_bg_{ac}-M_{acb}-D_cg_{ab}+M_{abc})
\end{equation}

\noindent Vu ce que valent les crochets, les symboles les plus simples sont les $\Gamma_{ijk}$, et comme les $e_i$ et
les $e_\mu$ sont orthogonaux, on obtient l'équation suivante
\begin{equation}
\Gamma_{i j}^{k}=\frac{1}{2}\sum_l(\partial_ig_{jl}+\partial_jg_{il}-\partial_lg_{ij})g^{lk}.
\end{equation}
Nous nommerons cet ensemble de composantes le \emph{champ de gravitation vertical}; en abrégé $\Gamma_F$.\\
De manière intrinsèque, il s'agit d'une famille paramétrée par $W$ de connexions de Levi-Civita sur les fibres de $E_F$.\\

\noindent De même
\begin{equation}
\Gamma_{\mu \nu}^{\lambda}=\frac{1}{2}\sum_\rho(D_\mu g_{\nu\rho}+D_\nu g_{\mu \rho}-D_\rho g_{\mu\nu})g^{\rho\lambda},
\end{equation}
sera nommé \emph{champ de gravitation horizontal}. Le long de $W$ c'est le champ de gravitation
usuel d'Einstein dans l'espace-temps de dimension quatre, cf. \cite{einstein1952principle}. On le note symboliquement $\Gamma_H$. C'est une
connexion partielle sur le fibré $M$ au-dessus de $X$, pas un champ de tenseurs.\\

\noindent A côté de ces connexions $\Gamma_F$ et $\Gamma_H$, la gravitation sur $X$ fait intervenir quatre champs de tenseurs:\\
\noindent le \emph{champ de déformation
verticale}, appelé $D_F$, dont les composantes sont les $D_ig_{\mu\nu}$ et qui est une $1$-forme verticale à valeur tenseur
symétrique d'ordre deux sur le fibré $M$, i.e. une section de $E_F^{*}\otimes S^{2}M^{*}$;\\
\noindent le \emph{champ de déformation horizontale}, appelé $D_H$, dont les composantes sont les $D_\mu g_{ij}$,
et qui est une section de $M^{*}\otimes S^{2}E_F^{*}$;\\
\noindent le champ de jauge, ou \emph{champ de connexion}, noté $C$ ou $C_F$, de composantes $D_ia_\mu^{j}=M_{i\mu}^{j}$, qui est
une section de $M^{*}\otimes End(E_F)$;\\
\noindent le \emph{champ de monodromie}, noté $M$ ou $M_H$, de composantes $M_{\mu\nu}^{j}$, qui est une section de $\Lambda^{2}M^{*}\otimes E_F$.\\
Rappelons que l'on a
\begin{align*}
M_{\mu\nu}^{j}&=D_\mu a_\nu^{j}-D_\nu a_\mu^{j}\\
&=\partial_\mu a_\nu^{j}-\partial_\nu a_\mu^{j}+\sum_ka_\mu^{k}\partial_ka_\nu^{j}-\sum_ka_\nu^{k}\partial_ka_\mu^{j}.
\end{align*}
Pour signaler le champ $C_F$, nous introduisons la notation suivante:
\begin{equation*}
C_{\mu i}^{j}=M_{i\mu}^{j}=-M_{\mu i}^{j}.
\end{equation*}
On a donc
\begin{equation*}
C_{\mu i}^{j}=\partial_{i}a_{\mu}^{j}.
\end{equation*}

\noindent Remarque: la division du champ de gravitation de dimension douze $\Gamma_{ab}^{c}$  en six champs vise à mieux
comprendre la Physique à quatre dimensions, cependant il faut garder en tête que la frontière entre ces six champs n'est
pas bien nette; par exemple dans $\Gamma_H$ interviennent les dérivées $D_\lambda g_{\mu\nu}$ où $D_F$, $C_F$ et $M_H$ sont présents
à travers les $a_\lambda^{i}D_ig_{\mu\nu}$. De même dans $D_H$ interviennent les $a_\mu^{k}\partial_kg_{ij}$, où se cachent
$C_F$, $M_H$ et $\Gamma_F$. Les degrés de liberté vraiment indépendants sont ceux des trois potentiels $g_{\mu\nu}$, $a_\mu^{i}$
et $g_{ij}$ que nous noterons respectivement $G_H$, $M$ et $G_F$.\\

\noindent Le champ $M_H$, mesurant la non-intégrabilité du sous-fibré $M$, se trouve
exprimé à côté du champ $D_F$ dans les composantes suivantes de $\Gamma$:
\begin{equation}
\Gamma_{\mu \nu}^{i}=\frac{1}{2}M_{\mu\nu}^{i}-\sum_j\frac{1}{2}\partial_j g_{\mu\nu}g^{ij},
\end{equation}
\begin{equation}
\Gamma_{\mu i}^{\nu}=\Gamma_{i \mu}^{\nu}=\frac{1}{2}\sum_\lambda g^{\lambda\nu}(\partial_i g_{\mu\lambda}-M_{\mu\lambda i}),
\end{equation}
on a
\begin{equation}
\sum_{j,\rho}g_{\rho \nu}g^{ij}\Gamma_{\mu j}^{\nu}=-\Gamma_{\mu \nu}^{i}.
\end{equation}
Remarquons que l'on a
\begin{equation}
M_{\mu\nu}^{i}=\Gamma_{\mu\nu}^{i}-\Gamma_{\nu\mu}^{i},
\end{equation}
et
\begin{equation}
D_i g_{\mu\nu}=\Gamma_{\mu\nu i}+\Gamma_{\nu\mu i}=\sum_jg_{ij}(\Gamma_{\mu\nu}^{j}+\Gamma_{\nu\mu}^{j}),
\end{equation}
On peut donc dire que $M_H$ et $D_F$ décrivent respectivement les parties anti-symétriques et symétriques
du tenseur $\Gamma_{\mu\nu}^{i}$. Notons que la connexion complète ne se transforme pas comme un tenseur,
mais, compte tenu de la décomposition du fibré tangent $T(X)$ que nous avons faite, certaines composantes
forment des tenseurs quand-même.\\

\noindent D'un autre côté, les champs $C_F$ et $D_H$ apparaissent dans les composantes suivantes:
\begin{equation}
\Gamma_{\mu i j}=\frac{1}{2}D_\mu g_{ij}-\frac{1}{2}M_{\mu ji}+\frac{1}{2}M_{\mu ij},
\end{equation}
\begin{equation}
\Gamma_{i \mu j}=\frac{1}{2}D_\mu g_{ij}-\frac{1}{2}M_{\mu ji}-\frac{1}{2}M_{\mu ij},
\end{equation}
\begin{equation}
\Gamma_{i j \mu }=-\frac{1}{2}D_\mu g_{ij}+\frac{1}{2}M_{\mu ji}+\frac{1}{2}M_{\mu ij}.
\end{equation}
D'où l'on tire
\begin{equation}
\Gamma_{i\mu j}=-\Gamma_{i j \mu}=-\Gamma_{j i \mu}=\Gamma_{j \mu i};
\end{equation}
\begin{equation}
M_{\mu ij}=\Gamma_{\mu ij}-\Gamma_{i\mu j}=\Gamma_{\mu ij}+\Gamma_{ij\mu};
\end{equation}
et
\begin{equation}
D_\mu g_{ij}=\Gamma_{\mu i j}+\Gamma_{i \mu j}.
\end{equation}
Si bien que $C_F$ et $D_H$ correspondent respectivement à des parties antisymétriques et
symétriques du tenseur $\Gamma_{\mu ij}$ par rapport aux deux indices $(\mu, i)$.\\
On a aussi
\begin{equation}
D_\mu g_{ij}=\Gamma_{\mu i j}+\Gamma_{\mu j i}.
\end{equation}
et
\begin{equation}
M_{\mu ij}=\Gamma_{\mu ij}-\Gamma_{\mu j i}.
\end{equation}
Si bien que, par rapport aux indices $i,j$, $C_F$
et $D_H$ décrivent respectivement la partie antisymétrique et la partie
symétrique du tenseur $\Gamma_{\mu ij}$.\\

\noindent On a aussi
\begin{equation}
\Gamma_{i j}^{\lambda}=-\frac{1}{2}D_\mu g_{ij}g^{\mu\lambda}+\frac{1}{2}M_{\mu ji}g^{\mu\lambda}+\frac{1}{2}M_{\mu ij}g^{\mu\lambda};
\end{equation}
\begin{equation}
\Gamma_{\mu i}^{j}=\frac{1}{2}D_\mu g_{ik}g^{kj}-\frac{1}{2}M_{\mu ki}g^{kj}+\frac{1}{2}M_{\mu i}^{j};
\end{equation}
\begin{equation}
\Gamma_{i \mu }^{j}=\frac{1}{2}D_\mu g_{ik}g^{kj}-\frac{1}{2}M_{\mu ki}g^{kj}-\frac{1}{2}M_{\mu i}^{j}.
\end{equation}
d'où l'on tire
\begin{equation}
C_{\mu i}^{j}=\Gamma_{i\mu}^{j}-\Gamma_{\mu i}^{j}.
\end{equation}

\noindent Du point de vue des symboles de Christoffel, la répartition des champs est faite de deux singletons $\Gamma_H$, $\Gamma_F$ et deux
paires $(D_F,M_H)$, $(C_F,D_H)$, qui correspondent respectivement aux symboles tout grec $\Gamma_{\mu\nu}^{\lambda}$, tout latin $\Gamma_{ij}^{k}$,
deux fois plus grecs que latins $(\Gamma_{\mu\nu}^{i}, \Gamma_{\mu i}^{\nu},\Gamma_{i \mu}^{\nu})$ et deux fois plus latins que grecs
$(\Gamma_{i j}^{\mu}, \Gamma_{i \mu}^{j},\Gamma_{\mu i}^{j})$.\\
\indent Il apparaît que du point de vue des potentiels (qui expriment tous la métrique sur $X$, et qui sont constitués par une métrique
$g_{ij}$ sur le feuilletage $E_F$, une distribution d'espaces orthogonaux $M$, et une métrique $g_{\mu\nu}$ sur cette distribution), il y a trois paires analogues,
de champs dérivés, qui sont $(\Gamma_H,D_F)$, $(\Gamma_F,D_H)$ et $(M_H,C_F)$.\\
Ainsi sont faites les composantes de la connexion totale de Levi-Civita $\Gamma_X$ en dimension douze, écrites dans une décomposition de l'espace-temps
adaptée aux explorations d'un voyageur de dimension trois au cours du temps.\\

En suivant ce découpage géométrique, la courbure de $\Gamma_X$ se décompose en $12$ termes cinétiques linéaires en les champs
et $56$ morceaux quadratiques en les champs.\\
\noindent En effet, il y a trois $2$-formes extérieures de courbures à valeurs $End(T(W)\oplus E_F)$ qui interviennent:
\begin{align}
\Omega_{\mu \nu}&=\nabla_\mu \nabla_\nu-\nabla_\nu \nabla_\mu-\nabla_[e_\mu,e_\nu],\\
\Omega_{i \mu}&=\nabla_i \nabla_\mu-\nabla_\mu \nabla_i-\nabla_[e_i,e_\nu]=-\Omega_{\mu i},\\
\Omega_{ij}&=\nabla_i \nabla_j-\nabla_j \nabla_i
\end{align}
En développant chacune, on obtient les composantes $R_{ a b d}^{c}$ où les deux premiers indices sont les indices
de forme différentielle suivant $dx^{a}\wedge dx^{b}$ et où les deux suivants sont les indices matriciels.\\
En tenant compte de l'antisymétrie de la matrice totale, il y a pour chacun des trois types $(a,b)$ quatre sortes de paires d'indices $(c,d)$,
et pour chacune il y a quatre
sommes de produits deux par deux de champs $\Gamma_{ef}^{g}$ en plus du terme cinétique $D_a\Gamma_{b d}^{c}-D_b\Gamma_{a d}^{c}$. A cela
il convient d'ajouter les termes venant des crochets de Lie; il y en a huit en tout.
Ce qui fait bien douze et cinquante six, respectivement. \\
Les douze composantes sont
\begin{multline}
R_{\mu\nu \kappa}^{\lambda}=D_\mu \Gamma_{\nu\kappa}^{\lambda}-D_\nu \Gamma_{\mu\kappa}^{\lambda}
+\sum_\rho \Gamma_{\nu \kappa}^{\rho}\Gamma_{\mu\rho}^{\lambda}- \sum_\rho\Gamma_{\mu \kappa}^{\rho}\Gamma_{\nu\rho}^{\lambda}\\
+\sum_l \Gamma_{\nu \kappa}^{l}\Gamma_{\mu l}^{\lambda}- \sum_l\Gamma_{\mu \kappa}^{l}\Gamma_{\nu l}^{\lambda}-\sum_jM_{\mu\nu}^{j}\Gamma_{j\kappa}^{\lambda}.
\end{multline}
\begin{multline}
R_{\mu\nu \kappa}^{i}=D_\mu \Gamma_{\nu\kappa}^{i}-D_\nu \Gamma_{\mu\kappa}^{i}
+\sum_\rho \Gamma_{\nu \kappa}^{\rho}\Gamma_{\mu\rho}^{i}- \sum_\rho\Gamma_{\mu \kappa}^{\rho}\Gamma_{\nu\rho}^{i}\\
+\sum_l \Gamma_{\nu \kappa}^{l}\Gamma_{\mu l}^{i}- \sum_l\Gamma_{\mu \kappa}^{l}\Gamma_{\nu l}^{i}-\sum_jM_{\mu\nu}^{j}\Gamma_{j\kappa}^{i}.
\end{multline}
\begin{multline}
R_{\mu\nu i}^{\kappa}=D_\mu \Gamma_{\nu i}^{\kappa}-D_\nu \Gamma_{\mu i}^{\kappa}
+\sum_\rho \Gamma_{\nu i}^{\rho}\Gamma_{\mu \rho}^{\kappa}- \sum_\rho\Gamma_{\mu i}^{\rho}\Gamma_{\nu \rho}^{\kappa}\\
+\sum_l \Gamma_{\nu i}^{l}\Gamma_{\mu l}^{\kappa}- \sum_l\Gamma_{\mu i}^{l}\Gamma_{\nu i}^{\kappa}-\sum_jM_{\mu\nu}^{j}\Gamma_{j i}^{\kappa}.
\end{multline}
\begin{multline}
R_{\mu\nu j}^{i}=D_\mu \Gamma_{\nu j}^{i}-D_\nu \Gamma_{\mu j}^{i}
+\sum_\rho \Gamma_{\nu j}^{\rho}\Gamma_{\mu\rho}^{i}- \sum_\rho\Gamma_{\mu j}^{\rho}\Gamma_{\nu\rho}^{i}\\
+\sum_l \Gamma_{\nu j}^{l}\Gamma_{\mu l}^{i}- \sum_l\Gamma_{\mu j}^{l}\Gamma_{\nu l}^{i}-\sum_kM_{\mu\nu}^{k}\Gamma_{kj}^{i}.
\end{multline}
\begin{multline}
R_{\mu j \kappa}^{\lambda}=D_\mu \Gamma_{j\kappa}^{\lambda}-D_j \Gamma_{\mu\kappa}^{\lambda}
+\sum_\rho \Gamma_{j \kappa}^{\rho}\Gamma_{\mu\rho}^{\lambda}- \sum_\rho\Gamma_{\mu \kappa}^{\rho}\Gamma_{j\rho}^{\lambda}\\
+\sum_l \Gamma_{j \kappa}^{l}\Gamma_{\mu l}^{\lambda}- \sum_l\Gamma_{\mu \kappa}^{l}\Gamma_{j l}^{\lambda}-\sum_kM_{\mu j}^{k}\Gamma_{k\kappa}^{\lambda}.
\end{multline}
\begin{multline}
R_{\mu j \kappa}^{i}=D_\mu \Gamma_{j\kappa}^{i}-D_j \Gamma_{\mu\kappa}^{i}
+\sum_\rho \Gamma_{j \kappa}^{\rho}\Gamma_{\mu\rho}^{i}- \sum_\rho\Gamma_{\mu \kappa}^{\rho}\Gamma_{j\rho}^{i}\\
+\sum_l \Gamma_{j \kappa}^{l}\Gamma_{\mu l}^{i}- \sum_l\Gamma_{\mu \kappa}^{l}\Gamma_{j l}^{i}-\sum_kM_{\mu j}^{k}\Gamma_{k\kappa}^{i}.
\end{multline}
\begin{multline}
R_{\mu j i}^{\kappa}=D_\mu \Gamma_{j i}^{\kappa}-D_j \Gamma_{\mu i}^{\kappa}
+\sum_\rho \Gamma_{j i}^{\rho}\Gamma_{\mu\rho}^{\kappa}- \sum_\rho\Gamma_{\mu i}^{\rho}\Gamma_{j\rho}^{\kappa}\\
+\sum_l \Gamma_{j i}^{l}\Gamma_{\mu l}^{\kappa}- \sum_l\Gamma_{\mu i}^{l}\Gamma_{j l}^{\kappa}-\sum_kM_{\mu j}^{k}\Gamma_{k i}^{\kappa}.
\end{multline}
\begin{multline}
R_{\mu j k}^{i}=D_\mu \Gamma_{j k}^{i}-D_j \Gamma_{\mu k}^{i}
+\sum_\rho \Gamma_{j k}^{\rho}\Gamma_{\mu\rho}^{i}- \sum_\rho\Gamma_{\mu k}^{\rho}\Gamma_{j\rho}^{i}\\
+\sum_l \Gamma_{j k}^{l}\Gamma_{\mu l}^{i}- \sum_l\Gamma_{\mu k}^{l}\Gamma_{j l}^{i}-\sum_lM_{\mu j}^{l}\Gamma_{l k}^{i}.
\end{multline}
\begin{multline}
R_{i j \kappa}^{\lambda}=D_i \Gamma_{j\kappa}^{\lambda}-D_j \Gamma_{i\kappa}^{\lambda}
+\sum_\rho \Gamma_{j \kappa}^{\rho}\Gamma_{i \rho}^{\lambda}- \sum_\rho\Gamma_{i \kappa}^{\rho}\Gamma_{j\rho}^{\lambda}\\
+\sum_l \Gamma_{j \kappa}^{l}\Gamma_{i l}^{\lambda}- \sum_l\Gamma_{i \kappa}^{l}\Gamma_{j l}^{\lambda}.
\end{multline}
\begin{multline}
R_{i j \kappa}^{l}=D_i \Gamma_{j\kappa}^{l}-D_j \Gamma_{i\kappa}^{l}
+\sum_\rho \Gamma_{j \kappa}^{\rho}\Gamma_{i \rho}^{l}- \sum_\rho\Gamma_{i \kappa}^{\rho}\Gamma_{j\rho}^{l}\\
+\sum_m \Gamma_{j \kappa}^{m}\Gamma_{i m}^{l}- \sum_m\Gamma_{i \kappa}^{m}\Gamma_{j m}^{l}.
\end{multline}
\begin{multline}
R_{i j l}^{\kappa}=D_i \Gamma_{j l}^{\kappa}-D_j \Gamma_{i l}^{\kappa}
+\sum_\rho \Gamma_{j l}^{\rho}\Gamma_{i \rho}^{\kappa}- \sum_\rho\Gamma_{i l}^{\rho}\Gamma_{j\rho}^{\kappa}\\
+\sum_m \Gamma_{j l}^{m}\Gamma_{i m}^{\kappa}- \sum_m\Gamma_{i l}^{m}\Gamma_{j m}^{\kappa}.
\end{multline}
\begin{multline}
R_{i j k}^{l}=D_i \Gamma_{j k}^{l}-D_j \Gamma_{i k}^{l}
+\sum_\rho \Gamma_{j k}^{\rho}\Gamma_{i \rho}^{l}- \sum_\rho\Gamma_{i k}^{\rho}\Gamma_{j\rho}^{l}\\
+\sum_m \Gamma_{j k}^{m}\Gamma_{i m}^{l}- \sum_l\Gamma_{i k}^{m}\Gamma_{j m}^{l}.
\end{multline}

\noindent Nous désignerons les douze cas suivant l'ordre d'apparition dans la liste des formules ci-dessus.\\
Les termes quadratiques correspondent à des combinaisons des couplages des champs
$\Gamma_H,M,\Gamma_F,C,D_F,D_H$. On constate que tous les couples de champs
apparaissent au moins une fois sauf le couple $\Gamma_H,\Gamma_F$. La liste des termes quadratiques dans les douze termes est la suivante:\\
1) $\Gamma_H\Gamma_H$ deux fois et $(M\pm D_F)(M\pm D_F)$ trois fois;
2) $\Gamma_H(M\pm D_F)$ deux fois et $(M\pm D_F)(C\pm D_H)$
trois fois;
3) également $\Gamma_H(M\pm D_F)$ deux fois, et $(M\pm D_F)(C\pm D_H)$ trois fois;
4) $(M\pm D_F)(M\pm D_F)$ deux fois, $(C\pm D_H)(C\pm D_H)$ deux fois et $(M\pm D_F)(C\pm D_H)$ une fois;
5) $\Gamma_H(M\pm D_F)$ deux fois et
$(M\pm D_F)(C\pm D_H)$ trois fois;
6) une seule fois $(M\pm D_F)(M\pm D_F)$, $\Gamma_H (C\pm D_H)$, $(M\pm D_F) \Gamma_F$ et deux fois $(C\pm D_H)(C\pm D_H)$;
7) pareil $(M\pm D_F)(M\pm D_F)$, $\Gamma_H (C\pm D_H)$, $(M\pm D_F) \Gamma_F$ une fois, et $(C\pm D_H)(C\pm D_H)$ deux fois;
8) deux fois $(M\pm D_F)(C\pm D_H)$ et trois fois $\Gamma_F(C\pm D_H)$;
9) $(M\pm D_F)(M\pm D_F)$ deux fois et $(C\pm D_H)(C\pm D_H)$ deux fois;
10) deux fois $(M\pm D_F)(C\pm D_H)$ et
deux fois $(C\pm D_H)\Gamma_F$;
11) également deux fois $(M\pm D_F)(C\pm D_H)$ et
deux fois $(C\pm D_H)\Gamma_F$;
12) deux fois $(C\pm D_H)(C\pm D_H)$ et deux fois $\Gamma_F\Gamma_F$.\\
En particulier, on remarque que les termes de courbure $R_{ab \mu}^{j}$ et $R_{ab j}^{\mu}$ ont
le même contenu d'interactions, ce qui est bien naturel, compte tenu de l'orthogonalité $g_{\mu i}=0$.
Il y a donc plutôt neuf coordonnées différentes que douze, et trente six cas de degré deux en les $\Gamma$ plutôt que quarante huit,
mais il semble préférable de procéder systématiquement en ne tenant pas trop compte
des redondances pour mener les calculs qui suivent.\\
Pour alléger, nous noterons $H$,$F$,$M$,$C$ au lieu de $\Gamma_H$,$\Gamma_F$,$M\pm D_F$,
$C\pm D_H$ respectivement. Nous voyons que $HH$ et $FF$ arrivent deux fois, $MM$ arrive neuf fois et $CC$ arrive dix fois,
$MC$ arrive seize fois, $HM$ arrive six fois,  $FC$ arrive sept fois,  et $HC$ et $FM$ arrivent deux fois. On trouve bien
$56$ termes au total.\\

\subsection{L'inertie et les tenseurs énergie-moments}

Les équations d'Einstein portent sur le tenseur d'Einstein qui s'exprime simplement à partir du
tenseur de Ricci, obtenu par contraction de dualité:
\begin{equation}
R_{a d}=\sum_b R_{ a b d}^{b}
\end{equation}
C'est un tenseur symétrique. La courbure scalaire est sa trace
dans une base orthonormée
\begin{equation}
R=\sum_a R_{ a a}
\end{equation}
L'équation d'Einstein dans le vide est l'annulation du tenseur de divergence covariante nulle qui s'écrit:
\begin{equation}
S_{ab}=R_{ab}-\frac{1}{2}Rg_{ab}.
\end{equation}
Cf. \cite{einstein1952principle}, \cite{hawking1973Largescale}.\\

\noindent Dans notre cas, le tenseur $S_{ab}$ se présente en trois composantes.\\
\indent La composante $S_{\mu\nu}$ fait intervenir des termes $R_{\mu \lambda \nu}^{\lambda}$
et des termes $R_{\mu i \nu}^{i}$; donc les tenseurs qui y sont
impliqués sont du type $1$ et $6$, c'est-à-dire $R_{\mu\nu\kappa}^{\lambda}$ et $R_{\mu i \nu}^{j}$.
Pour la composante $S_{\mu i}$ il faut tenir compte des termes $3$, $5$, $8$ et $10$, c'est-à-dire $R_{\mu\nu i}^{\lambda}$,
$R_{\mu i \nu}^{\lambda}=-R_{i \mu \nu}^{\lambda}$, $R_{\mu i j}^{k}$ et $R_{ i j \mu}^{k}$. Et pour $S_{ij}$
il faut tenir compte des termes du genre $7$ et $12$, c'est-à-dire $R_{\mu i j}^{\kappa}=-R_{i \mu j}^{\kappa}$ et
$R_{ijk}^{l}$.\\
Par conséquent $R_{\mu\nu}$ renferme des interactions du type $HH$, $MM$, $CC$, $HC$, $FM$; $S_{\mu i}$ des interactions du
type $HM$, $MC$, $CF$; $S_{ij}$ des interactions du type $CC$, $MM$, $HC$, $FM$ et $FF$.\\

\noindent Analysons de plus près les composantes du tenseur $R_{\mu\nu}$:
la composante $\sum_\lambda R_{\mu\lambda\nu}^{\lambda}$ induit le tenseur de Ricci à quatre dimension le long de $W$, mais
elle est étendue dans $X$ tout entier autour de $W$, en un tenseur porté par
la distribution de $4$-plans $M$; c'est de là que provient la trace de courbure moyenne. Nous divisons cette composante
en deux parties, une qui ne fait intervenir que des symbôles d'indices tous grecs, et une qui fait intervenir des symbôles avec
des indices grecs et latins mélangés:\\
Par définition,
\begin{equation}
R^{(0)}_{\mu\nu}=D_\mu \Gamma_{\lambda\nu}^{\lambda}-D_\lambda \Gamma_{\mu\nu}^{\lambda}
+\sum_{\rho,\lambda} \Gamma_{\lambda\nu }^{\rho}\Gamma_{\mu\rho}^{\lambda}- \sum_{\rho,\lambda}\Gamma_{\mu \nu}^{\rho}\Gamma_{\lambda\rho}^{\lambda};
\end{equation}
\begin{equation}
R^{(1)}_{\mu\nu}=\sum_{j,\lambda}\Gamma_{\lambda\nu}^{j}\Gamma_{\mu j}^{\lambda}-\sum_{j,\lambda}\Gamma_{\mu\nu}^{j}\Gamma_{\lambda j}^{\lambda}
-\sum_{j,\lambda}M_{\mu\lambda}^{j}\Gamma_{j \nu}^{\lambda}.
\end{equation}
La deuxième composante de $R_{\mu\nu}$ est $\sum_iR_{\mu i \nu}^{i}$; nous la noterons $R^{(2)}_{\mu\nu}$; elle se développe en
\begin{multline}
R^{(2)}_{\mu\nu}= \sum_iD_\mu \Gamma_{i\nu}^{i}-D_i \Gamma_{\mu\nu}^{i}+
\sum_{i \kappa} (\Gamma_{i \nu }^{\kappa}\Gamma_{\mu \kappa}^{i}- \Gamma_{\mu \nu}^{\kappa}\Gamma_{i\kappa}^{i})\\
+\sum_{i k} (\Gamma_{i \nu}^{k}\Gamma_{\mu k}^{i}- \Gamma_{\mu \nu}^{k}\Gamma_{i k}^{i})-\sum_{i k}M_{\mu i}^{k}\Gamma_{k \nu}^{i}.
\end{multline}
Notons qu'\emph{a priori} la symétrie du tenseur total $R_{\mu\nu}$ n'implique pas la symétrie séparée d'une des composantes $R^{(i)}_{\mu\nu}$.\\

\noindent \textbf{Lemme 1}:
\begin{multline}
R^{(1)}_{\mu\nu}=\frac{3}{4}\sum_{\kappa,\lambda,i,j}M_{\mu\kappa}^{i}g_{ij}M_{\nu\lambda}^{j}g^{\kappa\lambda}-\frac{1}{4}\sum_{\kappa,\lambda,j}(M_{\mu\kappa}^{j}D_j g_{\nu\lambda}g^{\kappa\lambda}+M_{\nu\kappa}^{j}D_j g_{\mu\lambda}g^{\kappa\lambda})\\
-\frac{1}{4}\sum_{\lambda\kappa jk} g^{jk}D_kg_{\lambda\nu}g^{\lambda\kappa}D_jg_{\mu\kappa}+\frac{1}{4}\sum_{\lambda\kappa jk}g^{j k}D_k g_{\mu\nu}D_j g_{\lambda\kappa}g^{\lambda\kappa}\\
-\frac{1}{4}\sum_{\kappa,\lambda,j}M_{\mu\nu}^{j}(D_jg_{\lambda\kappa})g^{\lambda\kappa}.
\end{multline}
\noindent \emph{Démonstration}: en utilisant la convention sommatoire d'Einstein, et l'antisymétrie des $M_{ab}$, on a
\begin{multline*}
\Gamma_{\lambda\nu}^{j}\Gamma_{\mu j}^{\lambda}=\frac{1}{4}(M_{\lambda\nu}^{j}-D_kg_{\lambda\nu}g^{kj})(g^{\lambda\kappa}(D_jg_{\mu\kappa}-M_{\mu\kappa j}))\\
=\frac{1}{4}M_{\nu \lambda}^{j}g^{\lambda\kappa}M_{\mu\kappa}^{k}g_{jk}-\frac{1}{4}g^{jk}D_kg_{\lambda\nu}g^{\lambda\kappa}D_jg_{\mu\kappa}\\
+\frac{1}{4}M_{\mu\kappa}^{k}D_kg_{\lambda\nu}g^{\lambda\kappa}-\frac{1}{4}M_{\nu\lambda}^{j}D_jg_{\mu\kappa}g^{\lambda\kappa}.
\end{multline*}
\begin{multline*}
-\Gamma_{\mu\nu}^{j}\Gamma_{\lambda j}^{\lambda}=-\frac{1}{4}(M_{\mu\nu}^{j}-D_k g_{\mu\nu}g^{kj})g^{\lambda\kappa}(D_jg_{\lambda\kappa}-M_{\kappa\lambda j})\\
=-\frac{1}{4}M_{\mu\nu}^{j}M_{\lambda\kappa j}g^{\lambda\kappa}+\frac{1}{4}g^{j k}D_k g_{\mu\nu}D_j g_{\lambda\kappa}g^{\lambda\kappa}\\
-\frac{1}{4}M_{\mu\nu}^{j}(D_jg_{\lambda\kappa})g^{\lambda\kappa}-\frac{1}{4}g^{\lambda\kappa}M_{\lambda\kappa j} g^{kj}D_kg_{\mu\nu}.
\end{multline*}
et
\begin{equation*}
-M_{\mu\lambda}^{j}\Gamma_{j\nu}^{\lambda}=\frac{1}{2}M_{\mu\lambda}^{j}g^{\lambda\kappa}M_{\nu\kappa}^{k}g_{k j}
-\frac{1}{2}M_{\mu\lambda}^{j}g^{\lambda\kappa}D_j g_{\nu\kappa}.
\end{equation*}
Le premier et le dernier terme du membre de droite de la seconde équation sont nuls à cause de l'orthogonalité
entre matrices symétriques et matrices anti-symétriques pour la forme de Killing
$Tr(XY)$. Le lemme s'en déduit par addition.\\

\noindent En faisant appel à l'identité $\partial_z \ln det G=Tr((\partial _zG)G^{-1})$, appliquée
à la matrice $G_H$ des $g_{\mu\nu}$, en considérant $M_H$ et $D_F=\nabla_F(G_H)$ comme des $8$-vecteurs
de matrices $4\times 4$, et en utilisant le symbole $\odot$ pour le composé du produit scalaire en dimension $8$
et de la composition des matrices $4\times 4$, on peut réécrire le lemme $1$ en raccourci de la manière suivante:
\begin{multline}
R^{(1)}=\frac{3}{4}M_H\odot M_H - \frac{1}{4}(M_H\odot D_F+D_F\odot M_H) +\frac{1}{4}D_F\odot D_F\\+\frac{1}{4}D_F.\nabla_F \ln det G_H
-\frac{1}{4} M_H.\nabla_F \ln det G_H.
\end{multline}
Tous les termes sont symétriques en $\mu$ et $\nu$, sauf le dernier des six qui est anti-symétrique.\\

\noindent \textbf{Lemme 2}:
\begin{multline}
R^{(2)}_{\mu\nu}=\frac{1}{2}\sum_{ij}D_i((D_j g_{\mu\nu})g^{ij})-\frac{1}{4}\sum_{\kappa\lambda i j}g^{\lambda\kappa}g^{ij}D_ig_{\nu\lambda}D_jg_{\mu\kappa}\\
+\frac{1}{4}\sum_{\kappa\lambda i j}g^{\lambda\kappa}(D_ig_{\mu\kappa}M_{\nu\lambda}^{i}+M_{\mu\lambda}^{i}D_j g_{\nu\kappa})+\frac{1}{2}
\sum_{i j k}D_jg_{\mu\nu}g^{kj}\Gamma_{i k}^{i}\\
+\frac{1}{2}\sum_{i j k}(C_{\mu i}^{k}(D_\nu g_{kj})g^{ij}
+C_{\nu i}^{k}(D_\mu g_{jk})g^{ij})+\frac{1}{4}\sum_{i j k l}(D_\mu g_{lk})g^{il}(D_\nu g_{ij})g^{jk}\\
+\frac{3}{4}\sum_{i j k l}C_{\mu j}^{l}g_{lk}C_{\nu i}^{k}g^{ji}+\frac{1}{4}\sum_{i k}C_{\mu i}^{k}C_{\nu k}^{i}\\
+\frac{1}{2}\sum_i(D_\mu C_{\nu i}^{i}+D_\nu C_{\mu i}^{i})+\sum_{i\lambda}\Gamma_{\mu\nu}^{\lambda}C_{\lambda i}^{i}\\
-\frac{1}{4}\sum_{\kappa \lambda i}g^{\kappa\lambda}M_{\mu\kappa}^{i}M_{\nu\lambda i}\\
+\frac{1}{2}(D_\mu D_\nu+ \sum_\kappa\Gamma_{\mu\nu}^{\kappa}D_\kappa) (\ln det G_F)-\frac{1}{2}\sum_{ik}\Gamma_{i k}^{i}M_{\mu\nu}^{k}.
\end{multline}

\noindent \emph{Démonstration}: en utilisant la convention sommatoire d'Einstein, on a
\begin{align*}
D_\mu \Gamma_{i\nu}^{i}&=\frac{1}{2}D_\mu((D_\nu g_{ik})g^{ik}-M_{\nu k i}g^{ik}-M_{\nu i}^{i})
=\frac{1}{2}D_\mu((D_\nu g_{ik})g^{ik})-D_\mu M_{\nu i}^{i}\\
&=\frac{1}{2}D_\mu((D_\nu g_{ik})g^{ik})+D_\mu C_{\nu i}^{i};
\end{align*}
\begin{equation*}
-D_i\Gamma_{\mu\nu}^{i}=-\frac{1}{2}D_iM_{\mu\nu}^{i}+\frac{1}{2}D_i((D_jg_{\mu\nu})g^{ij});\quad\quad\quad\quad\quad\quad\quad\quad\quad\quad\quad\quad\quad\quad\quad\quad
\end{equation*}
\begin{align*}
\Gamma_{i \nu }^{\kappa}\Gamma_{\mu \kappa}^{i}&=\frac{1}{4}g^{\lambda\kappa}(D_ig_{\lambda\nu}-M_{\nu\lambda i})(M_{\mu\kappa}^{i}-(D_jg_{\mu\kappa})g^{ij})\\
&=\frac{1}{4}g^{\lambda\kappa}(D_ig_{\mu\kappa}M_{\nu\lambda}^{i}+D_ig_{\nu\kappa}M_{\mu\lambda}^{i})
-\frac{1}{4}g^{\lambda\kappa}(g^{ij}D_ig_{\lambda\nu}D_jg_{\mu\kappa}+M_{\mu\kappa}^{i}M_{\nu\lambda i});
\end{align*}
\begin{multline*}
- \Gamma_{\mu \nu}^{\kappa}\Gamma_{i\kappa}^{i}=\frac{1}{2}\Gamma_{\mu\nu}^{\kappa}((D_\kappa g_{ik})g^{ik}-M_{\kappa k i}g^{ki}-M_{\kappa  i}^{i})\\
=\frac{1}{2}\Gamma_{\mu\nu}^{\kappa}(D_\kappa g_{ik})g^{ik}+\Gamma_{\mu\nu}^{\kappa}C_{\kappa  i}^{i};
\end{multline*}
\begin{multline*}
\Gamma_{i \nu}^{k}\Gamma_{\mu k}^{i}=\frac{1}{4}((D_\nu g_{ij})g^{jk}-M_{\nu j i}g^{jk}-M_{\nu i}^{k})
((D_\mu g_{lk})g^{il}-M_{\mu l k}g^{il}+M_{\mu k}^{i})\\
=\frac{1}{4}(D_\mu g_{lk})g^{il}(D_\nu g_{ij})g^{jk}+\frac{1}{4}(D_\nu g_{ij})g^{jk}M_{\mu k}^{i}
-\frac{1}{4}(D_\nu g_{ij})g^{il}g^{jk}M_{\mu l k}\\-\frac{1}{2}(D_\mu g_{lk})g^{il}M_{\nu i}^{k}
+\frac{1}{4}M_{\nu ji}g^{jk}M_{\mu lk}g^{il}-\frac{1}{4}M_{\nu ji}g^{jk}M_{\mu k}^{i}\\
-\frac{1}{4}M_{\nu i}^{k}M_{\mu k}^{i}+\frac{1}{4}M_{\nu i}^{k}M_{\mu lk}g^{il}\\
=\frac{1}{4}(D_\mu g_{lk})g^{il}(D_\nu g_{ij})g^{jk}+\frac{1}{2}C_{\nu i}^{k}(D_\mu g_{lk})g^{il}-\frac{1}{4}C_{\nu i}^{k}C_{\mu k}^{i}
+\frac{1}{4}C_{\nu i}^{k}C_{\mu lk}g^{il};
\end{multline*}
\begin{equation*}
-\Gamma_{\mu \nu}^{k}\Gamma_{i k}^{i}=\frac{1}{2}((D_jg_{\mu\nu})g^{kj}-M_{\mu\nu}^{k})\Gamma_{i k}^{i};\quad\quad\quad\quad\quad\quad\quad\quad\quad\quad\quad\quad\quad\quad\quad\quad\quad\quad
\end{equation*}
\begin{align*}
-M_{\mu i}^{k}\Gamma_{k \nu}^{i}&=\frac{1}{2}C_{\mu i}^{k}((D_\nu g_{kj})g^{ij}-M_{\nu jk}g^{ij}-M_{\nu k}^{i})
\quad\quad\quad\quad\quad\quad\quad\quad\quad\quad\quad\quad\quad\quad\\
&=\frac{1}{2}C_{\mu i}^{k}(D_\nu g_{kj})g^{ij}+\frac{1}{2}C_{\mu i}^{k}g^{ij}C_{\nu j}^{l}g_{kl}+\frac{1}{2}C_{\mu i}^{k}C_{\nu k}^{i}.
\end{align*}
\noindent Si on rassemble les termes en $D_H=\nabla_H(G_F)$ dans ces formules, on obtient une première partie de la composante $R^{(2)}$:
\begin{multline*}
R^{(2,1)}_{\mu\nu}=\frac{1}{2}(D_\mu ((D_\nu g_{ik})g^{ik})
+\frac{1}{2}\Gamma_{\mu\nu}^{\kappa}(D_\kappa g_{ik})g^{ik}+\frac{1}{4}(D_\mu g_{lk})g^{il}(D_\nu g_{ij})g^{jk}\\
+\frac{1}{2}C_{\nu i}^{k}(D_\mu g_{lk})g^{il}+\frac{1}{2}C_{\mu i}^{k}(D_\nu g_{kj})g^{ij}.
\end{multline*}
On peut réécrire cette partie en utilisant l'identité $\partial_z \ln det G=Tr((\partial _zG)G^{-1})$ appliquée
à la matrice $G_F$ des $g_{ij}$:
\begin{equation*}
R^{(2,1)}_{\mu\nu}=\frac{1}{2}(D_\mu D_\nu+\Gamma_{\mu\nu}^{\kappa}D_\kappa) (\ln det G)+\frac{1}{2}C_{\mu i}^{k}(D_\nu g_{kj})g^{ij}+\frac{1}{2}C_{\nu i}^{k}(D_\mu g_{lk})g^{il}.
\end{equation*}
On obtient ainsi les deux premières lignes de la formule du lemme 1; que l'on peut raccourcir en considérant $C=C_F$ et $D_H=\nabla_H(G_F)$ comme des $4$-vecteurs
de matrices $8\times 8$, et en utilisant le symbole $\odot$ pour le composé du produit tensoriel en dimension $4$
et de la forme de Killing des matrices $8\times 8$:
\begin{equation*}
R^{(2,1)}_{\mu\nu}=\frac{1}{2}(D_\mu D_\nu+\Gamma_{\mu\nu}^{\kappa}D_\kappa) (\ln det G)+\frac{1}{2}(C\oslash D_H+D_H\oslash C)_{\mu\nu}.
\end{equation*}
\indent Une seconde partie de $S^{(2)}$ est obtenue en rassemblant les termes en $D_F=\nabla_F(G_H)$:
\begin{multline*}
R^{(2,2)}_{\mu\nu}=\frac{1}{2}D_i((D_j g_{\mu\nu})g^{ij})-\frac{1}{4}g^{\lambda\kappa}g^{ij}D_ig_{\nu\lambda}D_jg_{\mu\kappa}
\\+\frac{1}{4}g^{\lambda\kappa}(D_ig_{\mu\kappa}M_{\nu\lambda}^{i}+D_ig_{\nu\kappa}M_{\mu\lambda}^{i})+\frac{1}{2}(D_jg_{\mu\nu})g^{kj}\Gamma_{i k}^{i}.
\end{multline*}
On peut la réécrire symboliquement:
\begin{equation*}
R^{(2,2)}_{\mu\nu}=\frac{1}{2}\triangle_F g_{\mu\nu}+\frac{1}{4}(D_F\odot M+M\odot D_F)_{\mu\nu}-\frac{1}{4}(D_F\odot D_F)_{\mu\nu}+\frac{1}{2}D_F.Tr(\Gamma_F).
\end{equation*}
Les termes qui restent donnent
\begin{multline*}
R^{(2,3)}_{\mu\nu}=\frac{3}{4}C_{\mu l}^{m}g_{mk}C_{\nu i}^{k}g^{li}+\frac{1}{4}C_{\mu i}^{k}C_{\nu k}^{i}\\
-\frac{1}{4}g^{\kappa\lambda}M_{\mu\kappa}^{i}M_{\nu\lambda i}-\frac{1}{2}D_iM_{\mu\nu}^{i}
+D_\mu C_{\nu i}^{i}+\Gamma_{\mu \nu}^{\kappa}C_{\kappa i}^{i}
-\frac{1}{2}M_{\mu\nu}^{k}\Gamma_{i k}^{i}.
\end{multline*}
Par ailleurs on a
\begin{multline*}
D_\mu C_{\nu i}^{i}-\frac{1}{2}D_i M_{\mu \nu}^{i}=\partial_\mu \partial_ia_\nu^{i}+a_{\mu}^{j}\partial_j\partial_ia_{\nu}^{i}\\
-\frac{1}{2}(\partial_i\partial_\mu a_\nu^{i}-\partial_i\partial_\nu a_\mu^{i}
+\partial_ia_\mu^{j}\partial_ja_\nu ^{i}-\partial_ia_\nu^{j}\partial_ja_\mu ^{i}
+a_\mu^{j}\partial_i\partial_ja_\nu^{i}-a_\nu^{j}\partial_i\partial_ja_\mu^{i})\\
=\frac{1}{2}\partial_i\partial_\mu a_\nu^{i}+\frac{1}{2}\partial_i\partial_\nu a_\mu^{i}
+\frac{1}{2}a_\mu^{j}\partial_j\partial_ia_\nu^{i}+\frac{1}{2}a_\nu^{j}\partial_j\partial_ia_\mu^{i};
\end{multline*}
donc
\begin{multline*}
R^{(2,3)}_{\mu\nu}=\frac{3}{4}C_{\mu l}^{m}g_{mk}C_{\nu i}^{k}g^{li}+\frac{1}{4}C_{\mu i}^{k}C_{\nu k}^{i}\\
-\frac{1}{4}g^{\kappa\lambda}M_{\mu\kappa}^{i}M_{\nu\lambda i}
+\frac{1}{2}(D_\mu D_iC_\nu^{i}+D_\nu D_iC_\mu^{i})+\Gamma_{\mu \nu}^{\kappa}C_{\kappa i}^{i}
-\frac{1}{2}M_{\mu\nu}^{k}\Gamma_{i k}^{i}.
\end{multline*}
Ce qui se réécrit symboliquement
\begin{multline*}
R^{(2,3)}_{\mu\nu}=\frac{3}{4}(CG_F\oslash CG_F^{-1})_{\mu\nu}+\frac{1}{4}(C\oslash C)_{\mu\nu}\\
-\frac{1}{4}(M\odot M)_{\mu\nu}
+\frac{1}{2}(D_\mu(\nabla_F.C_\nu)+D_\nu(\nabla_F.C_\mu))+\Gamma_{\mu \nu}^{\kappa}TrC_{\kappa}
-\frac{1}{2}M_{\mu\nu}^{k}Tr\Gamma_{k}.
\end{multline*}
D'où le lemme $2$.\\

\noindent Par ailleurs on a
\begin{equation*}
\Gamma_{i k}^{i}=\frac{1}{2}\sum_{ij}(\partial_ig_{jk}-\partial_jg_{ik}+\partial_kg_{ij})g^{ij}
=\frac{1}{2}\sum_{ij}(\partial_kg_{ij})g^{ij};
\end{equation*}
d'où
\begin{align*}
\frac{1}{2}\sum_{ik}\Gamma_{i k}^{i}M_{\mu\nu}^{k}&=\frac{1}{4}\sum_kM_{\mu\nu}^{k}\partial_k\ln det (G_F)\\
&=\frac{1}{4}(D_\mu D_\nu-D_\nu D_\mu)\ln det (G_F),
\end{align*}
si bien que
\begin{equation*}
\frac{1}{2}D_\mu D_\nu (\ln det G_F)-\frac{1}{2}\sum_{ik}\Gamma_{i k}^{i}M_{\mu\nu}^{k}
=\frac{1}{4}(D_\mu D_\nu+D_\nu D_\mu)\ln det (G_F).
\end{equation*}
On en déduit la formule suivante pour la somme:\\

\noindent \textbf{Proposition 1}:
\begin{multline}
R^{(1)}_{\mu\nu}+R^{(2)}_{\mu\nu}=\frac{1}{2}\sum_{ij}(D_i(g^{ij}D_j g_{\mu\nu})+\sum_{k}\Gamma_{i k}^{k}g^{ij}D_jg_{\mu\nu})\\
+\frac{1}{4}\sum_{\lambda\kappa jk}g^{j k}D_k g_{\mu\nu}g^{j k}D_j \ln det G_H-\frac{1}{2}\sum_{\kappa\lambda i j}g^{\lambda\kappa}g^{ij}D_ig_{\nu\lambda}D_jg_{\mu\kappa}\\
+\frac{1}{2}\sum_{i j k}g^{ij}(C_{\mu i}^{k}D_\nu g_{kj}+C_{\nu i}^{k}D_\mu g_{jk})+\frac{1}{4}\sum_{i j k l}(D_\mu g_{lk})g^{il}(D_\nu g_{ij})g^{jk}\\
+\frac{3}{4}\sum_{i j k l}C_{\mu j}^{l}g_{lk}C_{\nu i}^{k}g^{ji}+\frac{1}{4}\sum_{i k}C_{\mu i}^{k}C_{\nu k}^{i}\\
+\frac{1}{2}\sum_i(D_\mu C_{\nu i}^{i}+D_\nu C_{\mu i}^{i}+\sum_{\lambda}\Gamma_{\mu\nu}^{\lambda}C_{\lambda i}^{i})\\
+\frac{1}{2}\sum_{\kappa \lambda i}g^{\kappa\lambda}M_{\mu\kappa}^{i}M_{\nu\lambda i}\\
+\frac{1}{4}(D_\mu D_\nu+D_\nu D_\mu+2\sum_\kappa\Gamma_{\mu\nu}^{\kappa}D_\kappa)(\ln det G_F)\\
-\frac{1}{4}\sum_{\kappa,\lambda,j}M_{\mu\nu}^{j}D_j\ln det G_H
\end{multline}

\noindent En raccourci cela donne la formule suivante
\begin{multline*}
R^{(1)}_{\mu\nu}+R^{(2)}_{\mu\nu}=\frac{1}{2}\triangle_FG_H-\frac{1}{2}D_F\odot D_F+\frac{1}{2}D_F.Tr\Gamma_F+\frac{1}{4}D_F.\nabla_F\ln det G_H\\
+\frac{1}{2}(C_F\oslash D_H+D_H\oslash C_F)+\frac{1}{4}D_H\oslash D_H+\frac{3}{4}CG_FCG_F^{-1}+\frac{1}{4}C\oslash C\\+\frac{1}{2}(D_\mu Tr C_\nu+D_\nu Tr C_\mu)+\Gamma_H.Tr(C_F)
+\frac{1}{2}M\odot M\\+\frac{1}{4}(D_\mu D_\nu+D_\nu D_\mu+2\sum_\kappa\Gamma_{\mu\nu}^{\kappa}D_\kappa)(\ln det G_F)-\frac{1}{4}M_H.\nabla_F\ln det G_H.
\end{multline*}

\noindent Le seul terme non-symétrique en $(\mu,\nu)$ est le dernier, mais cela s'explique par la présence de
dérivées secondes $D_\mu D_\nu \ln det G_H$  dans la partie $R^{(0)}_{\mu\nu}$ du tenseur de Ricci:\\

\noindent \textbf{Lemme 3}:
\begin{multline}
R^{(0)}_{\mu\nu}=\frac{1}{2}D_\mu D_\nu \ln det G_H+\frac{1}{4}\sum_{\lambda\kappa jk}D_k g_{\mu\nu}g^{j k}D_j \ln det G_H\\
+\frac{1}{4}\sum_{\lambda\rho\kappa\omega}(D_\lambda g_{\mu\rho}-D_\rho g_{\mu\lambda})g^{\rho\kappa}g^{\lambda\omega}(D_\kappa g_{\nu\omega}-D_\omega g_{\nu\kappa})\\
+\frac{1}{4}\sum_{\lambda\rho\kappa\omega}D_\mu g_{\lambda\rho}g^{\rho\kappa}D_\nu g_{\kappa\omega}g^{\omega\lambda}
-\frac{1}{4}\sum_{\lambda\rho\kappa\omega}g^{\rho\kappa}g^{\lambda\omega}D_\kappa g_{\lambda\omega}(D_\mu g_{\rho\nu}+D_\nu g_{\rho\mu}).
\end{multline}

\noindent \emph{Démonstration}: toujours avec la convention de sommation d'Einstein, on a
\begin{multline*}
D_\mu \Gamma_{\lambda\nu}^{\lambda}-D_\lambda \Gamma_{\mu\nu}^{\lambda}+\Gamma_{\lambda\nu}^{\kappa}\Gamma_{\mu\kappa}^{\lambda}-\Gamma_{\mu\nu}^{\kappa}\Gamma_{\lambda\kappa}^{\lambda}
\\=\frac{1}{2}D_\mu((D_\lambda g_{\nu\kappa}+D_\nu g_{\lambda\kappa}-D_\kappa g_{\nu\lambda})g^{\kappa\lambda})
-\frac{1}{2}D_\lambda((D_\mu g_{\nu\kappa}+D_\nu g_{\mu\kappa}-D_\kappa g_{\nu\mu})g^{\kappa\lambda})\\
+\frac{1}{4}(D_\lambda g_{\nu\rho}+D_\nu g_{\lambda\rho}-D_\rho g_{\lambda\nu})g^{\rho\kappa}
(D_\mu g_{\kappa\omega}+D_\kappa g_{\mu\omega}-D_\omega g_{\mu\kappa})g^{\omega \lambda}\\
-\frac{1}{4}(D_\mu g_{\nu\rho}+D_\nu g_{\mu\mu\rho}-D_\rho g_{\mu\nu})g^{\rho\kappa}
(D_\lambda g_{\kappa\omega}+D_\kappa g_{\lambda\omega}-D_\omega g_{\lambda\kappa})g^{\omega \lambda}\\
=\frac{1}{2}D_\mu D_\nu \ln det G_H+\frac{1}{4}D_\mu g_{\kappa\omega}g^{\rho\kappa}D_\nu g_{\rho\lambda}g^{\omega\lambda}
+\frac{1}{4}g^{\kappa\rho}D_\kappa \ln det G_H D_\rho g_{\mu\nu}\\
+\frac{1}{4}(D_\lambda g_{\nu\rho}-D_\rho g_{\lambda\nu})
(D_\kappa g_{\mu\omega}-D_\omega g_{\mu\kappa})g^{\rho\kappa}g^{\omega \lambda}\\
-\frac{1}{4}(D_\mu g_{\nu\rho}+D_\nu g_{\mu\mu\rho})d_\kappa g_{\omega\lambda}g^{\kappa\rho}g^{\omega\lambda}.
\end{multline*}

\noindent Dans la somme des trois $R^{j}_{\mu\nu}$ on fait donc apparaître l'expression symétrique
\begin{equation*}
\frac{1}{2}D_\mu D_\nu \ln det G_H-\frac{1}{4}\sum_{\kappa,\lambda,j}M_{\mu\nu}^{j}D_j\ln det G_H=
\frac{1}{4}(D_\mu D_\nu+D_\nu D_\mu)\ln det G_H.
\end{equation*}
Ce terme se place plus naturellement du côté des termes d'inertie de $R^{(0)}_{\mu\nu}$ que du côté des
termes d'énergie-moment de la somme $R^{(1)}_{\mu\nu}+R^{(2)}_{\mu\nu}$. C'est également le cas pour d'autres termes qui
apparaissent dans $R^{(1)}_{\mu\nu}+R^{(2)}_{\mu\nu}$. Nous introduisons donc un découpage
un peu arbitraire, il faut le reconnaître, mais pratique pour parler de la théorie effective en dimension quatre
que nous cherchons à décrire:\\

\noindent \textbf{Définition 1)}: la partie d'inertie de $R_{\mu\nu}$ correspond à la somme suivante:
\begin{multline}
R^{(4)}_{\mu\nu}=\frac{1}{4}(D_\mu D_\nu+D_\nu D_\mu)\ln det G_H+\frac{1}{2}\sum_{\lambda\kappa jk}D_k g_{\mu\nu}g^{j k}D_j \ln det G_H\\
+\frac{1}{2}\sum_{ij}(D_i(g^{ij}D_j g_{\mu\nu})+\sum_{k}\Gamma_{i k}^{k}g^{ij}D_jg_{\mu\nu})-\frac{1}{2}\sum_{\kappa\lambda i j}g^{\lambda\kappa}g^{ij}D_ig_{\nu\lambda}D_jg_{\mu\kappa}\\
+\frac{1}{4}\sum_{\lambda\rho\kappa\omega}(D_\lambda g_{\mu\rho}-D_\rho g_{\mu\lambda})g^{\rho\kappa}g^{\lambda\omega}(D_\kappa g_{\nu\omega}-D_\omega g_{\nu\kappa})\\
+\frac{1}{4}\sum_{\lambda\rho\kappa\omega}D_\mu g_{\lambda\rho}g^{\rho\kappa}D_\nu g_{\kappa\omega}g^{\omega\lambda}-\frac{1}{4}\sum_{\lambda\rho\kappa\omega}g^{\rho\kappa}g^{\lambda\omega}D_\kappa g_{\lambda\omega}(D_\mu g_{\rho\nu}+D_\nu g_{\rho\mu});
\end{multline}
\noindent \textbf{Définition 2)}: la partie d'énergie-moment de $R_{\mu\nu}$ correspond au reste:
\begin{multline}
T^{(4)}_{\mu\nu}=\frac{1}{2}\sum_{\kappa \lambda i}g^{\kappa\lambda}M_{\mu\kappa}^{i}M_{\nu\lambda i}
+\frac{1}{4}\sum_{i j k l}(D_\mu g_{lk})g^{il}(D_\nu g_{ij})g^{jk}\\
+\frac{1}{4}(D_\mu D_\nu+D_\nu D_\mu+2\sum_\kappa\Gamma_{\mu\nu}^{\kappa}D_\kappa)(\ln det G_F)\\
+\frac{1}{2}\sum_{i j k}(C_{\mu i}^{k}(D_\nu g_{kj})g^{ij}
+C_{\nu i}^{k}(D_\mu g_{jk})g^{ij})
+\frac{3}{4}\sum_{i j k l}C_{\mu j}^{l}g_{lk}C_{\nu i}^{k}g^{ji}\\+\frac{1}{4}\sum_{i k}C_{\mu i}^{k}C_{\nu k}^{i}
+\frac{1}{2}\sum_i(D_\mu C_{\nu i}^{i}+D_\nu C_{\mu i}^{i}+\sum_{\lambda}\Gamma_{\mu\nu}^{\lambda}C_{\lambda i}^{i})
\end{multline}
Ou de façon symbolique
\begin{multline}
T^{(4)}_{\mu\nu}=\frac{1}{2}M\odot M+\frac{1}{4}D_H\oslash D_H+\frac{1}{4}\triangle_H^{\nabla}\ln det G_F\\
+\frac{1}{2}(C_F\oslash D_H+D_H\oslash C_F)+\frac{3}{4}C_FG_FC_FG_F^{-1}\\
+\frac{1}{4}C_F\oslash C_F+\frac{1}{2}(D_\mu TrC_{\nu}+D_\nu TrC_{\mu}+\Gamma_{\mu\nu}^{\lambda}TrC_{\lambda})
\end{multline}

\noindent Rappelons que dans $D_\mu$ intervient $a_\mu^{i}$, donc le prétendu terme d'inertie contient d'autres
champs que $\Gamma_H$. De même la décision de mettre $D_F$ presqu'entièrement du côté de l'inertie n'est pas
très fondée; elle se justifiera surtout par l'approximation qui néglige $D_F$ et considère $G_H$ plate.\\
On voit bien que la frontière entre les $R^{(i)}$ est floue. Un certain arbitraire nous incite à mettre
certains termes du côté de l'inertie et certains autres du côté des tenseurs énergie-moment de matière. Ceci n'aurait
sans doute pas surpris Einstein, qui pensait que le $T_{\mu\nu}$ manquait d'interprétation géométrique, car nous voyons
ici que, même venant entièrement de la géométrie, il demeure un peu flou.\\
Nous verrons plus loin que les conventions actuelles introduisent une constante dépendant des unités en facteur de la
partie $T^{(4)}_{\mu\nu}$ pour obtenir les tenseurs d'énergie-impulsions usuels. Cette constante sera en $GeV^{-2}$ et
sera tellement petite qu'elle justifiera un peu mieux la division entre inertie et le reste.\\

\noindent Pour passer au tenseur d'Einstein $S_{\mu\nu}$, on retire à chaque terme $A_{\mu\nu}$ de $R_{\mu\nu}$
la moitié de sa trace, notée $\frac{1}{2}Ag_{\mu\nu}$, où
\begin{equation}
A=\sum_{\kappa\lambda}A_{\kappa\lambda}g^{\kappa\lambda}=\sum_{\lambda}A_\lambda^{\lambda}.
\end{equation}
Nous soulignons qu'il faut soustraire les demi-traces
pour $T^{(4)}_{\mu\nu}$ aussi, avant d'obtenir les tenseurs d'énergie-impulsions usuels.\\

\noindent Enfin, notons que $S_{\mu i}$ et $S_{ij}$ n'apportent pas directement de l'énergie au monde de dimension quatre
dans le sens ordinaire, car
l'énergie usuelle appartient par définition aux $T_{\mu\nu}$, cependant les six champs ou leurs potentiels contribuent à $S_{\mu\nu}$,
donc les équations venant de $S_{\mu i}$ et $S_{ij}$ influencent la distribution d'énergie. Tout le territoire qui enlace notre espace-temps
peut se manifester aux hautes énergies.\\
Par exemple, nous verrons que les constantes de couplages et les masses des particules accessibles en dimension quatre aux
divers ordres d'un déplacement infinitésimal dans la direction de $E_F$, dépendent de la courbure de $g_{ij}$, par conséquent
l'équation sur $S_{ij}$ influence ces constantes et ces masses, en même temps que les corrections quantiques et le flot de
renormalisation en fonction de l'échelle d'énergie. \\

\noindent Nous allons commencer à explorer jusqu'à l'ordre deux ou trois les dimensions verticales. Mais d'abord parlons de la matière, car l'espace-temps $X$
n'est pas vide; il est parcouru et habité par des fermions. Il est même probable que l'origine des tenseurs métriques est à chercher dans
la dynamique de ces fermions.\\

\subsection{Les spineurs et les unités}

\indent Nous introduisons un fibré $S$ de spineurs de Dirac complexes sur $X$; ce qui suppose que $X$ (ou $W$, ce qui revient au même) est
munie d'une structure spin, i.e. un
revêtement d'ordre deux non-trivial en chaque fibre de l'espace des repères tangents. En chaque point de la variété $X$ l'espace des
spineurs est un espace vectoriel complexe de dimension $64$, qui possède une décomposition en deux
sous-espace de spineurs de Weyl. En fait la structure des spineurs $S$ en dimension $12$ et signature de Lorentz
est la même que celle de la dimensionn $4$, à cause de la périodicité modulo $8$ des algèbres de Clifford
(cf. \cite{cartan1966theory}, \cite{deligne1999quantum}, \cite{adams1996lectures}).\\
Les équations de particules sur les spineurs sont les équations de Dirac: on commence par appliquer la dérivée
covariante relevée au fibré $E_S$ des spineurs, puis on transforme la partie co-vecteur en vecteur avec la métrique,
puis on applique la multiplication de Clifford (i.e. les matrices de Dirac en coordonnées); cela donne
un opérateur $D_S$ échangeant les chiralités de Weyl.\\
Le couplage avec la gravité se fait naturellement à travers le Lagrangien de Hilbert et Dirac:
\begin{equation}\label{lagrangienhilbertdirac}
\mathcal{L}=R + c_0\widetilde{\psi}D_S\psi
\end{equation}
où le tilda ici désigne le spineur conjugué à $\psi$ vu comme une forme linéaire. En effet si $\psi\in S^{+}$
alors $D_S\psi \in S^{-}$, mais $S^{-}=\overline{S^{+}}$ est son propre dual via la structure symplectique canonique $\omega_S$.
En fait le choix d'une métrique de Lorentz sur $X$ fixe la norme de $\omega_S$ mais pas sa phase; il y a donc un petit choix
supplémentaire à faire ici, qu'on peut incorporer à la constante $c_0$.\\
\indent Pour obtenir le système complet des équations semi-classiques, on utilise les équations d'Euler-Lagrange.
De façon équivalente, on introduit au second membre de l'équation d'Einstein
$S_{ab}=c_0T^{(s)}_{ab}$ le seul tenseur de degré deux naturellement associé à un champ de spineurs sans masses, c'est-à-dire
la transposée de l'application $\gamma:T(W)\rightarrow End(S)$ appliquée à la partie spineur de la $1$-forme $\psi^{*}\otimes\nabla \psi
+(\nabla \psi)^{*}\otimes \psi$; on obtient ainsi une forme bilinéaire symétrique sur $T(X)$:
\begin{equation}\label{energiemomentfermion}
T^{(s)}=(Id_{T^{*}(X)}\otimes ^{t}\gamma)(\psi^{*}\otimes\nabla \psi
+(\nabla \psi)^{*}\otimes \psi).
\end{equation}

\noindent La décomposition de $T(X)$ au voisinage de $W$ dans $X$ en une partie verticale tangente au fibres de $E_F$
et une distribution $M$, munie on l'a vu de la même métrique que $T(W)$, permet de décrire $S$ comme le
produit tensoriel du fibré $E\oplus \overline{E}$ des spineurs de Dirac usuels en dimension $4$ Lorentzienne (le facteur $E$ étant
un espace vectoriel complexe de dimension $2$ muni d'une forme symplectique, et $\overline{E}$ désignant
son espace complexe conjugué), par une somme $S_8=S_8^{+}\oplus S_8^{-}$ de deux espaces vectoriels réels
de dimension $8$.\\
\indent Décrivons les couplages entre les champs qui traduisaient $\Gamma_X$ et les champs de spineurs
$4D$ qui expriment le champ $\psi_X$.\\
Tous ces couplages viennent du tenseur énergie moment de $\psi_X$, et on les voit sur l'opérateur de Dirac,
puisque celui-ci commence avec la dérivation covariante de Levi-Civita $\nabla_X$, se poursuit par la métrique
inverse $\eta_X^{-1}$, qui envoie co-vecteurs
sur vecteurs et se termine par la multiplication de Clifford $\gamma_X$ (matrices de Dirac).\\
La première étape est
\begin{equation}
\nabla_S(\psi_4\otimes \psi_8)=\sum_\mu\nabla^{S}_\mu (\psi_4\otimes \psi_8) dx^{\mu}+\sum_i \nabla^{S}_i (\psi_4\otimes \psi_8) dx^{i}
\end{equation}
où l'exposant $S$ indique le relèvement spin de la connexion.\\
Comme les directions verticales suivant $E_F$ et les directions horizontales de $M$ sont orthogonales pour la métrique
ambiante, les composantes de $\nabla_\mu$ et $\nabla_i$ qui agissent sur les spineurs ne peuvent pas mélanger les indices grecs
et latins, on a donc \emph{a priori} quatre combinaisons des champs qui pourraient agir:\\
1) le $\Gamma_H$, formé des composantes $\Gamma_{\mu\nu}^{\lambda}$, entre dans la dérivée du facteur $S_{1,3}=E\oplus \overline{E}$
et laisse le facteur $S_8$ inchangé; notons que la métrique horizontale intervient aussi dans la dualité pour $\eta_H^{-1}$ sur $M$
et dans les matrices $\gamma_H$;\\
2) le $\Gamma_F$ au contraire, formé des composantes $\Gamma_{ i j}^{k}$, entre dans la dérivée du facteur  $S_8$ et ne touche pas à $S_{1,3}$;
la métrique verticale s'exprime à traversla dualité $\eta_F^{-1}$ et les
matrices gamma Euclidiennes de dimension $8$, qui sont de taille $32\times 32$;\\
3) la somme $C+(D_H/2)$, formée des composantes de $\Gamma$ avec deux indices latins, comme $\Gamma_{\mu i}^{j}$, agit naturellement par sa partie anti-symétrique sur
les indices latins au travers de la représentation spinorielle $\sigma_8$ de $so_8$ dans $S_8$, et par multiplication de
Clifford du vecteur dual de la partie forme sur $S_4$. Notons qu'il s'introduit également  une action scalaire naturelle de la trace de
$C$ et de $D_H$. Nous verrons plus loin que l'algèbre $sl_8$ aussi agit de façon non-triviale.\\
4) La somme $(D_F-M_H)/2$, formée des composantes avec deux indices grecs, comme $\Gamma_{i \mu}^{\nu}$, pourrait agir symétriquement à travers une projection
sur l'algèbre de Lie $so(1,3)$ dans $gl_4(\mathbb{R})$, mais aucune projection invariante pour l'action adjointe de $so(1,3)$
n'existe, par conséquent cette partie des champs n'agit pas directement sur les spineurs. Notons qu'en signature Euclidienne il y aurait une action puisque
l'algèbre de Lie de $so_4$ s'identifie aux matrices anti-symétriques. Par contre, y-compris en Lorentzien, la trace dans $D_F$ pourrait engendrer
\emph{a priori} une action sur les spineurs.\\

Rigoureusement, pour ne pas lier le choix du fibré des spineurs au choix de la métrique, nous devons introduire
un fibré vectoriel auxiliaire $E_V$  de base $X$,
muni d'une métrique de fibré Lorentzienne, et lui attacher un fibré de spineurs
$E^{V}_{S}$ en passant par un
revêtement à deux feuillets $\widetilde{R}_V$ de son fibré des repères orthonormées $R_V$ (en signature $(1,11)$).
Alors une métrique d'univers
est fournie par un \emph{vielbein}, noté $\epsilon$, qui est un isomorphisme du fibré $E_V$ sur le fibré tangent $T(X)$.
A l'aide du vielbein on rappatrie la connexion de Levi-Civita, contenant tous les champs bosoniques, sur $R_V$ donc
sur $\widetilde{R}_V$, et on peut s'en servir pour dériver une section de $E^{V}_{S}$, i.e. un champ de spineurs;
là, on utilise la métrique d'univers pour transformer la partie co-vecteur en vecteur et on réutilise le vielbein
pour transformer ce vecteur en section de $E_V$ afin de pouvoir appliquer la multiplication de Clifford au
niveau de $E_V$ et de $E^{V}_{S}$ et obtenir une nouvelle section de $E^{V}_{S}$.\\
\indent Dans cette approche le potentiel des champs donnés par la métrique $\eta_{ab}$ est le vielbein $\epsilon_\alpha^{a}$,
qui est une sorte de racine carrée de la métrique, puisque, si on considère $\eta$ comme un morphisme
de $T(X)$ vers $T^{*}(X)$, on a
\begin{equation}
\eta=^{t}\epsilon^{-1}\circ\eta_V\circ\epsilon^{-1}
\end{equation}
où $\eta_V$ désigne la métrique fixée sur $E_V$ vue comme morphisme de $E_V$ vers $E^{*}_V$.\\
On a donc
\begin{equation}
\eta^{-1}=\epsilon\circ\eta_V^{-1}\circ(^{t}\epsilon),
\end{equation}
et en coordonnées
\begin{equation}
g^{ab}=\epsilon^{a}_\alpha g^{\alpha\beta}\epsilon_\beta^{b}
\end{equation}
Les degrés de liberté supplémentaires introduits pas $\epsilon$ sont les automorphismes
de $E_V$; ils proviennent naturellement des automorphismes spinoriels de $E_S$.\\
\indent Si $\omega$ est la forme d'une connexion linéaire sur le fibré tangent localement
trivialisé, la forme à valeurs $End(E_V)$ de la connexion sur $E_V$ qui s'en déduit est
\begin{equation}
\omega_V=\epsilon^{-1}(\omega)\epsilon+\epsilon^{-1}d\epsilon.
\end{equation}
On ne considèrera pas la connexion $\omega_V$ comme une variable indépendante; elle sera toujours
contrainte à être le pull-back de la connexion de Levi-civita par le vielbein.\\

\indent Dans notre situation, où un voisinage de $W$ dans $X$ (qu'on peut supposer être $X$ pour le moment)
est fibré par $E_F$ au-dessus de $W$ et où la métrique est donnée
par un champ de métriques Riemanniennes $g_{ij}(x,v)$ sur les fibres, une distribution $M(x,v)$ de
sous-espaces transverses et une métrique Lorentzienne sur cette distribution, il est indiqué de prendre pour $E_V$
le pull-back d'un fibré le long de $W$, ainsi la métrique de Lorentz sur $E_V$ est constante le long des
fibres, ainsi que les matrices de Dirac entre $E_V$ et $End(E^{V}_{S})$. En particulier $E_V$ contient
$E_F$ de manière canonique.\\
Une particularité des morphismes de vielbein mérite d'être signalée: si on utilise le repère mobile
des vecteurs $e_i$ tangents aux fibres et des vecteurs $e_\mu$ tangents à la distribution $M$, le
vielbein n'a pas de partie mixte, c'est-à-dire que l'on $\epsilon_\alpha^{a}=0$ si les deux indices
ne sont pas du même type, où $\alpha$ est dit du type $i$ et noté $\iota,\kappa,...$ s'il est tangent
aux fibres aussi, et du type $\mu$ et noté plutôt $\sigma,\tau,...$ s'il est orthogonal aux fibres
dans le modèle Lorentzien. Cette particulrité vient du fait que $M$ est orthogonal aux fibres de $E_F$.\\

\indent A propos des \emph{unités physiques}. Les quantités qui dépendent de la métrique sur $X$, ou $W$ ou $E_F$ ou $M$, ont une
dimension physique suivant la longueur, qui est définie par la façon dont elles changent lorsqu'on dilate
l'unité de longueur en un point. Par exemple, si on change tous les vecteurs de bases en les multipliant par $\lambda$,
il est naturel que la longueur d'un vecteur soit divisée par $\lambda$; on convient que toute quantité qui
change en étant divisée par $\lambda^{d}$ a une dimension $d$, et on écrit qu'elle est homogène à $L^{d}$. Ainsi
les coordonnées $u^{a}$ d'un vecteur donné sont comme $L$, mais les fonctions $g_{ab}$, quelque soit
la nature grecque ou romaine des indices, sont en $L^{0}$, car le carré de la longueur du vecteur est
donné par la somme des $g_{ab}u^{a}u^{b}$. Donc la dimension d'un tenseur n'est pas visible sur le nombre des
indices en haut et en bas. Les coefficients de Christoffel $\Gamma_{ab}^{c}$ ou $\Gamma_{abc}$, provenant d'une dérivation de $g$ par
rapport aux coordonnées sont de dimension $L^{-1}$ et la courbure est de dimension $L^{-2}$, de même le
tenseur de Ricci $R_{ab}$, la courbure scalaire $R$ et le tenseur d'Einstein $S_{ab}$. \\
\indent En particulier, le tenseur $S_{\mu\nu}$ est en $L^{-2}$. Nous avons séparé les composantes $R^{(4)}_{\mu\nu}$
venant de la métrique sur $M$, métrique horizontale, des autres composantes, formant $T^{(4)}_{\mu\nu}$. Nous constatons que toutes les composantes
sont d'origine géométrique, elles devraient donc posséder la même dimension. Cependant Einstein a écrit son équation
sous la forme $S^{(4)}_{\mu\nu}=\kappa T_{\mu\nu}$, où $T_{\mu\nu}$ est le tenseur énergie-moment de la matière, et plus généralement pour toutes les autres
interactions que la gravitation, en introduisant une constante, dite \emph{gravitationelle}, $\kappa$ qui a la dimension $L^{2}$.
Plus précisément, on a $\kappa=-8\pi G_N /c^{4}$,
où $G_N$ est appelée constante de Newton, et $c$ est la vitesse de la lumière. Il en résulte que la dimension de $T_{\mu\nu}$
est $1/L^{4}$. Une justification de ce décalage de dimension est que les densités d'énergie les plus courantes ont cette dimension là. Par exemple la
composante $T_{00}$ du tenseur associé à un flux de matière a la dimension d'une énergie divisée par un volume, or la dimension
de l'énergie, comme celle d'une composante d'impulsion $p_\mu$ doit être en $L^{-1}$, d'où le $L^{-4}$ en divisant par le volume $L^{3}$.
De même, le champ électromagnétique $F_{\mu\nu}$ est en $1/L^{2}$, puisqu'il s'obtient en dérivant une $1$-forme, qui est
le potentiel élctromagnétique $A_\mu$, de dimension $1/L$, or sa contribution au $T_{\mu\nu}$ est la composée
$F_{\mu\lambda}F^{\lambda}_{\nu}$ privée de sa trace, elle est donc en $1/L^{4}$. Notons que, si le cas du flux de matière est
lié à la dimension
quatre d'espace-temps, à cause de la division par un volume, ce n'est pas vrai pour le champ électromagnétique, qui a cette forme en toute dimension, mis à part
la trace. Dans la théorie que nous proposons, nous introduirons la constante d'Einstein et considérerons donc que $T_{\mu\nu}$
est en $L^{-4}$, mais nous devons souligner il n'y a
pas de raison \emph{a priori} d'attribuer arbitrairement des dimensions autres que $1/L^{2}$ aux composantes des $S_{ab}$ qui
s'interpétrent comme une énergie venant de la matière ou des interactions. Sauf peut-être pour les termes venant des fermions,
qui ont des charges et provoquent des courants.\\
\indent En effet, avec l'électromagnétisme, une nouvelle unité apparaît \emph{a priori} pour la charge électrique,
disons $C$, comme charge ou Coulomb. Le champ électrique et le champ magnétique ont une dimension
$C^{-1}L^{-2}$ donc l'énergie-moment du champ électromagnétique est en $C^{-2}L^{-4}$ et le courant électrique
en $CL^{-3}$. La constante $\epsilon_0$ de l'équation de Coulomb, qui dit que la charge entourée par une
surface est égale au flux du champ électrique multiplié par $\epsilon_0$, est en $C^{2}$. Cependant cette
constante possède une valeur expérimentale, de l'ordre de $10^{-10}$. Il est donc possible de l'utiliser
pour supprimer $C$,
comme on fait avec la vitesse $c$ de la lumière dans le vide pour ramener l'unité de temps à une unité de
longueur. Nous verrons que ceci s'interprète géométriquement en théorie de jauge, où \emph{a priori} une unité
serait nécessaire pour calculer dans l'algèbre de Lie du groupe de jauge, mais où cette unité est canoniquement
définie lorsque le groupe de jauge a un sens et qu'il est compact, ce qui est le cas du cercle $U_1$,
groupe de jauge de l'électromagnétisme.\\
\indent Pour les spineurs $\psi$, au-dessus de $X$ ou $W$, le quadri-vecteur $j$ du courant associé est fait sans dérivée et arrive au second membre d'une dérivée du champ $F$;
comme $j$ est en $L^{-3}$ et que le courant est de degré deux en $\psi$, il est naturel d'attribuer à $\psi$ une dimension
en $L^{-3/2}$. Ceci donne bien du $1/L^{4}$ pour le tenseur énergie-moment matériel \eqref{energiemomentfermion}. Notons que ces unités ne dépendent pas de
la dimension $4$ d'espace-temps non-plus, car une dérivée d'une courbure $F_{ab}$ est toujours en $L^{-3}$. Ceci suffirait à justifier que les
$T_{ab}$ de la matière soient en $1/L^{4}$, et qu'il faille introduire entre $S_{ab}$ et $T_{ab}$ une constante $\kappa$ ayant la dimension $L^{2}$.\\
Remarquons que le courant fait intervenir une dualité symplectique arbitraire $\omega_A^{B}$ entre spineurs: si $\psi=(\psi_+,\psi_-)$
on a $j=Re^{t}\gamma(\omega(\psi_+)\otimes \psi_-)$. Or $\omega_A^{B}$ pourrait avoir une dimension; cependant, au
moins en dimension quatre ou douze, la métrique $g$ est naturellement
proportionnelle à $\omega$, d'où une dimension $L^{0}$ pour $\omega$. De même, l'application $\gamma$ de Dirac n'apporte
pas de dimension supplémentaire puisque par définition
les carrés des matrices $\gamma(u)$ sont les carrés des longueurs des vecteurs $u$.
On obtient donc une même dimension $L^{-3/2}$ pour les $\psi$ et les $\psi^{*}$.\\
Par ailleurs les covecteurs d'espace-temps donnent naturellement des formes hermitiennes sur l'espace des spineurs $S$; et
pour les covecteurs d'énergie positive, dont l'orthogonal est du genre espace, cela donne un produit scalaire Euclidien défini positif.
Il est remarquable que ce soit seulement en dimension trois d'espace que l'intégrale du carré du produit des
spineurs donne un nombre, dont l'interprétation physique est la probabilité de présence d'une particule fermionique.
Dans toute autre dimension, il y a bien un produit scalaire invariant Lorentzien, défini positif sur les sections
de $S$ mais il ne correspond pas au volume Euclidien usuel.\\
La constante $c_0$ du Lagrangien \eqref{lagrangienhilbertdirac} est la constante $\kappa$ d'Einstein (\cite{einstein1952principle}). Dans les unités habituelle de
la physique des particules $c=\hbar=1$, l'unité de longueur est le $GeV^{-1}$, la constante $\kappa$ est en $GeV^{-2}$, et elle est très petite, de l'ordre
de $10^{-38} GeV^{-2}$; ce qui donne une idée de l'échelle à laquelle les effets gravitationnels
deviennent prépondérant, c'est $10^{-19} GeV^{-1}$, appelée \emph{longueur de Planck}. \\
En prenant cette longueur pour unité, on retirerait toutes les dimensions. Nous ne le ferons pas, car
la dimension $L$ est essentielle pour la discussion du développement limité des équations de champs
au voisinage de $W$, et aussi parce que l'interprétation physique comme théorie effective repose sur
des grandeurs d'énergie et de masse mesurées en $GeV$.\\
La petitesse de $\kappa$ justifie l'existence d'une Physique des particules à notre échelle, sans trop d'effet de la
gravitation à quatre dimensions, où on peut négliger aussi les effets de la déformation verticale $D_F$, mais où il faut tenir compte des effets
du champ de déformation horizontal $D_H$, de la
connexion $C$, de la monodromie $M$ et de la gravitation verticale $\Gamma_F$. \\

\indent Ceci nous invite à négliger les effets gravitationnels $4D$ dans un premier temps, où
l'on considère les trois premiers ordres du développement limité des équations des champs,
car on sait que à cet ordre, les énergies mises en jeu dans les interactions électromagnétiques, faibles et
fortes sont inférieures à $10000GeV$, ce qui laisse un facteur $10^{15}$ avant de trouver des difficultés dues
à la courbure $S^{(4)}_{\mu\nu}$. Cependant nous allons
constater qu'il ne faut pas négliger les effets gravitationnels transverses $8D$, car ils vont contribuer
aussi à la Physique en dessous de dix milles $GeV$. On apperçoit là une raison de la division de l'espace en un tangent
et un transverse le long de $W$, c'est que $W$ a très peu de courbure par rapport aux fibres. Notons que ceci contraste
fortement avec les hypothèses habituelles d'une théorie de membranes, cf. \cite{maartens2015standard}.\\
La possibilité de s'appuyer sur
les représentations unitaires irréductibles du groupe de Poincaré pour décrire les états des particules en
dimension $4$ repose sur la relative platitude du $g_{\mu\nu}$ à nos échelles d'énergie. Il n'en vas pas de
même pour l'univers à douze dimensions générique, parce que l'on ne sait pas encore comment
bien traiter la
gravitation du point de vue quantique.\\

\indent Ecrivons en coordonnées locales au voisinage d'un point de $W$, sans approximation pour le moment,
l'opérateur de Dirac $D_{S}$ sur les sections du fibré pull-back $E^{S}_V$, trivialisé par les $e_i$ du long de $W$.
D'abord, dans $X$ les petites lettres latines du début de l'alphabet désignent les coordonnées d'espace-temps,
les petits lettres du début de l'alphabet grec font référence au fibré auxiliaire Lorentzien $E_V$. Les grandes
lettres $I,J,...$ concernent pour le moment les spineurs $S_{12}$ associés au repère tangent de $X$ et les grandes lettres $A,B,...$ concernent les
spineurs $S_{12}$ associés aux repères venant du modèle Lorentzien $V$.
Il vient alors
\begin{multline}
D_S(\psi)^{C}=\sum_{\alpha,a,b,B}\gamma_{\alpha,B}^{C}
(\epsilon^{-1})^{\alpha}_b\eta^{ab}(\partial_a\Psi^{B}\\+\sum_{J}(\epsilon^{-1})_J^{B}\sum_{A}((\sum_I\Gamma_{a,I}^{J}\epsilon_A^{I}+\partial_a\epsilon_A^{J})\Psi^{A}).
\end{multline}

Maintenant tenons compte de la fibration de l'univers autour de $W$. Les petits indices latins $i,j,k,...$
et les petits indices grecs $\lambda,\mu,\nu,...$ du milieu de l'alphabet renvoient au repère mobile sur $X$ dépendant
de la métrique (mais cela ne gêne pas car le contenu géométrique de $\epsilon$ demeure, il s'agit uniquement d'une écriture
pratique); les indices $\eta,\iota,\kappa,...$ et $\alpha,\beta,\gamma,...$ renvoient à la décomposition correspondante de $V$;
les grandes lettres $I,J,K,H,...$ et $A,B,C,...$ sont pour les spineurs qui sont respectivement côté $F$ et côté $M$
dans $X$, et les lettres droites $\mathbf{I},\mathbf{J},\mathbf{K},\mathbf{H},...$ et $\mathbf{A},\mathbf{B},\mathbf{C},...$
sont pour les spineurs correspondant dans le modèle Lorentzien. A présent les spineurs ont un double exposant
du type $(\mathbf{A},\mathbf{J})$.\\
Comme les indices $\alpha,a,b,I,J,A,B,C$ peuvent à présent prendre chacun deux types différents on doit attendre
$2^{7}=128$ composantes dans l'écriture développée de $D_S$. Heureusement, le choix des repères et la sectorisation
des matrices gamma de Dirac réduit le nombre de composantes à quatre:
\begin{multline}
(D_S(\psi))^{\mathbf{C},\mathbf{K}}=\\\sum_{\alpha,\mu,\nu,\mathbf{B}}\gamma_{\alpha,\mathbf{B}}^{\mathbf{C}}
(\epsilon^{-1})^{\alpha}_\nu \eta^{\mu\nu}(\partial_\mu\Psi^{\mathbf{B},\mathbf{K}}+
\sum_{B}(\epsilon^{-1})_B^{\mathbf{B}}\sum_\mathbf{A}(\sum_{A}\Gamma_{\mu,A}^{B}\epsilon_\mathbf{A}^{A}+\partial_\mu\epsilon_\mathbf{A}^{B})\Psi^{\mathbf{A},\mathbf{K}})\\
+\sum_{\alpha,\mu,\nu,\mathbf{J}}\gamma_{\alpha,\mathbf{J}}^{\mathbf{K}}
(\epsilon^{-1})^{\alpha}_\nu \eta^{\mu\nu}(\partial_\mu\Psi^{\mathbf{C},\mathbf{J}}+
\sum_{J}(\epsilon^{-1})_J^{\mathbf{J}}\sum_{\mathbf{I}}(\sum_I\Gamma_{\mu,I}^{J}\epsilon_\mathbf{I}^{I}+\partial_\mu\epsilon_\mathbf{I}^{J})\Psi^{\mathbf{C},\mathbf{I}})\\
+\sum_{\iota,i,j,\mathbf{B}}\gamma_{\iota,\mathbf{B}}^{\mathbf{C}}
(\epsilon^{-1})^{\iota}_jg^{ij}(\partial_i\Psi^{\mathbf{B},\mathbf{K}}+\sum_{B}(\epsilon^{-1})_B^{\mathbf{B}}
\sum_{\mathbf{A}}(\sum_A\Gamma_{i,A}^{B}\epsilon_\mathbf{A}^{A}+\partial_i\epsilon_\mathbf{A}^{B})\Psi^{\mathbf{A},\mathbf{K}})\\
+\sum_{\iota,i,j,\mathbf{J}}\gamma_{\iota,\mathbf{J}}^{\mathbf{K}}
(\epsilon^{-1})^{\iota}_jg^{ij}(\partial_i\Psi^{\mathbf{C},\mathbf{J}}+\sum_{J}(\epsilon^{-1})_J^{\mathbf{J}}
\sum_{\mathbf{I}}(\sum_I\Gamma_{i,I}^{J}\epsilon_\mathbf{I}^{I}+\partial_i\epsilon_\mathbf{I}^{J})\Psi^{\mathbf{C},\mathbf{I}})
\end{multline}

Si on tient compte des ordres de grandeur, i.e. le fait que le champ de gravitation horizontal peut être négligé dans
le développement limité à l'ordre $n$ inférieur à $5$ en les coordonnées transversales,
toutes les parties dynamiques où $\psi$ est dérivé doivent
être conservées mais seuls le deuxième et le quatrième termes font intervenir des champs $\Gamma$ et $\partial\epsilon$
qui ne sont pas négligeables. De plus, seuls les troisième et quatrième termes dynamiques subissent des variations
à ces ordres là.\\

\subsection{Développement limité; équations de Dirac et Yang-Mills-Higgs}

Analysons ordre par ordre en la section $v$ du fibré vetoriel $E_F$ comment se manifestent les divers champs aux yeux des habitants
du monde apparent $W$.\\

La série de Taylor en $F$ des champs de spineurs le long de $W$ est simple; elle transforme l'unique champ $\psi_X$
en une série de spineurs de Dirac sur $W$ de dimension $4$ à valeurs dans le fibré
$S^{*}(E_F^{*})\otimes S_8$ des polynomes sur les fibres à valeurs dans $S_8$:
\begin{equation}
\psi(x,v)=\psi^{(0)}(x)+\sum_iv^{i}\psi_i^{(1)}(x)+\sum_{ij}v^{i}v^{j}\psi_{ij}^{(2)}(x)+...
\end{equation}
Puisqu'à ce jour, l'expérience n'a montré que trois générations de fermions, aller au-delà de l'ordre trois
serait prématuré.\\
Remarque: à première vue il y aurait des familles à huit, puis soixante-quatre groupes de fermions, mais, comme
nous allons le montrer les transformation de jauges ne laissent qu'un groupe à chaque génération en fixant un
vecteur $v$ dans chaque fibre, grâce à un mécanisme analogue à celui qu'ont décrit Brout, Englert et Higgs.\\

\noindent De même le développement de Taylor de $G_H$ est
\begin{equation}
g_{\mu \nu}(x,v)=g^{(0)}_{\mu \nu}(x)+\sum_kg^{(1)}_{k \mu \nu}(x)v^{k}+\sum_{k_1 k_2}g^{(2)}_{k_1 k_2 i \mu \nu}(x)v^{k_1}v^{k_2}+ ....;
\end{equation}
d'où l'on déduit celui de $D_F$:
\begin{equation}
D_{i\mu \nu}(x,v)=D_{i\mu \nu}(x)+\sum_kD_{k i \mu \nu}(x)v^{k}+\sum_{k_1 k_2}D_{k_1 k_2 i \mu \nu}(x)v^{k_1}v^{k_2}+ ....
\end{equation}

\noindent La métrique verticale $G_F$ se développe également:
\begin{equation}
g_{i j}(x,v)=g^{(0)}_{i j}(x)+\sum_kg^{(1)}_{k i j}(x)v^{k}+\sum_{k_1 k_2}g^{(2)}_{k_1 k_2 i j}(x)v^{k_1}v^{k_2}+ ....;
\end{equation}
d'où l'on déduit le développement de $D_H$:
\begin{equation}
D_{\mu i j}(x,v)=D_{\mu i j}(x)+\sum_kD_{\mu i j k}(x)v^{k}+\sum_{k_1 k_2}D_{\mu i  j k_1 k_2 }(x)v^{k_1}v^{k_2}+ ....
\end{equation}

\noindent Par difféomorphismes nous avons fixé la métrique verticale le long de $W$, et en imposant l'annulation des $\Gamma_{k j}^{i}$
en plus, ce qui fixe la métrique dans la fibre à l'ordre un, car $\partial_i g_{jk}=(\nabla _i e_j| e_k)+(e_j|\nabla _i e_k)$,
donc, en utilisant un difféomorphisme ambiant fixant $W$, on assure que le développement de $D_H$ ne commence qu'à l'ordre un.\\

\noindent En fait, il est bien connu que, dans ce cas, le tenseur de courbure $R_{ijk}^{l}$ permet d'exprimer simplement
le tenseur métrique (cf. \cite{cartan1946leccons}, \cite{green1987superstring}):
\begin{equation}\label{metriqueverticalecourbure}
g_{i j}(x,v)=g^{(0)}_{i j}(x)-\frac{1}{3}R_{ikjl}(x)v^{k}v^{l}-\frac{1}{6}(D_mR_{ikjl})(x)v^{m}v^{k}v^{l}+...;
\end{equation}
où l'on a posé $R_{ijkl}=\sum_nR_{ijk}^{n}g_{nl}$, et où $g^{(0)}_{i j}(x)$ est la matrice identité pour tout $x$.\\

\noindent On en déduit un développement des $D_\mu g_{ij}$ et un développement des symboles de Christoffel
\begin{equation}
\Gamma_{i j}^{k}(x,v)=\sum_{l}\Gamma_{i j; l}^{k}(x)v^{l}+...+\frac{1}{n!}\sum_{k_1,...,k_n}\Gamma_{i j; K}^{k}(x)v^{K}+o(\| v\|^{n}
\end{equation}
Si bien que $\Gamma_F$ intervient
à l'ordre $1$ par le polynome à valeurs connexions verticales unitaires et sans torsion
\begin{equation}
\sum_{l}\Gamma_{i j; l}^{k}(x)v^{l}
\end{equation}

\noindent Remarque importante pour la suite: il reste encore beaucoup de degrés de liberté donnés par les
difféomorphismes aux ordres supérieurs, mais il en reste aussi à l'ordre $1$, en effet toute transformation de
jauge pour le groupe orthogonal $SO_8$ des fibres respecte la condition que nous venons de dire: $g_{ij}$ fixée
à l'ordre un compris. C'est le point de départ de la seconde section.\\

Afin de déterminer ce que font $M$ et $C$, on doit faire un développement limité vertical à l'ordre $n$ des $a_\mu^{i}$,
qui rappelons-le, s'annulent en $0$ dans $\mathbb{R}^{8}$:
\begin{multline}
a_\mu^{i}(x,v)=\sum_ka_{\mu k}^{i}(x)v^{k}+\sum_{k,l}a_{\mu k l}^{i}(x)v^{k}v^{l}+...\\+\sum_{k_1,...,k_n}a_{\mu K}^{i}(x)v^{K}+o(\| v\|)^{n}.
\end{multline}
Nous obtenons une suite de champs de tenseurs sur $W$, qui sont des co-vecteurs (indice $\mu$) à valeurs vecteurs (indice $i$)
de formes multi-linéaires symétriques (indices $k_j$) sur
le fibré $E_F$, considéré comme
la restriction du fibré tangent vertical $T_V(E_F)$ à $W$. Avec les $(D_\mu g_{ij})_K$, ce seront les générations de bosons de la théorie de monodromie non-linéaire
qui étend la
théorie de jauge non-abélienne aux ordres supérieurs en la variable transversale $v$. \\

\indent La fonction $a_{\mu k}^{i}=\partial_ka_\mu^{i}$ n'est autre que la restriction de la composante $C_{\mu k}^{i}$ du
champ de connexion $C_F$ à $W$. De façon plus générale, le développement de $C_{\mu k}^{i}(x,v)$ est celui
de $a_\mu^{i}(x,v)$ décalé d'un cran:
\begin{equation}
C_{\mu k}^{i}(x,v)=a_{\mu k}^{i}(x)+\sum_{l}a_{\mu k l}^{i}(x)v^{l}+...\\+\sum_{l_1,...,l_n}a_{\mu k L}^{i}(x)v^{L}+o(\| v\|)^{n}.
\end{equation}

Il y a aussi un développement du champ de monodromie correspondant,
\begin{equation}
M_{\mu\nu}^{i}=\sum_{m=1}^{n}\sum_{k_1,...,k_m}M_{\mu\nu K}^{i}(x)v^{K}+o(\| v\|^{n}.
\end{equation}
En tenant compte de l'expression $D_\mu=\partial_\mu+\sum_ka_\mu^{k}\partial_k$, les deux premiers termes sont les suivants: à l'ordre $1$ en $v$,
le coefficient de $v^{k}$ est
\begin{equation}\label{courburedecf}
M_{\mu\nu k}^{i}=\partial_\mu a_{\nu k}^{i}-\partial_\nu a_{\mu k}^{i}+\sum_j(a^{j}_{\mu k}a_{\nu j}^{i}-a^{j}_{\nu k}a_{\mu j}^{i});
\end{equation}
qui correspond à la formule d'une courbure de connexion principale le long de $W$. Puis, à l'ordre $2$ en $v$ le coefficient
de $v^{k_1}v^{k_2}$ est
\begin{multline}
M_{\mu\nu k_1 k_2}^{i}=\partial_\mu a_{\nu k_1 k_2}^{i}-\partial_\nu a_{\mu k_1 k_2}^{i}
+\sum_j(a_{\mu k_1 k_2}^{j}a_{\nu j}^{i}-a_{\nu k_1 k_2}^{j}a_{\mu j}^{i})\\
+\sum_j(a^{j}_{\mu k_1}a_{\nu j k_2}^{i}-a^{j}_{\nu k_1}a_{\mu j k_2}^{i})
+\sum_j(a_{\mu k_2}^{j}a_{\nu j k_1}^{i}-a_{\nu k_2}^{j}a_{\mu j k_1}^{i}).
\end{multline}
C'est une $2$-forme à valeurs dans le fibré des morphismes d'un fibré vectoriel dans un autre, donc
pas directement une courbure. Cependant, on a, même en dehors de $W$, la formule suivante
\begin{multline}\label{courburedecf}
M_{\mu\nu k}^{i}(x,v)=D_\mu C_{\nu k}^{i}(x,v)-D_\nu C_{\mu k}^{i}(x,v)\\
+\sum_j(C^{j}_{\mu k}(x,v)C_{\nu j}^{i}(x,v)-C^{j}_{\nu k}(x,v)C_{\mu j}^{i})(x,v);
\end{multline}
où, de manière intrinsèque, $C_F$ définit une $1$-forme à valeurs dans le fibré $End(E_F)$:
\begin{equation}
C_F=\sum_\mu \sum_{i,k}C_{\mu k}^{i}(x,v)dx^{\mu}e_i\otimes e^{k}
\end{equation}

\noindent Remarque: le couplage avec la métrique n'est
pas le simple couplage avec la métrique à l'ordre zéro, parce qu'à chaque ordre en $v$ dans $E_F$
la déformation $\partial_i g_{\mu\nu}$ et ses dérivées successives interviennent \emph{a priori} pour évaluer
$g_{\mu\nu}(x,v)$:
\begin{equation}
g_{\mu\nu}(x,v)=g_{\mu\nu}(x)+\sum_kD_{i\mu\nu}(x)v^{i}+\frac{1}{2}\sum_{i,j}\partial_j D_{i\mu\nu}(x)v^{i}v^{j}+...
\end{equation}
Le champ de Levi-civita $\Gamma_H$ à l'ordre zéro est le champ de gravitation en dimension quatre proposé par Einstein;
pour avoir les ordres suivants on développe
\begin{multline}
\Gamma_{\mu\nu}^{\lambda}=\frac{1}{2}\sum_\rho g^{\rho \lambda}(D_\mu g_{\nu\rho}+D_\nu g_{\mu\rho}-D_\rho g_{\mu\nu})\\
=\frac{1}{2}\sum_\rho g^{\rho \lambda}(\partial_\mu g_{\nu\rho}+\partial_\nu g_{\mu\rho}-\partial_\rho g_{\mu\nu}\\
+\sum_i (a_\mu^{i}\partial_ig_{\nu\rho}+a_\nu^{i}\partial_ig_{\mu\rho}-a_\rho^{i}\partial_ig_{\mu\nu})).
\end{multline}
Par ailleurs
\begin{equation}
\partial_ig^{\rho\lambda}=-\sum_{\kappa\chi}g^{\rho\kappa}\partial_i g_{\kappa\chi}g^{\chi \lambda},
\end{equation}
et
\begin{multline}
\partial^{2}_{ij}g^{\rho\lambda}=-\sum_{\kappa\chi}g^{\rho\kappa}\partial^{2}_{ij} g_{\kappa\chi}g^{\chi \lambda}\\
+\sum_{\alpha\beta\kappa\chi}g^{\alpha \rho}\partial_j g_{\alpha\beta}g^{\beta\kappa}\partial_i g_{\kappa\chi}g^{\chi \lambda}\\
+\sum_{\alpha\beta\kappa\chi}g^{\rho\kappa}\partial_i g_{\kappa\chi}g^{\chi\alpha}\partial_j g_{\alpha\beta}g^{\beta\lambda}.
\end{multline}
Donc, à l'ordre $1$ en $v$, on a
\begin{multline}\label{deformationgravitation}
\Gamma_{\mu\nu}^{\lambda}(x,v)=\Gamma_{\mu\nu}^{\lambda}(x)+
\frac{1}{2}\sum_kv^{k}\sum_\rho g^{\rho\lambda}
(\partial_\mu D_{k \nu\rho}(x)+\sum_ia_{\mu i}^{k}D_{i\nu\rho}(x)\\+\partial_\nu D_{k \mu\rho}(x)+\sum_ia_{\nu i}^{k}D_{i\mu\rho}(x)
-\partial_\rho D_{k \mu\nu}(x)+\sum_ia_{\rho i}^{k}D_{i\mu\nu}(x))\\-\frac{1}{2}\sum_kv^{k}\sum_\kappa D_{k \kappa}^{\lambda}(x)\Gamma_{\mu\nu}^{\kappa}(x)+o(\| v\|).
\end{multline}
On reconnait sur $D_F$ l'action naturelle de la connexion $C$, ainsi que celle de $\Gamma_H$.\\
A l'ordre suivant, il s'ajoute trois termes en $v^{i}v^{j}$:
\begin{equation}
\frac{1}{2}\sum_\rho g^{\rho\lambda}(\nabla_\mu^{C}(D)_{ij \nu \rho}+...)-\frac{1}{2}\sum_\kappa D_{ij \kappa}^{\lambda}\Gamma_{\mu\nu}^{\kappa}
+\frac{1}{2}\sum_\rho g^{\rho\lambda}(\nabla_{\mu i}^{C}(D)_{j \nu \rho}+...),
\end{equation}
où le dernier terme représente la dérivation covariante pour l'extension de la connexion $C$ au second ordre.\\

\noindent Cette complication ne jouera aucun rôle pour l'instant, ca nous allons restreindre notre étude au cas où
toutes les dérivées de $G_H$ sont négligeables:\\

\noindent \emph{Axiome d'approximation plate}: les $\partial_\kappa g_{\mu\nu}$ et les $\partial_ig_{\mu\nu}$ sont supposées
s'annuler à un ordre suffisamment élevé pour ne pas intervenir aux énergies qu'on considère.\\

\noindent Ainsi, le terme $S^{(4)}_{\mu\nu}$ disparait de l'équation d'Einstein, cependant on ne peut pas en déduire
que le tenseur d'énergie-impulsion $T_{\mu\nu}$ est nul, même en y incluant la partie venant des fermions, à cause de
la constante $c_0$ en $L^{2}$ qui se trouve devant lui et qui est presque nulle. Les équations dynamiques se déduisent alors
de la contrainte de divergence covariante nulle (conservation de l'énergie et de l'impulsion):
\begin{equation}
\sum_\nu D_\nu T_\mu^{\nu}=0.
\end{equation}

A cette contrainte nous en ajouterons une deuxième qui ne porte sur le choix
des coordonnées, qui fixe une unité de longueur, et qui simplifie les calculs, en suivant Einstein
\cite{einstein1952principle}, 1916, 1919:\\

\noindent \emph{Choix unimodulaire}: les coordonnées sont choisies pour que le déterminant total $det G$
soit égal à $-1$.\\

\noindent Ainsi le terme en dérivée logarithmique de $det G_F$ se compense avec le terme en dérivée logarithmique de $det G_H$
et disparait de $T_{\mu\nu}$. Comme l'a souligné Einstein en 1916, le choix unimodulaire est pratique mais ne réclame pas une hypothèse en plus,
il est légitime dans une théorie covariante pour tous les difféomorphismes; cependant, évidemment, le choix unimodulaire réduit la covariance aux
difféomorphismes préservant le volume, par conséquent dans la suite nous n'utiliserons pas d'autres transformations.
Pour le rapport avec la
constante cosmologique, qui correspond à une théorie strictement plus générale, avec une constante en plus,
voir \cite{henneaux1989cosmological}, \cite{ngvadam1991unimodular}, \cite{finkelstein2001unimodular},\cite{smolin2009unimodular},
\cite{bufalo2015unimodularquantum}. \\

\noindent Remarque: une contrainte plus faible qui simplifie autant les calculs sans fixer d'unité de longueur
ou de volume, consiste à demander seulement que $det G$ soit constant; on pourrait la nommer \emph{choix équimodulaire}.\\

\noindent \emph{Autres choix}: de la même façon, le choix d'un fibré $E_F$ et le choix de coordonnées le long de
$W$ garantissant que la métrique $g_{ij}$ coïncide avec l'identité jusqu'au premier ordre transversal, est autorisé par
la covariance générale, et restreint d'autant le groupe des difféomorphismes.\\

Compte tenu de ces simplifications, calculons le tenseur $T^{(4)}_{\mu\nu}$ à l'ordre inférieur ou égal à $2$ en $v$.\\

\noindent Les dérivées $D_\mu g_{ij}$ étant d'ordre deux, à l'ordre deux on a
\begin{multline}
T^{(4)}_{\mu\nu}\thicksim \frac{1}{2}\sum_i(D_\mu C_{\nu i}^{i}+D_\nu C_{\mu i}^{i})
+\frac{3}{4}\sum_{i j k l}C_{\mu j}^{l}g_{lk}C_{\nu i}^{k}g^{ji}+\frac{1}{4}\sum_{i k}C_{\mu i}^{k}C_{\nu k}^{i}\\
+\frac{1}{2}\sum_{i j k}(C_{\mu i}^{k}(D_\nu g_{kj})g^{ij}
+C_{\nu i}^{k}(D_\mu g_{jk})g^{ij})
+\frac{1}{2}\sum_{\kappa \lambda i}g^{\kappa\lambda}M_{\mu\kappa}^{i}M_{\nu\lambda i}
\end{multline}
La première ligne donne des termes non-nuls à partir de l'ordre zéro, la seconde ligne n'en donne qu'à partir de
l'ordre deux:
\begin{multline}\label{energietrace}
\frac{1}{2}\sum_i(D_\mu C_{\nu i}^{i}+D_\nu C_{\mu i}^{i})\thicksim \frac{1}{2}\sum_i(\partial_\mu C_{\nu i}^{i}
+\partial_\nu C_{\mu i}^{i})\\+\frac{1}{2}\sum_{ik}v^{k}(\partial_\mu \partial_kC_{\nu i}^{i}
+\partial_\nu \partial_kC_{\mu i}^{i})+\frac{1}{2}\sum_{ijk}v^{k}(C_{\mu k}^{j}\partial_jC_{\nu i}^{i}+C_{\nu k}^{j}\partial_jC_{\mu i}^{i})\\
+\frac{1}{2}\sum_{ikl}v^{k}v^{l}(\partial_\mu \partial_k\partial_lC_{\nu i}^{i}+\partial_\nu \partial_k\partial_lC_{\mu i}^{i})
+\frac{1}{2}\sum_{ijkl}v^{k}v^{l}(C_{\mu k}^{j}\partial_j\partial_lC_{\nu i}^{i}+C_{\nu k}^{j}\partial_j\partial_lC_{\mu i}^{i})\\
+\frac{1}{2}\sum_{ijk}v^{k}v^{l}
(\partial_lC_{\mu k}^{j}\partial_jC_{\nu i}^{i}+\partial_lC_{\nu k}^{j}\partial_jC_{\mu i}^{i})+...;
\end{multline}
où tous les $C_{\mu k}^{i}$ et leurs dérivées sont des fonctions de $x$ seul, le long de $W$.\\

\begin{multline}\label{energielorentzcouplage}
\frac{3}{4}\sum_{i j k l}C_{\mu j}^{l}g_{lk}C_{\nu i}^{k}g^{ji}+\frac{1}{4}\sum_{i k}C_{\mu i}^{k}C_{\nu k}^{i}\thicksim
\frac{3}{4}\sum_{i j}C_{\mu i}^{j}C_{\nu i}^{j}+\frac{1}{4}\sum_{i j}C_{\mu i}^{j}C_{\nu j}^{i}\\
+\frac{3}{4}\sum_{i j k}v^{k}(\partial_kC_{\mu i}^{j}C_{\nu i}^{j}+C_{\mu i}^{j}\partial_kC_{\nu i}^{j})
+\frac{1}{4}\sum_{i j k}v^{k}(\partial_kC_{\mu i}^{j}C_{\nu j}^{i}+C_{\mu i}^{j}\partial_kC_{\nu j}^{i})\\
+\frac{3}{4}\sum_{i j k l}v^{k}v^{l}(\partial_k\partial_lC_{\mu i}^{j}C_{\nu i}^{j}+C_{\mu i}^{j}\partial_k\partial_lC_{\nu i}^{j})
+\frac{1}{4}\sum_{i j k l}v^{k}v^{l}(\partial_k\partial_lC_{\mu i}^{j}C_{\nu j}^{i}+C_{\mu i}^{j}\partial_k\partial_lC_{\nu j}^{i})\\
+\frac{3}{2}\sum_{i j k l}v^{k}v^{l}\partial_kC_{\mu i}^{j}\partial_lC_{\nu i}^{j}
+\frac{1}{2}\sum_{i j k l}v^{k}v^{l}\partial_kC_{\mu i}^{j}\partial_lC_{\nu j}^{i}\\
+\frac{1}{4}\sum_{i j k l m}v^{k}v^{l}R_{ i k j l}C_{\mu i}^{m}C_{\nu j}^{m}-\frac{1}{4}\sum_{i k l m n}v^{k}v^{l}R_{ m k n l}C_{\mu i}^{m}C_{\nu i}^{n}
\end{multline}

\begin{multline}\label{energielineaire}
\frac{1}{2}\sum_{i j k}(C_{\mu i}^{k}(D_\nu g_{kj})g^{ij}
+C_{\nu i}^{k}(D_\mu g_{jk})g^{ij})\\\thicksim
-\frac{1}{6}\sum_{i j k l}v^{k}v^{l}(C_{\mu i}^{j}\partial_\nu R_{i k j l}+C_{\nu i}^{j}\partial_\mu R_{i k j l}).
\end{multline}
Et enfin
\begin{equation}\label{energieyangmills}
\frac{1}{2}\sum_{\kappa \lambda i}g^{\kappa\lambda}M_{\mu\kappa}^{i}M_{\nu\lambda i}\thicksim
\frac{1}{2}\sum_{\kappa \lambda i k l}v^{k}v^{l}g^{\kappa\lambda}M_{\mu\kappa k}^{i}M_{\nu\lambda l}^{i}.
\end{equation}\\

\indent Jusqu'ici les coordonnées $v^{j};j=1,...,8$ sont arbitraires, elles ne servent qu'à décrire le développement limité
des champs et des équations génératrices, par conséquent l'application qui envoie le vecteur $v$ de
coordonnées $v^{1},...,v^{8}$ sur $\sum_iv^{i}C_{\mu i}^{j}$ s'identifie très bien avec la matrice $C_{\mu i}^{j}$.
De même l'appplication qui à $v^{1},...,v^{8}$ associe $\sum_iv^{i}M_{\mu \lambda i}^{j}$ s'identifie très bien avec la matrice $M_{\mu \lambda i}^{j}$.\\
Nous admettrons donc que le contenu en "particules" de la théorie engendrée par le tenseur d'Einstein correspond aux champs $C$, $M$,
$\Gamma_H$, $\Gamma_F$, $D_F$ et $D_H$, vus comme coefficients de la série de Taylor, indépendamment du déplacement infinitésimal $v$.\\
\indent Cependant il arrivera bientôt que $v$ prenne une valeur déterminée; et alors un terme comme $v^{i}v_jC_{\mu i}^{j}$ représentera
un co-vecteur ordinaire et un terme $v^{k}v_iM_{\mu\kappa k}^{i}$ représentera
une $2$-forme différentielle scalaire. Les termes correspondants de $T_{\mu\nu}^{(4)}$ s'interprèteront comme des termes d'interaction
entre les champs $C$, $M$, etc. et un nouveau champ $v$. Nous expliquerons pourquoi il est légitime d'identifier ce nouveau
champ "de déplacement" au champ de Higgs $H$, à valeurs dans le fibré auxiliaire $E_F$, et nous poserons $H=M_0v$.\\

\indent L'équation \eqref{energieyangmills} fait intervenir les composantes de $v$; ce n'est donc pas celle du tenseur énergie-impulsion
de Yang-Mills du champ $C$, qui s'écrit comme la trace
\begin{equation}\label{energieyangmillsstandard}
T_{\mu\nu}^{YM}=\frac{1}{2}\sum_{\kappa \lambda i k}g^{\kappa\lambda}M_{\mu\kappa k}^{i}M_{\nu\lambda i}^{k}.
\end{equation}
Cependant, pour des fluctuations gaussiennes réduites centrées de $v$ autour d'un équilibre, comme cela arrive au premier ordre pour
un champ de Higgs standard, la moyenne de \eqref{energieyangmills} donne bien \eqref{energieyangmillsstandard}.\\
C'est ainsi que tous les termes de $T_{\mu\nu}$ seront traduits: premièrement, avec $v=H/M_0$ un terme d'interaction avec le "nouveau champ"
$H$, deuxièmement, en moyenne, comme un terme intrinsèque des champs $C$, $M$, etc. Remarquons toutefois que les termes quadratiques
intrinsèques pourraient être trouvés directement en partant de l'interprétation des champs comme des particules dans $W$, générateurs des
équations de champs libres plus un couplage minimal, selon les règles usuelles de la Théorie Quantique des Champs, et que les termes
de degré un ou deux sans dérivée pourraient être négligés, comme ne contribuant pas à la dynamique.\\

\indent Donc, si on soustrait la trace on reconnaît dans l'équation \eqref{energieyangmills} le tenseur
d'énergie-impulsion de Yang-Mills
pour une théorie de jauge associée au groupe linéaire $GL_8(\mathbb{R})$. Toutefois, à ce stade, la contrainte sur
$g_{ij}$ le long de $W$ n'autorise comme groupe de jauge que le sous-groupe orthogonal $SO_8(\mathbb{R})$, on doit donc décomposer les matrices
$C_{\mu i}^{j}$ en leurs parties symétriques $S_{\mu i}^{j}$ et antisymétriques $A_{\mu i}^{j}$. Seule la partie
antisymétrique peut être considérée comme une forme de connexion principale,
l'autre partie doit être considérée comme une sorte de champ de Higgs, qui est en l'occurence une $1$-forme à valeurs dans un fibré associé
au fibré principal. Par orthogonalité le terme \eqref{energieyangmills} se découple en
un terme quadratique en la courbure de $A$, qui donne l'équation de Yang-Mills ordinaire, et en plusieurs termes
quadratiques en $S$ et $B=Tr(S)$
et leurs dérivées
covariantes horizontales suivant $A$, qui donnent des équations de Higgs du type de Hitchin, cf. \cite{atiyahhitchin1988geometrymonopoles}.\\
Nous verrons que les particules $W$ et $Z$, ainsi que le photon, pour une part au-moins, proviennent naturellement de ces champs de Higgs-Hitchin.
Il ne faut donc pas les confondre avec le champ de Higgs $H$ qui fut observé en $2012$, dont l'origine est plus subtile, faisant intervenir
dans notre modèle le déplacement $v$ dans $X$ et deux des champs de fermions, comme nous allons le voir, mais ceci est l'histoire des chapitres suivants.\\

\noindent Les termes de \eqref{energietrace} concernent la trace de la partie symétrique $C_{\mu i}^{i}$, qui sera le champ noté
traditionnellement $B$.
Le premier et le troisième termes sont linéaires; ce sont les dérivées covariantes de cette trace, plus un terme bilinéaire
en la partie symétrique $S$ et la dérivée verticale de la trace. De même, le second et le cinquième forment
la dérivée covariante de la dérivée verticale de la trace, plus un terme bilinéaire en la partie symétrique
et la dérivée seconde verticale de la trace. Le quatrième terme débute la dérivée covariante de la dérivée
seconde verticale de la trace, et le sixième et dernier est quadratique dans les dérivées premières verticales
de la partie symétrique. Les termes de \eqref{energietrace} son donc, soit linéaires en la dérivée
suivant $A$ des deux champs
de Higgs-Hitchin que sont la trace $Tr^{(0)}S^{(1)}$ et sa dérivée verticale $Tr^{(0)}S^{(2)}$, soit bilinéaires en ces champs de Higgs et
celui qui est donné par la partie symétrique $S^{(1)}$ et sa dérivée verticale $S^{(2)}$.\\
\indent Du point de vue du champ de Higgs $H$, on a des termes $\nabla B$, $BH$, $BBH$, $BHH$ et $BBHH$.\\

\noindent \emph{Remarque sur les unités}: dans le système d'unités naturelles où $\hbar=c=1$, la densité Lagrangienne doit être
en $L^{-4}$ pour que l'action soit sans unité, et les charges en facteur devant sans unités également; un champ scalaire comme $H$ doit
donc posséder la dimension d'une
énergie ou d'une masse $M=L^{-1}$ afin que sn terme cinétique soit en $L^{-4}$. Or le déplacement $v$ semblerait devoir être de dimension $L$. En fait, non, la dimension naturelle
de $v$ est $L^{0}$, car la métrique le long de $W$ est fixée. L'avantage est aussi que les dimensions des autres champs, à tous
les ordres, restent inchangées. Par conséquent la donnée d'une masse $M$ est nécessaire pour rapprocher $H$ et $v$, de sorte que
l'on posera $H=Mv$. Dans le Lagrangien de $H$ on a en plus un terme de masse $M_H^{2}HH$, donc \emph{a priori} il ne faut pas
confondre $M$ et $M_H$. Il faudra donc songer, chaque fois qu'on remplace un terme de degré $d$ en $v$ par un terme de degré $d$
en $H$, qu'il s'introduit automatiquement un facteur $M^{-d}$ devant. Par exemple $AHHA$ doit être lu $M^{-2}AHHA$.\\

\noindent Les huit premiers termes de \eqref{energielorentzcouplage} sont bilinéaires en les coefficients des parties symétriques
$S^{(1)}$, $S^{(2)}$, $S^{(3)}$, jusqu'à la dérivée verticale seconde, et bilinéaires aussi en les coefficients des
parties antisymétriques $A^{(1)}$, $A^{(2)}$, $A^{(3)}$. Seul le dernier terme de \eqref{energielorentzcouplage} est cubique, il fait intervenir
la métrique verticale $G_F$ à l'ordre deux, i.e. la courbure de la métrique $g_{ij}$, dérivée verticale de $\Gamma_F$, la partie antisymétrique $S^{(1)}$
et la partie symétrique $A^{(1)}$, toutes les autres composantes donnant zéro. Il est donc du type $R_F AS$.\\
\noindent Dans \eqref{energielorentzcouplage}, la contribution de $A^{(1)}$ aux équations du mouvement,
i.e. l'annulation de la divergence du tenseur, est assurée par l'équation de Lorentz:
\begin{equation}\label{lorentzjauge}
\sum_\mu \partial_\mu A^{\mu}=0.
\end{equation}
Dans la suite nous adopterons cette équation. Ceci laisse encore la liberté des transformations de jauge verticales,
i.e. le choix d'une base de $F$ en dimension huit.\\
\noindent Cependant il faut rappele que, au bout du compte, la théorie doit être quantifiée, et que toutes les sortes
de fixages de jauge donnent lieu à un calcul co-homologique, comme $BRST$ et $B-V$. Cela se traduit par l'apparition
de nouveaux champs auxiliaires $\omega$, $\overline{\omega}$, $\xi$, $\overline{\xi}$, ..., et de nouveaux couplages.\\
\indent Du point de vue du champ de Higgs $H$, on a des termes $AA$, $SS$, $AHA$, $SHS$, $AHHA$, $SHHS$, $ARS$, $AHHRS$.\\

\noindent Les termes de \eqref{energielineaire} introduisent de nouveaux vecteurs de Higgs-Hitchin $\partial_\mu g_{ij}$
et un couplage, avec la partie symétrique $S^{(1)}$.\\
\indent Du point de vue du champ de Higgs $H$, ceci introduit des termes $BHHR$, $SHHR$, $AHHR$.\\

En résumé, en plus de la connexion $A^{(1)}$, il est apparu neuf familles de champs de Higgs vectoriels, qui sont $S_0^{(1)}$, $TrS^{(1)}$, $S_0^{(2)}$,
$TrS^{(2)}$, $S_0^{(3)}$, $TrS^{(3)}$, $A^{(2)}$, $A^{(3)}$ et la courbure $R_F$, si on tient compte de la répartition de $S^{(1)}$,
$S^{(2)}$ et $S^{(3)}$ en parties sans trace $S_0^{(i)}$ et partie de trace pure $TrS^{(i)}$, où $i$ va de $1$ à $3$.
Pour l'indice $1$ on a une seule forme matricielle, pour l'indice $2$ on en a huit, pour l'indice $3$ on en a $64$.
Nous désignerons collectivement ces champs par la lettre $\phi$, et nous noterons $\phi_\mu$ ses composantes suivant les axes de
coordonnées $x^{\mu}$.\\
Le champ $H$, pour le moment n'intervient que dans des couplages, avec un degré inférieur ou égal à deux, mais bien
évidemment, aux ordres plus grand, on trouvera des termes en $H$ de degré arbitraireemnt grand, il n'y a aucune raison de
s'arrêter à l'ordre $4$, ce qui réclamera une
discussion serrée pour la renormalisation, cf. Weinberg, qui explique pourquoi il faut s'attendre à cela dans la suite
du modèle standard.\\

\noindent La meilleure façon de trouver les équations aux dérivées partielles des champs classiques
est de déterminer un Lagrangien à partir du tenseur d'énergie-impulsion et d'en prendre les équations
d'Euler-Lagrange. La règle pour les Lagrangiens d'une théorie co-variante générale, qui ne font pas intervenir
les dérivées de la métrique est la suivante (cf. \cite{landau1971theoriedeschamps} chapitre 11):
\begin{equation}
T_{\mu\nu}=\frac{1}{\sqrt{-g}}\frac{\delta(\mathcal{L}\sqrt{-g})}{\delta g^{\mu\nu}}.
\end{equation}
où $-g$ désigne l'opposé du déterminant de la métrique, $g=det G_W$.\\
Là, en chaque point les fonctions $g^{\mu\nu}$, comme les $g_{\mu\nu}$, sont indépendantes, mais sous la contrainte unimodulaire,
ce n'est pas le cas. Pour tenir compte de cette contrainte nous faisons appel à l'équation de son gradient:
\begin{equation}
g_{\mu\nu}= g\frac{\delta(g^{-1})}{\delta g^{\mu\nu}}=-g^{-1}\frac{\delta(g)}{\delta g^{\mu\nu}}
\end{equation}
Cette formule nous dit que $T_{\mu\nu}$ doit être considéré modulo les tenseurs de la forme $f(x)g_{\mu\nu}(x)$;
en particulier les multiples des traces peuvent être ajoutés ou retranchés à volonté.\\

\noindent Dans des coordonnées où le déterminant de $G_W$ est constant, le tenseur énergie-moment et le Lagrangien sont
reliés par l'équation plus simple:
\begin{equation}
T_{\mu\nu}=\frac{\delta\mathcal{L}}{\delta g^{\mu\nu}}.
\end{equation}
Vérifions qu'en appliquant cette formule au Lagrangien de Yang-Mills, avec $M$ antisymétrique, on trouve bien le terme $\frac{1}{2}M\odot M$:
\begin{align}
\frac{\delta\mathcal{L}}{\delta g^{\mu\nu}}&=\frac{1}{8}\frac{\delta \sum_{i \lambda\kappa} M_{\lambda\kappa}^{i}M^{\lambda\kappa i}}{\delta g^{\mu\nu}}\\
&=\frac{1}{8}\frac{\delta \sum_{i \lambda\kappa \rho\sigma} M_{\lambda\kappa}^{i}M_{\rho\sigma i}g^{\lambda\rho}g^{\kappa\sigma}}{\delta g^{\mu\nu}}\\
&=\frac{1}{2}\sum_{i \kappa} M_{\mu\kappa}^{i}M_{\nu i}^{\kappa}
\end{align}

\noindent Le tenseur $T^{YM}_{\mu\nu}$ ordinaire du champ de Yang-Mills s'obtient en soustrayant $\frac{1}{8}T_\lambda^{\lambda}g_{\mu\nu}$,
de manière à obtenir un tenseur de trace nulle, caractéristique d'une théorie de particules de masse nulle, invariante
par les transformations dites conformes, qui changent la métrique en la multipliant par une fonction strictement positive arbitraire.
Mais le fait de fixer la valeur de $detG$ égale à $1$ ne permet pas trop de considérer de telles transformations.\\

\noindent Symboliquement, compte tenu de la contrainte de jauge de Lorentz, le lagrangien bosonique classique à l'ordre deux transverse s'écrit
\begin{equation}
\mathcal{L}_B(A,\phi)=YM(A)+\epsilon (div_W(A))^{2}+d^{A}\phi.d^{A}\phi+L(d^{A}\phi)+V(\phi);
\end{equation}
où $L$ est linéaire et où, à cet ordre, $V(\phi)$ est un polynôme de degré deux en les composantes de $\phi$.\\
Du point de vue des ordres en $H$, tous les termes sauf le premier démarrent à l'ordre zéro.\\
A l'ordre deux on a vu que
\begin{multline}
V(\phi)=HB+HBB+HHB+HHBB+HAA+HSS+HHAA\\+HHSS+HHARS+HHB\nabla R+HHA\nabla R+HHS\nabla R\\
+HHM_AM_A+HHM_SM_S+HHM_BM_B;
\end{multline}
où $M_A,M_S,M_B$ désignent les champs de monodromie associés aux champs de connexions $A,S,B$ respectivement,
c'est-à-dire leurs courbures.\\
Les seuls champs scalaires autre que $H$ viennent de la courbure verticale.\\

\indent Pour obtenir les
équations complètes du mouvement des champs, il faudra ajouter la contribution fermionique, sous la forme d'un courant quadratique
en les $\psi$, mais contenant aussi des termes en $A,S,B,R$ et $H$, comme $\psi A\psi$, $\psi S\psi$, $\psi B\psi$, $\psi H\psi$, $\psi HH\psi$, $\psi AH\psi$, $\psi SH\psi$,
$\psi BH\psi$, $\psi AHH\psi$, etc.\\
\indent Ecrivons donc l'opérateur de Dirac dans l'espace de dimension douze, décomposée en ses quatre secteurs, à l'ordre
inférieur ou égal à $2$ en la section transversale $v$.\\
\indent Selon les huit dimensions verticales, le fibré tangent et le fibré
auxiliaire coïncident; on peut donc utiliser les mêmes indices $i,j,k,...$ pour les vecteurs verticaux
et les mêmes indices $I,J,...$ pour les spineurs verticaux, qu'ils soient associés au fibré tangent ou
au fibré Euclidien. Alors l'opérateur de Dirac à l'ordre $2$, selon une section $v$ de $E_F$, indépendamment des spineurs auxquels on l'applique, s'écrit:
\begin{multline}
(D_S\psi)^{\mathbf{C},K}=\\
\sum_{\alpha,\mathbf{B},\mu,\nu}\gamma_{\alpha,\mathbf{B}}^{\mathbf{C}}(\epsilon^{-1})^{\alpha}_\nu\eta^{\mu\nu}\partial_\mu\Psi^{\mathbf{B},K}
+\sum_{i,\mathbf{B}}\gamma_{i,\mathbf{B}}^{\mathbf{C}}\partial_i\Psi^{\mathbf{B},K}
+\sum_{i,J}\gamma_{i,J}^{K}\partial_i\Psi^{\mathbf{C},J}\\
+\sum_{\alpha,\mathbf{B},\mu,\nu}\gamma_{\alpha,\mathbf{B}}^{\mathbf{C}}(\epsilon^{-1})^{\alpha}_\nu\eta^{\mu\nu}\sum_{k,I}v^{k}(\Gamma_{\mu,I;k}^{J}
+\partial_\mu\epsilon_{I;k}^{J})\Psi^{\mathbf{B},I}
+\sum_{i,J}\gamma_{i,J}^{K}\sum_{k,I}\Gamma_{i,I;k}^{J}v^{k}\Psi^{\mathbf{C},I}\\
+\frac{1}{2}\sum_{i,\mathbf{B}}\gamma_{i,\mathbf{B}}^{\mathbf{C}}\sum_{k,l}v^{k}v^{l}(\partial^{2}_{k,l}\epsilon_{j}^{i})\partial_j\Psi^{\mathbf{B},K}\\
+\sum_{\alpha,\mathbf{B},\mu,\nu}\gamma_{\alpha,\mathbf{B}}^{\mathbf{C}}(\epsilon^{-1})^{\alpha}_\nu\eta^{\mu\nu}\sum_{k,I}v^{k}v^{l}
(\Gamma_{\mu,I;k,l}^{K}+\partial_\mu\epsilon_{I;k,l}^{K})\Psi^{\mathbf{B},I}\\
+\frac{1}{2}\sum_{i,J,k,l}v^{k}v^{l}\gamma_{i,J}^{K}(\partial^{2}_{k,l}\epsilon_{j}^{i})\partial_j\Psi^{\mathbf{C},J}
+\sum_{i,J}\gamma_{i,J}^{K}\sum_{k,l,I}v^{k}v^{l}(\Gamma_{i,I;k,l}^{J}+\partial_i\epsilon_{I;k,l}^{J})\Psi^{\mathbf{C},I}
\end{multline}

\noindent La première ligne est d'ordre zéro en $v$; elle correspond à l'opérateur de Dirac usuel $D_S^{W}$ sans connexion le long de $W$,
conservant les chiralités verticales,
et à l'opérateur de Dirac vertical, échangeant les chiralités verticales.\\
La deuxième ligne est la somme des deux termes d'ordre un en $v$. Le premier est la contraction de Clifford $\mathbb{A}_{S}(v)$
de la connexion de Yang-Mills
antisymétrique $A^{(1)}$ relevée aux spineurs, donnant les bosons de jauge du premier ordre; il préserve la chiralité de dimension huit
et échange la chiralité en dimension quatre. Le second terme fait
intervenir la courbure de la métrique verticale le long de $W$; \emph{a priori} il respecte la chiralité longitudinale (le long de $W$)
et change la chiralité verticale (le long des fibres $F$), ce qui ne lui permet pas
de donner de la masse aux fermions; cependant si on fait agir $v$ (qui deviendra $H$) en changeant la chiralité verticale, on verra
plus tard qu'on peut obtenir
par composition avec la courbure une source de masse possible.\\
La troisième ligne, qui prend en compte le vielbein vertical au premier ordre où il apparaît, change la chiralité verticale à l'ordre deux,
et ici on ne peut pas compenser le changement puisque $HH$ préserve la chiralité verticale.\\
Par contre, la quatrième ligne, également d'ordre deux en $v$, ne change pas la chiralité verticale; elle représente l'action de Clifford $\mathbb{A}_S^{(2)}(v,v)$
du boson vecteur de Higgs $A^{(2)}$ au second ordre en $v$. Ce terme peut donner des interactions de Yukawa, et il serait \emph{a priori} capable
d'accorder des masses aux fermions, après brisure spontanée de symétrie, s'il était scalaire.\\
Le premier terme de la cinquième ligne change la chiralité verticale; ce terme représente une modification
de l'opérateur de Dirac vertical qui tient compte de la courbure de $g_{ij}$. Le dernier terme d'ordre
deux change aussi la chiralité verticale. \emph{A priori} aucun de ces termes ne peut donner des masses aux fermions.\\
Notons que si le vielbein est pris symétrique dans ces équations, ce qui peut toujours être fait, et les $\partial_\mu\epsilon$
n'interviennent pas.\\

\noindent Par conséquent, si on ne retient que le résultat de l'application de l'opérateur de Dirac à l'ordre inférieur à $2$
appliqué aux trois premières générations
de fermions, et si on ne s'intéresse qu'au secteur $S_8^{+}$ de la multiplicié des fermions à quatre dimensions, on trouve l'opérateur suivant:
\begin{multline}\label{diracordredeuxantisym}
D_S(\psi^{(0)}+\sum_kv^{k}\psi_k^{(1)}+\sum_{k,l}v^{k}v^{l}\psi_{k,l}^{(2)})=\\
D_S^{W}\psi^{(0)}+\sum_kv^{k}D_S^{W}\psi^{(1)}_k+\sum_{k,l}v^{k}v^{l}D_S^{W}\psi^{(2)}\\
+\mathbb{A}_{S}^{(1)}(v)\psi^{(0)}+\sum_kv^{k}\mathbb{A}_S^{(1)}(v)\psi^{(1)}_k+\sum_{k,l}v^{k}v^{l}\mathbb{A}_S^{(1)}(v)\psi^{(2)}_{k,l}\\
+\mathbb{A}_S^{(2)}(v,v)\psi^{(0)}+\sum_kv^{k}\mathbb{A}_S^{(2)}(v,v)\psi^{(1)}_k+\sum_{k,l}v^{k}v^{l}\mathbb{A}_S^{(2)}(v,v)\psi^{(2)}_{k,l}.
\end{multline}

\noindent Le lagrangien fermionique semi-classique à l'ordre deux transverse s'écrit symboliquement
\begin{equation}
\mathcal{L}_F(\psi,A,\phi)=(\psi^{(0)}+v.\psi^{(1)}+vv.\psi^{(2)})D_S(\psi^{(0)}+v.\psi^{(1)}+vv.\psi^{(2)});
\end{equation}
où $\phi$ désigne les composantes de Higgs-Hitchin antisymétriques.\\
En le développant on obtient des termes polynomiaux
en $H$ là où $v$ lui laisse la place, qui ne sont pas forcément renormalisables vu de $W$, car leur degré superficiel de divergence
peuvent être supérieurs à cinq, comme $\widetilde{\psi} HHA\psi$, ou $\widetilde{\psi} HHR\psi$.\\

\noindent A l'ordre trois arriverait un nouveau terme de Higgs vectoriel (Higgs-Hitchin), mais \emph{a priori} il faudrait aussi faire intervenir une
nouvelle génération de spineurs.\\

\section{Les trialités de Higgs}

\subsection{Brisure complexe}

Le succès de l'équation de Dirac pour décrire la fonction d'onde de l'élec-tron ne fait pas de doute.
Pourtant quelque chose a toujours chagriné certains physiciens et certains géomètres, c'est que le fibré des
spineurs ne soit défini qu'à isomorphisme près. Plus tard nous parlerons du problème en
dimension quatre, mais pour le moment examinons le problème pour les huit dimensions fibrées de $E_F$.\\
\indent Soit $\Phi$ une forme quadratique inversible sur un espace vectoriel réel $F$, de dimension $n$; un espace de
spineurs $S$ est n'importe quel espace vectoriel complexe de dimension $2^{m}$ où $m$ est la partie entière de
$n/2$, muni de deux applications linéaires; la première, notée $\gamma$, va de $F$ dans $End(S)$ et satisfait à
$\gamma(x)\gamma(x)=\Phi(x)Id_S$ pour tout $x$ dans $F$; la seconde, notée $\sigma_*$ est une représentation linéaire de l'algèbre de Lie
$so(\Phi)$ dans $sl(S)$ et satisfait à $\gamma(X.x)=\sigma_*(X)\gamma(x)-\gamma(x)\sigma_*(X)$ pour tout $x$ dans
$F$ et tout $X$ dans $so(\Phi)$. C'est un fait que
$(S,\gamma,\sigma_*)$ est défini à isomorphisme unique près,
et que (sauf en signature $(1,1)$ ou $1$ ou $-1$), l'image de $so(\Phi)$ par $\sigma_*$ s'intègre en une représentation
d'un revêtement double non-trivial de la composante connexe de l'identité dans le groupe $SO(\Phi)$,
nommé groupe des spineurs, et noté $Spin(\Phi)$.\\

\indent Supposons que la dimension de $F$ soit paire. Une ancienne construction d'un espace de spineurs pour $\Phi$
(cf. \cite{cartan1966theory}, \cite{chevalley1997algebraic}, \cite{penrose1984spinors}, \cite{penrose1988spinors}, \cite{atiyah1975classical}) repose sur le choix d'une
structure complexe sur $F$, compatible avec la métrique;  elle consiste à prendre pour espace $S$ l'algèbre extérieure
complexe de $F$, qui est bien de dimension $2^{m}$ sur $\mathbb{C}$, et à définir
$\gamma$ (matrices de Dirac) et $\sigma_*$ (représentation spinorielle infinitésimale) à l'aide du produit extérieur, du produit intérieur et de la dualité
hermitienne. Cependant une manière un peu plus géométrique de faire la même chose est de complexifier $F$,
de choisir une \emph{polarisation} de $F$,
c'est-à-dire une décomposition $F=V'\oplus V"$ en deux sous-espaces $V'$ et $V"$ qui sont \emph{isotropes}, i.e.
tels que la restriction de $\Phi$ à ces sous-espaces soit
identiquement nulle (notons que sur $\mathbb{C}$ une telle décomposition existe toujours), de poser $S=\Lambda^{*}(V")$, et de définir
$\gamma(v'+v")(s)=\iota(v')s+v"\wedge s$ (produit intérieur plus produit extérieur). Pour définir le produit intérieur, on
identifie $V'$ au dual complexe de $V"$, en posant $\langle v',v"\rangle=B_{\Phi}(v',v")$, où $B_{\Phi}(x,y)$ est égal à
$\Phi(x+y)-\Phi(x)-\Phi(y)$. Il est alors
facile de vérifier que $\sigma_*$ est définie par
$\sigma_*(v_1\wedge v_2)=\gamma(v_2)\gamma(v_1)-2^{-m}tr(\gamma(v_2)\gamma(v_1))Id_S$. Le vecteur $1$ dans $\Lambda^{*}(V")$
s'appelle \emph{vecteur vide} de $S$. Par définition, le sous-espace pair $\Lambda^{ev}(V")$ est $S^{+}$ et le sous-espace
impair $\Lambda^{od}(V")$ est $S^{-}$; tous deux sont stables par le groupe $Spin(\Phi)$.\\
Lorsque $F$ sur $\mathbb{R}$ est définie positive, comme dans le cas qui nous intéresse en dimension $8$, les polarisations
qui sont stables par conjugaison dans $F\otimes \mathbb{C}$ correspondent aux graphes $V"$ des structures complexe de $F$
compatibles avec $\Phi$, vues comme applications de $F$ dans $iF$; en prenant pour $V'$ l'espace conjugué de $V"$ dans $F\otimes \mathbb{C}$.
Cf. \cite{weil20varietes}. Il revient donc au même de choisir un sous-espace
isotrope maximal dans $F\otimes \mathbb{C}$ ou une structure complexe compatible sur $F$. La projection de l'espace réel $F$
sur $V'$ parallèlement à $V"$, est un isomorphisme de la structure $J$ de $F$ avec la structure $i$ de $V'$,
celle qui est induite par $F\otimes \mathbb{C}$. Une autre façon de dire est que $V'$ et $V"$ sont
les sous-espaces propres de $J$ associés aux valeurs propres $i$ et $-i$ respectivement. On appelle \emph{positives} les structures complexes définissant la même
orientation que celle choisie sur $F$.\\
Pour chaque sous-espace totalement isotrope maximal $Z$ de $V\otimes \mathbb{C}$, il existe une unique droite complexe $\mathbb{C}u_Z$ dans $S$
telle que $z\in Z$ soit équivalent à $\gamma(z)u_Z=0$, cf. \cite{chevalley1997algebraic} p. 136. L'orbite de $u_Z$ pour le groupe spin complexe est
l'ensemble des conjugués $gZg^{-1}$ de $Z$. Un tel spineur $u_Z$ se nomme \emph{spineur pur}. Il appartient nécessairement à $S^{+}$ ou
à $S^{-}$. Nous considérerons les spineurs purs positifs comme des points dans l'espace projectif complexe $\mathbb{P}(S^{+})$,
car rien ne permet de choisir une phase meilleure qu'une autre pour un tel spineur.\\
Nous appellons \emph{spineurs purs positifs} les éléments de l'orbite
de la classe de $1$ dans $\mathbb{P}(S^{+})$ sous le groupe $Spin$ Euclidien, qui est un sous-groupe compact maximal de $Spin_{\mathbb{C}}(\Phi)$.
L'application qui à un spineur pur $s_Z$ associe
le sous-espace $V_Z$ de $F\otimes\mathbb{C}$ formés des vecteurs $v$ tels que $\gamma(v)s_Z=0$ établit une bijection
entre les spineurs purs positifs et les structures complexes positives sur $F$. Cf. \cite{chevalley1997algebraic}, pp. 135-140, pour
la version algébrique, la version réelle s'en déduit aisément. \\
L'orbite de $1$ sous le groupe compact $Spin(\Phi)$ dans l'espace projectif $\mathbb{P}(S^{+})$ s'identifie à l'espace homogène symétrique
$SO_{2n}/U_n$. Le groupe $\widetilde{U_n}$ qui relève $U_n$ dans $Spin_{2n}$ ne stabilise que la direction
complexe de $1$. Seul le sous-groupe $SU_n$ stabilise $1$ lui-même.\\
\indent Donc, pour tout fibré de spineurs positifs complexes en dimension paire, associé à un fibré principal en groupe
$Spin(\Phi)$ muni d'une connexion principale, il est possible d'appliquer le groupe
de jauge associé  pour ramener toute structure complexe positive de $F$ sur une
particulière qu'on a choisie. De cette façon, on munit le fibré $E_F$ lui-même d'une
structure complexe. Il devient ainsi hermitien positif, et le groupe de structure spin se brise sur
un revêtement à deux feuillets $\widetilde{U_m}$ du groupe unitaire.\\

\indent Un modèle concret est fourni par la structure complexe $J_0$ sur $\mathbb{R}^{2n}$ qui est définie par $J_0(e_k)=e_{n+k}$, pour
$k$ variant de $1$ à $n$; sa matrice est
\begin{equation}
J_0=
\begin{pmatrix}
0&-1\\
1&0
\end{pmatrix}
\end{equation}

L'espace $S_0^{+}$ complexe associé est $\Lambda^{ev}(F")$, où $F"$ est formé des vecteurs de la forme $v+iJ_0v=(x,y,-iy,ix)$ dans $\mathbb{C}^{n}\oplus i\mathbb{C}^{n}$.
Le long de $F"$ on a $J_0(w)=-iw$, si bien que la structure complexe naturelle sur $F"$ est celle de $\overline{\mathbb{C}^{n}}$.
De même l'espace $S_0^{-}$ est $\Lambda^{od}(F")$.\\
L'espace $F'$ qui est associé au vecteur $1$ dans $S=\Lambda^{*}(F")$ est formé des vecteurs $v-iJ_0v=(x,y,iy,-ix)$. \\
\indent Le stabilisateur de $J_0$ dans l'algèbre de Lie $o_{2n}$ est l'algèbre de Lie $u_n$ des matrices antisymétriques qui
commutent avec $J_0$, son orthogonal $\mathfrak{p}_n$ pour la métrique de Killing est l'espace tangent en $J_0$ de la variété des matrices
$J$ à la fois antisymétriques et orthogonales, donc $X$ est dans $u_n$ et $Y$ dans $\mathfrak{p}_n$ si et seulement si
\begin{equation}
X=-^{t}X,\quad J_0X-XJ_0=0,\quad Y=-^{t}Y,\quad J_0Y+YJ_0=0.
\end{equation}
D'où
\begin{equation}
X=
\begin{pmatrix}
A&-B\\
B&A
\end{pmatrix}
\quad
Y=
\begin{pmatrix}
A'&B'\\
B'&-A'
\end{pmatrix}
\end{equation}
avec
\begin{equation}
A=-^{t}A,\quad B=^{t}B,\quad A'=-^{t}A',\quad B'=-^{t}B'.
\end{equation}
Pour $u_n$ on retrouve l'écriture matricielle des nombres complexes.\\

\noindent La dimension de $u_n$ sur $\mathbb{R}$ est $n^{2}$; celle de $\mathfrak{p}_n$ est $n(n-1)$. En
particulier lorsque $n=4$, on trouve que $u_4$ est de dimension $16$ et que $\mathfrak{p}$ est de dimension $12$.\\
L'action adjointe de $u_n$ sur $\mathfrak{p}_n$ est donnée par:
\begin{equation}
ad(A,B)(A',B')=([A,A']-(BB'+B'B),[A,B']+(BA'+A'B)).
\end{equation}

\noindent On retrouve la représentation de $u_n$ par les matrices $n\times n$ anti-hermitiennes en posant
$X=A+iB$, pour que $^{t}\overline{X}=-X$. Alors $\mathfrak{p}_n$ s'identifie aux matrices $n\times n$ antisymétriques
complexes, et l'action de $u_n$ dessus correspond à l'action naturelle du groupe unitaire sur $\Lambda^{2}(\mathbb{C}^{n})$ induite
par l'action du groupe linéaire $GL_n(\mathbb{C})$,
\begin{equation}\label{actionlinantisym}
g^{*}\omega=g\omega^{t}g,
\end{equation}
puisqu'au niveau de l'algèbre de Lie, la formule \eqref{actionlinantisym} donne,
\begin{equation}
X.Y=XY+Y^{t}X=XY-Y\overline{X};
\end{equation}
et que dans notre cas, on vérifie bien
\begin{multline}
(A+iB)(A'+iB')-(A'+iB')(A-iB)\\=AA'-A'A-BB'-B'B+i(AB'-B'A+BA'+A'B).
\end{multline}

Si on restreint la représentation à $su_n$, et si $n=2m$ est lui-même un nombre pair; le choix d'une $n$-forme non-nulle décompose l'espace $\Lambda^{m}$
en une partie autoduale $\Lambda^{m}_+$ et une partie anti-autoduale $\Lambda^{m}_-$, qui sont deux représentations complexes d'égales dimensions
mais inéquivalentes de $SU_n$. \\

\noindent Appliquons cette construction générale au fibré des spineurs verticaux associé à $E_F$ de rang $8$ au-dessus de
$W$ de dimension quatre. Cela ramène le groupe $Spin_8$ au groupe $\widetilde{U_4}$ et définit une structure hermitienne complexe
sur $E_F$.\\

\subsection{Choix d'une trialité}

Les espaces de spineurs $S_8^{+}$ et $S_8^{-}$ possèdent des structures spéciales. Par exemple,
il existe sur chacun d'eux une structure réelle $r$ et une forme quadratique $B$ définie positive invariante sous l'action
du groupe des spineurs compact $Spin_8$, et covariantes pour les matrices $\gamma$ (cf. \cite{chevalley1997algebraic}, \cite{adams1996lectures},\cite{deligne1999quantum}).
Ces structures existent en toute
dimension congrue à $0$ modulo $8$ pour la signature Euclidienne. Elles ne sont pas tout à fait uniques. Il est important pour nous de décrire
précisément les choix dont ces structures dépendent, mais commençons par fixer un $r$ et un $B$ (ils sont tous équivalents), et notons $S_8^{+}$ et $S_8^{-}$
les espaces de spineurs réels munis de cette structure Euclidienne.\\
\indent L'exclusivité de la dimension $8$ (Euclidienne ou pas) est qu'il existe une
\emph{forme de trialité} sur la somme $\Delta=F\oplus S_8^{+}\oplus S_8^{-}$, c'est-à-dire une application trilinéaire $T$ spin invariante,
à valeurs dans $\mathbb{R}$. Dans le cas Euclidien, son image est $[-1,1]$ sur le produit des sphères $S^{7}$ pour les métriques $\Phi$ sur les
vecteurs et $B$ sur les spineurs; de plus,
pour toute paire de points dans les spères de deux de ces espaces, la restriction de $T$ à la troisième sphère n'est pas
identiquement nulle. A la multiplication près par $-1$ la trialité est unique, une fois que les métriques et les structures
réelles ont été choisies sur $S_8^{+}$ et $S_8^{-}$. Cf. \cite{adams1996lectures}.\\
La forme $T$ est facile à construire: il suffit de considérer l'application $\gamma$ qui envoie $F$ dans les
homomorphismes de $S_8^{+}$ dans $S_8^{-}$ comme un élément
\begin{equation}
\Gamma\in F^{*}\otimes (S_8^{+})^{*}\otimes S_8^{-};
\end{equation}
et alors, l'application de la dualité associée à la forme $B$ sur $ S_8^{-}$ donne $T$.\\
Mais il y a mieux encore, c'est le \emph{principe de trialité} remontant à Cayley, cf. \cite{chevalley1997algebraic} p.181:\\
il existe un automorphisme $\tau$ de $\Delta$ d'ordre trois envoyant $F$ sur $S_8^{+}$, $S_8^{+}$ sur $S_8^{-}$ et $S_8^{-}$ sur $F$,
préservant la forme quadratique et la forme de trialité.\\
\noindent \emph{Démonstration}: soit $s_8$ un élément de norme $1$ de $S_8^{+}$;
nous considérons l'application $\gamma_8^{+}$ qui envoie tout $v\in F$ sur $\gamma(v)s_8$ dans $S_8^{-}$; c'est une
bijection, car $\gamma(v)\gamma(v)s_8=\Phi(v)s_8$ n'est nulle que si $v=0$. On peut donc considérer l'involution
$\mu_8$ de $\Delta$ qui coïncide avec $\gamma_8^{+}$ et son inverse respectivement sur $F$ et sur $S_8^{-}$ et
qui coïncide avec la réflexion dans l'hyperplan orthogonal à $s_8$ dans $S_8^{+}$. De même, à un
élément $t_8$ de $S_8^{-}$ de norme $1$ on peut associer l'involution $\nu_8$ de $\Delta$ qui étend la réflexion
dans le plan orthogonal à $t_8$ dans $S_8^{-}$  en l'application qui
à $v$ dans $F$ associe $\gamma(v)t_8$ dans $S_8^{+}$. Alors l'application $\tau=\mu_8\circ \nu_8$ est une
trialité. (On vérifie que $\nu_8^{-1}\mu_8\nu_8=\mu_8^{-1}\nu_8\mu_8$.)\\
\indent Ainsi les trialités dépendent d'une paire d'éléments de norme $1$ dans $S_8^{+}$ et $S_8^{-}$.\\

\noindent Ayant choisi une structure complexe positive sur $F$, on dispose d'un sous-espace isotrope privilégié $F'$ de $F\otimes \mathbb{C}$
tel que
$F\otimes \mathbb{C}=F'\oplus F"$ où $F"=\overline{F'}$, et il vient
\begin{equation}
S_8^{+}\otimes \mathbb{C}=\Lambda^{0}(F")\oplus\Lambda^{2}(F")\oplus \Lambda^{4}(F").
\end{equation}
et
\begin{equation}
S_8^{-}\otimes \mathbb{C}=\Lambda^{1}(F")\oplus\Lambda^{3}(F").
\end{equation}

\noindent \emph{Première observation}: les formes quadratiques complexes spin invariantes $B$ sur $S^{+}=S_8^{+}\otimes \mathbb{C}$ resp.
$S^{-}=S_8^{-}\otimes \mathbb{C}$ forment un espace vectoriel complexe de dimension $1$ (ce qui
donne $2$ pour la somme $S=S^{+}\oplus S^{-}$). Elles sont toutes obtenues de la
façon suivante (cf. \cite{chevalley1997algebraic}, pp.142-143): on choisit un élément non-nul $\Omega$ de $\Lambda^{4}(F")$ et pour un spineur
$s$ dans $S^{+}$ ou dans $S^{-}$, on prend pour $B(s)$ la composante suivant $\Omega$ de $C(s)s$
où $C$ désigne \emph{l'antiautomorphisme principal} de l'algèbre de Clifford, i.e. pour $s=\gamma_1...\gamma_r$,
on a $C(s)=\gamma_r...\gamma_1$.\\
Observons que le carré de la norme de $\Omega$ est contraint par la métrique
sur $F$, qui donne une forme volume dans $\Lambda^{8}_\mathbb{R}(F)$, et donc une forme volume
dans $\Lambda^{8}_\mathbb{C}(F)$. Alors l'identification de
$S_8\otimes \mathbb{C}$ avec $\Lambda^{*}(F")$ fait apparaître un module d'unité de volume naturel
dans $\Lambda^{4}(F")$. Pour fixer la phase de $\Omega$, on fait appel à la structure réelle $r$, et on
impose que $B$ soit réelle et définie positive sur $S_8^{+}$ et sur $S_8^{-}$.\\

\noindent La structure réelle invariante sur les spineurs est unique à une phase près dans chaque
secteur, pair et impair.\\
En effet une structure réelle
est donnée par une involution $\mathbb{C}$-antilinéaire (cf. \cite{adamslectures}), donc le composé de deux
structures $I',I"$ est un endomorphisme $\mathbb{C}$-linéaire qui commute avec l'action de $Spin_8$; comme les deux
représentations spin, demi-spineurs pairs ou impairs, sont irréductibles, sur chacune d'entre-elles, ce composé est une homothétie $zId$, donc $I"=zI'$; comme $I'$
est antilinéaire, on a $Id=zI'\circ zI'=z\bar{z}I'I'=z\bar{z}Id$ donc $z$ est de module $1$.
Réciproquement, supposons que $I'$ soit une structure réelle invariante dans $S_8^{\pm}\otimes \mathbb{C}$;
changer $I'$ en $I"=-I'$ échange l'ensemble des points fixes, appelé sous-espace réel avec l'espace propre
associé à $-1$, appelé espace imaginaire pur. Les deux sont invariants par le groupe compact $Spin_8$.
Plus généralement changer $I'$ en $zI'$, avec $z=\exp(i\theta)$, où $\theta$ est un nombre réel, revient à faire
tourner l'espace des spineurs réels de l'angle $\theta/2$. Le groupe compact $Spin_8$ préserve un par un
tous les espaces de dimension réelle huit qu'on obtient en choisissant une structure réelle.\\

\noindent Par conséquent, si on connaît un sous-espace de spineurs réels, les autres s'en déduisent par changement
de phase uniforme, un pair et un impair.\\

\noindent \textbf{Définitions}: l'espace $F"$ est muni d'un produit scalaire hermitien défini positif canonique, qui se déduit
de la structure Euclidienne $\Phi$ de $F$ et de la structure complexe $J$ de $F$:
\begin{equation}
\langle v"| w"\rangle=\frac{1}{2}(v|w)-\frac{i}{2}(Jv|w)
\end{equation}
où nous avons posé comme d'habitude $v"=(v+iJv)/2,w"=(w+iJw)/2$.\\
Il est facile de vérifier que $\langle v"| w"\rangle$ est le conjugué de $\langle w"| v"\rangle$ et que
cette expression est $\mathbb{C}$-linéaire en $w"$ et anti-$\mathbb{C}$-linéaire en $v"$.\\
Le produit scalaire hermitien s'étend fonctoriellement en un produit scalaire hermitien sur l'algèbre extérieure $S=\Lambda^{*}(F")$.
Dans la droite $\Lambda^{max}(F")$, nous choisissons un élément $\Omega$ de norme $1$.\\
Sur l'espace $S$ est définie une \emph{étoile de Hodge} anti-$\mathbb{C}$-linéaire $*"$, par la formule
suivante: quelque soit $u\in\ S$,
\begin{equation}
*"s\wedge u=\langle s| u\rangle \Omega
\end{equation}
Dans cette dimension $4$ pour $F"$, il se trouve que $*"\circ*"$ vaut $Id$ sur les multivecteurs pairs, et
vaut $-Id$ sur les multivecteurs impairs. On définit l'application anti-$\mathbb{C}$-linéaire $I_0$, en
prenant $*"$ sur les multivecteurs de degrés $1$ et $2$, et en prenant l'opposée $-*"$ sur les multivecteurs
de degrés $0$, $3$ et $4$.
De sorte que $I_0\circ I_0=Id$ définit une structure réelle sur $S$.\\

\noindent \textbf{Proposition 2}: l'application $I_0$ commute avec l'action spinorielle de l'algèbre de
Lie des rotations Euclidiennes; par conséquent $I_0$ définit l'opérateur de conjugaison d'une structure réelle spin invariante sur $S_{\mathbb{C}}$.
Les multivecteurs auto-duaux-conjugués, de la forme $s+I_0s$, de $\Lambda^{odd}(F")$ (resp. $\Lambda^{ev}(F")$)
forment un espace réel $S_8^{-}$ (resp. $S_8^{+}$)
spin invariant. \\

\noindent \emph{Démonstration}: soit $e_1,...,e_8$ une base orthonormée de $F$, telle que, pour tout indice
$k$ impair on ait $J(e_k)=e_{k+1}$. La représentation spin infinitésimale réelle $\sigma_*$ est engendrée par les
matrices $\gamma(e_j)\gamma(e_k)$, où $j\neq k$; le bivecteur $e_k\wedge e_j$ correspond à la demi-somme $\frac{1}{2}\gamma(e_j)\gamma(e_k)$
(cf. Cartan, Lawson et Michelsson). Or nos matrices de Dirac sont données par la formule polarisée,
\begin{equation}
\gamma(v)s=v"\wedge s+v'\lrcorner s,
\end{equation}
donc, pour toutes les paires d'indices $j,k$, il vient
\begin{multline}\label{actionspin}
2\sigma_*(e_j\wedge e_k)s\\=(e"_j\wedge e"_k)\wedge s+(e'_j\wedge e'_k)\lrcorner s+e"_j\wedge (e'_k\lrcorner s)-e"_k\wedge (e'_j\lrcorner s)
\\+(e'_j\lrcorner e"_k)s.
\end{multline}
Supposons d'abord que $e"_j$ et $e"_k$ soient colinéaires, par exemple $j=k+1$, auquel cas $e'_j=ie'_k$ et $e"_j=-ie"_k$.
Par ailleurs, on a $e'_k\lrcorner e"_k=1/2$, donc
\begin{equation}\label{actionspincomplexe}
2\sigma_*(e_{k+1}\wedge e_k)s=-2ie"_k\wedge (e'_k\lrcorner s)+\frac{i}{2}s.
\end{equation}
Sur un vecteur $e"_l$, cela donne $ie"_l/2$, sauf si $l$ vaut $k$ ou $k+1$, auquel cas on trouve respectivement
$-ie"_k/2$ et $e"_k/2=-ie"_{k+1}/2$.\\
Sur un trivecteur $*e"_k$ ou $*e"_{k+1}$ on trouve le même multiplié par $i/2$, et sur un trivecteur $e"_k\wedge e"_m\wedge e"_n$,
on trouve ce même trivecteur multiplié par $-i/2$.\\
Sur une constante $c$, on obtient $ic/2$, et sur le quadrivecteur $\Omega_0=e"_1\wedge ...\wedge e"_7$ on trouve $-i\Omega_0/2$.\\
Enfin sur un bivecteur $e"_m\wedge e"_n$, on trouve $ie"_m\wedge e"_n/2$, sauf si un indice vaut $k$ ou $k+1$, par exemple
si $m=k$, où on trouve $-i e"_m\wedge e"_n/2$.\\
Dans tous les cas l'application $\varphi=2\sigma_*(e_{k+1}\wedge e_k)$ commute avec l'application antilinéaire $I_0$.\\
En effet, si $l$ diffère de $k$ et de $k+1$,
\begin{equation}
\varphi (I_0(e"_l))=\varphi(*"e"_l)=-\frac{i}{2}*"e"_l=*"(\frac{i}{2}e"_l)=I_0(\varphi(e"_l)),
\end{equation}
et
\begin{multline}
\varphi (I_0(*"e"_l))=\varphi(e"_l)=\frac{i}{2}e"_l\\=\frac{i}{2}I_0*"e"_l=I_0(-\frac{i}{2}*"e"_l)=I_0(\varphi(*"e"_l)).
\end{multline}
Par contre, lorsque $l$ est égal à $k$,
\begin{equation}
\varphi (I_0(e"_k))=\varphi(*"e"_k)=\frac{i}{2}*"e"_k=*"(-\frac{i}{2}e"_k)=I_0(\varphi(e"_k)),
\end{equation}
et
\begin{multline}
\varphi (I_0(*"e"_k))=-\varphi(e"_k)=\frac{i}{2}e"_k\\=-\frac{i}{2}I_0*"e"_k=I_0(\frac{i}{2}*"e"_k)=I_0(\varphi(*"e"_k)).
\end{multline}
Sur une constante réelle $c$, on a $I_0(c)=-c\Omega$, donc
\begin{equation}
I_0\varphi c=I_0(ic/2)=-iI_0(c/2)=i\Omega/2=-\varphi(\Omega/2)=\varphi I_0(c).
\end{equation}
Le résultat sur les $4$-vecteurs $c\Omega$ s'en déduit puisque $I_0\circ I_0=Id$.\\
Pour un bivecteur $e"_m\wedge e"n$, où aucun indice ne vaut $k$ ou $k+1$, on a
\begin{equation}
I_0(\varphi (e"_m\wedge e"n))=I_0(\frac{i}{2}e"_m\wedge e"n)=-\frac{i}{2}I_0(e"_m\wedge e"n)=\varphi(I_0(e"_m\wedge e"_n)),
\end{equation}
car $I_0(e"_m\wedge e"_n)=*"(e"_m\wedge e"_n)$ est pur avec un indice $k$ ou $k+1$.\\
Le cas où $m$ ou $n$ vaut $k$ ou $k+1$ s'en déduit, pour la même raison d'involutivité.\\
\indent Supposons à présent que $e"_j$ et $e"_k$ ne soient pas colinéaires. Dans ce cas le dernier terme de la formule \eqref{actionspin}
s'annule. Lorsque $e"_l$ n'appartient pas au sous-espace vectoriel complexe engendré par $e"_j,e"_k$, l'action de $2\sigma_*(e_j\wedge e_k)$
donne le trivecteur pur
$e"_j\wedge e"_k\wedge e"_l$, et sur le trivecteur pur $*"e"_l=\varepsilon e"_j\wedge e"_k\wedge e"_m$, on trouve
$-\varepsilon e"_m$. On trouve donc à nouveau que $\varphi=2\sigma_*(e_j\wedge e_k)$
commute avec $I_0$; en effet
\begin{equation}
I_0\varphi (e"_l)=-*"(e"_j\wedge e"_k\wedge e"_l)=-\varepsilon e"_m,
\end{equation}
et
\begin{equation}
\varphi I_0(e"_l)=\varphi (*"e"_l)=\varepsilon \varphi(e"_j\wedge e"_k\wedge e"_m)=-\varepsilon e"_m.
\end{equation}
Dans l'autre cas, si $k=l+1$ par exemple, l'action sur $e"_l$ donne $ie"_j/2$, et l'action sur $*e"_l=\varepsilon e"_j\wedge e"_m\wedge e"_n$
donne
\begin{equation}
-\varepsilon e"_k\wedge e"_m\wedge e"_n/2=i\varepsilon e"_l\wedge e"_m\wedge e"_n/2=-i*e"_j/2.
\end{equation}
Ceci démontre la proposition pour les spineurs impairs. Passons au cas pair.\\
Continuons de noter  $\varphi=2\sigma_*(e_j\wedge e_k)$.\\
Pour une constante réelle $c$; on a $I_0\varphi(c)=c I_0 e"_j\wedge e"_k=c\varepsilon e"m\wedge e"_n$, où les indices $m,n$ sont tous deux différents de $j$ et $k$,
et où $\varepsilon=\pm 1$, mais on a aussi
\begin{equation}
\varphi I_0(c)=\varphi(-c\Omega)=c\varepsilon e"_m\wedge e"_n,
\end{equation}
en vertu des signes dans la formule \eqref{actionspin}.\\
On en déduit la relation de commutation pour les $4$-vecteurs.\\
Venons-en aux bivecteurs. Si les indices $m,n$ sont tous deux différents de $j$ et $k$, l'action de $\varphi$ sur $e"_m\wedge e"_n$ donne $\varepsilon\Omega$,
où $\varepsilon$ est le signe de l'orientation de la base $(e"_j,e"_k,e"_m,e"_n)$ par rapport à $\Omega$
mais sur $\varepsilon e"_j\wedge e"_k=*"(e"_m\wedge e"_n)$ l'action de $\varphi$ donne $-\varepsilon$, qui est bien égal à $I_0(\varepsilon\Omega)$
grâce au retournement du signe en degré congru à $0$ modulo $4$. On en déduit la relation de commutation si l'ensemble $\{m,n\}$ est
égal à l'ensemble $\{j,k\}$.\\
Si $m=j$, et si $n$ est différent de $k$ ou $k+1$, on a $\varphi(e"_k\wedge e"_n)=\frac{1}{2}e"_j\wedge e"_n$; il existe
un indice $l$ tel que $*"(e"_j\wedge e"_n)=\varepsilon e"_k\wedge e"_l$, où $\varepsilon=\pm 1$ est le rapport
entre $e"_j\wedge e"_n\wedge e"_k\wedge e"_l$ et $\Omega$, alors $I_0\varphi(e"_k\wedge e"_n)=\frac{\varepsilon}{2}e"_k\wedge e"_l$.
Mais alors $I_0(e"_k\wedge e"_n)=-\varepsilon e"_j\wedge e"_l$, car la permutation qui mène de $k,n,j,l$ à $j,n,k,l$ est impaire;
donc
\begin{equation}
\varphi I_0(e"_k\wedge e"_n)=-\varepsilon \varphi(e"_j\wedge e"_l)=+\frac{\varepsilon}{2}e"_k\wedge e"_l.
\end{equation}
C.Q.F.D.\\

\noindent \textbf{Notations, définitions}: si on identifie $\Lambda^{2}(F)$, comme il se doit, à l'algèbre de Lie du groupe spin compact,
qui est aussi celle du groupe orthogonal, les
éléments $e_{j}\wedge e_{j+1};j=1,3,5,7$ engendrent une sous-algèbre de Lie commutative maximale (sous-algèbre de Cartan) $\mathfrak{t}_4$
de $spin_8=so_8$. Comme on vient de le voir au cours de la démonstration précédente, le spectre des doubles $2\sigma_*(e_{j}\wedge e_{j+1})$
est fait de $\pm i/2$, par conséquent il faut multiplier $e_{j}\wedge e_{j+1}$ par $8\pi$ pour obtenir que l'exponentielle vaille $1$.
Comme base canonique du \emph{réseau nodal} $\Gamma_0$ du groupe $SO_8=SO(\Phi)$ de $F$, nous choisissons donc les
$4\pi e_{j}\wedge e_{j+1};j=1,3,5,7$.
Alors (cf. \cite{bourbaki1981groupes456}, \cite{bourbaki1982groupes9}) le
réseau nodal $\Gamma_1$ du groupe spin $Spin_8=$ de $F$ est engendré par les sommes et différences de ces éléments;
il coïncide avec les éléments de $\Gamma_0$ dont la somme des coordonnées est paire; en particulier, il contient les vecteurs
$8\pi e_{j}\wedge e_{j+1};j=1,3,5,7$.\\
Pour respecter la tradition, nous noterons $H_j=\frac{2}{i}e_{j}\wedge e_{j+1}$ et $K_j=2i\pi H_j$, pour les indices $j=1,3,5,7$.
Les $K_j$ engendrent $\Gamma_0$ sur $\mathbb{Z}$ et engendrent $\mathfrak{t}_4$ sur $\mathbb{R}$, alors que les $H_j$ engendrent la
forme réelle déployée $\mathfrak{h}_4=i\mathfrak{t}_4$ de l'algèbre complexifiée $t_4\otimes \mathbb{C}$. \\

\noindent \emph{Seconde observation}: la donnée d'une métrique Euclidienne sur $F$, d'une structure complexe
compatible positive $J$ sur $F$ \emph{via} $(F', F")$, d'une
métrique spinorielle compatible $B$ \emph{via} un quadrivecteur unité $\Omega$ et d'un
élément $t$ de norme $1$ dans $S^{-}$ permet de construire une trialité complexe, compatible avec la métrique spinorielle.\\

\noindent En effet, $t$ se décompose en un vecteur $u"$ de $F"$ et un trivecteur $t"$ de $\Lambda^{3}(F")$ qui peut
toujours s'écrire $t"=u"_1u"_2u"_3.1$ au sens des matrices gamma, ou $u"_1\wedge u"_2\wedge u"_3$ au sens de l'algèbre
extérieure, ce qui revient au même dans la sous-algèbre anti-commutative $\Lambda^{*}(F")$ de l'algèbre de Clifford; nous l'écrivons
$u"_1u"_2u"_3$. Par hypothèse on doit avoir $B(u"+t")=C(u"+t")(u"+t")=\Omega$, mais $C(u")=u"$ et $C(t")=u"_3u"_2u"_1$ donc
\begin{multline}
B(u"+t")\\=\iota(\Omega)(u"u"+u"_3u"_2u"_1u"_1u"_2u"_3+u"_3u"_2u"_1u"+u"u"_1u"_2u"_3)\\
=2u"_3u"_2u"_1u".
\end{multline}
Ce produit est purement de degré quatre, et il doit valoir $\Omega$ pour que $B(t)=1$. En particulier,
les quatre vecteurs $u"_3,u"_2,u"_1,u"$ doivent être linéairement indépendants. Notons que l'on aurait pu se
restreindre aux éléments $t$ tels que $t"$ soit orthogonal à $u"$; ce qui ne dépend pas du choix de $\Omega$.
Volontairement nous ne le faisons pas pour le moment, afin de garder libre ce paramètre supplémentaire le plus longtemps possible.\\
\indent Posons $u_0=(u"+\overline{u"})/\|u"+\overline{u"}\|_F$;
c'est un vecteur de norme $1$ dans $F$, on peut donc appliquer $\gamma(u_0)$ à $t$ et on obtient un élément $s$ de norme $1$
pour $B$ dans $S^{+}$. On définit $\mu_8$ à partir de $s$ et $\nu_8$ à partir de $t$. Alors, suivant Chevalley, $\tau=\mu\nu$
est une trialité, respectant $\Phi\otimes \mathbb{C}$ et $B$.\\
\indent Le spineur négatif (ou impair) $t$ n'est pas complétement déterminé par le choix de $u_0$; en fait si on
décompose $u_0=u_0"+u_0'$ suivant $F"$ et $F'$, i.e. $u"_0=(u_0+iJu_0)/2$, $u'_0=(u_0-iJu_0)/2$, et
si l'on note $V_0$ le trivecteur unité orienté de l'espace à trois dimensions complexes dans $F$
orthogonal à $u_0$, que l'on décompose également en $V_0=V"_0+V'_0$ suivant les trivecteurs de type $(0,3)$
et $(3,0)$, et qu'en plus on choisisse arbitrairement un nombre complexe non-nul $\lambda$, un nombre réel $\beta$
modulo $2\pi$, et
un bivecteur  $a"$ de type $(0,2)$, dont le support appartient au support de $V"_0$, alors
\begin{equation}
t=u"+t",\quad u"=\lambda u"_0,\quad t"=\frac{1}{\lambda}(e^{-i\beta}V"_0+u_0"\wedge a"),
\end{equation}
convient également pour fonder une trialité.\\
En effet,
\begin{multline}
B(t)\Omega=u"t"-t"u"=2u"t"=2e^{-i\beta}u"_0V"_0+2u"_0u"_0a"\\
=2e^{-i\beta}u"_0V"_0,
\end{multline}
Or le carré de la norme de $u"_0V"_0$ est $1/4$ puisque $u_0V_0$ se décompose en quatre morceaux
orthogonaux de même norme pour la métrique de Kähler du complexifié de $F$, dont le morceau ci-dessus.\\
On voit donc qu'en choisissant
\begin{equation}
\Omega=2e^{-i\beta}u"_0V"_0
\end{equation}
comme base de $\Lambda^{4}(F")$
on a un spineur $t$ de norme $1$ (i.e. $B(t)=1$), d'où l'on peut partir pour faire toute la construction.
La phase $\beta$ est arbitraire.\\
On trouve alors
\begin{align}
s&=\gamma(u_0)t=u"_0\wedge u"+u"_0\wedge t"+\iota(u_0')u"+\iota(u_0')t"\\
&=0+\frac{1}{\lambda}e^{-i\beta}u"_0V"_0+\frac{1}{\lambda}u"_0\wedge u_0"\wedge a"+\frac{\lambda}{2}.1+\frac{1}{2\lambda}a"\\
&=\frac{\Omega}{2\lambda}+\frac{1}{2}(\lambda.1+\frac{a"}{\lambda})
\end{align}
On vérifie qu'on a bien $B(s)=1$.\\

\noindent \emph{Troisième observation}: si on veut tenir compte comme il se doit des structures réelles,
afin d'obtenir une trialité pour le groupe spin compact, il faut que $t$ appartienne à un des sous-espaces
réels invariants. D'après la proposition $2$, cela équivaut à dire que $\lambda$ est de module $1$ et que
$a"$ est nul; on a alors
\begin{equation}
t=\lambda u"_0+\overline{\lambda}e^{-i\beta}V"_0.
\end{equation}
En effet $\overline{\lambda}e^{-i\beta}V"_0=I_0(\lambda u"_0)$ et la proposition $2$ dit que toute
structure réelle spin invariante s'écrit $\exp(-i\beta)I_0$. Comme avant nous confondons la structure réelle
et la conjugaison qui lui est associée, cf. \cite{adamslectures}.\\
Notons qu'on doit prendre le même déphasage $\beta$ de $I_0$ sur les parties paires et impaires des
spineurs si l'on veut que la trialité définie par $t$ respecte les trois structures réelles, sur
$S_8^{+}$, $S_8^{-}$ et $F_8$.\\
La signification de la phase beta est donc le choix d'une structure réelle $I'$ sur tous les spineurs;
elle retentit sur le choix de la métrique $\exp(-i\beta)\Omega_0$.\\
La signification de $\lambda$ est une rotation du vecteur de référence $u_0$ dans le plan de $F$
engendré par $u_0$ et $Ju_0$. \\

\noindent Dans la suite on pose $u_0=e_8$, le nombre complexe $\lambda$ est de module $1$ et $a"$
disparaît complétement (jusqu'à nouvel ordre) parce qu'il doit valoir $0$ pour respecter les structures spin
réelles.\\

\noindent Attention: si l'on choisit sur $F"$ la mesure unité $\Omega_0=2u"_0V"_0$, alors, quelquesoit $\lambda$ de module $1$,
la conjugaison pour laquelle les spineurs $t$ et $s$ ci-dessus sont réels est $-e^{-i\beta}I_0$.\\
En effet, la définition de $*"$ fait que $V"_0=-*"u"_0$, par conséquent
\begin{equation}
\overline{\lambda}e^{-i\beta}V"_0=-\overline{\lambda}e^{-i\beta}*"(\lambda u"_0)=-e^{-i\beta}*"(\lambda u"_0)
=-e^{-i\beta}I_0(\lambda u"_0),
\end{equation}
et $\Omega_0=*"1=-I_0(1)$, par définition de $I_0$ en degré zéro, donc
\begin{equation}
\overline{\lambda}e^{-i\beta}\Omega_0=-\overline{\lambda}e^{-i\beta}I_0(1)=-e^{-i\beta}I_0(\lambda.1).
\end{equation}

\noindent Remarques: dans la présentation habituelle de la trialité réelle (cf. \cite{adams1996lectures}), on part d'un espace de spineurs réels
$S_8=S_8^{+}\oplus S_8^{-}$, munie d'un produit scalaire défini positif invariant $B$. On choisit un vecteur $u_0$
de norme $1$ dans $F_{8}$. Le stabilisateur de $u_0$ dans le groupe $Spin_8$ compact est un groupe $Spin_7$.
L'action de ce sous-groupe $Spin_7$ est transitive sur les sphères de dimension $7$
de $S_8^{-}$. En effet, déjà le groupe $Spin_5$ est transitif
sur les $S_7$ de sa représentation spinorielle de dimension complexe $4$, qui s'identifie à l'espace $S_8$ des spineurs réels en dimension $7$.
Cf. \cite{adams1996lectures}. Le fixateur d'un élément $t_0$ dans la sphère $S_7$ de rayon $1$ de $S_8^{-}$ est un groupe compact $G_2$. Et si on impose en plus de fixer une
structure complexe positive $J_0$ sur $F$, on trouve un $SU_3$ dans $G_2$. Ce $G_2$ est le groupe des automorphismes
de l'algèbre de Cayley qui est associée à la trialité et a pour unité $u_0=1$; le $SU_3$ est le sous-groupe qui
commute avec l'opérateur $J_0$ représentant $i$ dans l'algèbre.
L'espace homogène $G_2/SU_3$ paramètrisant les sous-algèbres complexes,
est isomorphe à une sphère $S^{6}$. Et l'espace homogène $Spin_7/G_2$
décrivant les structures d'octaves sur $F$ compatibles avec la structure Euclidienne est une sphère de
dimension $7$.\\
Variante: une structure d'octonions sur $F_8$ est choisie. Elle permet d'identifier un $G_2$ dans le $Spin_7$
fixant l'unité $u_0=1$. Alors, fixer un autre point $e_7$ dans la sphère de dimension $6$ de rayon $1$ dans
l'orthogonal de $u_0$, détermine un $SU_3$ dans $G_2$. Cela détermine la structure complexe $J_0$
telle que $J_0e_7=u_0$.\\
\indent Mais nous constatons que ces manières de faire loupent l'interprétation complexe de $\tau$,
parmi les spineurs pairs, le vecteur imaginaire $1$, qui sera le neutrino, le vecteur $\Omega$ qui sera l'électron, et les deux
espaces de quarks qui vont en résulter grâce à $\tau$, ainsi que le cercle $U_1$ de la phase $\beta$,
parce que la structure réelle est fixée trop tôt. Par conséquent, dans toutes ces approches,
fixer une trialité fait apparaître un $SU_3$, mais seul le choix des coordonnées complexes fait
apparaître les éléments du modèle standard, en particulier un $U_1$ qui peut s'interpréter comme celui de l'électromagnétisme.\\
Notre choix de coordonnées n'était pas fait dans ce but, il était fait pour identifier un modèle des spineurs
à partir des vecteurs, en passant par les sous-espaces isotropes maximaux de $F\otimes \mathbb{C}$. Il a donc
consisté à choisir deux éléments spinoriels, un $J_0$ et un $t$, sans avoir à choisir arbitrairement une représentation
spin. Cette stratégie est apparentée à un choix twistoriel, suivant Penrose (cf. \cite{penrose1988spinors}), puisqu'elle commence avec le choix d'un point
dans l'espace des structures complexes positives, or cet espace de structures complexes est un espace homogène de dimension douze, fait d'une famille
bien définie de sous-groupes conjugués entre eux dans le groupe $Spin_8$, qui lui, est canoniquement défini par $F_8$, donc
par la géométrie de l'espace-temps.\\

\noindent Le plongement $\tau$ de $F\otimes \mathbb{C}$ dans $S^{+}$ envoie $u_0$ sur $-s$ (à cause de la réflexion dans le $\mu$).\\
Pour le vecteur $e_7=-Je_8$ (à ne pas confondre avec $-ie_8$ ou $ie_8$, puisqu'il engendre une droite complexe
différente dans $F\otimes \mathbb{C}$), on a
\begin{align}
\tau(e_7)&=-ie"_8\wedge t+i\iota(e'_8)t=-iu"_0\wedge t+i\iota(u'_0)t\\
&=-iu"_0\wedge u"-iu"_0\wedge t"+i\iota(u_0')u"+i\iota(u_0')t"\\
&=0-\frac{i}{\lambda}e^{i\beta}u"_0V"_0+i\frac{\lambda}{2}.1\\
&=-i\frac{\Omega}{2\lambda}+\frac{i}{2}\lambda.1.
\end{align}
Et pour les six vecteurs $e_i$ d'une base orthonormée
quelconque de l'orthogonal du plan de $e_7$ et $e_8$, on a
\begin{align}
\tau(e_i)&=e"_i\wedge t+\iota(e'_i)t\\
&=\lambda e"_i\wedge u"_0+0+\frac{1}{\lambda}e^{-i\beta}\iota(e'_i)V"_0\\
&=\lambda e"_i\wedge u"_0+\frac{1}{\lambda}e^{-i\beta}\iota(e'_i)V"_0.
\end{align}

\noindent Nous appelerons \emph{quarks électriques} les bivecteurs complexes de type $(0,2)$ de la forme
\begin{equation}
q"(v")=2\lambda v"\wedge u"_0
\end{equation}
et
\emph{quarks magnétiques} les bivecteurs complexes de type $(0,2)$ qui ont la forme
\begin{equation}
q'(v')=2\lambda^{-1}e^{-i\beta}\iota(v')V"_0
\end{equation}
pour $v'$ dans $V'_0$. Les normalisations des modules sont faites pour que les deux formules donnent des plongements isométriques
du sous-espace $V_0$ de l'espace Euclidien $F$ dans l'espace $S_8^{+}$ réel. \\
Si bien qu'on peut écrire
\begin{equation}
\tau(e_j)=\frac{q"_j}{2}+\frac{q'_j}{2}.
\end{equation}

Le plongement de $F$ par $\tau\circ\tau$ se décrit ainsi: l'image de $u_0$ est le
vecteur $-t$, et l'application sur l'orthogonal est définie par
\begin{align}
\tau^{2}(e_i)&=e"_i\wedge s+\iota(e'_i)s\\
&=\frac{\lambda}{2} e"_i+\iota(e'_i)\frac{\Omega}{2\lambda}\\
&=\frac{\lambda}{2}e"_i+\frac{1}{2\lambda}\iota(e'_i)\Omega
\end{align}\\

\indent D'autre part, comme on peut le lire dans le livre de Claude Chevalley, toute trialité $\tau$ permet de décrire un automorphisme
extérieur $j$ d'ordre trois du groupe $G_0$ (que ce soit la version algébrique complexe ou que ce soit
la version compacte du groupe des spineurs $Spin_8$). Le groupe quotient $Out(G_0)/Int(G_0)$ est isomorphe
au groupe symétrique $\mathfrak{S}_3$, l'image de $j$ dedans est un cycle d'orde trois. Il y a donc deux classes de
conjugaison d'automorphismes
d'ordre trois de $G_0$ (ou de $Spin_8$); celle de $j$ et celle de $j^{-1}=j^{2}$.\\
Soit $\sigma$ la somme des trois représentations de $G_0$ dans la somme directe $\Delta$ de $F_8, S_8^{+}$ et $S_8^{-}$, et
$g\in G_0$, alors $j(g)$ est l'unique élément de $G_0$ tel que
\begin{equation}
\tau\circ \sigma(g)=\sigma(j(g))\circ \tau.
\end{equation}
Ainsi la paire $(\tau,j)$ réalise une équivalence extérieure des trois représentations
inéquivalentes de dimension huit du groupe des spineurs $G_0$.\\
La trialité $\tau$ apparaît donc comme une quantification de l'automorphisme $j$.\\

\noindent Partant de là, décrivons l'automorphisme $j_*$ de l'algèbre de Lie de $spin_8$
induit par $j$.\\
Le plus simple est de calculer $j^{-1}_*$, en partant des vecteurs $v$ de $F$ et en faisant appel aux formules
de la proposition $2$, pour satisfaire à la formule suivante:
\begin{equation}
\sigma_*\circ j^{-1}_*(X)(v)=\tau^{-1}\circ \sigma_*(X)\circ \tau (v)
\end{equation}\\

\noindent \textbf{Proposition 3}: dans la base ordonnée $(H_7=-2ie_7\wedge e_8,
H_5=-2ie_5\wedge e_6, H_3=-2ie_3\wedge e_4, H_1=-2e_1\wedge e_2)$ de l'algèbre de Lie réelle
$\mathfrak{h}_{4}$ dans $iso_8$, la restriction de $j^{-1}_*$ s'écrit $\frac{1}{2}T_8$, où
\begin{equation}\label{matricethuit}
T_8=
\begin{pmatrix}
-1&-1&-1&-1\\
\quad 1&\quad 1&-1&-1\\
\quad 1&-1&\quad 1&-1\\
\quad 1&-1&-1&\quad 1
\end{pmatrix}
\end{equation}

\noindent \emph{Démonstration}: on commence par remarquer que l'effet sur $\tau(e_8)$ et $\tau(e_7)$ de
$2\sigma_*(e_{k+1}\wedge e_k)$ est le même quelque soit $k=1,3,5,7$, et donne la rotation de $e_8$ vers
$e_7$ en revenant à $F$ par $\tau^{-1}$. Puis on examine l'effet sur $\tau(e_j)$ lorsque $j$ varie de
$1$ à $6$ en utilisant toujours la formule \eqref{actionspincomplexe}.\\

\noindent C'est un plaisir de vérifier que $T_8^{3}=8Id$, donc $(\frac{1}{2}T_8)^{3}=Id$.\\
\noindent La restriction de $j_*$ a pour matrice $\frac{1}{2}T_8^{2}=\frac{1}{2}^{t}T_8$, où
\begin{equation}\label{matricejetoile}
^{t}T_8=
\begin{pmatrix}
-1&\quad 1&\quad 1&\quad 1\\
-1&\quad 1&-1&-1\\
-1&-1&\quad 1&-1\\
-1&-1&-1&\quad 1
\end{pmatrix}
\end{equation}\\

\indent Introduisons la base duale $x_1,x_3,x_5,x_7$ des $H_j$ dans le dual $\mathfrak{h}_{4}^{*}$, qui s'identifie ainsi à $\mathbb{R}^{4}$;
une base de racines simples de $so_8\otimes \mathbb{C}$ est fournie par les co-vecteurs suivant:
\begin{equation}
\alpha_1=-x_1-x_3,\quad \alpha_2=x_5-x_3,\quad \alpha_3=x_5-x_7,\quad \alpha_4=x_5+x_7.
\end{equation}
Alors $j_*$ induit l'automorphisme d'ordre trois de $\mathfrak{h}_{4}^{*}$ qui envoie $\alpha_1$ sur $\alpha_3$,
$\alpha_3$ sur $\alpha_4$ et $\alpha_4$ sur $\alpha_1$, en laissant fixe $\alpha_2$, et nous retrouvons ainsi, à peu de
choses près, la présentation
usuelle de l'automorphisme d'orde trois du diagramme de Dynkin de $so_8$, tel qu'il est décrit par exemple dans
Bourbaki, Groupes et algèbres de Lie, chapitres 4,5,6; p 257, \cite{bourbaki1981groupes456}.\\
Par linéarité, on en déduit l'action de $j_*$ sur l'ensemble des $24$ racines de $so_8$,
qui sont les combinaisons $\epsilon_j x_j +\epsilon_k x_k$, lorsque $j$ et $k$ prennent toutes les valeurs impaires
possibles entre $1$ et $7$, et où $\epsilon_j$ et $\epsilon_k$ valent $\pm 1$ indépendamment
l'un de l'autre. Et de là
on déduit l'action de $j_*$ sur l'algèbre de Lie $so_8(\mathbb{C})$, en considérant la base de Chevalley des $X_\alpha$ et les vecteurs
$H_\alpha$ de $\mathfrak{h}_4$, lorsque $\alpha$ parcourt l'ensemble des racines. L'action sur $so_8(\mathbb{R})$ s'en déduit en prenant la base de
Chevalley standard de l'algèbre de Lie compacte $u_\alpha=X_\alpha+X_{-\alpha}$, $v_\alpha=i(X_\alpha+X_{-\alpha})$. Cf. Bourbaki, Lie 9, \cite{bourbaki1982groupes9}.\\

\noindent Remarque: les éléments $2iH_{j};j=1,3,5,7$, n'appartiennent pas seulement à l'algèbre de Lie de $\mathfrak{t}_4$;
étant tous de carré $-Id$, ils appartiennent aussi au groupe $Spin_8$, et l'automorphisme d'ordre trois du groupe $Spin_8$ agit dessus comme le prescrit
l'automorphisme linéaire $j_*$ représenté par la matrice transposée de $T_8$.\\

Sur $\mathfrak{h}_4$ nous retenons la structure Euclidienne positive associée à la base $H_j;j=1,...,7$, de même sur $\mathfrak{h}_4^{*}$ avec
la base duale des $x_j;j=1,...,7$, si bien que les
carrés des longueurs des racines et co-racines sont tous égaux à $2$. Sur $\mathfrak{t}_4=i\mathfrak{h}_4$ la métrique est
définie négative; on y retrouve la métrique de Killing de $so_8(\mathbb{R})$, qui fait que, pour $\alpha$ parcourant l'ensemble des racines positives,
les $u_\alpha$ et $v_\alpha$ sont deux à deux orthognaux, de carrés égaux à $-2$.\\

\indent L'algèbre de Lie $\mathfrak{u}_4$ du stabilisateur de $J_0$ est somme directe de l'algèbre de Lie $\mathfrak{su}_4$,
dont les douze co-racines $H_\beta$ sont les vecteurs $i(H_j-H_k); k\neq j$
de $\mathfrak{t}_4$, et du centre $\mathfrak{z}_4$, engendré par
\begin{equation}
Z_4=i(H_1+H_3+H_5+H_7).
\end{equation}
L'algèbre $\mathfrak{su}_4$ est somme du tore $\mathfrak{t}_4$ et des douze droites réelles engendrées par les $u_\beta$ et $v_\beta$,
lorsque $\beta$ est positive, i.e. $j< k$.
Si bien que l'orthogonal $\mathfrak{p}_4$ de $\mathfrak{su}_4$ dans $so_8$, est la
somme des douze droites engendrées par les  vecteurs $u_\gamma$ et $v_\gamma$, associés aux six co-racines $i(H_j+H_k); k\neq j$
et à leurs opposées.\\
\indent L'algèbre de Lie $\mathfrak{u}_4$ est représentée par l'ensemble des
matrices de Clifford qui peuvent s'écrire dans une base orthornomée convenable sous
la forme suivante
\begin{equation}\label{matriceuquatre}
\varphi=\sigma_*(\varphi_1 e_1\wedge e_2+\varphi_2 e_3\wedge e_4+\varphi_3 e_5\wedge e_6+\varphi_4 e_7\wedge e_8).
\end{equation}

\indent  En ces termes le stabilisateur $\widetilde{U_t}$ du spineur impair spécial $t_0$ est décrit par le lemme suivant:\\

\noindent \textbf{Lemme 4}: l'algèbre de Lie $\mathfrak{u}_t$ stabilisant $t_0$ est faite des matrices de
la forme $A-\frac{2}{i}Tr(A)\gamma_7\gamma_8$ où $A$ est l'algèbre $\mathfrak{u}_3$ des matrices antihermitiennes
opérant sur l'espace de dimension trois complexe engendré par $e_1,e_3,e_5$. Autrement dit, dans \eqref{matriceuquatre},
on doit avoir $\varphi_4=\varphi_1+\varphi_2+\varphi_3$, ou encore, en termes des poids du tore contenant $A$ et $\gamma_7\gamma_8$,
on doit avoir $x_7=x_1+x_3+x_5$ (ou $-x_1-x_3=x_5-x_7$).\\

\noindent \emph{Démonstration}: une telle matrice $R$ s'écrit dans une base orthonormée bien choisie de
$F$ complétant $(e_7,e_8)$ sous la forme \eqref{matriceuquatre} qui précéde. Son action sur $e_8$ a été calculée au cours de
la preuve de la proposition $2$. En appliquant à $e"_8$ on trouve donc
\begin{equation}
R(e"_8)=\frac{i}{4}(-\varphi_1-\varphi_2-\varphi_3+\varphi_4)e_8.
\end{equation}
La condition sur la trace équivaut à l'identité $\varphi_4=\varphi_1+\varphi_2+\varphi_3$, qui est la condition
nécessaire et suffisante pour stabiliser $e_8$, i.e. donner $0$.\\

\noindent En particulier, le centre $\mathfrak{z}_t$ de $\mathfrak{u}_t$
est fait des homothéties $i\omega Id_3$, avec $\omega\in \mathbb{R}$
quelconque, auxquelles on ajoute $12\omega \sigma_*(e_7\wedge e_8)$; il est engendré par
\begin{equation}
Z_t=i(H_1+H_3+H_5+3H_7),
\end{equation}

\indent Dans la suite, sauf grave inconvenance, nous noterons $SU_3$ le groupe des éléments de déterminant $1$ dans $\widetilde{U_t}$.\\
\indent Ce groupe agit sur l'espace $F$ comme les transformations spéciales
unitaires qui fixent point par point le plan de $e_7$ et $e_8$. Puisque $e_8=J_0e_7$, cela donne un représentation isomorphe
à la somme d'une représentation triviale complexe $\mathbb{C}$ et d'une représentation fondamentale $\mathbb{C}^{3}$.\\
Nous en déduisons que l'action de $SU_3$ sur $S_8^{-}$ et sur $S_8^{+}$ se décompose de la même façon, grâce au lemme suivant:\\

\noindent \textbf{Lemme 5}: l'automorphisme $j$ préserve $SU_3$ dans $Spin_8$. Donc les trois représentations
de $SU_3$ sur $F$, $S_8^{+}$ et $S_8^{-}$ sont équivalentes.\\

\noindent \emph{Démonstration}: les deux racines simples canoniques de $su_3$ dans $so_8$ sont $x_1-x_3$ et $x_3-x_5$,
et elles sont invariantes par $T_8/2$.\\

\noindent Attention: il n'est pas vrai que $j_*$ préserve $\mathfrak{u}_4$, ni même $\mathfrak{u}_t$.\\

\indent Les formules \eqref{actionspincomplexe} permettent de trouver directement les décompositions précises des représentations
spin de $SU_3$: pour $S_8^{-}$
le plan de $u"_0$ et de $V"_0$ est préservé point par point et l'espace de dimension six orthogonal est isomorphe
à un $\mathbb{C}^{3}$; pour $S_8^{+}$ le plan de $1$ et de $\Omega_0$ est préservé point par point et l'espace de dimension six
des quarks est isomorphe
à un $\mathbb{C}^{3}$ fondamental.\\
\indent Les structures complexes naturelles $SU_3$-invariantes sur les spineurs réels sont celles qui se déduisent de $J_0$
par la trialité: dans $S_8^{-}$, on a
\begin{equation}
J^{+}(1)=i.1;\quad J^{+}(\Omega)=-i\Omega;\quad J^{+}(q")=-iq";\quad J^{+}(q')=iq';
\end{equation}
dans $S_8^{+}$, on a, pour $u"$ orthogonal hermitien à $u"_0$ et $V"$ orthogonal hermitien à $V"_0$:
\begin{multline}
J^{-}(u"_0)=iu"_0;\quad J^{-}(V"_0)=-iV"_0;\\ J^{-}(u")=-iu";\quad J^{-}(V")=iV";
\end{multline}
Toutefois, n'importe quel changement de signe effectué séparément pour les leptons et les quarks donne
aussi une structure complexe $SU_3$ invariante.\\

\noindent Nous constatons que sur l'espace $Q'$ (resp. $Q"$)  des quarks magnétiques (resp. électriques), vu comme
sous-espace du complexifié $S_8^{+}\otimes \overline{\mathbb{C}}=\Lambda^{ev}F"$, la struture complexe invariante par trialité $J^{+}$
coïncide avec la structure induite naturelle (resp. son opposée). Il en va de même pour $-J^{-}$ sur les orthogonaux de $u"_0$ et $V"_0$
dans $S_8^{-}\otimes \overline{\mathbb{C}}=\Lambda^{od}F"$, et pour $J_0$ sur les sous-espaces $F'$ et $F"$ dans les vecteurs $F\otimes \mathbb{C}$.\\
En particulier, c'est pour la structure $J^{+}$ et pas pour la structure induite que le groupe $SU_3$ agit de façon équivalente
sur les deux espaces de quarks, $Q'$ et $Q"$.\\

\subsection{L'action des bosons vecteurs et les accords de Yukawa}

\indent Rappelons que les représentations linéaires complexes irréductibles de $SU_3$ et de l'algèbre de Lie $su_3$ sont toutes
isomorphes à un des sous-espaces $D(p,q)$ qui est formé par les tenseurs
séparément symétriques qui sont annulés
par toutes les contractions dans $T(p,q)=S^{p}(\mathbb{C}^{3})\otimes S^{q}((\mathbb{C}^{3})^{*})$. L'espace $D(p,q)$
s'identifie à l'espace des polynômes $P$ sur $\mathbb{C}$ en les variables $x'_1,...,x'_p, x"_1,...,x"_q$
satisfaisant à $\sum_{k,l}\partial'_{k}\partial"_{l}P=0$. Cf. \cite{bourbaki1982groupes9}, p.126. Le couple $(p,q)$ est le poids
dominant de la représentation dans la base usuelle des poids fondamentaux. Deux représentations $D(p,q)$ et $D(r,s)$
avec $(r,s)\neq (p,q)$ ne sont pas isomorphes.
Par contre $D(p,q)$ et $D(q,p)$ sont duales l'une de l'autre (et aussi conjuguées l'une de l'autre), et elles sont extérieurement équivalentes en passant par
l'auto-morphisme extérieur (unique modulo les intérieurs): $g\mapsto \overline{g}$. \\
Les représentations irréductibles réelles sont les $D(p,p)$, comme, par exemple, la représentation adjointe pour $p=1$, ou bien les
points fixes $(D(p,q)\oplus D(q,p))_{\mathbb{R}}$ de la conjugaison de $D(p,q)\oplus D(q,p)$, pour $p\neq q$.\\

\noindent En ces termes, les représentations de $SU_3$ dans les espaces Euclidiens réels $F$, $S_8^{+}$ et $S_8^{-}$
sont toutes trois complexes pures du type $D(0,0)\oplus D(1,0)$.\\

\indent D'autre part $SU_3$ agit sur tous les éléments de l'algèbre de Lie de $GL_8(\mathbb{R})$ par crochet. Or cette algèbre
permet de décrire tous les champs $C$ et $M$, que nous avons divisés en une partie antisymétrique $A,M_A$, une partie symétrique sans trace $S,M_S$,
et une trace globale $B,M_B$.\\

En particulier,
les éléments de $so_8$ étant données par les matrices anti-symétriques, le supplémentaire
orthogonal $\mathfrak{p}_4$ de $u_4$ dans $so_8$ est fait des matrices complexes antisymétriques $A'+iB'$ vues plus haut; il forme la représentation $\Lambda^{2}$
de $SU_4$, qui se casse en $\Lambda_+^{2}$ et $\Lambda_-^{2}$, autoduales et anti-autoduales, toutes deux de dimension
complexe trois, mais non-équivalentes. Pour $SU_3$ ces deux représentations deviennent respectivement équivalentes à la représentation standard
$\mathbb{C}^{3}=D(1,0)$
et à sa duale $\overline{\mathbb{C}^{3}}=D(0,1)$.\\
\indent Quant aux sept champs réels orthogonaux à $u_3$ dans $u_4$, un est stabilisé par l'action adjointe de $U_3$, et donne
donc une représentation triviale réelle $D(0,0)_{\mathbb{R}}$, alors que les six autres
forment une représentation complexe isomorphe à la représentation fondamentale $\mathbb{C}^{3}=D(1,0)$, comme le montre le calcul
matriciel direct. \\

\indent L'algèbre $so_8$ contient un plan réel neutre pour $SU_3$, engendré par le centre de $u_4$ et par
le commutant de $u_3$ dans $u_4$; comme générateur privilégié du premier nous retenons $Z_4=i\sum H_j$ et comme
générateur privilégié du second nous retenons
\begin{equation}
Z_1=iH_7.
\end{equation}
Les directions $Z_1$, $Z_4$ et $Z_t$
son liées par la relation linéaire $Z_t-Z_4=2Z_1$.\\

\indent Du côté des endomorphismes symétriques de $F$, i.e. dans $\mathfrak{p}_8$, il arrive $36$ dimensions réelles.\\
\indent Afin de décrire la représentation de $U_4$ sur $\mathfrak{p}_8$, puis celle de $SU_3$, nous reprenons les notations matricielles du paragraphe
sur la brisure complexe. Les matrices anti-hermitiennes $4\times 4$ étaient notées
\begin{equation}
X=A+iB=
\begin{pmatrix}
A&-B\\
B&A
\end{pmatrix}
\end{equation}
où $A,B$ étaient $4\times 4$ réelles, avec $A$ antisymétrique et $B$ symétrique.\\
Nous décomposons les matrices $8\times 8$ symétriques réelles en sommes
\begin{equation}
\Sigma=\Sigma_1+\Sigma_2;\quad \Sigma_1=
\begin{pmatrix}
S_1&M_1\\
M_1&-S_1
\end{pmatrix}
,\Sigma_2=
\begin{pmatrix}
S_2&M_2\\
-M_2&S_2
\end{pmatrix};
\end{equation}
où $S_1,S_2,M_1$ sont des matrices $4\times 4$ symétriques réelles et où $M_2$ est une matrice $4\times 4$ antisymétrique réelle.\\
Le calcul de l'action adjointe des $X$ par crochets avec les $\Sigma$, démontre que si l'on pose $Q_1=S_1+iM_1$ et $H_2=S_2+iM_2$,
on obtient respectivement l'action naturelle de $u_4$ sur les tenseurs symétriques complexes de degré deux $S^{2}\mathbb{C}^{4}$
(irréductible de dimension $10$ sur $\mathbb{C}$) et la représentation naturelle de $u_4$ sur les formes hermitiennes, qui se décompose en une trace réelle
de dimension $1$ et une représentation adjointe $su_4$ de dimension réelle $15$.
\begin{multline}
ad(X)\Sigma_1\approx (A+iB)Q_1-Q_1(A-iB),\\ ad(X)\Sigma_2\approx (A+iB)H_1-H_1(A+iB).
\end{multline}
Les crochets entre des $\Sigma_1$ donnent des éléments de $u_4$:
\begin{equation}
[\Sigma_1,\Sigma'_1]=
\begin{pmatrix}
[S_1,S'_1]+[M_1,M'_1]&\{S_1,M'_1\}-\{M_1,S'_1\}\\
\{M_1,S'_1\}-\{S_1,M'_1\}&[S_1,S'_1]+[M_1,M'_1]
\end{pmatrix}
\end{equation}
où les accolades désignent l'anti-crochet $\{M,M'\}=MM'+M'M$.\\
Les crochets entre des $\Sigma_2$ donnent aussi des éléments de $u_4$:
\begin{equation}
[\Sigma_2,\Sigma'_2]=
\begin{pmatrix}
[S_2,S'_2]-[M_2,M'_2]&[S_2,M'_2]+[M_2,S'_2]\\
-[S_2,M'_2]-[M_2,S'_2]&[S_2,S'_2]-[M_2,M'_2]
\end{pmatrix}
\end{equation}
Les blocs diagonaux y sont antisymétriques et égaux et les blocs anti-diagonaux y sont symétriques et opposés.\\
Et les crochets des $\Sigma_1$ avec des $\Sigma_2$ donnent des tenseurs antisymétriques du type des $Y$ dans $\Lambda^{2}\mathbb{C}^{4}$,
que nous avons nommés $U,V$:
\begin{equation}
[\Sigma_2,\Sigma_1]=
\begin{pmatrix}
[S_2,S_1]+\{M_2,M_1\}&[S_2,M_1]-\{M_2,S_1\}\\
[S_2,M_1]-\{M_2,S_1\}&-[S_2,S_1]-\{M_2,M_1\}
\end{pmatrix}
\end{equation}
Les blocs diagonaux y sont antisymétriques et opposés et les blocs anti-diagonaux y sont antisymétriques et égaux.\\

\indent Lorsqu'on restreint ces représentations à $SU_3$, les $\Sigma_1$ du $S^{2}\mathbb{C}^{4}$ se cassent en
une représentation $S^{2}\mathbb{C}^{3}=D(2,0)$,
complexe irréductible de dimension $6$, une représentation fondamentale $\mathbb{C}^{3}=D(1,0)$,
et une triviale complexe $\mathbb{C}=D(0,0)$. La partie hermitienne $u_4^{H}$ des $\Sigma_2$ se casse en une adjointe $u_3^{H}$, réelle
de dimension $8$, qu'on peut noter $D(1,1)_{\mathbb{R}}$, une trace réelle $D(0,0)_{\mathbb{R}}$, et une $D(1,0)$. Ce
qui fait bien les $8+1+6=15$ dimensions réelles de $u_4^{H}$.\\
Au total, on trouve donc
\begin{equation}
\mathfrak{p}_8(\mathbb{R})=su_3^{H}\oplus S^{2}\mathbb{C}^{3}\oplus 2\mathbb{C}^{3}\oplus \mathbb{C}\oplus 2\mathbb{R}.
\end{equation}
En notant avec les poids dominants, cela donne
\begin{equation}
\mathfrak{p}_8(\mathbb{R})=D(1,1)_{\mathbb{R}}\oplus D(2,0)_{\mathbb{C}}\oplus 2 D(1,0)_{\mathbb{C}}\oplus  D(0,0)_{\mathbb{C}}\oplus  2D(0,0)_{\mathbb{R}}
\end{equation}

\indent Les deux facteurs triviaux réels sont donnés par la trace totale $B_0=Id_8$, et par une trace partielle $B_1$, que
nous prenons égale à $1$ sur $e_7$ et sur $e_8$, et à $0$ sur l'orthogonal dans $F$. \\
Et enfin le facteur $D(0,0)_{\mathbb{C}}$ est engendré par deux éléments de $\mathfrak{p}_8$, qui avec $Z_1$, engendrent l'espace
des endomorphismes sans trace
du plan $(e_7,e_8)$ qui fixent point par point l'orthogonal de ce plan dans $F$.\\
Matriciellement, on peut écrire les blocs suivant les directions $e_7,e_8$:
\begin{equation}
Y=
\begin{pmatrix}
0&1\\
1&0
\end{pmatrix};\quad
X=
\begin{pmatrix}
1&0\\
0&-1
\end{pmatrix};\quad
B_1=
\begin{pmatrix}
1&0\\
0&1
\end{pmatrix};\quad
2Z_1=
\begin{pmatrix}
0&-1\\
1&0
\end{pmatrix}
\end{equation}
De sorte que $[Y,X]=4Z_1$. Il est naturel d'identifier la matrice $xX+yY$ à $x+iy\in \mathbb{C}$, comme représentation
des similitudes, d'où le $D(0,0)$ complexe.\\

Il existe une façon naturelle de faire agir les $64$ champs de bosons sur les fermions.\\
\indent Elle consiste à prendre l'action de $spin_8$ telle qu'elle se présente, sans la tordre, sur les espaces Euclidiens réels de dimension huit
$S_8^{+}$ et $S_8^{-}$, de l'étendre $\mathbb{C}$-linéairement à l'espace complexifié  muni de la structure hermitienne
canonique, puis de prolonger cette action à l'espace $i\mathfrak{p}_8$, identifié au supplémentaire de $spin_8$ dans l'algèbre de Lie $u_8$.\\
Cette manière de faire correspond
au "truc unitaire" de Weyl, qui exhibe une bijection naturelle entre les représentations
irréductibles de dimensions finies holomorphes de $gl(E\otimes \mathbb{C})$ et les représentations unitaires irréductibles
de $u(E\otimes \mathbb{C})$ pour tout espace vectoriel réel Euclidien $E$ de dimension finie.\\
L'espace $i\mathfrak{p}$ donne les endomorphismes anti-hermitiens imaginaires purs, et $so(E)$ les
endomorphismes anti-hermitiens purement réels.\\

Les formules de la proposition $2$ décrivent les représentations spin; elles montrent en particulier que l'élément $2Z_1=2iH_7$
de $spin_8$ agit comme $-J^{+}$ sur $S_8^{+}\otimes \mathbb{C}$ et comme $J^{-}$ sur $S_8^{-}\otimes \mathbb{C}$.\\
C'est-à-dire que $2Z_1$ agit en multipliant par $-i$ sur $1$ et $Q'$ ainsi que sur $u"_0$ et $(V"_0)^{\perp}$, et en multipliant
par $i$ sur $\Omega$ et $Q"$ ainsi que  sur $V"_0$ et $(u"_0)^{\perp}$.\\

Mais, attention, en vertu de la formule \eqref{matricejetoile}, l'élément qui joue le rôle de $Z_1$ dans la représentation spin sur $S_8^{+}$,
c'est-à-dire celui qui commute avec la représentation de $U_3$, en étant orthognal au centre de $U_3$, est
\begin{equation}
j_*(Z_1)=-\frac{1}{2}Z_4.
\end{equation}
Il est donc égal à la somme des $\gamma_{k+1}\gamma_k$, qui agit sur les constantes comme $-i$,
sur $\Omega$ comme $+i$, et sur toutes les sortes de quarks comme $0$.\\

Nous retrouvons ainsi la même structure de représentation de $SU_3$ sur $S_8^{+}$ que celle sur $F=\mathbb{R}^{8}$.\\
En particulier, il existe bien deux opérateurs $X^{+},Y^{+}$ analogues aux $X,Y$ du $sl_2$ sur $e_7,e_8$, qui font ce que
les opérateurs $W_+$ et $W_-$ des interactions faibles font en échangeant le neutrino left et l'électron left, mais
par contre $X^{+}$ et $Y^{+}$ ne font rien avec
les quarks. Plus généralement, il est impossible de définir un opérateur commutant avec $SU_3$ qui échange les espaces
$Q'$ et $Q"$, car la représentation de $SU_3$ sur $Q'$ est du type $D(1,0)$ alors que sur $Q"$ elle est du type $D(0,1)$.
C'est ce qui arrive lorsqu'on complexifie la représentation réelle sous-jacente à une représentation complexe (cf.
\cite{adamslectures}).\\
Or c'est un fait d'expérience physique bien établi que les créations des deux
sortes de quarks constituent des représentations isomorphes, toutes les deux du type $D(1,0)$. Ceci s'oppose à l'identification
de $Q"$ dans $S_8^{+}\otimes \mathbb{C}$ avec les quarks observés.\\

\indent La solution du paradoxe vient des principes de reversibilité du temps et de positivité de l'énergie, qui
entraînent que toutes les particules de spin $1/2$ en dimension Lorentzienne quatre
arrivent par paires complexes d'une particule et d'une anti-particule, portant des structures conjuguées.
Par conséquent, à côté de $S_8^{+}$ complexe, et \emph{a priori} à côté de $S_8^{-}$ complexe avant sa pétrification,
il est nécessaire d'introduire
les espaces conjugués $\overline{S_8^{+}}$ et $\overline{S_8^{-}}$.\\
Pour les secteurs leptoniques, l'espace $\overline{S_8^{+}}$ correspond aux anti-neutrinos et aux positrons, etc. mais pour le
secteur des quarks, la moitié, au contraire, va correspondre à des particules, c'est la moitié $\overline{Q"}$ des conjugués de nos
"quarks électriques", qui eux en fait sont des anti-particules. C'est en tout cas ce que dit l'expérience en révèlant que les
vrais quarks sont tous du même type, $D(1,0)$.\\

Nous procéderons de même avec les spineurs impairs, en retenant sur $(u"_0)^{\perp}$ dans $F"$
la structure complexe conjuguée de celle qui se présente naturellement.\\

\noindent \textbf{Définitions}: la division particule-antiparticule et lepton-quark invite à redéfinir les espaces de multiplicités des spineurs
de Dirac, en posant
\begin{multline}
L=\mathbb{C}1+\mathbb{C}\Omega;\quad \overline{L}=\overline{\mathbb{C}}\overline{1}+\overline{\mathbb{C}}\overline{\Omega};\\
Q=Q'\oplus \overline{Q"};\quad \overline{Q}=\overline{Q'}\oplus Q";\\
M_8^{+}=L\oplus Q;\quad \overline{M_8^{+}}=\overline{L}\oplus \overline{Q}.
\end{multline}
Les espaces sans barre donnent les directions des particules, ceux avec barre donnent les directions des anti-particules associées.\\
Il y a une répartition en tous points analogue pour $S_8^{-}\oplus \overline{S_8^{-}}$ donnant $M_8^{-}$ et $\overline{M_8^{-}}$.
La lettre $M$ est pour multiplicité, car c'est ainsi que l'espace $M_8^{+}$ (et d'une façon plus cachée $M_8^{-}$) apparaissent
dans notre univers $W$ à quatre dimension.\\

\noindent Remarque: cette redistribution des quarks n'aurait pas beaucoup de sens dans une théorie Euclidienne, puisqu'elle s'appuie
sur la nécessité de doubler les degrés de liberté des particules fermioniques.\\

\noindent \textbf{Définitions}:
l'opérateur $W_+$ (resp. $W_-$) est l'endomorphisme $\mathbb{C}$-linéaire
de $M_8^{+}$ qui est défini par les formules suivantes, pour tout vecteur réel $v$ de $F$, décomposé en
$v'+v"$ dans $F'\oplus F"$:
\begin{gather}
W_+(\overline{v"}\wedge \overline{u"_0})=\exp(-i\beta)\iota(v')V"_0,\quad W_+(\iota(v')V"_0)=0\\
W_+(1)=-\exp(-i\beta)\Omega,\quad W_+(\Omega)=0,\\
W_-(\overline{v"}\wedge \overline{u"_0})=0,\quad W_-(\exp(-i\beta)\iota(v')V"_0)=-v"\wedge u"_0;\\
W_-(1)=0,\quad W_-(\exp(-i\beta)\Omega)=1.
\end{gather}
On définit de même les opérateurs conjugués, correspondant aux anti-particules de $W_{\pm}$, agissant
dans $M_8^{-}$.\\

\noindent Remarque: nous pourrions considérer à la place des opérateurs $W_{\pm}$ et $\overline{W_{\pm}}$,
des applications anti-$\overline{C}$-linéaires de $S_8^{+}$ (ou de $S_8^{-}$), mais cela
irait trop contre les habitudes et provoquerait sans doute pas mal de fautes de signes, en plus de celles
qu'il doit déjà y avoir.\\

\noindent Rappelons que $q"$ désigne un élément générique de $Q"$, donc de la forme $v"\wedge u"_0$ pour $v"\in F"$,
et $q'$ désigne un
élément générique de $Q'$, donc de la forme $v'\lrcorner V"_0$ pour $v'\in F'$.\\
Nous noterons aussi $\nu"$ (comme neutrino) un nombre complexe quelconque, vu comme élément de $\Lambda^{0}F"$, et $\varepsilon"$
(comme électron) tout multiple complexe de $\Omega$ dans $\Lambda^{4}F"$.\\

\noindent Le crochet $[W_+,W_-]=W_+W_--W_-W_+$ satisfait à
\begin{multline}
[W_+,W_-](\overline{q"})=\overline{q"},\quad [W_+,W_-](q')=-q',\quad [W_+,W_-](\nu")=-\nu",\\ [W_+,W_-](\varepsilon")=\varepsilon".
\end{multline}
C'est-à-dire que, si l'on note $Z'_1$ l'opérateur sur $M_8^{+}$ qui vaut $-i/2$ sur $\Lambda^{0}F"$ et $Q'$
et qui vaut $i/2$ sur $\Lambda^{4}F"$ et $\overline{Q"}$, on a
\begin{equation}\label{crochetzcomplexe}
[W_+,W_-]=2iZ'_1.
\end{equation}

\noindent Remarque: on constate que la définition de $Z'_1$ reproduit exactement celle de $Z_1$; la nuance est que sur $\overline{Q"}$
on devrait s'attendre à ce que l'opérateur donné par $Z_1$ soit $\overline{Z_1}$, ce qui donnerait la multiplication par $-i/2$
comme sur $Q'$. Or nous n'avons pas eu le choix en calculant le commutateur de $W_+$ et $W_-$.
Il est crucial d'avoir $Z'_1$ et pas $\overline{Z_1}$, pour que l'algèbre de $u_8$ soit respectée
sur le secteur $X,Y,Z_1$ à travers son action sur les fermions.\\

\noindent Les opérateurs anti-hermitiens associés sont définis par:
\begin{equation}
W_1=\frac{W_++W_-}{\sqrt{2}},\quad W_2=\frac{i(W_+-W_-)}{\sqrt{2}},
\end{equation}
de sorte que $\sqrt{2}W_+=W_1-iW_2$ et $\sqrt{2}W_-=W_1+iW_2$.\\

\noindent Il vient
\begin{equation}
[W_+,W_-]=i[W_1,W_2],
\end{equation}
par conséquent
\begin{equation}\label{crochetzunitaire}
[W_1,W_2]=2Z'_1.
\end{equation}
Donc $W_1,W_2,Z'_1$ forment une algèbre de Lie $su_2$, et $W_+,W_-,H_7$ une algèbre de Lie $sl_2$, toutes deux dans leurs
vêtements habituels.\\
Ce sont respectivement des sous-algèbres de l'algèbre $su_8$ de $M_8^{+}$ et de l'algèbre
$sl_8(\mathbb{C})$ du même espace.\\
\indent Comme les particules $W_+$ et $W_-$ vont être chargées et que la théorie électrofaible va reposer sur elles,
nous pensons à elles comme à des cariatides.\\

\noindent L'action de $W_{\pm}$ sur les fermions échange $\Omega$ et $1$ et échange les deux
familles de quarks en commutant avec l'action de $SU_3$, de la seule façon possible à deux constantes près, complexes de module $1$,
une pour les leptons, une pour les quarks.\\
L'opérateur $Z'_1$ prolongeant le $Z_1$ standard sur $L$ à l'espace $Q$, il n'y pas de problème à considérer qu'il est le représentant
du générateur $Z_1$ dans $spin_8=o_8$, vu dans l'algèbre $\mathfrak{u}_8$ de l'espace $M^{+}$. D'ailleurs nous le noterons également $Z_1$
lorsqu'il n'y aura pas de risque de confusion avec $j_*(Z_1)=-Z_4/2$.\\
Sur $S_8$ l'opérateur $Z_4$ coïncide avec $4Z_1$ sur $L$ et vaut $0$ sur
$Q'$ et $Q"$; le plus simple est de garder cette formule sur $L$ et $Q'$ et de prendre $0$ sur $\overline{Q"}$ aussi,
comme sur $Q"$;
c'est ce que nous ferons. Alors le générateur $Z_3$ du $U_3$ standard, qui vaut naturellement $Z_4-Z_1$, sera donc égal à $3Z'_1$ sur
l'espace $L$ des leptons et à $-Z'_1$ sur l'espace $Q^{+}$ des quarks dans $M$.\\
\noindent L'opérateur $B_0$ sera encore pris comme l'identité sur $M_8$ et l'opérateur $B_1$ comme l'identité sur les leptons
et zéro sur les quarks.\\

Notons $\rho^{+}$ la représentation de $\mathfrak{u}_3\times \mathfrak{u}_2\times \mathfrak{u}_1$
construite ci-dessus, avec le $s\mathfrak{u}_3$ standard, le $s\mathfrak{u}_2$ de $W_1,W_2,Z_1$, le $Z_3$ centre de $\mathfrak{u}_3$, le $iB_1$ centre de $\mathfrak{u}_2$
et le $iB_0$ engendrant $\mathfrak{u}_1$, et notons $\mathfrak{u}_8$ l'algèbre de Lie $spin_8\oplus i\mathfrak{p}_8$ qui contient naturellement
$\mathfrak{u}_3\times \mathfrak{u}_2\times \mathfrak{u}_1$.\\

\noindent \textbf{Proposition 4}: il n'existe pas d'extension de $\rho^{+}$ en une représentation d'une sous-algèbre de Lie
de $\mathfrak{u}_8$ qui contienne $\mathfrak{u}_3\times \mathfrak{u}_2\times \mathfrak{u}_1$, et qui soit stable par l'action adjointe
de $s\mathfrak{u}_3$ dans $\mathfrak{u}_8$.\\

\noindent \emph{Démonstration}: commençons par considérer un élément du $u_6=u_3\oplus i\mathfrak{p}_3$; celui-ci
doit commuter avec $u_2$ et $u_1$, en particulier il commute avec $B_1$, donc il doit envoyer $Q$ dans $Q$ et
$L$ dans $L$. Il commute également avec $Z'_1$, ce qui impose qu'il respecte séparément $Q'$ et $\overline{Q"}$;
mais alors le fait qu'il commute avec les $W$ dit qu'il appartient à $u_3$.\\
Ce $u_6$ est invariant par l'action adjointe de $u_3$ et le supplémentaire orthogonal de $u_6\times u_2\times u_1$
est l'espace
vectoriel complexe $V$ de dimension $12$ qui correspond aux trois blocs $2\times 2$ des deux dernières colonnes
autre que $u_2$. Les crochets avec un élément $U$ de $u_2$ correspondent à la multiplication de chaque bloc par
$U$.\\
En tant que représentation de $SU_3$, $V$ est isomorphe à $3D(1,0)\oplus D(0,1)$, deux facteurs venant de $o_8$ et deux autres venant
de $i\mathfrak{p}_8$. Les commutateurs des éléments de $V$ appartiennent à $u_6$, ceux qui appartiendrait à une
extension $\rho$ de $\rho^{+}$ à un sous-espace $V_\rho$  de $V$, devraient donc avoir leurs commutateurs dans $su_3$; mais nous avons vu que les crochets des
éléments du type $\Sigma_1$ ($M_1$ symétrique) avec des éléments du type $\Sigma_2$ ($M_2$ antisymétrique) ne tombent pas dans $su_3$, si bien que $V_\rho$ ne peut
contenir des éléments que d'un seul type, ce qui est incompatible avec l'action de $u_2$, donc $\rho$ ne pourrait
pas étendre $\rho^{+}$. Par conséquent $V_\rho$ ne peut contenir que des éléments de $o_8$, dont les commutateurs sont
dans $u_3$, c'est-à-dire des éléments du $u_4$ standard, correspondant au facteur $D(1,0)$ restant, i.e. une dernière colonne
de trois nombres complexes. Mais ceci est incompatible avec l'action de $u_2$ puisque les crochets avec ceux-ci donneraient
un autre type d'éléments.\\

\noindent Remarque: nous allons voir que, mis à part une combinaison particulière des traces $B_0,B_1$, les trois vecteurs $X,Y,Z_1$,
et bien entendu $su_3$, aucun des $19$
autres champs de bosons de type $A$ et aucun des $33$ autres champs de bosons de type $S$ que nous venons de décrire ne correspond à une particule observée jusqu'ici.
Cinquante-deux opérateurs possibles sur les fermions sont cachés, ou interdits.\\
La proposition ci-dessus en donne une raison plausible. Les seules particules qu'elle n'interdirait pas sont attachées
à des champs abéliens, qui seraient deux composantes non-détectées issues de $iB_0,iB_1,Z_1,Z_4$.\\
Pour quelle raison celles-ci restent cachées? Sont elles trop lourdes? Ou bien sont elles exclues par un autre principe de conservation?\\

Les $52$ particules cachées n'ont pas d'interaction décelable avec la matière spinorielle. Elles constituent de bons candidats pour
faire partie de la \emph{matière noire}, décelable à l'échelle des galaxies. D'ailleurs, dans la théorie que nous présentons la matière noire ne risque pas de manquer, car
chaque ordre de déplacement en $v$ dans $X$ autour de $W$ apporte de nouveaux champs.\\
Notons que les champs bosoniques qui n'interagissent pas avec la matière fermionique, même découplés, interagissent
certainement avec le boson de Higs et acquièrent sans doute des masses de cette manière.\\

\subsection{Répartition des charges et modèle standard}

En plus des gluons de $su_3$, on sait qu'il doit y avoir un champ de vecteurs abélien de masse nulle, le photon.
Il se distingue dans le modèle standard comme étant celui qui préserve le champ
de Higgs. Nous allons proposer bientôt un scenario d'apparition d'un champ de Higgs scalaire dans notre modèle, mais
pour le moment, \emph{a priori} le
mécanisme de brisure de symétrie invoque un champ de spin $1/2$, le $t_0$, ou le $u"_0$. Regardons donc lequel des
champs abéliens en présence préserve ce spineur particulier.\\

\noindent Parmi les combinaisons linéaires des charges $iB_0,iB_1,Z_1,Z_4$, plusieurs stabilisent le $u"_0$; toutes sont
des candidates à la charge électrique.\\

\noindent \textbf{Définition}:  $S_4^{-}$ est le sous-espace vectoriel complexe de dimensionce deux de $S_8^{-}$ qui est
engendré par le vecteur $u"_0$ dans $\Lambda^{1}(F")=F"$ et par le trivecteur $V"_0$ dans $\Lambda^{3}(F")$.\\
On retrouve cet espace à l'identique dans $M_8^{-}$.\\
Tous les éléments $t$ qu'on a considèrés pour le fixage de jauge appartiennent à cet espace $S_4^{-}$. En particulier les cercles
orbites de
structures réelles contenant un point de $S_4^{-}$ demeurent dans $S_4^{-}$.\\

\noindent $S_4^{-}$ est le petit jardin qui sert de scène au mécanisme classique de Higgs:\\

\noindent Dans $S_4^{-}$,
les opérateurs $\mathbb{C}$-linéaires $W_{\pm}$ satisfont aux formules suivantes
\begin{align}\label{actionwimpaire}
W_+(\overline{u"_0})=\exp(-i\beta)V"_0,\quad &W_+(\exp(-i\beta)V"_0)=0,\\
W_-(u"_0)=0,\quad &W_-(\overline{\lambda}\exp(-i\beta)V"_0)=-\lambda u"_0.
\end{align}
Le commutateur
$[W_+,W_-]$ agit comme $Id$ sur la direction de $u"_0$ et comme $-Id$ sur celle
de $V"_0$, donc les équations \eqref{crochetzunitaire} et \eqref{crochetzcomplexe} sont encore valides.\\

\noindent Dans $S_4^{-}$, les générateurs $2Z_1$ et $Z_4/2$ sont tous deux représentés
par les rotations anti-diagonales d'un angle droit: angle $\theta$ dans la droite complexe des $u"$ et $-\theta$ dans la droite
complexe des $V"$. Ceci se traduit par la multiplication par $i$ sur les $u"$ et $-i$ sur les $V"$.\\
Par ailleurs les traces $iB_0$ et $iB_1$ sont toutes deux représentées par l'homothétie de rapport $i$.\\
Il est donc naturel d'introduire les combinaisons de ces opérateurs qui laissent fixes les $u"$ et qui font tourner
les $V"$, ainsi que leurs combinaisons orthogonales, qui font tourner les $u"$ et laissent fixes les $V"$:
\begin{equation}
Q=i\beta_0B_0+i\beta_1B_1+2\alpha_1Z_1+\alpha_4\frac{Z_4}{2};
\end{equation}
où $\beta_0+\beta_1=\alpha_1+\alpha_4$.\\
\begin{equation}
Z=i\beta_0B_0+i\beta_1B_1-2\alpha_1Z_1-\alpha_4\frac{Z_4}{2}.
\end{equation}
Les opérateurs $Q$ sont tous candidats à la charge électrique, les champs de bosons qui expriment leurs dynamiques étant candidats
pour représenter le champ électromagnétique. De leur côté, les $Z_0$ sont les candidats au boson vecteur neutre massif.\\

Mais une seule combinaison $Q$ est la bonne, donnant du même coup le bon $Z_0$; c'est celle
qui attribue les charges connues aux particules observées venant de $M_8^{+}$, neutrinos, électrons, muons, taus, quarks $q'$,
quarks $\overline{q"}$ de toutes les saveurs. A ce moment là, nous faisons appel à l'expérience depuis plus de cent ans:  il faut trouver
respectivement $0$, $-1$, $2/3$ et $-1/3$. Plus loin nous proposerons une raison géométrique justifiant ces valeurs des charges,
mais pour le moment admettons les comme elles sont.\\

Le système d'équation à résoudre est le suivant:
\begin{multline}
\beta_0+\beta_1=\alpha_1+\alpha_4;\quad \beta_0+\beta_1+\alpha_1+\alpha_4=-1;\\
\beta_0-\alpha_1=\frac{2}{3};\quad \beta_0+\alpha_1=-\frac{1}{3}.
\end{multline}
Son unique solution est
\begin{equation}
\beta_0=\frac{1}{6};\quad \alpha_1=-\frac{1}{2};\quad \beta_1=-\frac{2}{3};\quad \alpha_4=0.
\end{equation}
Ce qui donne
\begin{equation}
Q=Q_0=\frac{i}{6}(B_0-4B_1)-Z_1;\quad Z=Z_0=\frac{i}{6}(B_0-4B_1)+Z_1.
\end{equation}
L'opérateur $B'=(B_0-4B_1)/3$ est égal à $-1$ sur les leptons et à $1/3$ sur les quarks.\\
Il est naturel de mettre partout des signes moins pour les quatre opérateurs $iB_0,iB_1,Z_1,Z_4$ lorsqu'on passe aux anti-particules.
Alors $B'_0$ donne $+1$ sur les partenaires des leptons, anti-neutrinos et positrons, etc. et donne $-1/3$ sur les anti-quarks.
On voit que le confinement observé pour les quarks et leurs combinaisons correspond à l'énoncé suivant:\\

\noindent \emph{Principe d'un confinement soft}: Sont observables à grande distance et à notre échelle les combinaisons de fermions telles que $B'$ soit un entier.\\
\noindent Cette condition implique que $Q_0$ et $Z_0$ prennent des valeurs entières également.\\

\noindent Remarque hard: le confinement des quarks repose bien plus probablement sur une propriété non-perturbative de la
théorie de jauge $SU_3$ des interactions fortes, actuellement très difficile à mettre en évidence théoriquement,
cependant la coïncidence avec le principe sur les charges mérite d'être signalée.\\

\indent Les charges des bosons résultent de la conservation de la charge totale dans les interactions, comme elles
apparaissent dans l'équation de Dirac, les gluons, le photon et le $Z_0$ sont électriquement neutres, mais $W_{+}$
et $W_{-}$ sont respectivement de charges $+1$ et $-1$. Elles satisfont donc au principe ci-dessus.\\

\noindent Un problème se pose alors: les opérateurs $B_0,B_1,Z_1,Z_4$ de $M_8^{-}$ préservent tous l'espace $S_4^{-}$, donc tout opérateur $Q$ tel que
$\beta_0+\beta_1=\alpha_1+\alpha_4$ appartient au stabilisateur de $u"_0$ dans $M_8^{-}$.\\
Si fixer $u"_0$ était la seule condition sur la jauge, nous devrions donc observer trois particules abéliennes de masse nulle et non-pas une seule.
Par suite il est nécessaire d'ajouter deux conditions de jauge pour trouver une théorie an accord avec l'expérience.\\
En vertu des équations de Yang-Mills-Higgs que nous avons trouvées en partant des équations d'Einstein, il est tentant d'attribuer une masse à $B_0$, qui
représente la trace totale $\sum_iC_{\mu,i}^{i}$. D'abord, il y a nombre de termes quadratiques en $B_0$ pour le justifier,
ais surtout, cela a un sens d'imposer à $B_0$ de fixer quelque chose puisque
la jauge d'Einstein préscrit des coordonnées ou le déterminant de la métrique est constant. Dans les fibres de $E_F$, le long de
$W$ la métrique est fixée, donc son discriminant également; si bien que l'action de $B_0$ sur le volume de $g_{\mu\nu}$
équivaut à son action sur le volume total.\\
Le second déterminant, celui qui correspond à $Z_4$, est plus caché; il ne s'agit
pas d'un déterminant réel mais d'un déterminant complexe dans un espace complexe de dimension quatre.
Nous imposerons donc aux coordonnées complexes choisies de garder une unité de volume
constante. Le sous-espace engendré par $e_7$ et $e_8$, dont le complexifié correspond par trialité à l'espace $S_4^{-}$, ne reçoit pas
de condition pour autant.\\
Les champs $iB_0+Z_4/2$ et $iB_1+2Z_1$ ne reçoivent pas de masses à travers le mécanisme de Higgs, ils forment une
base du supplémentaire de $Q_0$ dans l'espace des combinaisons satisfaisant à $\alpha_0+\alpha_1=\beta_1+\beta_4$.
Ils sont neutres, ils coïncident sur $L$; sur $Q$ le premier est la constante $iB_0$ et le second la diagonale $2Z_1=2Z'_1$.
Ils se peut qu'ils aquièrent des masses, comme nous venons de le supposer, mais il se peut aussi qu'ils se découplent
des autres champs; le plus honnête est d'avouer que l'uteur ne sait pas.\\

\noindent La charge $-1$ pour $\varepsilon"$ est également attribuée aux autres générations de leptons, le \emph{muon}, et le \emph{tau}.
La charge $0$ est attribuée aux générations des neutrinos associés.  Les quarks-éléctriques et les quarks-magnétiques, qui ont respectivement les charges
$-1/3$ et $+2/3$, correspondent
respectivement aux générations $(d,u)$, \emph{down} et \emph{up}, $(s,c)$, \emph{strange} et \emph{charme}, $(b,t)$ \emph{bottom} et \emph{top},
parfois nommée beauty et truth. \\

\indent Le "groupe de jauge auxiliaire" $SU_2\times U_1$,
où $SU_2$ est celui de $W_{\pm}$ et de $Z_1$, et où $U_1$ est celui de $iB'$, agit comme il faut
sur $L$ et sur $Q$, pour être en accord avec les expériences, et son action dans $S_4^{-}$ reproduit exactement le mécanisme de Weinberg.
Dans ce groupe le stabilisateur de $u"_0$ est le cercle engendré par $Q_0$.\\
Notre scenario permet donc de retrouver le modèle standard, comme il fut exposé par Weinberg, cf. \cite{weinberg1996quantum}.\\

\indent Pourtant il manque encore le champ de Higgs scalaire. Car c'est aussi un fait d'expérience maintenant bien établi,
que le champ de Higgs observé n'a pas de spin, il n'est pas fermionique, comme $t_0$,
ou $u"_0$, au contraire il est scalaire.\\
\indent Il est naturel dans notre approche identifiant multiplicités spinorielles et vecteurs transverses de se tourner
vers la section privilégiée, qui est $\overrightarrow{u}_0=e_8$ dans $E_F$. Ce sera lui, ou plutôt un multiple constant de celui-ci,
qui représentera le champ de Higgs $H$.\\
\emph{A priori} il semble que nous devrions lui attribuer huit dimensions à cause de $F$, ou au contraire deux
dimensions, à cause du plan $e_7,e_8$. Dans le discours standard, sont distinguées
les deux composantes complexes $\varphi^{0}$ et $\varphi^{+}$. C'est ce que nous allons reproduire.
Les quatre dimensions correspondront aux seuls degrés de liberté dynamiques disponibles.\\

Le champ de vecteurs $v$ qui exprimait un déplacement virtuel
dans l'espace $X$ inconnu autour de notre univers $W$, va ainsi devenir un champ physique, un champ scalaire, variant autour de son équilibre,
fixant une échelle et distribuant des masses à presque tout le monde.\\
\indent Le point crucial est que certaines particules vont rester sans masse, au moins les gluons et le photon,
et ceci à tous les ordes du développement. La raison selon notre approche, en est que les générateurs du $SU_3$ non-brisé et $Q_0$,
générateur du $U_1$ électromagnétique,
donnent zéro quand ils sont appliqués à la section $e_8$, et que par conséquent aucun terme de masse
venant de $H$, s'il correspond bien à $e_8$, ne s'introduit pour les générateurs de $SU_3$ et de $U_1$.\\
Aux champs correspondant à $iB_0+Z_4/2$ et à $iB_1+2Z_1$, ce n'est pas $H$ qui va donner des masses, cela peut être directement le Lagrangien,
pour $iB_0$ c'est tentant de le dire, ou cela peut provenir de
l'action sur les fibrés déterminants réels et complexes des fibres de $E_F$, comme nous l'avons évoqué.\\

Notons que le déplacement $v$ intervient partout dans le Lagrangien; une partie du potentiel de Higgs vient de là, et nous verrons comment une
autre partie s'y ajoute plus loin, mais sa dynamique propre, d'où
vient-elle?\\
\indent Or une autre caractéristique du modèle standard n'est pas présente dans notre théorie, il s'agit de
l'absence de la composante right (impaire) $\nu_R$ du neutrino.\\
\noindent Nous proposons donc l'hypothèse suivante, qui offre une dynamique au champ $H$ et fait disparaître
la composante impaire de la particule $\nu"$:\\

\noindent \emph{Bosonisation}: la composante impaire du neutrino transmet au champ de Higgs ses deux degrés de degrés de libertés complexes, si on tient
compte de l'antiparticule paire associée, qui elle, est paire.\\

La transmission se fait en trois temps, reprenant les trois étapes de la trialité. D'abord un champ scalaire $\varphi=(\varphi',\varphi")$ est défini par la variation des
phases de $\nu"$ et $\overline{\nu"}$:
\begin{equation}\label{bosonfermion}
\nu(x)=\varphi'(x)\nu"+\varphi"(x)\overline{\nu"};
\end{equation}
Ensuite, on identifie $\overline{\nu"}$ à l'étoile de $\nu"$ dans le plan des leptons, sans que cela porte
atteinte à l'électron $\Omega$, de façon à retrouver par trialité les deux composantes $u"$ et $V"$.
La première composante $\varphi'(x)$ donne la coordonnée sur l'axe complexe de $u"_0$ dans $S_4^{-}$, la
deuxième composante $\varphi"(x)$ donne la coordonnée sur l'axe de $V"_0$, décalée par la phase beta.\\
Nous proposons d'identifier $\varphi'$ à $\varphi^{0}$ et $\varphi"$ à $\varphi^{+}$.\\
C'est ici que l'on rejoint tout à fait le modèle standard.\\
Enfin, on interprête la droite complexe $\mathbb{C}u"_0$ comme le vecteur $e"_8$ dans le plan complexe qui
complexifie le plan réel $(e_7,e_8)$, et on identifie $H$ au déplacement
dans l'espace transverse. Là, il est vrai que nous nous écartons, en un certain sens, du modèle standard, et que nous
rejoignons l'autre origine possible des masses, qui correspond
au rayon d'une sphère dans l'espace, comme c'est le cas pour les trous noirs.\\
Le module de $\varphi^{0}$, qui est considérée par Weinberg comme le seul champ physique, correspond au rayon de
la sphère où se déplace $v$ et, comme nous allons l'expliquer, c'est lui qui donne l'échelle en $GeV$ de la théorie.\\

Andrzej Okninski a récemment remarqué que l'équation de Dirac pour un spineur $\nu$ de direction fixe donne
pour les coordonnées $\varphi$ un système d'équations de Duffin-Kemmer-Petiau, dit $DKP$, de spin zéro. Cf. \cite{okninski2011supersymmetric}, \cite{okninski2014neutrino}.\\
C'est cela qui va définir $H$ en tant que particule; c'est-à-dire associée à une représentation unitaire du groupe de Poincaré.\\
\noindent Le système $DKP$ est un système hyperbolique d'orde $1$ qui étend l'équation des ondes, et aussi l'équation de Klein-Gordon
si on introduit une masse. Avec des conditions initiales convenables les solutions de $DKP$ satisfont à l'équation de Klein-Gordon.\\
Une version simplifiée de la construction de Okninski va nous suffire: on suppose que le choix de jauge a fixé la partie impaire
ou right du neutrino $\nu"_R$ ainsi que la partie paire de l'anti-neutrino $\overline{\nu"}_L$, supposés tous les deux de masse nulle.
Cela détermine un co-vecteur de lumière $p_0$. On sait qu'on peut alors choisir les
coordonnées $x^{0},x^{1},x^{2},x^{3}$ sur $W$ de sorte que $p_0=(E,0,0,E)$; les plans $e_0,e_3$ et $e_1,e_2$ sont déterminés, ainsi
que les axes de lumière $e_0\pm e_3$ dans le premier plan, et la structure conforme $e_1+e_2$ dans le second plan.\\
Les coordonnées sur les spineurs sont choisies pour faire apparaître les chiralités $S=S^{+}\oplus S^{-}$ (spineurs de Weyl) dans $\mathbb{C}^{4}$; un
jeu de matrices de Dirac possible est le suivant (cf. cours I):
\begin{multline}
\Gamma_1=
\begin{pmatrix}
0& \sigma_0\\
-\sigma_0&0
\end{pmatrix}
\quad
\Gamma_2=
\begin{pmatrix}
0& -i\sigma_3\\
i\sigma_3&0
\end{pmatrix}\\
\Gamma_3=
\begin{pmatrix}
0& -i\sigma_1\\
i\sigma_1&0
\end{pmatrix}
\quad
\Gamma_0=
\begin{pmatrix}
0&\sigma_2 \\
-\sigma_2&0
\end{pmatrix}
\end{multline}
où les matrices $\sigma_j;j=1,2,3$ sont les matrices de Pauli:
\begin{equation}
\sigma_1=
\begin{pmatrix}
0& 1\\
1&0
\end{pmatrix}
\sigma_2=
\begin{pmatrix}
0& -i\\
i&0
\end{pmatrix}
\sigma_3=
\begin{pmatrix}
1& 0\\
0&-1
\end{pmatrix}
\end{equation}
et, par définition, la matrice $\sigma_0$ est l'identité.\\
Le sous-espace
propre de $S^{-}$ associé à la valeur propre zéro de $\gamma(p_0)$, qui est engendré par $\nu"_R$, est fait des
vecteurs $(a,0,b,0)$.  On a
\begin{equation}
E(\Gamma_0-\Gamma_3)=
\begin{pmatrix}
0&E(\sigma_2+i\sigma_1) \\
-E(\sigma_2+i\sigma_1)&0
\end{pmatrix}
=
\begin{pmatrix}
0&0&0&0 \\
0&0&2iE&0\\
0&0&0&0\\
-2iE&0&0&0\\
\end{pmatrix}
\end{equation}
Décomposons la fonction d'onde de la paire neutrino-antineutrino comme dans l'équation
\eqref{bosonfermion}; alors en appliquant l'opérateur de Dirac à $\nu$ on trouve l'opérateur différentiel $P(\partial)$ de degré $1$ qui suit:
\begin{equation}
(\varphi',\varphi")\mapsto
\begin{pmatrix}
-(\partial_1+i\partial_2)\varphi"\\
i(\partial_0+\partial_3)\varphi"\\
(\partial_1+i\partial_2)\varphi'\\
-i(\partial_0+\partial_3)\varphi'
\end{pmatrix}
\end{equation}
On voit qu'au signe près, c'est le même opérateur sur $\varphi'$ et $\varphi"$.
L'équation $P(\partial)\varphi'=0$ (ou $P(\partial)\varphi"=0$) dit que $\varphi'$ est
une fonction de $x_0-x_3$ holomorphe en $x_1+ix_2$. Un cas particulier est un paquet d'ondes
en $x_1-x_3$ indépendant de $x_1,x_2$. Surtout on voit que $\varphi'$ comme $\varphi"$
satisfont à
\begin{align}
\partial_0^{2}\varphi-\partial_3^{2}\varphi&=0;\\
\partial_1^{2}\varphi+\partial_2^{2}\varphi&=0.
\end{align}
Elles sont donc
solutions de l'équation des ondes
\begin{equation}
d^{*}d\varphi=(\partial_0^{2}-\partial_1^{2}-\partial_2^{2}-\partial_3^{2})\varphi=0.
\end{equation}
Ce que nous aurions également vu en appliquant le carré de l'opérateur de Dirac à $\nu(x)$.\\

Mais, au delà des solutions classiques, ce qui est plus pertinent pour nous, est que
le champ $\varphi=(\varphi',\varphi")$ possède une dynamique relativiste qui lui donne
un statut de particule scalaire de rang complexe deux.\\
Le saut, qui est moins évident à justifier, mais que nous proposons comme hypothèse, est que $\varphi(x)$ puisse être assimilé au complexifié du vecteur de déplacement
$v(x)$ contraint par le plan $e_7,e_8$, et que, en particulier, sa norme, ou l'espérance $v$ de sa norme dans le vide,
fixe une échelle d'observation, et commande également l'attribution des masses aux particules $W$ et $Z$
et aux fermions libres.\\
La synthèse des deux champs $\varphi(x)$ et $v(x)$ donnant le champ de Higgs, nous la désignerons par le symbole $H(x)$.
La lettre $H$ renvoie à Higgs bien entendu, mais aussi à Hâpy, divinité fertile du Nil et androgyne, parfois représentée sous la forme de deux jumaux,
ce qui convient bien.\\
Son potentiel, en particulier le terme d'ordre deux qui fournit une masse, provient en partie du Lagrangien de Hilbert
en dimension douze que nous avons développé, mais aussi du Lagrangien de Dirac, côté impair.\\

\noindent Le terme cinétique d'un champ régi par une équation du premier ordre doit être
en $L^{-3/2}$, comme pour le champ $\psi$ de spin $1/2$, mais $\varphi$ étant le coefficient devant un $\psi$ fixe
n'a pas d'unité; ce qui permet de l'identifier directement au déplacement $v$. Cependant, comme nous l'avons vu au
paragraphe $2.4$, le passage à $H$ demande de poser $H=M_0v$ donc $H=M_0\varphi$, où $M_0$ est une constante
qui a l'unité de l'inverse d'une longueur, comme une masse, mais qu'il ne faut pas confondre avec
la masse de $H$.\\

Dans la partie bosonique du Lagrangien se trouvent les termes $HHSS$ et $HHBB$ responsables d'une masse pour $W_{\pm}$
et $Z_0$ respectivement. Les masses interviennent là par leurs carrés, donc $M_W$ et $M_Z$ sont du premier ordre
en la valeur moyenne $H_0=M_0v_0$. Rappelons l'argument de Weinberg:\\
\noindent Les trois constantes arbitraires qui interviennent dans le calcul de $M_W,M_Z$ sont
$g,g',H_0$, respectivement la taille de la charge du $SU_2$ avant brisure, la taille de la charge du cercle $U_1$
et la valeur moyenne dans le vide du champ de Higgs. Dans notre approche on ne peut pas non-plus les prédire, la première est
le coefficient du terme de Yukawa pour $su_2$ dans $su_8$, la seconde le coefficient du terme central
modifié $B=B_0-4B_1$, la troisième donne l'échelle en $GeV$.\\
Il n'y a effectivement pas de raison que $g$ soit égale à $g'$ si la théorie est générique.\\
L'angle de Weinberg $\theta_W$ est défini par
\begin{equation}
\sin \theta_W=\frac{g'}{\sqrt{g^{2}+(g')^{2}}},\quad \cos \theta_W=\frac{g}{\sqrt{g^{2}+(g')^{2}}}.
\end{equation}
On a alors, cf. \cite{itzykson1980quantum}, \cite{weinberg1996quantum}:
\begin{equation}
M_W=\frac{1}{2}gH_0,\quad M_Z=\frac{M_W}{\cos \theta_W}
\end{equation}
Les valeurs estimées actuellement à partir des expériences sont
\begin{equation}
M_W=80.4 GeV,\quad M_Z=91.2 GeV
\end{equation}
Ce qui donne un $\theta_W$ pas très loin de $\pi/6$. En fait $\sin^{2}\theta_W\approx 0.24$.\\
La constante $H_0$ est expérimentalement donnée par la
constante de Fermi $G_F=1/v^{2}\sqrt{2}\approx 1.166 10^{-5} GeV^{-2}$; ce qui donne une valeur de
$H_0$ proche de $250 GeV$. Donc $g$ et $g'$ sont de l'ordre des unités.\\

Comme $H_0=M_0v_0$, il nous faut une information supplémentaire pour avoir une idée des ordres de grandeur de $M_0$ et
de $v_0$ séparément.\\
Pour cela, nous nous tournons vers les fermions. Là les suites des masses des leptons et celles des quarks suivent approximativement
une loi géométrique de raison $200$; nous verrons plus loin que cela s'explique bien en prenant $v_0=200^{-3/4}\approx 1/53$. Ceci donne une masse
$M_0$ de l'ordre de $13250 GeV= 13.25 TeV$.\\
Nous verrons aussi plus loin que les premiers termes du Lagrangien pouvant donner naturellement des masses
aux fermions sont quadratiques en $v$ et linéaires dans la courbure transversale, ils s'écrivent
$C\widetilde{\psi} Rvv\psi$, où $C$ est une constante possédant la dimension $L^{-1}$.
Mais, contrairement
au cas des bosons, ce n'est pas le carré de la masse mais la masse elle-même qui est proportionnelle au terme du Lagrangien,
donc $m\approx CRv_0^{2}\approx 4.10^{-4}CR$. Expérimentalement
la masse $m_e$ de l'électron est de l'ordre de $0.5 MeV$, ou $5. 10^{-4}GeV$, et celle des quarks $u,d$ de l'ordre de quelques $MeV$.
Donc les plus petites valeurs de $CR$ sont de l'ordre de l'unité, les suivantes de l'ordre des centaines. Nous verrons que pour le moment,
il n'y a pas d'évidence expérimentale pour des courbures plus grandes.\\
Notons que notre approche exige $v_0$ assez petit, ce qui implique $M_0$ plutôt grand, puisqu'expérimentalemnt $H_0$ est
de l'ordre des centaines de $GeV$, par suite, il n'est pas trop abusif de prétendre que notre modèle
explique assez bien l'importante différence qui existe entre les masses des bosons et celles des fermions.\\

\noindent Pour la suite, nous retenons que $v_0$ est à peu près égal à $1/53$, i.e. environ $0.02$.\\

\noindent Afin de déterminer plus exactement les masses des fermions, nous devons décrire plus précisément les termes de courbure $Rvv$
et leurs généralisation; nous le ferons dans la section suivante, intitulée résultats.\\
Notons que l'origine des masses des fermions dans notre théorie se distingue à peine de celle qui est présentée dans le
modèle standard, car les termes du Lagrangien fermionique $\mathcal{L}_F$ utilisant la courbure peuvent être considérés comme des termes de Yukawa.
Ce que nous ajoutons est donc plutôt l'hypothèse d'une origine géométrique, ou cosmologique, des termes de Yukawa.\\

Reste à dire d'où provient la masse $M_H$ du boson de Higgs lui-même; elle doit venir d'un monôme quadratique en $H$.
Le plus naturel dans notre approche est de faire appel au terme de Yukawa dans  $\mathcal{L}_F$ qui fait
intervenir le spineur $t$ gelé du secteur impair $S_8^{-}\oplus \overline{S_8^{-}}$ fixé par le choix de
jauge. Ce terme est de la forme $C_0\overline{t} HH t$; le
coefficient $C_0$ ayant la dimension d'une longueur pour respecter
la dimension $L^{-4}$ dans la densité Lagrangienne. Le terme dynamique, issu de la bosonisation de la partie
droite du neutrino $\nu"_R$, sécrit
$M_1H.\nabla H$, avec
une constante $M_1$ en $L^{-1}$; la masse au carré de $H$ est donc $M_H^{2}=C_0\overline{t_0}t_0$.
La valeur expérimentale de $M_H$ est $125 GeV$. Rien \emph{a priori} dans notre modèle ne permet de le prévoir. Remarquons
que si la valeur moyenne $t_0$ de $t$ se déduit de la valeur moyenne $H_0$ en prenant la puissance $3/2$, on trouve
$C_0$ de l'ordre de $10^{-3}$. Par contre, si $t_0$ se déduit de $v_0$, considérée comme longueur typique, en prenant la
puissance $-3/2$, on trouve une constante $C_0$ de l'ordre des unités.\\
Dans l'estimation ci-dessus, nous avons tenu compte de la masse au carré pour $H$, comme il se doit pour un boson, même si
l'équation de Klein-Gordon est remplacée par un système $DKP$ d'ordre $1$.\\

D'autres termes, de degrés plus élevés en $H$, doivent être pris en compte, comme $\lambda \overline{t} H^{4} t$; ils
forment le potentiel propre du champ de Higgs. Dans la phase considérée, il est préférable de stabiliser $H$ par son
potentiel; on choisit donc $\lambda$ de sorte que
l'équilibre moyen $H_0\approx 250 GeV$ corresponde au minimum du potentiel $M_H^{2}H^{2}-\lambda H^{4}+...$.\\

\section{Résultats}

La structure du \emph{modèle standard} de Weinberg et Salam, et le mécanisme de Brout-Englert-Higgs, dit $BEH$,
sous une forme presque standard, ont émergé canoniquement du seul principe
de covariance général, appliqué en dimension douze à l'équation de Dirac, couplée avec l'équation d'Einstein.\\
\indent Plus précisément, nous avons supposé que la sous-variété $W$ Lorentzienne de dimension quatre qui correspond
à notre univers, est quasiment plate par rapport aux huit dimensions orthogonales, et nous avons restreint le groupe
des difféomorphismes ambiants à ceux qui respectent $W$ et une structure fibrée normale $E_F$; nous avons également
fixé la métrique des fibres le long de $W$ à l'ordre un, et restreint le jet des difféomorphismes en conséquence. Les composantes
de la métrique et les spineurs ont été développés en série de Taylor transversalement à $W$, faisant apparaître
plusieurs générations de champs du point de vue de l'univers à quatre dimensions. Ensuite, nous
avons utilisé le groupe de jauge $Spin_8$ agissant sur les spineurs pour fixer une trialité, qui identifie les fibres de
$E_F$ aux deux sortes de spineurs, les pairs $S_8^{+}$ et les impairs $S_8^{-}$. La manière de fixer la trialité, en
choisissant un spineur pur complexe pair $1$ et un spineur réel de norme un impair $t_0$ a permis de geler tout le secteur impair.
De plus, une partie du secteur pair a été interprété
comme donnant des particules, une autre comme donnant des anti-particules, de façon à redéfinir une multiplicité $M_8$
pour les spineurs de Dirac vus en dimension quatre qui soit conforme à l'expérience. Ceci a fourni la décomposition usuelle des fermions
élémentaires, en neutrino, électron, et deux sortes de quarks, où $SU_3$ agit de la même façon. Toute cette structure se répétant à chaque ordre du développement
des champs le long de $W$.\\
\indent La structure résultante comporte un champ scalaire massif $H$ de dimension $4$ à la Weinberg, trois bosons vecteurs massifs
$W_+,W_-,Z_0$, neuf bosons sans masse, qui sont le photon et huit gluons, ainsi que plusieurs générations de fermions,
chacune avec deux leptons, un de charge électrique nulle, le neutrino, l'autre de charge $-1$, l'électron (ou le muon ou
le tau) et avec deux familles de trois quarks,
une de charge électrique $2/3$ ($u$ ou $c$ ou $t$), l'autre de charge $-1/3$ ($d$ ou $s$ ou $b$).\\
Le Lagrangien est celui de Yang-Mills-Higgs pour les bosons et celui de Dirac pour les fermions, les interactions
sont celles que prescrit le modèle standard, couplages minimaux, termes de Yukawa naturels polynomiaux. Cf. \cite{plehn2015lectures}, \cite{gardi2015lhc}, \cite{schorner2015large}.\\

Une interprétation originale du champ de Higgs $H$ en est ressortie: sa dynamique résulte de la bosonisation d'un spineur
disparaissant, qui est la composante impaire $\nu_R$ du neutrino, cette composante ayant été fixée par la jauge transverse $so_8$
en choisissant une \emph{trialité}, qui identifie vecteurs transverses de dimension $8$, spineurs pairs $S_8^{+}$
et spineurs impairs $S_8^{-}$; son potentiel propre $V(H)$  provient de termes de couplages de Yukawa avec le spineur impair de
référence qui a servi à fixer la trialité; ses interactions avec les champs de bosons proviennent du Lagrangien
de Einstein-Hilbert en dimension douze, développé au voisinage de l'espace-temps $W$, en identifiant le vecteur de déplacement
transverse $v$ avec $H/M_0$, où $M_0$ est un facteur d'échelle (que nous avons trouvé de l'ordre de $10^{4} GeV$). Le champ $H$ distribue leurs
masses aux bosons et aux fermions comme dans le scenario $BEH$ du modèle de Glashow-Weinberg-Salam. Mais la valeur moyenne de $H$
correspond également à un rayon dans l'espace-temps transverse, et les masses des bosons résultent de la courbure
de l'univers dans ces huit directions. Ainsi deux mécanisme possibles
d'apparition des masses se rejoignent, celui des trous noirs et celui de la brisure de symétrie du vide.\\

Comme $H$ sert au développement canonique des champs en dimension douze le long de l'univers à quatre dimensions, il apparaît
une suite de générations de particules nouvelles à tous les ordres en $v=H/M_0$. La structure algébrique des fermions se reproduit
à l'identique à tous les ordres de $v$, mais chaque ordre ajoute une complexité nouvelle aux bosons. Il arrive des champs
scalaires venant de la gravité transversale $\Gamma_F$, et des champs vectoriels de Higgs-Hitchin, correspondant
à des jets de connexions de tous les degrés venant des composantes croisées $4\times 8$ de la métrique en dimension $12$. Cependant
la structure algébrique des bosons de jauges $SU_3\times SU_2\times U_1$, demeure la même à tous les ordres.\\
A partir de l'ordre sept ou huit, on ne peut probablement plus négliger les effets de la gravitation horizontale $\Gamma_H$
en dimension quatre. Mais les expériences du moment se limitent à l'ordre deux ou trois. Cf. \cite{schorner2015large}.\\

Par conséquent, notre approche prédit beaucoup d'autres champs bosoniques que ceux du modèle standard, et \emph{a priori} ces nouveaux champs
possèdent des masses, même si on ne les connaît pas. Pourtant un principe d'exclusion (énoncé par la Proposition $4$) résultant de l'unitarité de l'action
sur les fermions, exclue les interactions de ces bosons avec les fermions. De sorte que les seuls bosons observables
directement, par leurs effets électro-faibles ou forts sur la matière standard, sont ceux qu'on connaît, avec leurs avatars aux ordres suivant en $v$.
Il est donc bien tentant d'identifier tous ces champs avec \emph{la matière noire}.\\
De plus, il n'est pas exclu que les interactions de ces champs supplémentaires avec $H$ et avec les autres champs standard soient observables
un jour.\\
\indent Notons que l'\emph{énergie noire} pourrait également trouver une place dans notre théorie, en effet, comme nous allons le montrer, celle-ci prédit un
équilibre entre fermions et bosons, sans supersymétrie complète, or il est connu que les contributions à l'énergie du vide
des fermions et des bosons peuvent théoriquement se compenser, et qu'alors l'ajout d'une constante cosmologique
dans le Lagrangien de Einstein-Hilbert suffit à rendre compte de l'énergie noire observée, cf. \cite{henneaux1989cosmological},
\cite{bousso2008cosmological}, \cite{dupays2013can}. En vrai, pour que bosons et fermions se
compensent, il faudrait qu'ils aient les mêmes masses, donc même la super-symétrie, dès lors qu'elle est brisée, ne convient pas.
Le calcul des spécialistes, à nos échelles d'énergie, donne un facteur $10^{60}$ entre la prédiction et l'expérience!
De même, dans notre théorie effective, la structure à chaque ordre est équilibrée mais
pas après brisure spontanée de symétrie, donc le problème grave semble persister. Cependant, dans l'approche que nous proposons, la théorie génératrice en
dimension douze se réduit à la gravitation pure, avec éventuellement une constante cosmologique, plus un spineur sans aucune
masse; ce sont les seuls champs de départ, donc à ce niveau l'énergie noire peut s'expliquer par la constante cosmologique seule. Par conséquent,
nous ne prétendons pas donner l'explication de l'énergie noire, mais plutôt offrir un cadre où, peut-être, il y a suffisamment d'éléments
nouveaux, dont l'égalité des degrés de liberté bosoniques et fermioniques, pour chercher une explication plausible de cette énergie.
Intégrer le modèle en dimension douze dans une théorie cosmologique serait un début.\\

Comme nous l'avons vu plus haut, les ordres de grandeur des masses compatibles avec notre modèle s'accordent assez
bien avec l'observation, notamment avec les rapports entre masses des bosons et masses des fermions.
Dans la suite de cette section, nous allons préciser ces rapports, en suggérant une explication des masses des fermions par
la forme précise des termes de courbure qui interviennent et par le jeu des ordres du développemnt canonique.
Ainsi, partant de l'hypothèse que les termes de Yukawa impliquent la courbure suivant les fibres de $E_F$, nous allons démontrer que\\
1) au premier ordre possible, la masse du neutrino est exactement $0$;\\
2) la plus petite masse non-nulle est attribuée à la direction de charge $-1$ des leptons, le facteur
$200$ qui fait passer de la masse de l'électron à celle du muon et de celle du muon à celle du tau étant
expliqué par la courbure et l'échelle des générations de fermions;\\
3) la forme de l'opérateur qui donne les masses pour les ordres successifs de leptons permet aux directions propres des masses de ne pas coïncider
avec les directions propres des saveurs, ceci s'exprime à travers la matrice $PMNS$ pour les trois neutrinos; mais, apparamment,
la direction de l'électron définit aussi sa masse, car il n'est pas observé de mélange entre les trois générations $\varepsilon$, $\mu$ et $\tau$;\\
4) génériquement, il n'y a pas de raison évidente pour que la masse des quarks soit constante sur un même triplet,
encore moins pour une même saveur; cependant il se pourrait que seule la moyenne soit observée, ou encore, que, dynamiquement au moins,
les masses des quarks aient convergée vers une valeur commune; notons d'ailleurs que les masses observées dépendent du
régime d'interaction forte, qui intègre tous les effets dynamiques;\\
5) en prenant les moyennes des masse des quarks des différentes saveurs, on trouve des phénomènes analogues à ceux trouvés pour les
leptons, c'est-à-dire que, comme pour les leptons chargés, il existe un rapport à peu près égal à $200$ d'une génération à l'autre, et que,
comme pour les neutrinos, les directions propres des masses des quarks ne
coïncident pas avec les saveurs, ce qui est exprimé par la matrice nommée $CKM$.\\
6) en nous appuyant sur l'hypothèse que la théorie est générique, nous proposons des matrices $PMNS$ et $CKM$ qui ne sont
pas loin, au-moins pour ce qui est des ordres de grandeur, de celles qui sont mesurées expérimentalement, avec tant d'ingéniosité
et de succès par les Physiciens.\\
Notre approche offre donc des éléments théoriques pour aborder la distribution des masses et la \emph{Physique des saveurs}.\\
Dans la dernière partie, en discussion, seront discutées les corrections de ces masses aux énergies plus hautes que la centaine de $GeV$.\\

\noindent Nous verrons aussi ce que nous pouvons déduire à propos de la structure et des masses des bosons cachés, qui constituent une
éventuelle matière noire.\\
Puis nous reviendrons sur la géométrie du groupe $U_1$ de l'électrodynamique, en essayant de comprendre le pourquoi
théorique des charges fractionnaires des quarks.\\
Dans la section d'après, nous examinerons les effets de la nature quantique des particules, et nous discuterons la renormalisation
de notre théorie. Nous verrons qu'il s'agit plutôt d'une suite de théories effectives, au sens de Wilson et de Weinberg.
Ces théories ne sont certainement pas renormalisables selon les critères de comptages de puissances, mais il est vraissemblable
qu'elles soient renormalisables suivant le flot de renormalisation, à échelle fixée en-dessous des effets gravitationnels,
en vertu de la covariance générale, suivant un schéma proche de celui qu'a développé Damiano Anselmi récemment pour le
modèle standard couplé à la gravitation d'Einstein.
Cependant, l'évolution des charges et des masses effectives en fonction de l'échelle d'énergie sera en grande partie due
à l'intervention aux ordres successifs du développement en $v=H/M_0$ de toutes les nouvelles composantes des
champs. Ainsi à chaque nouvel ordre de nouvelles constantes imprévisibles s'ajouteront aux précédentes.\\
Pourtant, quelque chose préserve une certaine intelligibilité de l'ensemble du système de particules, c'est que les
nouvelles constantes semblent rester du même ordre de grandeur, au coefficient d'échelle près, comme nous le montre l'analyse des masses
des particules connues jusqu'à présent.\\
Autrement délicat sera le passage aux très hautes énergies, car en principe la haute courbure de notre univers rendra difficile
l'analyse en termes de particules; mais rien d'interdit qu'un autre point de vue, sans particules limites usuelles,
ne puisse décrire ces échelles, comme cela se passe pour la théorie conforme des systèmes critiques.\\

Il faut garder en tête que le principal résultat de ce texte est une explication géométrique de la structure du Modèle
Standard des interactions électromagnétiques, faibles et fortes. Tout ce qui vient en plus, prédictions sur les masses,
explication des charges, repose sur une hypothèse supplémentaire, disant que les termes de Yukawa impliquant bosons et fermions
sont naturellement associés à la courbure transversale de l'univers, et néglige les effets radiatifs, pourtant certainement importants,
ainsi que les effets non-perturbatifs, également importants.\\

\subsection{Du côté des fermions. Matrices CKM et PMNS.}

\indent Du côté des fermions, tous les degrés de liberté $S_8^{-}$ ont disparu, mais tous les
degrés de liberté $S_8^{+}$ sont restés,
avec une division de l'espace complexifié de dimension huit,
en un axe électron $\varepsilon"$, un axe neutrino $\nu"$, et deux familles de quarks à trois dimensions,
une de charge $2/3$, type \emph{up}, écrits $q'$, et une de charge $-1/3$, type \emph{down}, écrits $\overline{q"}$.\\
Nous avons redéfini la structure complexe de $S_8^{+}\otimes \mathbb{C}$,  donnant l'espace $M_8^{+}$, de manière
à ce que la représentation de $SU_3$ y soit du type $2D(0,0)\oplus 2 D(1,0)$. L'action de $SU_2$ à travers $W,Z$ est celle du commutant
de $SU_3$. Comme $S_8^{+}\otimes \mathbb{C}$ l'espace $M_8^{+}$
possède une structure hermitienne canonique, venant du choix de $\Omega$, ou $\varepsilon"$.\\
\indent Il existe également une manière (presque) canonique d'identifier $M_8^{+}$ à l'espace vectoriel $F\otimes \mathbb{C}$,
d'où nous allons tirer une famille de couplages de Yukawa d'origine géométrique. Cette manière consiste à faire que $\nu"$
corresponde à $e_8$, $\varepsilon"$ à $e_7$, l'espace engendré par les vecteurs impairs $e_1,e_3,e_5$ à
$Q'$ et l'espace engendré par les vecteurs pairs $e_2,e_4,e_6$ à $\overline{Q"}$, ceci de façon compatible
avec l'action de $SU_3$. Alors la correspondance est déterminée à une phase près (un nombre complexe de
module $1$) sur $L$ et à une phase près sur $Q$.\\

\indent Les seuls termes scalaires qui puissent figurer naturellement à l'ordre $2$ en $v$ dans le Lagrangien des fermions, sont ceux qui
proviennent de la courbure verticale. Pour chaque génération de spineur $\psi$, chaque section $v$ de $E_F$, et chaque constante réelle $C$,
on peut introduire le terme de Yukawa suivant:
\begin{equation}\label{lagrangienmassefermions}
\mathcal{L}_R(\psi)=\frac{C}{i}\widetilde{\psi}^{a}\sum_{k,l}v^{k}v^{l}R_{kalb} \psi^{b}.
\end{equation}
(Remarque: $R_{kalb}=R_{akbl}$, mais nous préférons l'air calme à l'air accablé.)\\
Dans cette équation il faut comprendre que la matrice réelle $8\times 8$ d'indices $a,b$ formée avec la courbure et
le vecteur $v$,
opère sur l'espace de multiplicité des fermions $M_8^{+}$, où nous utilisons une base orthonormée
venant de la base orthonormée $(e_1,...,e_8)$ via l'identification choisie de $M_8^{+}$ avec $F\otimes \mathbb{C}$. Sans le $1/i$ cette matrice est symétrique;
le facteur imaginaire $1/i$ est là pour avoir un opérateur anti-hermitien $T_R$ dans $u_8$, agissant sur
$M_8^{+}$ en respectant la structure hermitienne. Cet opérateur est donc diagonalisable; ses valeurs
propres sont les \emph{courbures sectionnelles} des fibres $F_x$. \emph{A priori} elles peuvent être
positives ou négatives; nous supposerons qu'elles sont toutes positives pour ne pas rencontrer de difficultés
avec ce point, mais nous pourrions aussi bien modifier l'opérateur symétrique $vvR$, en prenant sa valeur absolue.
Ce qui compte est une association naturelle entre courbure et interaction ajoutée au Lagrangien. Celle-ci étant censée
traduire les équations d'Einstein et Dirac entre les champs en dimension douze que nous n'avons pas pris en
compte dans le tenseur énergie-moment en dimension quatre.\\
Bien entendu, nous choisissons pour $v$ la direction privilégiée qui correspond au champ de Higgs.\\

\noindent Remarques: 1) le terme \eqref{lagrangienmassefermions} doit posséder la dimension $L^{-4}$, or
$\psi$ a la dimension $L^{-3/2}$ et $R$ la dimension $L^{-2}$ donc la constante $C$ doit avoir la dimension
de la longueur $L$; par ailleurs $v$ doit
être transformé en $H/M_0$ afin de pouvoir s'interpréter comme un champ. Les masses qui ressortent de ce
Lagrangien semblent donc faire intervenir le carré $HH$, ce qui n'est pas le cas dans les termes de Yukawa du modèle standard.
Dans ce modèle, qui semble bien confirmé par les expériences récentes, le couplage avec les fermions est linéaire en $H$ et
quadratique en $\psi$. Nous pourrions songer à prendre une racine carrée de l'opérateur symétrique $vvR$, mais il est
plus simple et convenable de considérer qu'on a affaire à un développement autour de $H_0$, et de considérer
que $vvR$ est remplacé par $(v_0)uvR$, où le vecteur $u$ de norme $1$ est fixe.\\
2) Rappelons que dans le Lagrangien fermionique initial, aucun terme de peut donner de masses à l'ordre deux
à cause des échanges de chiralité; en particulier \eqref{lagrangienmassefermions} n'est pas présent, et il
n'a de sens qu'après le fixage de jauge résultant du choix de $t$ dans $M_8^{-}$. Ce spineur $t$ correspondant à $1$ (qui est aussi
la direction $\nu"$ du neutrino) et au vecteur $e_8$. Or, au premier ordre, l'équation d'Einstein portant sur $R_{ij}$ a pour second membre le
tenseur énergie-moment des fermions. Par conséquent, il se peut (c'est notre hypothèse) que la courbure $vvR=\sum_{k,l}v^{k}v^{l}R_{kalb}$
traduise une réponse dynamique aux variations de $H$ (ou $t$) autour de sa valeur moyenne. Ainsi, l'adaptation que
nous allons décrire entre les directions de $M_8^{+}$ et la courbure peut se comprendre mieux.\\

\noindent Que se passe-t-il d'une génération à la suivante? Les spineurs de la $n$-ième génération
apparaissent avec le facteur $v_0^{n}$, il est nécessaire de diviser par ce facteur pour trouver une section
de $M_8$ de norme un; comme les spineurs
ont une dimension $L^{-3/2}$, cette normalisation peut être interprétée comme un changement d'échelle d'espace qui multiplie
les longueurs par le facteur $v_0^{2n/3}$, et alors les courbures sectionnelles, faisant intervenir deux dérivations d'espace, sont
multipliées par $v_0^{-4n/3}$. Approximativement, et si la constante $C$ ne dépend pas de la génération,
cela signifie que d'une génération à la suivante, les
masses doivent être multipliées par $v_0^{-4/3}$. Puisque ce facteur est expérimentalement proche de $200$ (en négligeant
les irrégularités),
on trouve que $v_0$ doit être proche de $1/53$.\\
A part la valeur exacte, cette croissance de la masse s'accorde avec l'interprétation des masses comme des énergies
mises en jeu pour explorer l'univers à des échelles d'énergie de plus en plus grandes, correspondant à des distances
de plus en plus petites.\\

En particulier, si les courbures $R_{kalb}$, mesurées en $L^{-2}$, multipliées par $C$ qui est en $L$, sont de l'ordre des unités,
on trouve des masses de fermions au premier ordre en $v_0^{2}\approx 4.10^{-4} GeV$, ce qui est conforme à l'expérience.
Nous ne pouvons pas prétendre que cet argument ait valeur de prédiction, cependant il nous semble légitime de faire remarquer
que, dans notre schéma, la différence de trois ordres de grandeur entre masses des premiers bosons et masses des premiers fermions apparaît
naturellement comme un effet de la courbure sur les fermions, qui n'interviendra pour les bosons vecteurs que comme une perturbation de
l'interaction directe avec le champ de Higgs. \\

\indent \emph{A priori} le tenseur de courbure transversal $R_{kalb}$ dépend du point $x$ dans $W$, donc l'opérateur
$T_R$ qui en résulte sur les spineurs, pourrait ne pas être constant le long de $W$; ce qui prédirait des masses de fermions dépendantes du lieu. Remarque:
pour les bosons également, en fait, la courbure et $H$ peuvent être conjuguées dans le Lagrangien et donc modifier
les masses en fonction du lieu; nous reparlerons plus loin de cette intervention éventuelle de la courbure dans les masses des
bosons. Il se pourrait donc qu'à très grande échelle de distance dans $W$, les masses de toutes les particules varient de
façon non-négligeable. Cependant, en accord avec notre hypothèse de courbure nulle de $W$, disant que la géométrie intrinsèque varie
peu d'un lieu de $W$ à un autre, nous allons supposer dans la suite que $R_{kalb}(x)$ varie très peu en fonction de $x$,
en comparaison de ce qu'il fait transversalement, dans les directions des fibres. Par conséquent, dans la discussion qui
suit, nous faisons comme si $R_{kalb}$ ne dépendait pas de $x$. Par contre il est certainement utile de prendre en
compte le développemnt de $R_{kalb}$ en fonction de $v$:
\begin{equation}\label{devcourburesectionnelle}
R_{kalb}(v)=R_{kalb}+\sum_jR_{kalb;j}v^{j}+\sum_{j h}R_{kalb;j h}v^{j}v^{h}+...;
\end{equation}
ceci permet d'obtenir des termes de masses d'ordres supérieurs à $2$ en suivant la formule \eqref{lagrangienmassefermions}.\\
\indent En plus de choisir le vecteur $e_8$ égal au vecteur $u_0$ de norme $1$, qui répond à $t$ par trialité, nous
pouvons choisir une déformation de $v$ aux ordres supérieurs; par exemple à l'ordre trois, nous posons
\begin{equation}
v=v_0e_8+v_0^{2}\sum_j\alpha^{j}e_j+v_0^{3}\sum_,\beta^{j}e_j+...
\end{equation}
Alors l'opérateur associé au terme $\mathcal{L}_R$ de \eqref{lagrangienmassefermions} à l'ordre quatre, devient
\begin{multline}\label{massedeformation}
(T_R)_{ab}=\frac{C}{i}v_0^{2}(R_{8a8b}+v_0(\sum_k(R_{8akb}+R_{ka8b})\alpha^{k}+R_{8a8b;8})\\
+v_0^{2}(\sum_{j h}R_{8a8b;j h}\alpha^{j}\alpha^{h}+\sum_j(R_{8ajb}+R_{ja8b})\beta^{j}+R_{8a8b;88}+...)
\end{multline}
\indent Le choix d'une telle déformation est compatible avec le principe de  covariance
générale, car il revient à utiliser un difféomorphisme dont la dérivée le long le long de
$W$ est l'identité, mais dont les jets d'ordres plus grands ne sont pas nuls. Nous entrons ainsi dans un nouveau domaine, qui fait appel aux transformations
non-linéaires des coordonnées transverses, une sorte de théorie de jauge non-linéaire (cf. \cite{signori2008}).\\
\indent De cette discussion, il faut surtout retenir que des termes d'ordres supérieurs viennent modifier
l'opérateur simple $R_{8a8b}$.\\

\noindent Quelquesoit $a$ (resp. $b$) on a $R_{8a88}=0$ (resp. $R_{888b}=0$), donc $e_8$ est
dans le noyau de $T_R$. En suivant l'hypothèse de la relation naturelle entre terme de Yukawa et courbure sectionnelle, nous trouvons
que le neutrino au premier ordre possède bien une masse nulle.\\
Si nous nous limitions à cet ordre, nous retrouverions le modèle
standard, mais là, ce serait un peu un défaut, car les neutrinos sont connus aujourd'hui pour avoir des masses non-nulles.
Expérimentalement, la masse de $\nu_e$ est inférieure à $2.5 eV=2.5\times 10^{-6} MeV$, celle de $\nu_\mu$ est
de l'ordre de $170 keV= 0.17 MeV$ et celle de $\nu_\tau$ de l'ordre de $18 MeV$. (La règle du rapport de
$200$ n'est pas vérifiée, mais nous n'en sommes pas trop loin non-plus.)\\
Or notre approche permet de faire apparaître une
masse pour les neutrinos, si dans la formule \eqref{massedeformation}, les termes en
$v_0^{2}$ ou $v_0^{3}$ sont non-nuls.\\
Nous décidons malgré tout que la plus petite valeur propre de l'opérateur \eqref{massedeformation} correspond au
neutrino.\\
Les trois spineurs indépendants aux trois premiers ordres en $v$ représentant les neutrinos théoriques de saveur déterminée
sont $\nu_0$, $v_0\nu_1$, $v_0^{2}\nu_2$, où $\nu_0$, $\nu_1$, $\nu_2$ sont de norme $1$ dans la direction $e_8$.
Mais \emph{a priori} les trois directions propres des masses
aux trois ordres des générations de neutrinos pour l'opérateur de masse perturbé ne coïncident pas avec ces directions,
ils font intervenir d'autres directions dans $M_8^{+}$. Notons que cela ne fait pas encore apparaître une matrice de passage entre
les deux bases, puisque les nouveaux vecteurs sont seulement des triplets de vecteurs orthogonaux dans $M_8^{+}$, qui est de
dimension huit et concerne tous les fermions.\\

\noindent Le second vecteur de base, $e_7$, peut être choisi dans $F$, de norme $1$ et orthogonal à $e_8$, de manière à ce que $R_{8787}$
soit le plus petit possible en valeur absolue. On obtient ainsi que la direction de l'électron
est également, au premier ordre, une direction propre, celle qui correspond à la plus petite masse non-nulle possible, étant donné $e_8$.\\
Le fait que nous puissions décider que $e_7$ est dans une direction choisie perpendiculaire à $e_8$ repose
sur notre liberté de choix au début, sur la structure complexe arbitraire $J$ sur $F$.\\
\indent Notons que, pour la première fois depuis le début de notre étude, il apparaît naturellement des \emph{modules},
c'est-à-dire des invariants caractéristiques continus du système des particules, venant de la métrique en dimension douze
au voisinage de $W$. Ici ces invariants sont responsables des masses des fermions.
La plus importante dans les applications est sans doute celle de l'électron, mais il est connu également,
que toutes les autres masses ont des conséquences fondamentales sur  notre univers.\\
\indent Là aussi, pour les leptons chargés, \emph{a priori} les ordres supérieurs donnent une base différente pour les vecteurs
propres des masses, à cause des termes d'ordre plus grand que deux dans la formule de $T_R$. \\
\indent La loi des rapports de masses est bien vérifiée pour les leptons chargés. En effet,
la masse de l'électron est proche de $0.5 MeV$, celle du muon est de l'ordre de $100MeV$, et celle
du tau est estimée $1800MeV$. En fait, c'est la seule série possédant une telle régularité.\\

\noindent Le même scenario s'applique au quarks.\\
\noindent La masse de $u$ est située entre $1.5 MeV$ et $3.3 MeV$, et celle de
$d$  entre $3.5 MeV$ et $6 Mev$. A la génération suivante $c$ possède une masse de l'ordre de
$1200MeV$, donc plus que $200$ fois celle de $u$, bien que d'ordre comparable, par contre $s$ possède une masse de l'ordre de $100MeV$,
qui est inférieure d'un ordre de puissance de $10$ à ce qu'on attendrait, qui devrait être environ $1000 MeV$. Ensuite, pour
$t$ on a $170000MeV$ qui n'est pas trop loin de $200$ fois $1200$, mais pour $b$, on a $4200MeV$ qui est à nouveau
trop petit, cette fois-ci c'est même de deux ordres de grandeur. Par conséquent, la série des quarks $\overline{q"}$ que nous avons
nommés électriques, $d,s,b$, de charge $-1/3$, est beaucoup moins
régulière du point de vue des échelles que la série des quarks $q'$,
que nous avons nommés magnétiques, c'est-à-dire $u,c,t$, de charge $2/3$. Nous ne prétendons pas expliquer ce
phénomène avec notre modèle actuel.\\

\indent A présent, nous introduisons l'hypothèse qu'il est possible de séparer
les espaces engendrés par les vecteurs propres des masses des leptons de ceux des masses des quarks. Ceci se justifie,
au moins en partie, par la constatation expériméntale
que les leptons se distinguent des quarks par des masses sensiblement plus petites.\\

\noindent Cette hypothèse se traduit par l'existence de deux matrices de passage entre les bases théoriques des saveurs
et les directions propres des masses. Celle des leptons est a priori une matrice $6\times 6$ et celle des quarks une
matrice $18\times 18$. Mais dans les deux cas il y a une bonne raison pour ramener la matrice à une matrice $3\times 3$.\\
\indent Du côté des leptons, il se trouve que les directions propres pour l'électron, le muon et le tau
ne diffèrent pas sensiblement des directions théoriques.\\
Malgré de nombreux efforts, cf. la revue \cite{bernstein2013charged}, il n'y a pas d'indication expérimentale
de mélange des saveurs pour les leptons chargés. Les directions de l'électron, du muon et du tau théoriques, venant
pour nous de l'analyse linéaire, sont les directions propres des masses, ou en tout cas en sont très proches.\\
Dans notre approche géométrique, cela signifie que le terme de courbure est guidé par $e_7$ et non
par $e_8$ aux ordres supérieurs. Est-ce lié à la non-générescence de la plus petite masse non-nulle?
ou bien à la détermination par la direction $\Omega$ de la métrique spinorielle? \\
Il résulte de l'orthogonalité entre directions propres des masses, que les transformations non-triviales entre leptons ne portent que sur les neutrinos.\\
La matrice de passage dans ce cas est nommée $PMNS$, d'après Pontecorvo, Maki, Nakagawa et Sakata. Cf. \cite{suekane2015neutrino}.\\
Les masses des neutrinos sont responsables du phénomène important nommé oscillations des neutrinos: un neutrino de saveur prescrite
étant combinaison de trois neutrinos de masses différentes, sa fonction d'onde évolue comme une superposition de
trois oscillateurs indépendants; elle est donc une fonction quasi-périodique des trois neutrinos de saveurs pures.\\
\indent Du côté des quarks, les
courbures sectionnelles, correspondant à la diagonalisation de $R_{a8b8}$ dans une base orthonormée de l'espace orthogonal au plan $(e_7,e_8)$
fournissent six vecteurs $e_6,e_5,e_4,e_3,e_2,e_1$ de norme $1$, orthogonaux entre eux et orthogonaux à $e_7,e_8$,
associés à six nombres $R_{k8k8}$ \emph{a priori} tous différents. En particulier dans ce scenario, il n'y a
pas de raison pour que seulement deux nombres,
correspondant à $m_d$ masse du down et $m_u$ masse du up émergent.\\
Bien sur il se pourrait que les masses soient de multilicité trois; mais nous préfèrerions supposer que la
théorie est générique, et que les six masses distinguent effectivement les six quarks
les uns des autres. (Il faudrait alors leur trouver un petit nom, pourquoi pas lièvre, loir et chapelier fou?)
Suivant l'hypothèse de généricité, et en tenant compte des masses connues, nous aurions donc un premier quark, disons $u'$,
de l'ordre de $1 MeV$, puis un second $u"$, entre $1 MeV$ et $2 MeV$, disons $2$
pour fixer les idées et un troisième $u'''$, de masse $3 MeV$; nous leurs ferons correspondre les vecteurs
impairs $e_1, e_3, e_5$. De même nous aurions trois quarks de saveur $down$, nommés $d'$, $d"$, $d'''$, de masses
respectives $4 MeV$, $5 MeV$, $6 MeV$, toujours pour simplifier et fixer les idées, en restant à peu près en accord avec
les données expérimentales.\\
Le plus simple serait qu'aux générations suivants les masses des quarks correspondants aient des masses en rapport
$1/v$, $1/v^{2}$ etcetera avec celles-ci. C'est bien ce qui se produit grossièrement du côté de $u$ mais on ne peut
pas dire ça du côté de $d$.\\
\indent Cependant notre approche n'est pas non-plus incompatible avec le fait que les quarks d'un même type, électrique ou
magnétique, aient une masse commune.\\
Il se pourrait par exemple que les observations actuelles ne donnent que les moyennes de leurs masses et qu'il soit
légitime d'attribuer la valeur moyenne de ces masses à chaque génération de quark. Ainsi la symétrie $SU_3$ des quarks serait
plutôt effective qu'exacte. Une autre formulation de la même idée serait de dire que, conformément à la symétrie $SU_3$,
le terme de couplage viendrait de la courbure moyenne dans la $3$-direction $(e_1,e_3,e_5)$ pour les uns
et de celle dans la $3$-direction $(e_2,e_4,e_6)$ pour les autres. Ceci revient à prendre les moyennes des masses définies
par $R_{8a8b}$, mais c'est dit plus géométriquement.\\
Mais une autre possibilité, réellement différente, et plus satisfaisante, mais difficilement contrôlable à l'heure actuelle,
serait que, dynamiquement, en régime non-perturbatif d'interaction forte, les masses effectives
convergent vers une valeur commune. Variante: les contraintes apportées par les équations d'Einstein et Dirac couplées
dans le secteur transversal $E_F$ relient les symétries de la courbure aux symétries des fermions. Alors, comme c'est observé,
la symétrie $SU_3$ serait exacte.\\
\indent Quoiqu'il en soit, dans l'état actuel des choses, nous sommes obligés de faire comme si la symétrie non-brisée $SU_3$ préservait effectivement
les états propres des masses, d'où à nouveau une matrice unitaire $3\times 3$. Cette fois elle se note $CKM$, d'après Cabibbo, Kobayashi et Maskawa.
Cf. \cite{kuhr2013flavor}, T.Gershon dans \cite{gardi2015lhc}.\\
Comme pour les leptons de charge nulle et ceux de charge $-1$ (ou $+1$), il existe deux bases naturelles des espaces
de quarks $Q'$ et $Q"$, celle qui correspond aux
masses les mieux déterminées possibles, et celle qui correspond aux saveurs, c'est-à-dire
à l'action de $W_+$ et de $W_-$. La matrice de passage est la matrice $CKM$.\\

La théorie perturbative de la réduction des matrices hermitiennes en base unitaire, cf. \cite{kato2012perturbation}, prédit la forme des matrices $PMNS$ et
$CKM$. Le petit paramètre est $\epsilon=v_0\approx 0.02$. Les trois valeurs propres de la matrices non-perturbée
sont par orde croissant $\lambda_0$, $\lambda'_0$ et $\lambda"_0$; elles semblent dans les trois cas bien séparées
les unes des autres, au moins pour l'électron et les quarks. Les valeurs propres de la perturbation sont des séries en $\epsilon$. Les vecteurs propres également.
On s'attend donc à une matrice de passage $P$ proche de l'identité. Par exemple avec deux dimensions au lieu de trois, et
en supposant que les opérateurs sont  réels symétriques, cela fait
\begin{equation}
P=
\begin{pmatrix}
1+a_1\epsilon+a_2 \epsilon^{2}+...& b_1\epsilon+b_2 \epsilon^{2}+...\\
c_1\epsilon+c_2 \epsilon^{2}+...& 1+d_1\epsilon+d_2 \epsilon^{2}+...
\end{pmatrix}
\end{equation}
L'unitarité impose
\begin{equation}
a_1=d_1=0,\quad a_2=d_2=-\frac{1}{2}b_1^{2},\quad b_1=-c_1.
\end{equation}
Dans le cas de $CKM$, pour les quarks des deux premières générations, les expériences ont bien établi que $b_1\epsilon$ est
de l'orde de $0.22$, ce qui donne $b_1\approx 10$, et n'est pas incompatible avec notre modèle, étant donnée l'hypothèse
que les courbures sont de l'ordre des unités. Dans la suite, nous notons  $b_1\epsilon =\lambda$, comme tout le
monde.\\
Admettant ce résultat pour les deux premières générations, l'introduction de la troisième génération, fait apparaître la forme
générique de la matrice unitaire $P_{CKM}$:
\begin{equation}
P_{CKM}=
\begin{pmatrix}
1-\frac{1}{2}\lambda^{2}& \lambda& A\lambda^{3}(\rho-i\eta))\\
-\lambda& 1-\frac{1}{2}\lambda^{2}& A\lambda^{2}\\
A\lambda^{3}(1-\rho-i\eta)& -A \lambda^{2}& 1
\end{pmatrix}
+O(\lambda^{4}).
\end{equation}
Le nombre $\eta$ a une très grande importance car, dans le modèle standard, c'est lui qui est responsable
de la violation de la symétrie $PC$, d'échange de parité et de conjugaison de charge, qui entraîne l'excès
de particules sur les anti-particules dans notre univers quotidien.\\
Si $T_R$ était réel et symétrique le terme de courbure sectionnelle \eqref{massedeformation}
n'expliquerait pas $\eta\neq 0$, car $P$ serait une matrice orthogonale. Mais rappelons nous que
l'identification de $F$ avec $M_8$ n'est possible qu'après complexification des deux espaces et qu'elle
possède deux ambiguités de phases (dont nous allons reparler à propos du cercle électromagnétique);
par conséquent le changement de base est seulement unitaire, et $\eta$ peut bien être différent de zéro.\\

\noindent Dans le cas des neutrinos, pour la matrice $PMNS$, le coefficient $\lambda$ est encore plus grand,
ainsi que le nombre $\eta$. Une bonne explication viendrait du fait que la matrice non-perturbée possède une valeur propre nulle
de multiplicité deux ou même trois, et que dans ce cas, la théorie prédit (cf. \cite{kato2012perturbation}) que les valeurs propres et
les vecteurs propres de la matrice perturbée ne sont pas des
séries en $\epsilon$ mais en $\epsilon^{1/2}$ ou en $\epsilon^{1/3}$. Nous trouverions donc un $\lambda$
de l'ordre de celui qui est observé, avec un $b_1$ de l'ordre de $1$ cette fois.\\

\subsection{Du côté des bosons}

\indent Les champs de spin entiers, au deuxième ordre en $v$, sont le champ scalaire de Higgs $H$, les composantes de la connexion
$C_{\mu i}^{j}$, ses dérivées verticales $\partial_kC_{\mu i}^{j}$ et $\partial_l\partial_kC_{\mu i}^{j}$, la courbure $M_{\mu\nu i}^{j}$
qui provient des dérivées horizontales de $C_{\mu i}^{j}$ ainsi que la courbure de la métrique verticale $R_{ikjl}$ et
ses dérivées premières verticales et horizontale.\\
\indent Parmi ces champs, en vertu de la prop. $4$, les seuls qui interagissent avec les fermions sont $H$, $W_{\pm}, Z_0$, $Z_4$,
$B_0$, $B_1$, et les gluons et photons du champ de jauge résiduel
$SU_3\times U_1$. Ces derniers sont sans masse, les premiers ont des masses qui ont déjà été discutées. Pour tous les autres, le plus naturel
est de supposer que les masses sont données par le couplage avec $H$.\\
Comme nous l'avons mentionné plus haut, les combinaisons de $iB_0$, $iB_1$, $2Z_1$ et $Z_4/2$ auxquelles $H$ ne donnent pas de masse,
par exemple $iB_0+Z_4/2$ ou $iB_1+2Z_1$ posent un problème particulier, car elles agissent sur les fermions. La première pourrait
être contrainte par le choix de coordonnées unimodulaire, or la seconde coïncide avec la première sur les leptons et agit
comme $2Z_1$, qui est la diagonale de $SU_2$ sur les quarks.\\

Si on omet la trace $B_0$, si on néglige les dérivées horizontales de $R$, et qu'on se reporte aux équation \eqref{energielorentzcouplage}, \eqref{energielineaire} et \eqref{energieyangmills},
le Lagrangien bosonique de degré deux en $v$ s'écrit
\begin{multline}\label{energielorentzcouplagedegredeux}
\mathcal{L}_2(C,\nabla C,\nabla\nabla C)
=\frac{3}{4}\sum_{i j k l}v^{k}v^{l}(\partial_k\partial_lC_{\mu i}^{j}C_{i}^{\mu j}+C_{\mu i}^{j}\partial_k\partial_lC_{i}^{\mu j})\\
+\frac{1}{4}\sum_{i j k l}v^{k}v^{l}(\partial_k\partial_lC_{\mu i}^{j}C_{j}^{\mu i}+C_{\mu i}^{j}\partial_k\partial_lC_{j}^{\mu i})\\
+\frac{3}{2}\sum_{i j k l}v^{k}v^{l}\partial_kC_{\mu i}^{j}\partial_lC_{i}^{\mu j}
+\frac{1}{2}\sum_{i j k l}v^{k}v^{l}\partial_kC_{\mu i}^{j}\partial_lC_{j}^{\mu i}\\
+\frac{1}{4}\sum_{i j k l m}v^{k}v^{l}R_{ i k j l}C_{\mu i}^{m}C_{j}^{\mu m}-\frac{1}{4}\sum_{i k l m n}v^{k}v^{l}R_{ m k n l}C_{\mu i}^{m}C_{i}^{\mu n}\\
+\frac{1}{2}\sum_{\kappa \lambda i k l}v^{k}v^{l}g^{\kappa\lambda}M_{\mu\kappa k}^{i}M_{\lambda l}^{\mu i}.
\end{multline}
\noindent L'orthogonalité des composantes symétriques sans traces et antisymétriques des matrices carré fait que ces deux familles
peuvent être traitées séparément.\\
La troisième ligne
donne un terme de masse pure pour $\nabla C$, et la cinquième décrit la masse possible des composantes de $M$.\\
Par contre, on voit que les états propres de masse n'ont pas de raison de respecter la divison entre $C$ et $\nabla\nabla C$.
On constate également que, comme nous l'avons annoncé plus haut, les courbures sectionnelles $R_{8a8b}$ contribuent aux masses des
bosons, à travers les
composantes de la paire $C,\nabla\nabla C$.\\

Les couplages des champs de bosons avec $H$ étant tous du même type, il est plus simple d'admettre que
les masses de tous les bosons massifs à la première génération sont du même ordre que celles de $W$ et $Z$, c'est-à-dire autour de
$100 GeV$.\\
Pour les générations suivantes on propose un raisonnement de changement d'échelle analogue à celui qui a été fait pour les fermions:
le champ de la $n$-ième génération est multiplié par $v_0^{n}$, comme il est de dimension $L^{-1}$, les longueurs sont
multipliées par $v_0^{n}$, donc la courbure multipliée par $v_0^{-2n}$, donc la partie de la masse au carré expliquée par la courbure
est multipliée par $v_0^{-2n}$, l'autre partie restant du même ordre. Si bien que d'une génération à la suivante un facteur
$53$ est appliqué. Ceci prédit, par exemple, que les $W,Z$ de la deuxième génération auront des masses de l'ordre de $5 TeV$.
Ce sont des valeurs d'énergie bientôt accessibles.\\

Les particules connues représentent
$20$ pour cent de la masse de l'univers observable.
Or en comptant le nombre des bosons visibles, $3$ par rapport aux $55=64-9$ possibles, on a moins
que $1/5$, mais il faut tenir compte des fermions, en particulier des plus lourds, qui font $3$ fois
$2$ égal $6$, d'où un rapport $9/55$, qui s'approche de $1/5$. Par conséquent notre suggestion pour la
matière noire n'est pas déraisonnable.\\

Reste la question de l'énergie noire, qui représente $70$ pour cent de l'énergie de l'univers accessible, selon les estimations actuelles.
L'argument qui s'oppose
à une simple explication par la constante cosmologique positive est le calcul de l'énergie du vide quantique sur la
base du modèle standard, donnant une valeur gigantesque à cause de l'excès de fermions par rapport aux bosons. Il est connu que la supersymétrie,
rétablissant l'équilibre entre fermions et bosons, règlerait ce problème en prévoyant une énergie du vide nulle, si elle n'était
pas brisée (cf. \cite{henneaux1989cosmological},
\cite{bousso2008cosmological}). Or dans notre
théorie il y a au départ pour chaque génération, $64$ directions réelles bosoniques, les $9$ de $SU_3\times U_1$ plus les trois de $SU_2$ et les $52$
étranges, ce qui vient évidemment de la dimension de $gl_8$, d'où $128$ degrés de libertés pour les particules,
et il y a $128$ degrés de liberté réels fermioniques, puisque
les quatre dimensions complexes de Dirac en dimension quatre se multiplient par les $16$ réelles des spineurs en dimension huit,
pour donner $64$ dimensions complexes. Ce compte est modifié par la bosonisation des quatre degrés de liberté du neutrino impair
(right) en un boson de Higgs, ce qui fait $68$ degrés de liberté bosoniques contre $60$ degrés de liberté fermioniques, de plus
il faut avouer que le décompte exact des masses suivant les hypothèses faites dans notre modèle ne donnerait
pas l'équilibre, la balance penchant du côté des bosons. Cependant
l'équilibre n'est pas loin d'être atteint quand même. Alors la constante $\Lambda$ pourrait bien expliquer
l'inflation.
Mais il se pourrait surtout qu'à très grande échelle, ce soit le modèle de départ exact qui compte le plus, or
ce modèle est la gravitation avec un spineur de Dirac sans masse,
donc, à nouveau en ajoutant une constante cosmologique $\Lambda$, on retrouverait la meilleure explication connue.\\
\indent Signalons que des modèles s'appuyant sur des dimensions supplémentaires de l'univers, comme le fait notre modèle, ont été
proposés pour expliquer l'énergie noire, cf. \cite{appelquist1983quantum}, \cite{dupays2013can}.\\

\subsection{Le cercle électromagnétique}

\indent Le choix de la charge électrique sur les quarks est resté mystérieux jusqu'ici, d'un point de vue géométrique. Pourquoi ce $1/3$ si peu compatible avec
le groupe de jauge usuel?\\
\indent Nous démontrons dans ce paragraphe que l'opérateur de charge électrique $Q_0$ qui a été retenu est l'unique opérateur sans masse qui engendre le groupe circulaire
de plus petite longueur dans un sous-quotient de $U_8$. Il se trouve en plus que ce quotient, qui est $PU_6\times PU_2$,
est celui qui contient l'opérateur de masses des fermions. Ceci donne une explication possible des charges fractionnaires avec des $1/3$ pour les quarks.\\
Nous obtenons aussi de cette manière, une jolie formule de $Q_0$ et $Z_0$ basée sur la géométrie des racines de $\mathfrak{u}_8$, $\mathfrak{u}_6$ et $\mathfrak{u}_2$.\\
Une autre formule donne une belle interprétation de $Q_0$ en terme des ambiguités de phase des structures complexes et réelles des spineurs.\\

\indent Afin de décrire l'algèbre de Lie $\mathfrak{u}_8$, nous introduisons une base orthonormée $x_k;k=1,...,8$
du dual $\mathfrak{h}_8^{*}$ de l'espace Euclidien $\mathfrak{h}_8$. Le système de racines de $\mathfrak{su}_8$ est l'ensemble des
$x_j-x_k;j\neq k$; les racines simples usuelles sont les
$x_{k+1}-x_k;k=1,...,7$. Comme base duale des $x_k$ dans $\mathfrak{h}_8$ on a les vecteurs $H_k;k=1,...,8$.\\
\indent Le sous-espace $\mathfrak{h}_4$, diagonale de $so_8$, est défini par l'annulation des coordonnées paires $x_2,x_4,x_6,x_8$.
Les racines de $so_8$ sont les combinaisons $\pm x_{2k+1}\pm x_{2l+1}$ où $k,l$ sont distincts et où les $\pm$ varient
indépendamment l'un de l'autre; cet ensemble est le système de racines $D_4$, clé de la trialité.\\
On introduit les cinq réseaux principaux de $\mathfrak{h}_4$
(cf. \cite{bourbaki1981groupes456} p. 200 et suivantes), qui correspondent aux groupes nodaux (groupes fondamentaux des tores maximaux) des cinq groupes
d'algèbre de Lie $\mathfrak{h}_4$ associés au système des racines $D_4$; la correspondance étant donnée en associant $2 i\pi v$
à un vecteur $v$ de $\mathfrak{h}_8$. Le réseau $L_0$ est fait des vecteurs à coordonnées entières,
il donne le groupe nodal de $SO_8$ standard; $L_2$ est la somme de $L_0$ et
du groupe cyclique engendré par le vecteur $H^{+}=\frac{1}{2}(H_1+H_3+H_5+H_7)$, il donne le groupe nodal du quotient $PSO_8$
de $SO_8$ par son centre $\{\pm 1\}$; $L_0^{+}$ (resp. $L_0^{-}$) est le sous-groupe de $L_2$ engendré par $H_3$, $H_5$, $H_7$
et par $H^{+}$ (resp. $H^{-}=\frac{1}{2}(H_1+H_3+H_5-H_7)=H^{+}-H_7)$, il correspond au groupe nodal du groupe $j(SO_8)$ (resp. $j^{2}(SO_8)$,
groupe que nous noterons $SO_8^{+}$ (resp. $SO_8^{-}$);
$L_1$ est le sous-groupe de $L_0$ formé par les vecteurs dont la somme des quatre coordonnées est paire,
il correspond au groupe nodal du revêtement $Spin_8$. \\
Les groupes nodaux correspondant seront respectivement notés $\Gamma_0$, $\Gamma_0^{+}$, $\Gamma_0^{-}$, $\Gamma_1$ et $\Gamma_2$.\\
Le groupe quotient $L_2/L_1$ est isomorphe à $\mathbb{Z}/2\mathbb{Z}\times \mathbb{Z}/2\mathbb{Z}$, les quatre classes
étant celles de $0$, $H^{+}$, $H^{-}$, $H_1$, mais $H_1$ ici peut être remplacé par n'importe lequel des $H_{2k+1}$,
ou par toute somme d'un nombre impair de ces vecteurs, en particulier $H^{+}+H^{-}=H_1+H_3+H_5$.\\
Les trois groupes $L_0=L_0^{0}$, $L_0^{+}$ et $L_0^{-}$ sont contenus dans $L_2$ et contiennent $L_1$, donc les quotients
$L_2/L_0^{\varepsilon}$ et $L_0^{\varepsilon}/L_1$ sont isomorphes à $\mathbb{Z}/2\mathbb{Z}$ quelque soit $\varepsilon=0,+,-$.\\
L'automorphisme d'ordre trois $j_*$ de $so_8$ que nous avons préféré envoie $H^{-}$ sur $H_7$, $H^{+}$ sur $-H^{-}$ et $H_7$ sur
$-H^{+}$.\\
\indent Le réseau $L=Q_8^{\vee}$ de $\mathfrak{h}_8$ qui correspond au le groupe nodal de $U_8$ est celui qui est engendré par les vecteurs $H_k;k=1,...,8$.
Pour avoir la diagonale $\mathfrak{v}_8$ de $\mathfrak{h}_8$ qui correspond à $su_8$,
on impose que la somme des coordonnées $x_k$ soit nulle.
Soit $Q_8$ le réseau dans $\mathfrak{v}_8^{*}$ engendré par les racines de $su_8\otimes \mathbb{C}$, le réseau dual dans
$\mathfrak{v}_8$ est noté $P_8^{\vee}$;
il est constitué par les vecteurs dont les différences de coordonnées sont entières. Il contient $L$.
Notons
\begin{equation}
y_8=x_1+x_2+...+x_8,\quad Y_8=H_1+H_2+...+H_8.
\end{equation}
Alors $P_8^{\vee}$ est engendré par $H'_1=H_1-Y_8/8$ (ou par n'importe quel élément $H'_k=H_k-Y_8/8$). De même
le réseau $P_8$ des poids de $su_8$ (formes à valeurs entières sur $\Gamma_0$) est engendré par $Q_8$ et par l'élément $x'_1=x_1-\frac{1}{8}y_8$
ou n'importe quel élément $x'_k=x_k-\frac{1}{8}y_8$.\\
Le centre $C_8$ de $SU_8$ est le groupe cyclique des racines huitième de l'unité; il est contenu dans le centre de $U_8$,
constitué par les matrices scalaires $zId$ où $z$ est de module un. Un générateur privilégié est $\exp(2i\pi H'_1)$;
On a un isomorphisme canonique entre $C_8$ et le quotient $P_8^{\vee}/L$; ce qui correspond au fait que
$SU_8$ est simplement connexe.\\

Tout ceci s'étend aux groupes $U_n$ pour tous les entiers naturels $n$. En particulier deux sous-groupes de $U_8$ méritent notre attention en plus de $SO_8$,
et de ses transformés par trialité; il s'agit de $U_6$
qui contient le $U_3$ des gluons et de $U_2$ qui contient $W_{+}$, $W_{-}$ et $Z_1$. Leurs diagonales (algèbres de Cartan, racines, etc.)
se décrivent de la même manière que pour
$\mathfrak{u}_8$; les élements qui remplacent $y_8$ sont respectivement $y_6=x_1+x_2+...+x_6$ et $y=x_7+x_8$. Le groupe cyclique d'orde huit est
remplacé par un groupe cyclique d'ordre six et un groupe d'ordre deux respectivement. Celui d'ordre six dans
$P_6^{\vee}/Q_6^{\vee}$ est engendré par la
classe de $H_1-Y_6/6$ et celui d'ordre deux dans $P_2^{\vee}/Q_2^{\vee}$  est engendré par la
classe duale de $H_7-Y/2=\frac{1}{2}(H_7-H_8)$.\\

Sur l'espace vectoriel complexe de dimension huit $M_8^{+}$, on sait faire agir $H_7$ par $Z'_1/i$, faisant
$1/2$ sur $\overline{Q"}$ et $\Omega$, $-1/2$ sur $Q'$ et $1$. On sait aussi que
$B_0$ est l'identité et $B_1$ le projecteur sur les leptons. Il est naturel d'identifier $Y_8$
à $B_0$, donc il est naturel de supposer que $Y_6$ est $Id$ sur $Q^{+}$ et $0$ sur $L$ et $Y$ est $0$ sur $Q^{+}$ et $Id$ sur $L$.\\

\noindent \textbf{Lemme 5}: en faisant agir $H_7$ à travers la représentation spinorielle paire de $spin_8$ sur $S_8^{+}$, on a
\begin{align}
\frac{Q_0}{i}&=\frac{Y_6}{6}-\frac{Y}{2}-H_7,\\
\frac{Z_0}{i}&=\frac{Y_6}{6}-\frac{Y}{2}+H_7.
\end{align}\\

\noindent \emph{Démonstration}: on sait que $Q_0/i=(B_0-4B_1)/6-H_7$ et que $Z_0/i=(B_0-4B_1)/6+H_7$;
il suffit donc de remplacer $B_0$ par $Y_8=Y_6+Y$ et $B_1$ par $Y$.\\

Ce qui est remarquable dans ces formules, est que les trois monômes sont des plus hauts poids de représentations
fondamentales, le premier celui de $U_6$ sur l'espace des quarks seuls, le second celui de $U_2$ sur l'espace des leptons seuls,
le troisième celui de $U_1$, étant l'action diagonale du cercle engendré par le demi-crochet $Z'_1$ de $W_+$ avec $W_-$, agissant à la fois
sur les quarks et sur les leptons, comme $1/2$ sur $\Omega$ et $\overline{Q"}$, comme $-1/2$ sur $1$ et $Q'$.\\

Dans la théorie quantique, il est important de considérer le groupe $S^{1}$ engendré par $2\pi Q_0$; comme $Q_0/i$ n'appartient
pas à $L$ ce groupe n'est pas contenu dans
$U_8$; seulement un revêtement à six feuillets de $S^{1}$ est naturellement contenu dans $U_8$.\\
Le groupe naturel où on peut intégrer $i(Y_6/6-Y/2)=iB'$ est le produit du groupe quotient $PU_6=U_6/C_6$ et du groupe
quotient $PU_2=U_2/C_2=SO_3\times U_1$. Son groupe fondamental est isomoprphe à $(\mathbb{Z}/12\mathbb{Z})\times \mathbb{Z}$.\\

Nous constatons un remarquable accord de la charge électrique avec la masse, qui elle aussi fait appel à un élément de $U_6\times U_2$,
à savoir $T_R$.\\
Surtout, le choix de $PU_6\times PU_2$ est également le seul possible du point de vue de l'opérateur principal des masses
à cause de l'ambiguité des phases sur $L$ et $M$ résultant de l'identification de la multiplicité $M_8$ avec
l'espace transverse complexifié $F\otimes \mathbb{C}$.\\

\noindent A présent, nous nous proposons d'établir que, parmi tous les choix possibles d'un opérateur
de charge pour une particule de masse nulle (i.e fixant le choix de Higgs), le choix minimal du point de vue
géométrique du cercle engendré est celui
de l'électromagnétisme connu, attribuant une charge $0$ au neutrino, $-1$ à l'électroon et $-1/3$ et $2/3$
aux deux familles de quarks.\\

\indent Rappelons que le fixage de jauge menant à $SU_3$ et à $SU_2$ laissait une grande ambiguité sur le choix de la particule
sans masse qui pourrait être le photon. Nous avions un espace vectoriel réel de dimension trois de candidats: les combinaisons
$i\beta_0B_0+i\beta_1B_1+2\alpha_1Z_1+\frac{1}{2}\alpha_4 Z_4$ de $iB_0$, $iB_1$, $Z_1$ et $Z_4$,
satisfaisant à la condition $\beta_0+\beta_1=\alpha_1+\alpha_4$.\\
Une autre base naturelle de l'espace engendré par $iB_0$, $iB_1$, $Z_1$ et $Z_4$ est $Y_6$, $Y$, $Z_4/2$ et $Z_6=2Z_1-Z_4/2$,
car les deux premiers vecteurs sont les projections sur $L$ et $Q$ respectivement, et les deux suivantes sont les opérateurs
$i\sigma_z$ canoniques dans $L$ et $Q$ respectivement (diagonale $-1,1$ dans la base $1,\Omega$ ou $Q',Q"$).
Si nous réécrivons la combinaison de la façon suivante:
\begin{equation}
Q=iz\frac{Y_6}{6}+tZ_6+iy\frac{Y}{2}+x\frac{Z_4}{2},
\end{equation}
nous trouvons que l'annulation de la masse se traduit par la condition
\begin{equation}
y=2x.
\end{equation}
Cette condition ne porte donc que sur les leptons; elle dit que le neutrino est neutre (bien sur puisqu'il fixe la jauge), et que $y$
est la charge de l'électron. A ce stade il n'y a pas de raison que cette charge soit égale à $-1$.\\

En admettant que $W_+$ et $W_-$ aient des charges bien définies, il s'ensuit que la différence entre la charge des $q'$ et
celle des $\overline{q"}$ doit être égale à $y$, d'où l'équation $t=-y/2$.\\
Nous sommes ramenés à deux constantes arbitraires:
\begin{equation}
Q=iz\frac{Y_6}{6}-y\frac{Z_6}{2}+iy\frac{Y}{2}+y\frac{Z_4}{4}.
\end{equation}

A présent, nous faisons l'hypothèse que $Q$ engendre un sous-groupe compact dans un quotient de $U_6\times U_2$, cela équivaut
à demander que $Q$ appartienne au réseau des poids de ce groupe, donc que $z$ et $y$ soient des nombres entiers.
La solution minimale étant $z=\pm1$, $y=\pm1$.\\

Au signe près, qui est une convention, cela fait deux solutions possibles: $z=1$,$y=-1$ et $z=-1$, $y=-1$.\\
La seconde correspond simplement à l'échange de $Q'$ et de $\overline{Q"}$ et donc à un choix symétrique de $W_{\pm}$.\\
Nous aurions alors un opérateur $Q=Q'_0$ donné par
\begin{equation}
\frac{Q'_0}{i}=\frac{Y_6}{6}+\frac{Z_6}{2}-\frac{Y}{2}-\frac{Z_4}{4}=\frac{Y_6}{6}-\frac{Y}{2}-\frac{Z_4}{2}+Z_1.
\end{equation}
Et au fond, on n'y verrait que du feu, il suffirait de permuter partout $u$ et $d$, $c$ et $s$, $t$ et $b$.\\
\indent Pour les conséquences physiques, la chose la plus importante pour les valeurs des charges électriques
des quarks, est qu'elles s'accordent avec l'observation de tous
les mésons et baryons connus, en particulier le proton $p=(uud)$ de charge $+1$ et le neutron $n=(ddu)$ de charge nulle, ce qui
imposait déjà que les charges $c'$ et $c"$  des $q'$ et des $\overline{q"}$ satisfaisaient à $2c'+c"=1$, $2c"+c'=0$, d'où $c'=2/3$
et $c"=-1/3$.\\

\noindent Voyons à présent une interprétation Galoisienne de $Q_0$.\\

\indent D'autres groupes unitaires de dimension $1$ sont de nature Galoisienne. En effet, le choix d'une trialité a fait
apparaître des paramètres de phase arbitraires $\lambda$ et $\beta$, et il est légitime que l'ambiguité de chaque phase soit
prise en charge par un groupe
de jauge.\\

\indent Le paramètre $\lambda$ intervient dans la formule de $t$ et dans nos définitions des fermions. Il représente un
changement de base dans le plan $e_7,e_8$. Son action naturelle donne
les charges suivantes: $+1$ sur le neutrino $\nu"=\lambda.1/2$, $-1$ sur l'électron $\varepsilon"=\Omega/2\lambda$ (où
$\Omega= \exp(-i\beta)u"_0\wedge V"_0$), $+1$
sur les $q"=\lambda v"\wedge u"_0$ et $-1$ sur les $q'=\lambda\lambda^{-1}\exp(-i\beta)v'\lrcorner V"_0$. Cela  définit un générateur $Z_\lambda$, qui coincide avec $-H_7$ sur
le plan des leptons et avec $H_7$ sur l'espace de dimension six des quarks. On a $Z_\lambda=Z'_1-\frac{1}{2}Z_4$; ce qui fait que
$Z_\lambda$ n'apporte pas beaucoup d'information nouvelle.\\

\indent L'autre paramètre est la phase $\beta$ qui détermine la structure réelle de $S_8^{+}$ et $S_8^{-}$, et du même coup
la métrique spinorielle retenue sur $S_8$, visible sur $\Omega$. Elle correspond donc un générateur $Z_\beta$ qui agit comme
$-1$ sur $\Omega$ (ou $\varepsilon"$), $-1$ également sur les quarks magnétiques $q'$, et $0$ sur $1$ (ou $\nu"$) et sur
les quarks électriques $\overline{q"}$.
Cela vient de nos conventions, car $\Omega$ et les $q'$
ont en facteur le nombre $\exp(-i\beta)$.\\
Du côté de $S_8^{-}$ ce sont $V"$ et l'orthogonal de $u"_0$ qui ont une charge $Z_\beta$ égale à $-1$.
L'action déplace le spineur fixé $t_0$ sur son orbite circulaire.\\

\indent Le groupe circulaire qui traduit la graduation d'algèbre extérieure sur $\Lambda^{*}(F\otimes \mathbb{C})$ peut aussi être pris en compte; son
générateur $N'$ donne $0$ sur les constantes, $-1$ sur les vecteurs de $F"$, $+1$ sur les vecteurs de $F'$, etcetera. Nous décidons d'avoir des charges
négatives du côté de $F"$ et des charges positives du côté de l'algèbre extérieure de $F'$.\\
Ce qui fait $0$ pour le neutrino, $-4$ pour l'électron, $-1$ pour les quarks électriques et $+1$ pour les quarks magnétiques.\\
On étend l'action de $N'$ à $S_8^{-}$ par l'identité, de sorte qu'il respecte trivialement l'orbite de chaque $t$.\\

\noindent Les générateurs $N'$ et $Z_\beta$ donnent une formule très simple pour les charges élecriques des particules:
\begin{equation}\label{chargetiers}
Q_0=\frac{1}{3}N'-\frac{1}{3}Z_\beta
\end{equation}
Comme $N'$ vaut zéro sur $S_8^{-}$, l'opérateur $Q_0$ respecte l'orbite de $t$; mais il la fait tourner trois fois
trop lentement pour qu'elle se ferme lorsque $\beta$ change d'un multiple entier de $2\pi$.
On voit clairement sur la formule \eqref{chargetiers} la division par trois. L'angle du groupe engendré par $Q_0$ sera noté $\theta/3$.\\\\
Un autre intérêt de cette formule  est de montrer que le cercle électromagnétique
fait tourner la structure réelle de $S_8$. \\
\noindent Les ensembles de trois structures réelles obtenues en translatant $\theta$ d'un multiple de $2\pi$ semblent
devoir jouer un certain rôle.\\

\subsection{Au-delà du modèle standard}

Les résultats que nous venons d'exposer n'entrent apparamment pas en contradiction avec les observations
expérimentales; ils recouvrent largement les prédictions du modèle standard, même s'ils s'en écartent sur
certains points, comme l'origine fermionique de la brisure de symétrie, la nature du $SU_2$ et celle du boson de Higgs.\\
\indent Le principal intérêt de notre approche est de donner une explication géométrique unifiée de toutes les composantes
du modèle standard de Glashow, Weinberg et Salam, en ajoutant une explication possible des ordres de grandeurs des
masses, des matrices $PMNS$ et $CKM$, des termes de Yukawa et du boson de Higgs.\\

\noindent Cependant nous pouvons aussi déduire de notre approche quelques prédictions pour aller au-delà du modèle standard:\\

\noindent 1) on doit attendre plusieurs nouvelles générations de fermions, analogues aux trois précédentes, un lepton neutre,
une lepton de charge $-1$, une famille de trois quarks de charge $-1/3$ et une famille de trois quarks de charge $2/3$.
Pour leurs masses, en se fiant à la régularité de la série électron et de la série up, et en utilisant la règle des $200$,
on peut prévoir que le prochain
analogue de l'électron, disons le  theton, puisque la lettre grecque theta a déjà été attribuée à une
particule mais qu'elle s'assortit trop bien avec la lettre tau en Mathématiques, aura une masse d'environ $350 GeV$, et on peut prévoir que
le suivant dans la série $u,c,t$, qu'on pourrait nommer $j$ comme justice, aura une masse d'environ $34 TeV$. Le neutrino
du theton devrait avoir une masse de l'ordre de $2 GeV$ et l'autre quark, que l'on pourrait nommer $h$ comme honnêteté
devrait avoir une masse de l'ordre de $8 TeV$. \\
Cette prédiction n'est pas bien maline et a sans doute été suggérée déjà. L'analogue pour les bosons n'est probablement pas plus nouveau:\\

\noindent 2) il devrait exister une seconde génération (au moins) de $W$ et $Z$, possédant les mêmes charges que ceux qui
son connus, mais avec des masses environ $53$ fois plus grandes, c'est-à-dire environ $5$ ou $6 TeV$.\\

\noindent 3) De nombreuses particules massives sans interaction avec les fermions devraient exister également
et constituer la majeure partie de la matière noire; à la première génération, elles correspondent aux générateurs de l'algèbre
de Lie $gl_8$, ceux qui sont antisymétriques, dans l'orthogonal de $u_3$ à l'intérieur de $so_8$, et ceux qui sont symétriques
et qui sont dans l'orthogonal du $sl_2$ à l'intérieur de $\mathfrak{p}_8$. Leurs masses devraient être de l'ordre de $100 GeV$.
A la seconde génération les masses devraient être de l'ordre de $6 TeV$.\\

\noindent 4) Compte tenu de ces nouveaux bosons, ou du modèle gravitationnel pur, l'énergie noire devrait être expliquée par une constante cosmologique positive.\\

\noindent 5) Le mécanisme donnant naissance au boson de Higgs, fixage de jauge sur un fermion dans un nouveau secteur,
bosonisation et identification à un champ de déplacement spatial en dimension douze, devrait posséder une forme de
signature identifiable.\\

\noindent 6) De même, le mécanisme responsable de la naissance de la lumière, le photon $Q_0$ vu comme combinaison de
deux traces de $gl_8$ et de deux racines de $u_4$ et $u_2$, devrait également pouvoir être identifié, ne serait-ce
que sous la forme des deux particules soeurs de la lumière, que sont $iB_0$ et $Z_4$, à côté du $Z_0$ qui est connu. Celles-ci devraient
être neutres et massives, sans que leurs masses proviennent du boson de Higgs. La charge de chaque fermion vis à vis de $B_0$
est égale à l'unité $1$, alors que vis à vis de $Z_4$ on aurait une sorte de double de $Z_0$ sur les leptons, et rien sur
les quarks.\\

\section{Discussion quantique}

\subsection{Théorie quantique des champs}

En Physique des particules, le premier principe est la covariance pour le groupe de Poincaré $G=V\ltimes Spin(\Phi)$,
disant que les états quantiques d'une particule élémentaire libre
sont les vecteurs de l'espace de Hilbert d'une représentation unitaire irréductible $\rho$
d'énergie positive et de spin fini de $G$. Le lien avec l'expérience repose
sur l'introduction d'un espace-temps $X$, localement modelé sur $V$, et sur l'identification des éléments de l'espace de
Hilbert $\mathcal{H}_\rho$ avec certaines sections lisses d'un fibré de spineurs-tenseurs au-dessus de $X$.
Lorsque l'espace-temps est plat, onn peut identifier $X$ à $V$, la transformation de Fourier permet de construire un produit scalaire invariant sur les
sections à support compact d'un tel fibré, tel que
l'espace séparé complété soit isomorphe à la représentation unitaire irréductible $\rho$ de $G$. Dire qu'une section appartient
à cet espace équivaut à dire qu'elle satisfait à certaines équations différentielles remarquables. Pour les spineurs (spin $1/2$), on trouve ainsi
l'équation de Dirac, avec ou sans masse, pour les vecteurs (spin $1$), on trouve l'équation de Maxwell et la condition de jauge de Lorentz,
pour les champs scalaires on trouve l'équation de Klein-Gordon.
De cette manière les équations aux dérivées partielles linéaires des champs classiques se déduisent des axiomes quantiques.
La parametrix de ces opérateurs qui correspond aux résultats expérimentaux en Physique des particules a été introduite par Stueckelberg et Feynman.
Elle stipule que les sources produisent des particules d'énergie positive qui se déplacent vers le futur et
des anti-particules d'énergie négative, qui se déplacent vers le passé, donc ne violent pas l'interdiction sur l'énergie
négative. En fait, le produit scalaire sur $\mathcal{H}_\rho$ apparaît comme la somme de la forme sesquilinéaire associée à cette parametrix,
et de la forme qui correspond à la règle renversée dans le temps. Donc la règle de propagation de Stueckelberg et Feynman est
une version raffinée de la représentation unitaire.\\
\indent A partir de là, il est possible de définir les espaces de Hilbert d'un nombre arbiraire de particules, sous forme d'espaces de Fock.
Ces espaces sont naturellement munis d'opérateurs de \emph{créations} et \emph{annihilations}, qu'on peut considéder comme des distributions
d'opérateurs associées aux champs, libres au début. Une redéfinition de l'état du vide des anti-particules permet de préciser comment la
positivité de l'énergie est sauvée. Cf. \cite{bogoliubov1960introduction}, \cite{itzykson1980quantum}, \cite{peskin1995introduction},
\cite{weinberg1995quantum}.\\

\indent Le second principe est celui de l'interaction minimale, formulé dans des termes variés par
Feynman, Tomonaga, Schwinger et Dyson, à la fin des années $40$ (cf. \cite{feynman1962quantumelectrodymaics}, \cite{tomonaga1946relativistically}).
Ce principe requiert, pour devenir
opératoire, un cadre conceptuel, qui aujourd'hui est celui de la Théorie Quantique des Champs, sous sa forme
Hamiltonienne ou Lagrangienne.
Dans ce cadre les interactions sont interprêtées comme des combinaisons de créations et annihilations de particules.
On dispose, pour chaque hypersurface $M$ du genre espace dans $X$ (ou plus simplement, pour chaque hyperplan du genre espace),
d'un espace de Fock $\mathcal{F}_M$ muni des opérateurs de créations et annihilations en restriction à $M$, qui satisfont
à des relations canoniques (cf. \cite{itzykson1980quantum}, \cite{peskin1995introduction},
\cite{weinberg1995quantum}), d'où résulte une algèbre $\mathcal{A}_M$ d'endomorphismes de $\mathcal{F}_M$.
Les interactions semi-classiques entre champs sont décrites par les termes d'un Lagrangien (ordre zéro en $\hbar$), reposant en général
sur les couplages tensoriels invariants naturels de plus petit degrés entre particules. Ces termes peuvent être interprêtés comme
les coefficients d'une forme différentielle sur
la variété $\mathcal{M}$ des hypersurfaces $M$, à valeurs dans le fibré $\mathcal{A}_M$.\\
Par exemple, si on a un
champ de spineurs de Dirac $\Psi$ et un champ de vecteurs $A$, vu comme comme connexion unitaire sur un fibré en droite
complexe, le couplage minimal dans le Lagrangien est du type $\widetilde{\Psi}A\Psi$; il donne en dimension quatre une combinaison linéaire de
monomes d'opérateurs de créations et d'annihilations
d'électrons, de positrons et de photons. L'équation d'évolution est alors un couplage des équations de Dirac et de Maxwell
avec second memebre.\\
Il en résulte, pour chaque chemin d'une hypersurface $M_1$ à une hypersurface $M_2$ dans $\mathcal{M}$, un système infini d'équations
différentielles linéaires à résoudre,
afin de calculer les amplitudes de transition entre états des espaces de Fock associés. Cette équation est l'expression
locale d'une connexion $\nabla_M$ qui déforme la connexion canonique $\nabla^{0}_M$ donnée par les équations à une particule. \\
L'accord des équations pour des chemins différents entre $M_1$ et $M_2$ est assuré par un théorème (ou axiome) de Tomonaga qui affirme
que $\nabla_M$, comme $\nabla^{0}_M$, est une connexion plate. (Le théorème en question dit en fait que cette condition équivaut au théorème de
conservation du tenseur énergie-moment semi-classique. Cf. \cite{tomonaga1946relativistically}, \cite{koba1950note}).\\
Calculer la résolvante de cette connexion équivaut à calculer les coefficients entre états de particules élémentaires de
la parametrix de Feynman pour l'équation incluant les interactions.
La matrice $\textbf{S}$ historique étant plutôt celle
qui décrit le transport asymptotique entre
les temps $-\infty$ et $+\infty$, cf. \cite{bogoliubov1960introduction}.\\
Cette résolution semble dans la plupart des cas ne pouvoir être effectuée que de façon perturbative, en supposant les charges
devant les termes d'interaction petites,
et elle donne lieu au calcul des intégrales de Feynman, canoniquement attachées à des graphes. On peut alors dire, avec 't Hooft et Veltman
(\cite{thooftveltman1973}), que toute la théorie repose sur un ensemble de \emph{règles de Feynman} qui associent des intégrales à des
graphes décorés. Mais, dans toute théorie intéressante, il apparaît  des quantités infinies, certaines intégrales sont divergentes et elles
exigent des régularisations. En fait, il ne s'agit pas que de calculs, la solution du problème réclame
un changement de point de vue: l'introduction des choix d'échelles d'énergie, puis une analyse asymptotique de compensations des régularisations
en tenant compte des relations entre diverses échelles. Ceci est le domaine des \emph{corrections radiatives}, le début de la théorie de
la renormalisation.\\
Heureusement, dans certains bons cas, dont le modèle standard fait partie, le développement perturbatif donne une théorie effective à
chaque échelle, correspondant à un Lagrangien nouveau,
et de nouveaux termes d'interactions
qu'il faut prendre en compte et il est possible d'assurer
la convergence du procédé, en compensant des quantités tendant vers l'infini par d'autres, de façon naturelle, tenant compte de la
covariance. Mais cela ne suffit pas pour déduire certains effets dynamiques quantiques essentiels,
comme les apparitions spontannées de masses, car ces effets proviennent de couplages forts, où
\emph{a priori} la théorie des perturbations ne dit rien. Cf. \cite{t1994under}, \cite{polyakov1987gauge}.\\

Notons que dans la théorie que nous avons proposée pour expliquer la structure
particulière du modèle standard, la Physique se découvre ordre par ordre en le déplacement $v_H$
le long de $W$ dans la variété $X$, et là, il est naturel que tout dépende de l'échelle d'énergie, un ordre étant un facteur $50$ environ si l'on compte
en $GeV$ (ce qui fait un facteur $200$ pour les masses des fermions élémentaires). Cependant, dans cette théorie, nous devons préciser le rapport
qu'il y a entre l'apparition des nouveaux champs à chaque ordre et l'effet déjà observé des échelles
en Physique des particules, tel qu'il est expliqué par un flot de renormalisation. Et là aussi, il faut distinguer entre
renormalisation perturbative, la seule qui soit méthodique aujourd'hui, et théorie quantique non-perturbative, la seule qui
puisse correspondre à la plupart des effets globaux observés à notre échelle, dans un régime de couplage fort.\\

Une notation élégante pour écrire la résolvante, ou la propagation une fois réordonnée, est l'intégrale fonctionnelle sur les champs $\phi$ en présence
pour une mesure complexe formelle:
\begin{equation}
U(M_2,M_1)=\int\mathcal{D}\phi e^{\frac{i}{\hbar}S(\phi)}
\end{equation}
où
\begin{equation}
S(\phi)=\int_W dx\mathcal{L}(\phi(x),\partial \phi(x), ...),
\end{equation}
est l'intégrale du Lagrangien classique, et où les conditions
au bord sur les hypersurfaces $M_1$ et $M_2$ correspondent aux coefficients de $U$.\\
Cette notation due à Feynman évoque celle de Leibnitz en dimension $1$, où il y a bien coïncidence entre
primitives et intégrales indéfinies, et sous-entend que la coïncidence s'étend en partie aux résolvantes des connexions linéaires,
même dans un cadre de dimension infinie. Sauf que $\mathcal{D}\phi$ doit être invariant par translation, or il
n'existe pas de telle mesure, donc seul le produit avec la phase peut avoir un sens, en particulier si $S$
est réduit à la partie libre $S_0$, cela évoque bien entendu une mesure Gaussienne, prolongée en co-variance
imaginaire pure. De nombreux efforts continuent d'être effectués pour définir correctement un tel objet.
Cf. \cite{wetterich2001effective}, \cite{costello2011renormalization}, \cite{delamotte2012introduction}, \cite{anselmi2015renormalization}.\\
La propagation peut aussi être écrite formellement comme un système infini d'équations intégro-différentielles, appelé
équation de Dyson-Schwinger:
\begin{multline}
\langle \emptyset|\frac{\delta S}{\delta \phi_0(x_0)} \phi_1(x_1)...\phi_n(x_n)|\emptyset\rangle\\
=\frac{\hbar}{i}\sum_{k=1}^{n}\langle \emptyset|\phi_1(x_1)...\delta_{0k}\delta^{(4)}(x_0-x_k)...\phi_n(x_n)|\emptyset\rangle
\end{multline}
A l'ordre zéro en $\hbar$ on trouve les équations de champs classiques; cette équation indique que les
corrections quantiques sont contenues dans les termes dits "de contact", aux points coincidents des champs. Cependant il
faut faire attention que le membre de gauche, faisant intervenir les seconds membres polynomiaux des équations de champs classiques,
comme ils viennent des dérivées du Lagrangien d'interaction, introduit des amplitudes qui ont des degrés supérieurs à ceux
du membre de droite, et qui demandent en plus à être régularisés (cf. \cite{itzykson1980quantum}). Cette écriture est
à comprendre "avant renormalisation". Pourtant, bien que l'intégrale fonctionnelle et l'équation de Dyson-Schwinger apparaissent
ainsi dans un cadre perturbatif,
elles ouvrent des perspectives non-perturbatives aussi. Par exemple la discrétisation Euclidienne de Wilson, qui contient une
régularisation ultra-violette évidente, ne fait pas de référence directe à un système de petit nombre de particules. \\
\indent Afin de mettre en lumière les aspects quantiques, la stratégie dominante est de passer à l'action dite "effective",
qui est une série en $\hbar$:
\begin{equation}
\Gamma(\phi)=S(\phi)+\hbar \Gamma_1(\phi)+ \hbar^{2} \Gamma_2(\phi)+...
\end{equation}
où le terme de degré $n$ représente les corrections à l'action classique venant des graphes de Feynman à $n$ boucles.\\
Formellement, la transformation qui fait passer de $S=\Gamma_0$ à $\Gamma$ est la composée d'une transformation de
Fourier, d'un logarithme et d'une transformation de Legendre; on introduit les sources $J=J_\phi$, duales des champs, alors
\begin{equation}
Z(J)=\int\mathcal{D}\phi e^{\frac{i}{\hbar}S(\phi)+i\int_W J.\phi dx};
\end{equation}
\begin{equation}
\frac{i}{\hbar}W(J)=\ln Z(J);
\end{equation}
\begin{equation}
\Gamma(\phi)=(W(J_\phi)-\int_W J_\phi.\phi dx)|_{\phi=\delta W(J_\phi)/\delta J}.
\end{equation}
La relation entre $J_\phi$ et $\phi$ dit aussi que $J_\phi=-\delta \Gamma(\phi)/\delta \phi$.\\
(Remplacer $S$ par $\Gamma$ dans le développement de l'intégrale fonctionnelle qui résulte du développemnt
en série de l'exponentielle revient à remplacer les graphes de Feynman par les graphes "propres", ou une-particule-irréductibles,
i.e. les graphes connexes qu'on ne décompose pas en coupant une seule arête.) Cf. \cite{itzykson1980quantum}, \cite{peskin1995introduction},
\cite{weinberg1995quantum}, \cite{weinberg1996quantum}, \cite{costello2011renormalization}, \cite{anselmi2015renormalization}.\\

\noindent Notons que toutes ces équations doivent être modifiées dans le cas de champs $\phi$ qui représentent des
fermions $\psi$, là où les espaces de Fock sont des sommes de puissances extérieures: les sources $J_\psi$
anticommutent entre elles et avec les $\psi$, les intégrales doivent être prises au sens de Berezin (cf. \cite{weinberg1996quantum}). Mais,
quitte à travailler dans un cadre d'algèbres graduées, qui est bien décrit par le super-calcul différentiel, la
méthode de développement de l'action s'étend formellement.\\

\noindent Le développement en $\hbar$ comptabilise le nombres de boucles; le développement perturbatif de la résolvante
se fait en fonction des charges $g_k$ placées
aux sommets des graphes. Il s'agit maintenant de définir ordre par ordre en $\hbar$ des séries en $g_k$
dont les termes sont finis, qui représentent les prédictions de la théorie à une certaine échelle $\mu$. \\
\indent Pour cela, en premier lieu on choisit une méthode de régularisation. Le standard actuel,
au moins pour les théories de jauges, et en particulier pour le modèle standard, est la régularisation dimensionnelle,
qui passe par des formules de propagateurs etc. où la dimension $4$ est remplacé par $4\pm \epsilon$, mais on doit également
considérer l'effet des énergies et fréquences prises en compte (cf. \cite{anselmi2014adler}, \cite{anselmi2015adler}). Pour donner une idée simple
de la renormalisation effective, nous partons d'une régularisation par troncature lissée au-delà de l'échelle $\Lambda$
(cf. \cite{wetterich2001effective}, \cite{delamotte2012introduction}).\\
Suivant Kadanoff et Wilson, cette régularisation est moyennée entre $\mu$ et $\Lambda$. Elle donne alors une
\emph{théorie effective} (au sens de Wilson) à l'échelle $\mu$ de la théorie $T$
régularisée à l'échelle $\Lambda$, que l'on note\\
\begin{equation}
T'_\mu=T(\mu,\Lambda;T_\Lambda)=R_{\Lambda,\mu}(T_\Lambda).
\end{equation}
Cette théorie s'obtient donc en négligeant les fréquences au-delà de $\Lambda$ et en sommant les fluctuations entre les fréquences d'ordres $\mu$ et $\Lambda$.\\
Certaines amplitudes de  $T'_\mu$ s'interprètent comme des constantes de couplage $g_i$ et des masses $m_i$.
Les coefficients $g_i^{0}$ et $m_i^{0}$ étant respectivement associés aux types de sommets et aux types d'arêtes des graphes d'interaction
de la théorie initiale. Ce sont  respectivement des charges et des masses généralisées.\\
Génériquement,
nous allons désigner ces constantes devant des termes monomiaux d'interaction des champs (y-compris les masses vues comme
des auto-interactions de degré deux),
par les symbôles $\lambda_i$.
La théorie $T$ est censée être décrite par un nombre fini de ces
constantes, allors que $T'_\mu$ requière en général plus de paramètres, et même, dans beaucoup de cas intéressants, de plus en plus
de paramètres au fur et à mesure que $\Lambda$ croît. C'est ce que Wilson a démontré sur réseau, et que l'on
constate facilement avec des modèles statistiques comme celui de Ising. Nous admettons donc la présence d'une infinité de $\lambda_i$.\\
\indent Les \emph{conditions de renormalisation} sont des contraintes imposées aux $\lambda_i$ à l'échelle $\mu$ pour que
une théorie effective soit compatible à un certain nombre de données expérimentales:\\
\begin{equation}
F_i(T'_\mu)=C_i;
\end{equation}
Cela implique des contraintes sur $T_\Lambda$ qui se traduisent par des équations fixant les $\lambda$ de $T_\Lambda$:\\
\begin{equation}
\lambda_{i,\Lambda}=\lambda_i(\lambda_{j,\mu};\Lambda;\mu).
\end{equation}
En général lorsque $\Lambda$ tend vers l'infini, les paramètres $\lambda_{i, \Lambda}$ tendent vers $0$ ou $\infty$.\\
En considérant plusieurs échelles $\mu$ entre deux valeurs d'énergies $\mu_0$ et $\mu_1$, on étend naturellement
ces fonctions, par exemple
\begin{equation}
\lambda_{\mu'}=\lambda^{\Lambda}(\lambda_\mu;\mu';\mu),
\end{equation}
qui expriment les paramètres d'une théorie effective $T'_{\mu'}$ équivalente à $T_\Lambda$ à l'échelle $\mu'$ en
fonction de la théorie effective en $\mu$.\\
Ces fonctions satisfont à une équation, dite de type Dyson-Shirkov:
\begin{equation}
\lambda_{\mu_2}(\lambda_{\mu_1}(\lambda_\mu;\mu_1;\mu);\mu_2;\mu_1)=\lambda_{\mu_2}(\lambda_\mu;\mu_2;\mu);
\end{equation}
où il est entendu que l'exposant $\Lambda$ doit figurer partout.\\
Alors on fait tendre $\Lambda$ vers l'infini.
Démontrer un théorème de renormalisation revient à démontrer la convergence de $T'_\mu$ satisfaisant
aux contraintes $F_i(T'_\mu)=C_i$ lorsque $\Lambda$ tend vers l'infini, ainsi que la convergence des
transformations $\lambda^{\Lambda}(\lambda_\mu;\mu_1;\mu)$. Pratiquement les amplitudes de $T'_\mu$
sont des parties finies de celles de $T_\Lambda$, convenablement adaptées à l'échelle $\mu$.\\
La confrontation avec l'expérience doit alors reposer sur les prédictions des amplitudes d'autres évènements que ceux
qui ont servi à fixer les couplages de base.\\
On dit que la limite $R_\mu T$ de $T'_\mu(\Lambda;T_\Lambda)$ lorsque $\Lambda$ tend vers l'infini est la théorie
$T$ renormalisée à l'échelle $\mu$.\\
\noindent Le groupe de renormalisation est la dynamique sur les amplitudes de $R_\mu T$ en fonction
de $\mu$. Une expression simplifiée de ce flot (qui est souvent un groupe à un paramètre) est l'évolution des
paramètres $\lambda_i$
en fonction de $\mu$, entre deux bornes $\mu_0< \mu_1$, qui peuvent être $0$ et $+\infty$.\\
\indent Par exemple, la constante de couplage $\alpha_s$ des interactions fortes, dont l'inverse est à
mettre devant le lagrangien des gluons, est de l'ordre de l'unité à notre échelle, mais elle
varie en fonction des énergies $\mu$ mises en jeu (en $GeV$) selon la formule suivante (Gross, Politzer, Wilczek)
\begin{equation}
\alpha_s(\mu)=\frac{C_0}{(33-2n_f)\ln(\mu/0.2)};
\end{equation}
où $n_f$ dénote le nombre de triplets de quarks, cf. \cite{weinberg1996quantum}, \cite{peskin1995introduction}.
La liberté asymptotique de $QCD$ dit que cette constante tend vers zéro lorsque $\mu$ tend
vers l'infini, on voit que cela laisse  \emph{a priori} la place pour huit générations de
quarks, dont les trois connues.
Par contre, ce n'est pas une mauvaise nouvelle qu'au delà de neuf générations la tendance se renverse
et que la théorie aux plus grandes énergies deviennent fortement couplée; en effet, dans notre approche,
neuf générations peuvent théoriquement se manifester bien avant que l'échelle de Planck ne soit atteinte,
et il serait inquiétant que les gluons soient asymptotiquement libre alors qu'ils proviennent de composantes
de la métrique d'univers, de plus en plus courbée.\\

\noindent Pratiquement, Wilson et Weinberg expliquent qu'il n'y a pas de théorème de renormalisation utile,
mais seulement des évidences que en-dessous d'une certaine énergie $\Lambda_0$ pour $T_\Lambda$, les amplitudes
de $T'_\mu$ contraintes par des $(g_\mu,m_\mu)$ décrivent bien les expériences faites autour de l'énergie $\mu$.
Particulièrement intéressant est le test des variations prédites pour les constantes de couplage $g_\mu$.
Ces variations dépendent du flot de renormalisation sur l'ensemble des paramètres.\\
\indent Toutefois, le résultat sur lequel les développements récents s'appuient, et qui demeure fondamental,
est la preuve de la renormalisation (au sens perturbatif usuel) du modèle standard, donnée par $^{t}$ Hooft,
Veltman, Lee et Zinn-Justin au début des années $70$. Cf. \cite{t1994under}, \cite{zinn1975renormalization}.\\

\subsection{Co-homologie BRST-BV et renormalisation}

Dans une théorie comme l'électrodynamique quantique, la chromodynamique quantique, ou le modèle standard, il est important que les
théories effectives à toutes les échelles soient compatibles avec l'invariance de jauge. Les espaces de Fock, la connexion
de Tomonaga,
la renormalisation, tout doit être revu sous cet éclairage, cf. \cite{t1994under}, \cite{polyakov1987gauge}. L'approche qui s'est montrée la plus efficace
au cours des années (depuis les années $70$ jusqu'à présent) est aussi la plus profonde conceptuellement;
elle dit que les équations quantiques, comme les états et les amplitudes qui en résultent, sont de nature homologique.
L'instrument précis pour mettre ce principe en forme a été découvert par Becchi-Rouet-Stora-Tyutin, c'est la
co-homologie $BRST$, \cite{becchi1976renormalization}, dont une extension élégante et géométrique adaptée à la notation des intégrales de chemin,
est la théorie de Batalin-Vilkovyski, $BV$; cf. \cite{cataneo1995}, \cite{costello2011renormalization}, \cite{weinberg1996quantum}. Du point de vue mathématique, les exposés de
Henneaux, 1985, 1990, \cite{henneaux1990lectures}, \cite{henneaux1985hamiltonian}, de Hennaux et teitelboim, \cite{henneauxteitelboim1992quantizationgauge},
et ceux de Barnich, Brandt et Henneaux 1995, \cite{barnich1995local},
sont très recommandables, ainsi que ceux de
J.Stasheff, \cite{stasheff1997homological}, \cite{stasheff2005poisson}.\\
\indent Plusieurs exposés récents, originaux sur bien des points, et rigoureux, développent la théorie perturbative
de la renormalisation en théorie quantique des champs dans le cadre $BV$, en particulier pour Yang-Mills et la gravitation; il faut
citer en particulier le livre $AMS$ de Kevin Costello \cite{costello2011renormalization}, et les travaux de
Kasia Rejzner, \cite{rejzner2011batalin}, et R.Brunetti, K. Rejzner, , K.Fredenhagen, \cite{brunetti2013quantum}. Dans ces derniers travaux la théorie $BV$ et la renormalisation
évitent l'intégrale de chemin et utilisent une approche par la théorie algébrique des champs.\\
Voir également les travaux de Connes et Kreimer, \cite{conneskreimer1999} et Connes et Marcolli, \cite{connes2008noncommutative},
qui éclaircissent considérablement le procédé de renormalisation en utilisant
les caractères d'algèbres de Hopf associées aux graphes de Feynman.\\

Comme il est expliqué par Henneaux et Teitelboim, \cite{henneaux1985hamiltonian}, \cite{henneauxteitelboim1992quantizationgauge}, l'idée de $BRST$ est de mener de front les deux aspects les plus
importants du point de vue dynamique (classique comme quantique), à savoir, satisfaire à des équations, et former
des quantités observables invariantes de jauge. \\
\indent Dans le cas spécial où $A$ est une algèbre de Poisson et où les équations dans $I$ qui définissent $B$ satisfont aux
relations d'une algèbre de Lie (on parle d'application moment), la co-homologie $BRST$ réunit les deux
aspects en définissant un opérateur $s$ de carré nul sur
un complexe bigradué. Alors le résultat des opérations de restriction et de quotient ensemble, est défini
par la co-homologie de ce double complexe. Lorsque les éléments de $I$ sont seulement en involution, une déformation naturelle
de $s$ rend le même service; elle porte le nom de $BV$. Cf. \cite{stasheff1997homological}, \cite{henneauxteitelboim1992quantizationgauge}, pour
la construction générale de $s$, qui appartient au domaine de l'algèbre homotopique. \\
\indent L'observation géométrique
de Batalin et Vilkovisky est qu'il existe une structure de Poisson graduée impaire (i.e. une structure de Gerstenhaber)
sur le complexe total $C$ des cochaines $BV$, notée par une parenthèse $(X,Y)$, et qu'il existe un élément $S$ de $C$
tel que, pour tout
$X$, on ait $s(X)=(X,S)$. Cet élément satisfait à l'équation quadratique $(S,S)=0$, dite équation maîtresse,
ou master equation classique.\\
L'idéal $I$ est défini par les dérivées ordinaires de $S$, donc passer au quotient par $I$ équivaut à
prendre la restriction au lieu critique $\Sigma$ de $S$. La fonction $S$ correspond à l'action du problème variationnel
engendrant les équations hamiltoniennes.\\
L'algèbre de Gerstenhaber qui intervient est une algèbre de $BV$ (cf. \cite{cataneo1995}, \cite{roger2009gerstenhaber}), ce qui signifie qu'il existe un
opérateur $\Delta$ d'ordre deux de degré $-1$, de carré nul tel que, pour tout triplet $(a,b,c)$ d'éléments gradués de
l'algèbre $\mathcal{A}$, on ait
\begin{multline}
\Delta(abc)=\Delta(ab)c+(-1)^{|a|}a\Delta(bc)+(-1)^{(|a|+1)|b|}b\Delta(ac)\\
-\Delta(a)bc-(-1)^{|a|}a\Delta(b)c-(-1)^{|a|+|b|}ab\Delta(c).
\end{multline}
Le crochet est alors défini par
\begin{equation}
(a,b)=(-1)^{|a|}(\Delta(ab)-\Delta(a)b-(-1)^{|a|}a\Delta(b)),
\end{equation}
et $\Delta$ est automatiquement une antidérivation pour ce crochet, i.e.
\begin{equation}
\Delta((a,b))=(\Delta(a),b)-(-1)^{|a|}(a,\Delta(b)).
\end{equation}
\indent Par définition, une quantification du système défini par $S$ est donnée par une série
$S_Q$ en $\hbar$ commençant par $S$,
\begin{equation}
S_Q=S+\hbar S_1+ \hbar^{2} S_2+...
\end{equation}
satisfaisant à l'équation maitresse quantique
\begin{equation}
(S_Q,S_Q)=2i\hbar \Delta S_Q,
\end{equation}
qui équivaut à
\begin{equation}
\Delta(\exp\frac{i}{\hbar}S_Q)=0.
\end{equation}
La fonction $S_Q$ est vue comme une fonction génératrice d'équations de Dyson-Schwinger.\\
A cette fonction est associée une déformation de l'opérateur $s$
en un opérateur $s_Q$, de carré nul aussi, dont la co-homologie est isomorphe à celle de $s$, mais pas de façon
évidente à décrire. La formule de $s_Q$ est
\begin{equation}
s_QX=(X,S_Q)-i\hbar \Delta X,
\end{equation}
Les quantités observables quantiques $\Psi$ sont définies comme les fonctions satisfaisant à
\begin{equation}
\Delta(\Psi e^{\frac{i}{\hbar} S_Q})=0;
\end{equation}
ce qui équivaut à
\begin{equation}
i\hbar\Delta(\Psi)=(\Psi, S_Q),
\end{equation}
ou encore $s_Q \Psi=0$.\\

Ce qui précède concerne toutes les dynamiques hamiltoniennes possédant une forme d'invariance de jauge.
Dans le cas de la théorie des champs, en dimension infinie et dans un cadre Lagrangien,
partant d'une famille de champs $\phi$, on ajoute des champs fermioniques de fantômes $\xi$,
et d'antifantômes $\xi^{*}$, pour tenir compte de l'invariance et des contraintes, et des multiplicateurs de Lagrange $\beta$ pour les termes de fixation
de jauge; l'ensemble de tous ces champs est noté $\Phi$, et chaque composante $\Phi_a$ est notée par un indice
$a\in A$. Pour trouver ue théorie $BV$, on double les variables $\Phi_a$ par des variables duales $\Phi_a^{*}$
qui sont interprétées comme des sources. Cf. \cite{anselmi2014adler}, \cite{anselmi2015adler}, \cite{weinberg1996quantum}, \cite{dewitt2003global}.
Nous les écrirons aussi $J_\phi$, $J_\xi$, $J_{\xi^{*}}$
et $J_\beta$ respectivement (Anselmi utilise la lettre $K$ au lieu de $J$). Un point fondamental pour les formules qui suivent
est le renversement de la parité Grassmannienne: les coordonnées duales des champs bosoniques doivent anti-commuter, celles
des champs fermioniques doivent commuter; on a donc affaire à un structure symplectique impaire (dite \emph{périplectrique})
sur un super-espace de dimension $(\infty|\infty)$, ici du type d'un co-tangent impair $\prod T^{*} \Xi$.
Cf. \cite{cataneo1995}, \cite{roger2009gerstenhaber}.\\
L'opérateur $BV$ canonique est donné par la formule du Laplacien de Poisson \emph{impair}:
\begin{equation}
\Delta X=\sum_a\frac{\delta^{L}}{\delta \Phi_a^{*}}\frac{\delta^{R}X}{\delta \Phi_a}.
\end{equation}
Le renversement des degrés fait que $\Delta\circ \Delta=0$. Le crochet de Poisson qui s'en déduit est
le crochet impair canonique.\\
Une action $S(\phi)$ invariante par un groupe de jauge se prolonge canoniquement en une fonction $S(\Phi,\Phi^{*})$
qui est solution de l'équation maitresse classique. Cf. \cite{anselmi1994removal}, \cite{anselmi1995more}, \cite{gomis1996nonrenormalizable}.\\
De plus, si l'action effective au sens des boucles, $\Gamma(\phi)$, est également invariante par le groupe
de jauge, elle se prolonge en une fonction $\Gamma(\Phi,\Phi^{*})$, qui sarisfait $(\Gamma,\Gamma)=0$ et
$\Delta\Gamma=0$. Il en résulte un opérateur $BRST$ $s_Q$, tel que $s_Q \Gamma=0$.\\
L'équation $(\Gamma,\Gamma)=0$ est souvent nommée équation de Zinn-Justin; elle traduit \emph{l'absence d'anomalies}.
Cette équation est considérée comme la clé de la renormalisation.
Des cas particuliers apparaissaient dans le travail de Kluberg-Stern et Zuber, 1975, \cite{kluberg1975renormalization}, \cite{kluberg1975bisrenormalization},
motivé par les découvertes toutes récentes alors de $BRST$
et de Zinn-Justin, \cite{becchi1976renormalization}, \cite{zinn1975renormalization}.\\
Cf. \cite{anselmi2014adler}, \cite{anselmi2015adler}, \cite{costello2011renormalization}.\\

La théorie de Batalin et Vilkovisky fournit actuellement le meilleur cadre algébrique et géométrique pour une théorie
quantique perturbative, mais il ne faut pas sous-estimer le travail
à faire pour définir à partir de là des quantités régularisées convenables et démontrer les compensations asymptotiques
effectives des termes divergents, qui assurent l'annulation des anomalies
à chaque ordre des boucles.\\
Immédiatement après les découvertes de Becchi, Rouet, Stora et Tyutin, les premiers auteurs qui aient
mené à bien une renormalisation en théorie de jauge basée sur des calculs de type homologique sont H.Kluberg-Stern et
J-B. Zuber, \cite{kluberg1975renormalization}, \cite{kluberg1975bisrenormalization}.
D'ailleurs Damiano Anselmi observe que le succès de la méthode $BRST$ ou $BV$ dépend d'une propriété d'annulation co-homologique, qu'il appelle conjecture de
Kluberg-Stern et Zuber, $KSZ$:\\
\noindent Soit $G_j;j\in J$ une famille de fonctionnelles \emph{locales} invariantes de jauge séparant les points du quotient de la
variété critique $\Sigma$ par l'action des
transformations de jauge; donc en particulier on a $(S,G_j)=0$, pour tout $j$ dans $J$; on suppose en plus que l'action classique s'écrive
\begin{equation}
S(\phi)=\sum_i\lambda_iG_i(\phi);
\end{equation}
pour des constantes $\lambda_j; j\in J$, qui contiennent les constantes de couplages, les masses, etc.\\
\indent Alors la condition $KSZ$ s'énonce ainsi: quelque soit la $0$-cochaine \emph{locale} $X$ (de degré fantômatique zéro);
\begin{equation}
sX=0\Leftrightarrow \exists a_i, \exists Y,\quad X=\sum_ia_iG_i+sY,
\end{equation}
où les $a_j$ sont fonctions des paramètres et où $Y$ est une fonctionnelle \emph{locale} de ddegré $-1$.\\
Dans les cas traités par Kluberg-Stern et Zuber, la théorie de jauge était renormalisable au sens de Dyson,
il n'y avait qu'un nombre fini de paramètres $\lambda_j$ non-nuls, mais Anselmi considère aussi le cas non-renormalisable
au sens usuel, où il y a un
nombre infini de $\lambda_j$ non-nuls.\\
En $1995$ Anselmi, \cite{anselmi1995more}, et Gomis et Weinberg, \cite{gomis1996nonrenormalizable}, montrent comment cette propriété $KSZ$ peut être utilisée de façon générale
pour ajuster les compensations des divergences et renormaliser une théorie sans rencontrer d'anomalies.
D'où l'existence de limites effectives invariantes de jauge, à tous les ordres en $\hbar$ et pour toutes les échelles
d'énegies $\mu$.\\
\indent Ces travaux confirment l'existence de la relation profonde entre l'extension de l'invariance de jauge au système
quantique et la possibilité d'une renormalisation.\\
\indent Selon les résultats de Barnich, Brandt et Henneaux, 1995, \cite{barnich1995local}, la propriété $KSZ$ est vérifiée par les théories de jauge
couplées à la gravitation, pourvu que le groupe de jauge soit semi-simple. Récemment, Anselmi a réussi à utiliser
une condition un peu plus faible, permettant entre autre d'inclure des facteurs abéliens $U_1$ dans le groupe
de jauge; ce qui lui a permis d'établir la renormalisation effective sans anomalie du modèle
standard couplé à la gravitation quantique, \cite{anselmi2014adler}, \cite{anselmi2015adler}. Même si la ligne à suivre est bien dégagée par Anselmi,
son travail est techniquement très délicat, en particulier à cause des contrôles de régularisation.\\

Pour mener à bien la renormalisation il est nécessaire de choisir de bonnes conditions fixant la jauge;
ce point est très élégamment traité dans la théorie $BV$, où les fixages de jauge correspondent à la
donnée de sous-variétés Lagrangiennes impaires (dont les variétés co-normales
des sous-variétés de la base $\Xi$ font partie). Or, après un choix de mesure globale, les fonctionnelles
peuvent être intégrées canoniquement sur toute variété Lagrangienne impaire, et selon un théorème de Albert Schwarz,
lorsque la fonctionnelle $X$ est un co-cycle, i.e. $\Delta X=0$,
l'intégrale de $X$ sur $L$ ne dépend que de la classe d'homologie de $L$, cf. \cite{cataneo1995}, \cite{roger2009gerstenhaber}. Sont ainsi définies, au moins formellement les
intégrales des fonctionnelles $X=\Psi \exp (\frac{i}{\hbar} S_Q)$, pour les actions satisfaisant à l'équation
maitresse quantique et pour les observables quantiques.\\
Les variétés Lagrangiennes considérées par Gomis, Weinberg et Anselmi sont du type des graphes:
\begin{equation}
\Phi^{*}=\frac{\delta \Psi (\Phi)}{\delta \Phi},
\end{equation}
avec,
\begin{equation}
\Psi(\Phi)=\int\sqrt{ -g}\xi^{*}.(G(\phi)+P(\phi,\partial_x)\beta);
\end{equation}
où $P$ est un opérateur différentiel, et $G$ une fonctionnelle des champs classiques; par exemple
$G(\phi)=\sum_\mu\partial^{\mu}A_\mu$ dans le cas de la jauge de Lorentz en théorie de yang-Mills.\\
\indent La première chose que nous devrons faire pour étendre les résultats de convergence d'Anselmi à notre théorie est
de décrire la variété lagrangienne impaire correspondant au fixage de jauge sur une trialité.\\

\subsection{La théorie suggérée au-delà}

Le Lagrangien de base en dimension douze est la somme de celui de Dirac et de celui de Einstein-Hilbert,
le couplage étant donné par la connexion de Levi-Civita de la métrique, et la métrique elle-même étant donnée
par un vielbein depuis le fibré Lorentzien $E_V$ qui est associé au fibré de spineurs $E_S$. Le groupe de covariance
locale est le produit
du groupe des difféomoprphismes de $X$ par un groupe de jauge $Spin_{1,11}$.\\
\indent Cependant la théorie quantique engendrée, même semi-classique, ne serait pas très informative, car
le long de la variété $W$ de dimension quatre qui figure notre espace-temps, le rapport entre
les courbures longitudinales et les courbures transversales est de l'ordre de $10^{-40}$. De plus
les éléments du modèle standard n'apparaissent pas dans le modèle de base, ils n'apparaissent que
dans dans le modèle où la symétrie de jauge est spontanément brisée. Par conséquent, le Lagrangien
et la covariance que nous proposons de retenir pour la quantification est un modèle réduit de celui
de dimension douze:\\
dans la théorie que nous proposons, $W$ est fixée, la métrique est fixée le long de $W$,
le fibré orthogonal $E_F$ est fixé aussi; tous les champs sont en dimension quatre; les champs de bosons correspondent
à des dérivées successives des six familles de champs $C_F$, $M_H$, $\Gamma_H$, $\Gamma_F$, $D_F$ et $D_H$
le long de $W$ dans $X$, plus un champ de Higgs $H$, qui est un multiple du champ de déplacement $v_H=H/M_0$,
servant à écrire le développement de Taylor; les champs de fermions sont les dérivées successives du champ
de Dirac en dimension douze sur $W$, on trouve donc le champ de Dirac usuel avec une multiplicité $S_8$.\\
Le groupe d'invariance locale que nous retenons est le produit du groupe des jets d'ordre infini
des difféomorphismes de $X$ préservant $W$ et les
fibres de $E_F$, qui induisent en plus une isométrie le long de $W$, par le groupe de jauge des spineurs
en $Spin_{1,3}\times Spin_8$.\\
En particulier, on ne souhaite plus toucher à la décomposition tensorielle des spineurs qui provient du choix de $E_F$.\\
Remarque: il est plus joli de garder les difféomorphismes de $W$ et le groupe de jauge horizontal pour les
degrés spinoriels en dimension quatre, mais tant qu'on n'approche pas de la longueur de Planck, ce n'est
pas important, on pourrait aussi bien ne considérer que l'invariance de Lorentz globale pour la dimension quatre.\\
Le champ $H$ est considéré comme un champ auxiliaire; il n'acquière de dynamique que par bosonisation après
brisure de symétrie.\\
Le Lagrangien bosonique est une série formelle en $H$, section du fibré vectoriel $E_F$, dont la formule condensée est
\begin{multline}
\mathcal{L}_B=\frac{1}{8}Tr (M_HM_H)+\frac{1}{8}Tr (\Gamma_HD_H)+\frac{1}{8}\nabla_H\ln det G_F\\
+\frac{1}{4}div^{\nabla}tr C_F+\frac{1}{4}Tr(C_FD_H)+\frac{3}{8}Tr(C_FG_FC_FG_F^{-1})+\frac{1}{8}Tr(C_FC_F).
\end{multline}
Seule la partie antisymétrique à l'ordre zéro de la connexion $C_F$ est considérée comme un champ de Yang-Mills $spin_8$, les autres
composantes de $C_F$ sont considérées comme des champs de matières vectoriels. Le couplage avec la gravitation
quantique en dimension quatre serait évident.\\
Les deux autres Lagrangiens de la matière sont le potentiel du champ scalaire de Higgs et la partie fermionique venant du Lagrangien
de Dirac sur les deux secteurs $S_8^{+}$ et $S_8^{-}$ avant brisure de symétrie. Les termes de Yukawa de tous les ordres
en $H$ sont ajoutés en faisant appel aux structures $u_8$ naturelles sur $S_8^{+}\otimes \mathbb{C}$ et $S_8^{-}\otimes \mathbb{C}$.\\

Un avantage de ce Lagrangien et de ce groupe est l'absence d'anomalie \emph{a priori}, puisque la symétrie entre les deux
chiralités des fermions est respectée, et que $spin_8$ est simple. Notons que cela peut expliquer pourquoi le
modèle standard tel qu'il est d'habitude n'a pas d'anomalie, ce qui repose pour le moment sur une étonnante suite de coincidences
numériques (cf. \cite{peskin1995introduction}). Il semble aussi que la condition $KSZ$ soit satisfaite. Ceci laisse espérer
que nous pourrons appliquer
les méthodes de D. Anselmi 2014, 2015; cependant, nous devons faire
attention aux propriétés un peu particulières de la brisure spontanée de symétrie que nous avons suivie,
donc de la nature étrangement mixte du champ $H$, qui se substitue à une composante fermionique.\\

\subsection{Supersymétrie douce et gravitation}

La clé de notre approche est la levée de l'indétermination sur la nature des spineurs, et la
possibilité d'identifier leur multiplicité à une partie de l'espace. La théorie de la gravitation y est
posée au départ comme dynamique pour la connexion spinorielle dans le Lagrangien de Dirac.\\

J.Scherk et J.Schwarz ont découvert qu'en théorie des cordes et super-cordes, l'équation d'Einstein s'impose
comme le premier ordre des conditions de renormalisation cf. \cite{green1987superstring}, \cite{deligne1999quantum}.\\
Est-ce-que le modèle dont nous sommes partis pourrait se voir comme une certaine limite de super-cordes?\\

Comme nous l'avons constaté, une sorte de symétrie entre fermions et bosons résulte du choix des
coordonnées, de manière compatible avec la covariance générale; cette symétrie se manifeste par le
même nombre de degrés de liberté
des particules effectives tant que la gravitation longitudinale ne compte pas; une autre propriété, qui
n'est d'habitude obtenue que dans certaines théories super-symétriques étendues, a lieu dans notre
théorie, c'est la relation étroite entre la masse
et la charge électrique des fermions, qui dépendent du même groupe $PU_6\times PU_2$ et de l'ordre sur les courbures sectionnelles
dans les huit directions transverses à l'espace-temps.
Comme nous l'avons dit, ces propriétés ouvrent des pistes pour étudier l'énergie noire et la matière noire.\\
D'autre part, beaucoup des ingrédients que nous avons rencontrés, comme la symétrie
effective $U_8$, la brisure spontanée sur $SU_3\times U_1$ se retrouvent en théorie de supergravité,
comme l'ont observé De Wit, Nicolai et Warner, \cite{de1986d}, \cite{nicolai19853}. Les développments récents
dans cette direction sont encore plus proches de nos résultats,
cf. Meissner et Nicolai, \cite{meissner2015standard}, Kleinschmidt et Nicolai \cite{kleinschmidt2015}.
Cependant, dans cette théorie supersymmétrique, le $SU_2$ que nous avons
identifié pour les interactions faibles ne se trouve pas, et la symétrie $SU_3\times U_1$ représente un secteur parmi d'autres,
alors que dans notre approche, il n'y a pas le choix. \\
Il serait tout de même intéresssant de mieux comprendre la relation entre les deux approches.\\
Pourtant notre théorie de départ n'a rien ou presque de supersymétrique. Le presque vient de ce que
la supergravité classique renferme nécessairement au-moins les équations d'Einstein et celles
de Dirac, et que ces deux équations couplées reposent sur l'ingrédient essentiel de la super-symétrie
globale au sens de Haag-Lopuszanski-Sohnius, qui est l'application de $S\otimes S$ dans l'espace de Lorentz $V$, transposée
de l'application fondamentale de $V$ dans $End(S)$.
Cf. \cite{wess1992supersymmetry}, \cite{weinberg2000quantum}, \cite{deligne1999quantum}. \\
Mais la super-symétrie introduit la version réelle positive de $\Gamma$, donc des fermions du type
de Majorana, ce qui n'est pas le cas ici.\\
Il est couramment admis que au niveau semi-classique, les équations super-symétriques
exigent l'introduction de variables anti-commutantes, mais, dans un travail récent avec Michel Egeileh (\cite{egeileh2016}),
nous montrons que ces équations sont équivalentes à des équations de champs usuelles, donc
l'écart entre super et ordinaire ne vient pas de là. Par contre l'écart est important dans la correspondance
précise qui est prédite entre bosons et fermions, en particulier dans la répartition en super-multiplets des
représentations du groupe de Poincaré qui définit les particules, or celle-ci n'est pas celle
que nous avons rencontrée, car nous n'avons pas de particule de spin $3/2$ et nous avons un excès de spin $1$
par rapport au spin $0$.\\

L'application $\gamma_V$ de Dirac est un morphisme entre les fibrés $E_V$ et $End(E_S)$.
Le choix d'un vielbein $\epsilon$ donne la métrique sur $X$ d'où on tire la forme de connexion métrique;
le composé $\gamma_V\circ \epsilon^{-1}$ permet de faire agir les vecteurs sur $S$, d'où résulte le couplage Einstein-Dirac.
On voit que l'image de ce composé est toujours la même dans $End(E_S)$.\\
Une variante serait donc de remplacer le vielbein par un champ plus général d'applications linéaires injectives $\gamma_X$
de $T(X)$ (ou $T^{*}(X)$) dans les endormorphisme d'un fibré $E_S$, de sorte que, pour tout couple $(\xi, \eta)$ de champs de
vecteurs sur $X$, on ait
\begin{equation}
\gamma_X(\xi)\gamma_X(\eta)+\gamma_X(\eta)\gamma_X(\xi)=2g(\xi,\eta)Id_S,
\end{equation}
avec $g$ de signature Lorentzienne. La différence avec $\epsilon_X$ viendrait de ce que le groupe de tous les automorphismes de $E_S$
agirait sur le champ $\gamma_X$.\\

Les applications $\gamma_X$ offrent des déformations naturelles de l'opérateur de Dirac, il est donc
naturel de chercher à justifier les équations d'Einstein dans ce cadre.\\

Le problème plus général de déduire la théorie d'Einstein, ou une modification de celle-ci, des équations pour la matière
a retenu l'attention de
nombreux chercheurs dans la passé (De Broglie, Heisenberg, Sakharov, ..., cf. \cite{wetterich2004gravity}).\\
On peut considérer que la théorie des twisteurs de R.Penrose (\cite{penrose1988spinors}) traite en particulier ce problème, en l'étendant
à l'ensemble des équations de champs.
Une autre approche fut développée par Amati et Veneziano \cite{amati1981dynamical}, \cite{amati1982unified}, qui montrèrent que la métrique d'univers
peut être  engendrée par les
fluctuations quantiques des fermions interagissant avec une infinité de champs de bosons auxiliaires.
Plus récemment, Wetterich, cf. \cite{wetterich2004gravity}, a développé une théorie voisine, où la métrique est un objet composé,
et où l'invariance de Lorentz est globale mais pas locale.
Dans un autre ordre d'idées, la théorie de Chamseddine et Connes (\cite{chamseddine1997spectral}, \cite{connes2008noncommutative}), se fonde sur
un principe d'action spectrale qui ne concerne que l'opérateur de Dirac, et déduit le Lagrangien d'Einstein et Weyl,
plus une série d'autres termes dérivés de la courbure.
Cette théorie est adaptée par ses auteurs pour contenir le Modèle Standard; un point en commun avec la théorie
que nous avons présentée est que les deux considèrent une suite de générations
de fermions. Cependant, pour Chamseddine et Connes, le spectre fermionique du modèle standard, qui correspond à une partie non-commutative de l'espace-temps,
est posé
au départ comme une donnée géométrique \emph{a priori} de l'univers. Cette structure n'est pas déduite dynamiquement d'une situation
plus générique, comme nous
l'avons fait, par "brisure de symétrie", pour identifier les multiplicités des spineurs à huit des
dimensions de l'univers, les charges et les masses de ces spineurs venant de la courbure dans ces directions.\\
\indent Qu'une telle structure, deux leptons, deux familles de trois quarks, reproduite plusieurs fois,
soit générique, à partir du moment où le champ de Dirac se propage en huit dimensions transversales,
provient du miracle des trialités, et que les opérateurs définissent une théorie de jauge non-abélienne est une
conséquence des équations d'Einstein, ou d'une déformation de celle-ci. Tout cela évoque bien l'hypothèse de Kaluza-Klein,
étendue à plus de cinq dimensions (cf. \cite{coquereaux1983geometry}, \cite{bailin1987kaluza}); incontestablement
l'idée d'étendre l'univers à plus de quatre dimensions vient de là,
cependant la grande différence entre la théorie de Kaluza-Klein et
la notre  est que nous n'avons fait aucune hypothèse de symétrie au départ. Nous partons d'un univers générique, au sens de René Thom.
Il est d'autant plus remarquable qu'il en résulte une structure rigide des fermions et un équilibre
entre fermions et bosons comparable à celui de la super-symétrie.\\
La réduction de la dimension douze à la dimension quatre que nous avons faite s'appuie sur la périodicité modulo huit, découverte par R. Bott à propos de l'homotopie
stables des groupes de rotations, étroitement liée à la périodicité des algèbres de Clifford et des spineurs (cf. \cite{atiyah1964clifford}, \cite{atiyah1975classical}),
et qui
est un élément important de la topologie des variétés différentiables de toutes dimensions localisées aux nombres premiers
impairs, comme l'a montré D. Sullivan (\cite{sullivan2005geometric}). Celui-ci a suggèré plus récemment que les champs quantiques et les cordes pourraient donner
un accès à une définition directe de ce type de localisation des variétés. Il est tentant d'aller explorer au-delà
de douze, de huit en huit, ce que pourrait révêler un univers possédant une infinité de dimensions.\\

\section{Remerciements}

L'auteur tient à remercier Jean-Bernard Zuber pour son intérêt, ses questions et ses
suggestions pertinentes et bienveillantes. Merci également à Bernard Julia pour ses
remarques très utiles. Le soutien de Michèle Bompard-Porte, Alain Chenciner et
Sylvie Paycha
fut extrèmement précieux tout au long de cette recherche, merci à eux.\\

L'auteur est conscient des nombreux défauts de son texte et de son style inhabituel, mais l'essentiel de
ce travail a été fait il y a un an exactement et le reste a été complété l'été dernier, donc il décide de
le rendre public, en espérant qu'il pourra malgré tout intéresser d'autres chercheurs et leur permettra
de réaliser de nouvelles découvertes dans le domaine des particules.\\

\bibliographystyle{abbrv}
\bibliography{biblio_jouets}
\nocite{*}